\documentclass[webpdf,modern,small]{oup-authoring-template}

\theoremstyle{thmstyleone}%
\theoremstyle{thmstyletwo}%
\theoremstyle{thmstylethree}%

\usepackage{setspace}
\usepackage{amsmath, amssymb, amsthm, bm}
\usepackage{booktabs}
\makeatletter

\newcommand{\runningtitle}{%
	From Economic Explanation to Performance Improvement in Machine-Learning Portfolios%
}

\def\ps@headings{%
	\let\@oddfoot\@empty
	\let\@evenfoot\@empty
	
	\def\@evenhead{%
		\vbox{%
			\hbox to \textwidth{%
				\fontsize{8bp}{10bp}\selectfont
				{\sffamilyfontbold{\selectfont\thepage}}%
				\hfill
				{\sffamilyfontitalic{%
						\fontshape{it}\selectfont
						\runningtitle
				}}%
			}%
			\vspace{8\p@}%
			\color{black!60}%
			\rule{\textwidth}{1\p@}%
		}%
	}%
	
	\def\@oddhead{%
		\vbox{%
			\hbox to \textwidth{%
				\fontsize{8bp}{10bp}\selectfont
				{\sffamilyfontitalic{%
						\fontshape{it}\selectfont
						\runningtitle
				}}%
				\hfill
				{\sffamilyfontbold{\selectfont\thepage}}%
			}%
			\vspace{8\p@}%
			\color{black!60}%
			\rule{\textwidth}{1\p@}%
		}%
	}%
	
	\def\titlemark##1{\markboth{##1}{##1}}%
}

\makeatother

\begin{document}
	
%	\journaltitle{Journal of Financial Econometrics}
%	
%	\DOI{DOI added during production}
%	\copyrightyear{2026}
%	\pubyear{2026}
%	\vol{XX}
%	\issue{x}
%	\access{Published: Date added during production}
%	\appnotes{Article}
	
	\pagestyle{headings}
	
	\firstpage{1}
	
	% ============================================================
	% Title
	% ============================================================
	
	\title[From Economic Explanation to Performance Improvement in Machine-Learning Portfolios]
	{From Economic Explanation to Performance Improvement in Machine-Learning Portfolios}
	
	% ============================================================
	% Authors
	% ============================================================
	
	%	\author[1,$\ast$]{Yunqi Liang}
	\author[]{Yunqi Liang}
	
	\address[]{
		\orgdiv{Department of Engineering Science},
		\orgname{University of Oxford}
				\orgaddress{
						\city{Oxford},
%						\postcode{OX1 2JD},
%						\country{United Kingdom}
						\country{UK}
					}
%				{Yunqi.Liang@reuben.ox.ac.uk}
	}
	
	\author[]{Stefan Zohren}
	
	\address[]{
		\orgdiv{Department of Engineering Science},
		\orgname{University of Oxford}
		\orgaddress{
				\city{Oxford},
%				\postcode{OX1 2JD},
%				\country{United Kingdom}
				\country{UK}
			}
%		{stefan.zohren@eng.ox.ac.uk}
	}
	
	\author[]{Mihai Cucuringu}
	
	\address[]{
		\orgdiv{Department of Mathematics},
		\orgname{University of California, Los Angeles}
		\orgaddress{
				\city{Los Angeles},
				\state{CA},
%				\postcode{90095},
				\country{USA}
%			{mihai@math.ucla.edu}
		}}
		
%			\corresp[]{
%					Corresponding author.
%					\href{mailto:Yunqi.Liang@reuben.ox.ac.uk}
%					{Yunqi.Liang@reuben.ox.ac.uk}
%				}
		\corresp[]{
			Corresponding author: Yunqi Liang.
			Address: Department of Engineering Science, University of Oxford,
			Parks Road, Oxford, OX1 3PJ, UK.
			Tel: +44 (0)7942497932.
			Email: \href{mailto:Yunqi.Liang@reuben.ox.ac.uk}
			{Yunqi.Liang@reuben.ox.ac.uk}
		}

	\clearpage
%	\thispagestyle{empty}
	
%	\begin{titlepage}
		
		\begin{center}
			
			\vspace*{1cm}
			
			{\LARGE\bfseries
				From Economic Explanation to Performance Improvement
				in Machine-Learning Portfolios
				\par
			}
			
			\vspace{1.2cm}
			
			{\large\bfseries Yunqi Liang\par}
			\vspace{0.10cm}
			Department of Engineering Science, University of Oxford, Oxford, UK
			
			\vspace{0.55cm}
			
			{\large\bfseries Stefan Zohren\par}
			\vspace{0.10cm}
			Department of Engineering Science, University of Oxford, Oxford, UK
			
			\vspace{0.55cm}
			
			{\large\bfseries Mihai Cucuringu\par}
			\vspace{0.10cm}
			Department of Mathematics, University of California, Los Angeles,\\
			Los Angeles, CA, USA\\
			Department of Statistics and 
			Oxford-Man Institute of Quantitative Finance,\\
			University of Oxford, Oxford, UK

			\vspace{1cm}
			
			{\large 1 September 2026}
			
		\end{center}
		
		%		\vfill
		
		\vspace{1cm}
		
		\noindent
		\textbf{Corresponding author:} Yunqi Liang\\
		Department of Engineering Science, University of Oxford\\
		Parks Road, Oxford OX1 3PJ, UK\\
		Tel: +44 (0)7942497932\\
		Email:
		\href{mailto:Yunqi.Liang@reuben.ox.ac.uk}
		{Yunqi.Liang@reuben.ox.ac.uk}
		
		\vspace{1cm}
		
%	\end{titlepage}
	
	\clearpage
	% ============================================================
	% Abstract Page
	% ============================================================
	
	\vspace{0.3cm}
	\begin{center}
		{\large\bfseries Abstract}
	\end{center}
	
	%	\vspace{0.3cm}
	
	\noindent
	To extend conventional model-level feature importance to economically meaningful portfolio outcomes and decisions, we propose a local first-order attribution-propagation framework that maps allocation-function feature exposures to local sensitivities of sequential portfolio performance metrics. We relate the propagated attributions to Shapley values defined directly for portfolio objectives and derive statistical inference for Sharpe-ratio attributions. The resulting attribution-guided feature-exposure intervention enables post-training adaptation, translating portfolio-metric explanations into out-of-sample decisions without retraining the underlying allocation model. We further introduce temporal stabilisation and cross-model averaging as complementary regularisation mechanisms. Empirically, the interventions yield statistically significant improvements in net Sharpe ratios across different model classes. 
	
	\vspace{0.5cm}
	
	\noindent
	\textbf{Keywords:}
	feature attribution,
	portfolio performance attribution,
	machine-learning portfolios,
	Shapley values,
	statistical arbitrage,
	feature-exposure intervention
	
	\vspace{0.3cm}
	
	\noindent
	\textbf{JEL Classification:}
	C45, C58, G11

%	\newpage
	\section*{Introduction}

	As machine-learning methods are increasingly used in portfolio construction, it has become more difficult to understand how complex allocation models transform input features or trading signals into portfolio decisions. Model-level explanation methods can identify which features influence model outputs, such as predictions or allocation weights, but this is not necessarily the information most relevant to an investor. 
%	Ultimately, investors care about realised portfolio outcomes, including returns, risk, turnover, transaction costs, and risk-adjusted performance. 
	Ultimately, investors care about realised portfolio outcomes, including returns, risk, turnover and transaction costs.
	A feature that is important for determining portfolio weights need not contribute positively to these outcomes. In particular, a signal may improve gross portfolio performance while simultaneously generating substantial turnover and transaction costs, so that its contribution to net performance becomes negative.
	
	This distinction creates a gap between explaining a machine-learning model and explaining the economic performance of the portfolio generated by that model. Existing machine-learning explanation tools primarily attribute model outputs to input features. In portfolio construction, however, model outputs are only an intermediate stage of the investment process. Features affect allocation weights, while changes in weights determine portfolio turnover. Allocations combined with asset returns generate portfolio returns, and the resulting return distribution determines mean return, volatility, and risk-adjusted performance. Once transaction costs are introduced, turnover further determines the wedge between gross and net performance. An economically relevant attribution framework should therefore explain not only how features affect allocation decisions, but also how these effects propagate through the portfolio-construction process to the performance metrics that investors ultimately seek to improve.
	
	A second limitation of conventional explainability analysis is that it is predominantly diagnostic. Feature attributions are typically computed after a model has been trained to explain its predictions, but they are rarely used directly to inform subsequent out-of-sample decisions. This is particularly restrictive in portfolio construction, where the primary objective is not interpretability itself but improved out-of-sample economic performance. Moreover, explanation methods are often tailored to particular model classes, making it difficult to determine whether economically useful feature information is specific to an individual allocation model or shared across different models. In practice, portfolio managers may employ several allocation models simultaneously, and feature contributions that are persistent through time or common across models may provide more robust information for subsequent investment decisions.
	
	Against this background, this paper addresses two related questions.
	
	\paragraph{\textbf{(1) How can portfolio metrics be attributed within a unified framework?}}
	
	Individual portfolio metrics can in principle be attributed separately using Shapley values or other feature-attribution methods. Such an approach, however, does not explicitly characterise how feature contributions are transmitted across economically linked portfolio quantities. For example, an attribution of the Sharpe ratio should reveal whether a feature's positive contribution arises primarily from a higher mean return, lower portfolio volatility, or a combination of the two. Similarly, attribution of the net Sharpe ratio should distinguish a feature's contribution to gross risk-adjusted performance from its effect on transaction costs through portfolio turnover. Treating allocation weights, turnover, returns, volatility, and Sharpe ratios as separate attribution problems obscures these economic channels.
	
	We therefore develop a local first-order attribution-propagation framework that characterises the sensitivities of sequential portfolio metrics to feature exposures in an allocation function. Starting from feature contributions to allocation weights, the framework propagates local effects through portfolio turnover, daily portfolio returns, mean returns, volatility, and ultimately gross and net Sharpe ratios. The resulting attribution provides a unified decomposition of the portfolio-construction process while retaining the relationships among its individual components. Because the framework is local, feature contributions can be evaluated at the feature-stock-date level and subsequently aggregated across stocks, trading dates, and evaluation periods.
	
	We further establish the theoretical relationship between the propagated first-order attributions and Shapley values defined directly for the target portfolio metrics. The latter provide a natural benchmark for assessing whether the local approximation correctly captures the direction and magnitude of feature contributions, but can be substantially more computationally demanding for nonlinear portfolio objectives. In addition, we derive statistical inference for Sharpe-ratio attributions, allowing us to assess whether estimated feature contributions are statistically distinguishable from sampling variation. The framework therefore provides not only an economically structured attribution of portfolio outcomes, but also a means of evaluating the reliability of those attributions.
	
	\paragraph{\textbf{(2) Can portfolio-metric attribution be translated from explanation into improved portfolio decisions?}}

%	A trained allocation function that performs well on its training sample need not remain optimal for a subsequent validation period or out-of-sample trading environment. Conditional on a fixed allocation function and the available input features, local portfolio-metric attribution can identify directions in feature exposure that are associated with improvements or deteriorations in a target objective. If these attribution structures contain a persistent component, information estimated before the out-of-sample trading period may provide useful guidance regarding which trading signals should receive greater or smaller exposure subsequently.
	
	A trained allocation function that performs well on its training sample need not perform equally well in a subsequent validation period or out-of-sample trading environment. Economically consistent feature-level attribution may therefore provide useful information for adapting feature exposures over time. Conditional on a fixed allocation function and the available input features, local portfolio-metric attribution can identify directions in feature exposure that are associated with improvements or deteriorations in a target objective. If these attribution structures contain a persistent component, information estimated before the out-of-sample trading period may provide useful guidance regarding which trading signals should receive greater or smaller exposure subsequently.
	
	This observation motivates an attribution-guided feature-exposure intervention. We use portfolio-metric attribution to modify the effective exposure of individual features entering the existing allocation rule. Features with persistently positive contributions to the target objective receive greater exposure, whereas features with persistently negative contributions receive lower exposure. The procedure therefore constitutes a form of post-training adaptation, translating attribution information into subsequent portfolio decisions while the underlying allocation model remains fixed.
	
	The effectiveness of such an intervention depends critically on the stability of the attribution signal. We consequently introduce two complementary regularisation mechanisms. First, temporal stabilisation aggregates previously available attribution information to reduce the influence of transitory feature contributions. Second, cross-model averaging combines attribution information from different allocation models. The latter is motivated by the possibility that feature contributions to economic outcomes contain a common component related to persistent feature characteristics or market conditions, beyond solely the idiosyncratic structure of a particular trained model. Cross-model attribution information can therefore provide an additional form of regularisation when constructing feature-exposure rules.
	
	The framework makes three main methodological contributions. First, it extends feature attribution from model outputs to economically relevant portfolio outcomes while preserving the propagation channels linking allocation weights, turnover, returns, volatility, transaction costs, and Sharpe ratios. Second, it connects the resulting local first-order attributions to Shapley values for the corresponding portfolio objectives and provides statistical inference for Sharpe-ratio feature contributions. Third, it converts attribution from a purely post hoc diagnostic into an input to subsequent portfolio decisions through feature-exposure interventions. Temporal stabilisation and cross-model averaging provide complementary mechanisms for regularising these interventions without requiring retraining of the underlying allocation model.
	
	We evaluate the framework in a statistical-arbitrage setting using two substantially different machine-learning allocation functions, extreme gradient boosting (XGB) and feed-forward neural networks (FFN). The analysis covers daily out-of-sample trading from 2006 to 2024. Starting from efficient SHAP-based feature attributions of residual portfolio weights, we propagate feature contributions through portfolio turnover and gross and net portfolio-performance metrics. Our baseline net-performance analysis incorporates transaction costs of 5 basis points per unit of turnover. We then compare the resulting first-order portfolio-metric attributions with numerically converged Shapley-value estimates and examine whether the attribution information available before each trading period can improve subsequent out-of-sample portfolio decisions.
	
	The empirical results provide evidence that the proposed attribution framework is both statistically reliable and economically informative. First, the local first-order Sharpe-ratio attributions are closely aligned with estimated Shapley values. Across hundreds of feature-year observations, the correlations between the proposed attributions and estimated Shapley values are approximately $0.76$ for XGB and $0.90$ for FFN for both gross and net Sharpe ratios. Approximately $10\%$ of feature-year attributions are statistically significant at the $5\%$ level for both gross and net Sharpe ratios, indicating that a non-negligible subset of feature contributions is distinguishable from noise.
	
	The attribution analysis also demonstrates that feature contributions to model outputs and portfolio outcomes are related but economically distinct. Features with large contributions to residual portfolio weights need not have correspondingly large contributions to turnover or risk-adjusted performance. Some relatively short-horizon or residual-based signals make limited contributions to residual weights but generate substantial turnover, particularly in certain stress periods. As a result, features that contribute positively to gross returns or the gross Sharpe ratio can contribute negatively to net performance once trading costs are taken into account. The propagation framework identifies these channels directly by decomposing the path from allocation decisions to turnover, gross performance, transaction costs, and net performance.
	
	We further find systematic structure in portfolio-level feature contributions across both time and allocation models. Turnover attributions are highly persistent across adjacent trading periods. Net Sharpe-ratio attributions display stronger temporal persistence than gross Sharpe-ratio attributions, particularly for the higher-turnover FFN strategy, consistent with the persistent contribution of turnover to transaction costs. At the same time, gross and net Sharpe-ratio attributions display positive cross-model similarity between XGB and FFN, despite substantial differences between the underlying allocation functions. These results suggest that portfolio-performance attributions contain an economically relevant component that extends beyond model-specific allocation mechanisms.
	
	Most importantly, the attribution information is useful for subsequent out-of-sample portfolio decisions. Using only information available prior to each trading period, we construct feature-exposure rules based on previous-period and temporally stabilised net Sharpe-ratio attributions, both at the model-specific level and by averaging information across models. The resulting interventions produce statistically significant improvements in net Sharpe ratios across the two allocation models. For FFN, the annualised net Sharpe ratio over 2006-2024 increases from $0.5750$ for the baseline portfolio to $0.6846$ using previous-year attribution information and to $0.8250$ using temporally stabilised attributions. The latter improvement is accompanied by a substantial reduction in average daily turnover, from $26.70\%$ to $13.02\%$. For XGB, stabilised cross-model attribution increases the net Sharpe ratio from approximately $0.6175$ to $0.7559$, with the improvement statistically significant at the $5\%$ level.
	
	The economic channels underlying these improvements differ across the two allocation models. For the higher-turnover FFN strategy, temporal stabilisation exploits the strong persistence of turnover-related attribution and improves net performance primarily by reducing portfolio turnover and transaction costs. For the lower-turnover XGB strategy, cross-model averaging provides additional attribution information from FFN and improves performance primarily through an increase in the gross Sharpe ratio while turnover remains relatively stable. Thus, the same portfolio-objective attribution framework can improve net performance through different economic channels depending on the characteristics of the underlying allocation model.
	
	Additional evidence confirms that net factor-adjusted alpha also improves using net Sharpe ratio attribution information. Under the stabilised cross-model feature-exposure rule, the annualised net factor-adjusted alpha increases from $1.19\%$ to $1.85\%$ for XGB and from $1.12\%$ to $2.23\%$ for FFN. In both cases, the attribution-guided portfolios exhibit statistically significant net alphas even though the corresponding baseline net alphas are statistically insignificant. Moreover, the attribution-guided interventions remain effective across a broad range of transaction-cost assumptions. Attributions estimated under higher transaction-cost rates tend to perform better in higher-cost out-of-sample environments, particularly for the higher-turnover strategy, while the benefits for the lower-turnover strategy are less sensitive to the assumed cost level.
	
	Overall, the results show that portfolio-level explainability can serve a role beyond model interpretation. By characterising how individual features affect economically relevant portfolio objectives and identifying persistent components of those contributions, attribution information can be transformed into implementable adjustments to subsequent portfolio decisions. The framework therefore connects explainability, statistical inference, and portfolio choice without requiring the underlying machine-learning allocation model to be retrained.
	
	The remainder of the paper is organised as follows. Section~1 reviews the related literature. Section~2 presents the methodology, including the statistical-arbitrage portfolio construction, the local first-order attribution-propagation framework, statistical inference for Sharpe-ratio attributions, and attribution-guided feature-exposure interventions. Section~3 presents the empirical results, including baseline portfolio performance, feature contributions to allocation decisions and portfolio metrics, validation of attribution reliability, cross-model attribution similarity and temporal persistence, and out-of-sample performance improvements from feature-exposure interventions. Section~5 summarises and concludes this work.

	\section{Related Literature}
	
	This paper relates to three strands of literature, including machine learning in asset pricing and portfolio construction, explainable machine learning and feature attribution, and portfolio-performance attribution and economic feature importance.
	
	\paragraph{\textbf{Machine learning in asset pricing and portfolio construction}}
	
	A growing literature applies machine-learning methods to return prediction and asset-pricing models. 
	\citet{GuKellyXiu2020} document substantial gains in predictive and economic performance from flexible nonlinear methods, particularly tree-based models and neural networks. 
	\citet{FreybergerNeuhierlWeber2020} similarly show that nonlinear methods help identify the incremental information contained in firm characteristics. 
	A related strand uses machine learning to model latent factors and conditional risk exposures. 
	\citet{KellyPruittSu2019} introduce instrumented principal component analysis, in which observable characteristics determine time-varying factor exposures, while \citet{GuKellyXiu2021} extend this framework using autoencoder neural networks to capture nonlinearities and interactions. 
	\citet{GiglioKellyXiu2022} provide a broader review of machine-learning approaches to expected returns, factor structures, risk premia, and stochastic discount factors.
	
	A complementary literature applies machine learning directly to portfolio construction. 
	\citet{LimZohrenRoberts2019} introduce Deep Momentum Networks, which jointly learn trading signals and position sizing while directly optimising the Sharpe ratio and accounting for turnover. 
	\citet{ZhangZohrenRoberts2020} extend direct Sharpe-ratio optimisation to multi-asset portfolio allocation, and \citet{ZhangZhangCucuringuZohren2021} develop a general end-to-end framework that maps inputs directly into portfolio weights under economically relevant objectives and constraints. 
	\citet{PohLimZohrenRoberts2021} formulate cross-sectional portfolio construction as a learning-to-rank problem, explicitly learning pairwise and listwise relations across securities. 
	More recently, \citet{GuijarroOrdonezPelgerZanotti2025} combine residual-based statistical arbitrage with deep-learning signal extraction and neural-network portfolio allocation. 
	These studies show that machine-learning models can move beyond intermediate return forecasting to target economically relevant portfolio objectives directly.

%	A growing literature applies machine-learning methods to return prediction and portfolio construction. \citet{GuKellyXiu2020} systematically compare a broad range of machine-learning methods and document substantial improvements in predictive and economic performance relative to conventional linear models. Tree-based methods and neural networks perform particularly well, reflecting their ability to capture nonlinearities and interactions among predictors. Their analysis also finds considerable agreement across model classes regarding the characteristics with the strongest predictive importance. \citet{FreybergerNeuhierlWeber2020} similarly show that flexible nonlinear methods are useful for identifying the incremental information contained in firm characteristics.
%	More recent work applies machine learning directly to portfolio construction and statistical arbitrage. \citet{GuijarroOrdonezPelgerZanotti2025} construct residual portfolios from U.S. equities, extract predictive signals using a convolutional transformer, and map these signals into portfolio allocations through a feed-forward neural network under risk-adjusted objectives. 
%%	Their framework demonstrates how flexible machine-learning methods can be integrated into both signal extraction and portfolio allocation.
	
	These studies establish the value of machine learning for return prediction and portfolio construction. Our paper addresses a complementary question, namely how individual input features contribute to the realised economic performance of the resulting portfolio once a machine-learning allocation model has been trained. 
%	A feature that is important for predicting returns or determining allocation weights need not contribute positively to portfolio performance once turnover, risk, and transaction costs are taken into account.
	
	\paragraph{\textbf{Explainable machine learning and feature attribution}}
	
	Our methodology builds on the literature on Shapley values and explainable machine learning. \citet{Shapley1953} introduced the Shapley value as an allocation rule based on the average marginal contribution of each participant across possible coalitions. \citet{LundbergLee2017} adapt this principle to machine learning through SHAP (SHapley Additive exPlanations), which decomposes individual model predictions into feature-level contributions.	
	Subsequent work develops efficient attribution methods for specific model classes. \citet{LundbergEtAl2020} provide exact SHAP algorithms for tree-based models and introduce SHAP interaction values for measuring pairwise feature interactions. For neural networks, \citet{ShrikumarGreensideKundaje2017} propose DeepLIFT, while \citet{SundararajanTalyYan2017} develop Integrated Gradients. More broadly, \citet{Molnar2025} surveys a wide range of model-specific and model-agnostic approaches to interpretable machine learning.
	
%	An important issue in feature attribution is dependence among predictors. \citet{AasJullumLoland2021} show that independence-based implementations of Kernel SHAP can produce misleading explanations when predictors are correlated and propose conditional approaches that account for feature dependence. This concern is particularly relevant in financial applications, where trading signals and firm characteristics can be strongly related. We therefore assess the reliability of the proposed portfolio-level attributions using Shapley-value benchmarks and interaction diagnostics.
%	The interaction structure underlying model predictions need not coincide with that governing economically relevant portfolio outcomes, such as the Sharpe ratio.
	
	A related literature studies interpretability in financial machine-learning models.
	\citet{WoodGiegerichRobertsZohren2021} use attention to assess variable and temporal importance in the Momentum Transformer, while \citet{TanRobertsZohren2023} apply SHAP to interpret spatio-temporal features in momentum strategies.
	An important issue in feature attribution is dependence among predictors. \citet{AasJullumLoland2021} show that independence-based implementations of Kernel SHAP can produce misleading explanations when predictors are correlated and propose conditional approaches that account for feature dependence. This concern is particularly relevant in financial applications, where trading signals and firm characteristics can be strongly related. We assess the reliability of the proposed portfolio-level attributions using Shapley-value benchmarks and separately examine interaction effects. The interaction structure underlying model predictions need not coincide with that governing economically relevant portfolio outcomes, such as the Sharpe ratio.
	
	The distinction between model-level importance and economic contribution is central to our analysis. Standard feature-importance measures, such as split-gain importance in boosted trees, quantify the relevance of a feature for model fitting but do not indicate whether its effect is beneficial for the investor's objective. SHAP-type methods provide signed local contributions and therefore offer a more informative starting point. Nevertheless, attribution to a model output is not equivalent to attribution to portfolio performance. 
%	A feature may contribute positively to allocation weights or gross returns while simultaneously increasing portfolio turnover and transaction costs, thereby reducing net performance.
	
	Our framework extends model-level attribution to economically relevant portfolio outcomes. Starting from feature contributions to the allocation function, we propagate local effects through portfolio weights, turnover, returns, volatility, and gross and net Sharpe ratios. The objective is therefore not only to explain how a feature affects a model output, but also to identify how that effect is transmitted to the portfolio-performance metric relevant to the investor.
	
	\paragraph{\textbf{Portfolio performance attribution and economic feature importance}}
	
	Our paper is most closely related to research that evaluates investment inputs directly in terms of portfolio outcomes. \citet{MoehleBoydAng2022} apply Shapley values to portfolio-performance attribution and decompose achieved portfolio performance into contributions from investment-style components. 
%	Their framework demonstrates how Shapley values can be used to attribute economically relevant portfolio outcomes.
	Our setting differs primarily in the object and granularity of attribution. Instead of focusing on broad investment-style components, we study individual predictive features within nonlinear machine-learning allocation functions. The proposed framework generates local feature contributions at the stock-date level and propagates them through sequential portfolio metrics before aggregation across stocks and time. This allows us to identify not only whether a feature contributes positively or negatively to a portfolio objective, but also the economic channel through which the contribution arises. In particular, the contribution to the net Sharpe ratio can be decomposed into effects on gross performance and transaction costs through turnover.
	
	Transaction costs could be central to the distinction between predictive and economic importance. \citet{DeMiguelEtAl2020} show that trading costs materially affect the economic value of firm characteristics in portfolio choice. \citet{JensenKellyMalamudPedersen2026} further develop the implementable efficient frontier and introduce an economic feature-importance measure within a machine-learning portfolio framework. Their analysis shows that transaction costs can substantially alter the economic value assigned to individual signals.
%	 particularly when persistent signals are compared with transient, high-turnover predictors.
	
	Our approach is complementary to these works. Their framework incorporates trading costs directly into the portfolio-learning problem and evaluates feature importance within an economically optimised allocation rule. We instead take an already-trained allocation function, optimised for the net Sharpe ratio, as given and measure the local contribution of its individual features to subsequent portfolio objectives. When these attribution structures exhibit sufficient persistence, they can be used to adjust feature exposures for next trading period without re-estimating the underlying allocation model. The resulting intervention therefore provides a form of post-training adaptation.
	
	Related work also studies interpretable portfolio models directly. \citet{BellKakhbodLettauNazemi2026} apply Explainable Boosting Machines to corporate bond return prediction, providing interpretable feature effects and interactions and highlighting the economic value of model transparency. Our approach addresses a different problem by retaining flexible allocation functions, whether inherently interpretable or not, while developing a common portfolio-level attribution framework that can be applied across model architectures.
	
	Overall, the existing literature provides methods for predicting returns with machine learning, explaining model outputs, and attributing economic importance of investment signals at a global level. Our paper connects these strands by tracing local feature effects from machine-learning allocation models through the portfolio-construction process and by using the resulting attribution information for subsequent portfolio decisions. The framework links allocation weights, turnover, returns, volatility, and gross and net Sharpe ratios within a unified first-order attribution system, relates these attributions to Shapley-value benchmarks and statistical inference, and uses temporally stabilised and cross-model attribution signals to construct feature-exposure interventions without retraining the underlying allocation model.

	\section{Methodology}
	
%	In methodology, we first give the methods for constructing the arbitrage portfolios. Next, we propose the local first-order objective attribution, introducing the relationship to shapley values, details of propagation from model output of allocation weights to portfolio emtrics including sharpe raito, and the statistical inference for the significance test of sharpe ratio attribution. Finally, we give the possible methods and regularity ways, such as temporal stabilisation and cross-model averaging, for using the portfolio metrics attribution to feature exposure interventions for trading portfolios.
	
	The methodology proceeds in three stages. First, we describe the construction of the arbitrage portfolios. Second, we introduce the local first-order objective-attribution framework, establish its relationship with Shapley values, and describe how feature contributions propagate from the allocation-model output, represented by residual portfolio weights, to economically relevant portfolio metrics, including turnover and the Sharpe ratio. We also derive statistical inference for assessing the significance of Sharpe-ratio attributions. Finally, we present the attribution-guided feature-exposure intervention framework and introduce temporal stabilisation and cross-model averaging as complementary regularisation mechanisms for translating portfolio-metric attributions into out-of-sample trading decisions.
	
	\subsection{Statistical arbitrage}
	
	Statistical arbitrage can be decomposed into three essential elements: (1) the identification of groups of similar assets to form arbitrage portfolios, (2) the extraction of time-series signals that capture temporary deviations in asset similarity, and (3) the design of trading policies that allocate capital across arbitrage portfolios based on the extracted signals. The methods used in each part are given in detail.
	
	\subsubsection{Arbitrage portfolios}
	According to \citet{GuijarroOrdonezPelgerZanotti2025}, we consider a panel of excess returns $R_{n, t}$, defined as the return minus risk-free rate of stock $n=1, \ldots, N$ at time $t=1, \ldots, T$. The excess return vector of all stocks at time $t$ is denoted by $ \mathbf{R}_{1:N, t}=\left(\begin{array}{lll}R_{1, t} & \cdots & R_{N, t}\end{array}\right)^{\top}$. We assume that asset returns can be modelled using a conditional factor structure:
	\begin{equation}
		R_{n, t}=\boldsymbol{\beta}_{n, t-1}^{\top} \mathbf{F}_t+\epsilon_{n, t}
	\end{equation}
	
	The $K$ factors $\mathbf{F} \in \mathbb{R}^{T \times K}$ capture the systematic risk, while the risk loadings $\boldsymbol{\beta}_{t-1} \in \mathbb{R}^{N \times K}$ are allowed to vary over time as functions of the information set at time $t-1$. 
	
	Without loss of generality, the factors can be treated as excess returns on traded portfolios. For these tradable factors, such as the market factor, they may be included directly into the return vector $\mathbf{R}_t$ for asset space. Thus, we define the arbitrage portfolio as residual portfolios $ \epsilon_{n, t} = R_{n, t}-\boldsymbol{\beta}_{n, t-1}^{\top} \mathbf{F}_t$, which is a linear conbination of all the assets in $\mathbf{R}_t$. As factors are traded assets, the arbitrage portfolios are themselves traded portfolios. 
%	More compactly, the vector of residual portfolios can be expressed as
%	\begin{equation}
%		\epsilon_t = R_{1:N, t} - \beta_{t-1} F_t = \Phi_{t-1} R_t
%	\end{equation}
%	for some weighting matrix $ \Phi_{t-1}$. 
	
	Arbitrage portfolios can be interpreted as trades relative to mimicking stock portfolios. The factor component of the residual corresponds to a well-diversified replicating portfolio with exposures closely aligned to the stock under consideration. In the case of the Fama-French factors, the mimicking portfolios are constructed to match the target stock’s sensitivities to the fundamental risk factors.

	%%%%%%%%%%%%%%%%%%%%%%%%%%%%%%%%%%%%%%%%%%%%%%%%%%%%%%%%%%%%
	\paragraph{\textbf{Factor mimicking portfolios and tradable residual portfolios}}
%	With a conditional factor model
%	$
%	\mathbf{R}_{1:N, t}
%	=
%	\boldsymbol{\beta}_{t-1}\mathbf{F}_t
%	+
%	\boldsymbol{\epsilon}_t,
%	$
%	where $\mathbf{R}_{1:N, t}$ is the cross-sectional vector of excess returns across the stock universe, $\mathbf{F}_t \in \mathbb{R}^{K}$ are systematic factors, and
%	$\boldsymbol{\beta}_{t-1} \in \mathbb{R}^{N \times K}$ are
%	time-varying factor loadings, 
	We recover tradable factor realisations as a linear combination of excess returns across the stock universe:
	\begin{equation}
	\hat{\mathbf{F}}_t
	= \mathbf{w}_{t-1}^{F} {}^{\top} \mathbf{R}_{1:N, t}=
	\left(
	\boldsymbol{\beta}_{t-1}^\top
	\boldsymbol{\beta}_{t-1}
	\right)^{-1}
	\boldsymbol{\beta}_{t-1}^\top
	\mathbf{R}_{1:N, t},
	\end{equation}
	with $ \mathbf{w}_{t-1}^{F}\in \mathbb{R}^{N \times K}$ as factor-mimicking weights, derived from known factor exposures $\boldsymbol{\beta}_{t-1}$.
	
	The formula represents the Ordinary Least Squares (OLS) estimation of factor returns within a cross-sectional asset pricing framework. The objective function is minimising the sum of squared residuals $\min_{\mathbf{F}_t} ||\mathbf{R}_{1:N, t} - \boldsymbol{\beta}_{t-1}\mathbf{F}_t||_2^2$. Assuming the matrix $\boldsymbol{\beta}_{t-1}^\top
	\boldsymbol{\beta}_{t-1}$ is non-singular, we solve for $\hat{\mathbf{F}}_t$.
	
	%%%%%%%%%%%%%%%%%%%%%%%%%%%%%%%%%%%%%%%%%%%%%%%%%%%%%%%%%%%%
%	\paragraph{Residual (factor-neutral) portfolios}
    \label{sec:factor_neutral_portfolios}
	
	According to \citet{Ross1976}, define the projection matrix:
	\begin{equation}
	\boldsymbol{\Phi}_{t-1}
	=
	\mathbf{I}
	-
	\boldsymbol{\beta}_{t-1}
	\left(
	\boldsymbol{\beta}_{t-1}^\top
	\boldsymbol{\beta}_{t-1}
	\right)^{-1}
	\boldsymbol{\beta}_{t-1}^\top,
	\end{equation}
	and the tradable residual portfolios are given by
	\begin{equation}
	\boldsymbol{\epsilon}_t
	= \mathbf{R}_{1:N, t} - \boldsymbol{\beta}_{t-1}\hat{\mathbf{F}}_t = \left(\mathbf{I}
	-
	\boldsymbol{\beta}_{t-1}\mathbf{w}_{t-1}^{F} {}^{\top} \right)  \mathbf{R}_{1:N, t} =
	\boldsymbol{\Phi}_{t-1}
	\mathbf{R}_{1:N, t}.
\end{equation}
%	The residual portfolios satisfy factor neutrality, meaning that they are orthogonal to the factor space:
%	\[
%	\boldsymbol{\beta}_{t-1}^\top
%	\boldsymbol{\epsilon}_t
%	=\boldsymbol{\beta}_{t-1}^\top \boldsymbol{\Phi}_{t-1}
%	\mathbf{R}_t= 
%	\mathbf{0}.
%	\]

	\subsubsection{Arbitrage signal}
	
	The arbitrage signal is derived from the time series of arbitrage portfolios and serves as the input to the trading policy. Specifically, we construct trading signals from the cumulative residuals over a specific lookback window.
	We apply the signal extraction to the time-series of the last $L$ lagged residuals of stock $n$, which we denote in vector notation as
	$
		\epsilon_{n, t-1}^L:=\left(\begin{array}{lll}
			\epsilon_{n, t-L} & \cdots & \epsilon_{n, t-1}
		\end{array}\right) .
	$
	The arbitrage signal function is a mapping $\boldsymbol{z} \in \boldsymbol{Z}$ from lagged residuals in $\mathbb{R}^L$ to a signal in $\mathbb{R}^p$, where $\boldsymbol{Z}$ defines an appropriate function space:
	\begin{equation}
		\boldsymbol{z}(\cdot): \epsilon_{n, t-1}^L \rightarrow z_{n, t-1}
	\end{equation}
	
	The signals $z_{n, t-1} \in \mathbb{R}^p$ for the arbitrage portfolio $n$ at time $t$ only depend on the time-series of lagged returns $\epsilon_{n, t-1}^L$. We extend this elementwise across all assets, $\boldsymbol{z}\left(\epsilon_{t-1}^L\right)=\left(\begin{array}{lll}z_{1, t-1} & \cdots & z_{N, t-1}\end{array}\right)=z_{t-1} \in \mathbb{R}^{p \times N}$ with $\epsilon_{t-1}^L=\left(\begin{array}{lll}\epsilon_{1, t-1}^L & \cdots & \epsilon_{N, t-1}^L\end{array}\right)  \in \mathbb{R}^{L \times N}$. The arbitrage signal $z_{n, t-1}$ is treated as a sufficient statistic for trading, which means it summarises all relevant information needed for portfolio decisions. In practice, these signals capture mean-reversion and reversal patterns with a small number of parameters. We assume that the first $L$ lagged returns provide a sufficient basis for estimating $z_{n, t-1}$. 
	
	The underlying motivation is that arbitrage opportunities represent temporary deviations from fair value. Thus, parametric models of mean reversion provide a natural signal. Trading strategies can be designed around the estimated speed of mean reversion and the current deviation from the long-run mean.
	
	%	\textbf{Ornstein-Uhlenbeck (OU) Process}
	
	We construct signals from last $L$ cumulative residuals $X^L_{n, t-1} = \sum_{l=1}^L \epsilon_{n, t-L-1+l}$ for each time index $t-1$ and the stock index $n$. For each stock, the cumulative residuals $X^L_t$ are assumed to be the discrete realisations of a continuous-time process. Following the influential papers by \citet{AvellanedaLee2010} and \citet{YeoPapanicolaou2017}, we model the cumulative residuals $X^L_t$ as an Ornstein-Uhlenbeck (OU) process
	\begin{equation}
		d X^L_t=\kappa\left(\mu-X^L_t\right) d t+\sigma d W_t
	\end{equation}
	where $d W_t$ is a standard Brownian motion. The OU dynamics provide a parsimonious framework for capturing mean reversion in arbitrage signals, characterised by a constant speed of reversion $\kappa$, long-term mean level $\mu$, and volatility $\sigma$. These OU parameters $(\kappa_{n, t-1}, \mu_{n, t-1}, \sigma_{n, t-1})$ could be used as components of trading signals $z_{n, t-1}$.
	
	%	As the analytical solution of the above stochastic differential equation is
	%	$$
	%	X_{t+\Delta t}=\left(1-e^{-\kappa \Delta t}\right) \mu+e^{-\kappa \Delta t} X_t+\sigma \int_t^{t+\Delta t} e^{-\kappa(t+\Delta t-s)} d W_s
	%	$$
	%	for any $\Delta t$, we can without loss of generality set $\Delta t=1$, and estimate the parameters $\kappa, \mu$ and $\sigma$ from the AR(1) model
	%	$$
	%	X_{t+1}=a+b X_t+e_t, \quad e_t \sim N(0, \sigma_e^2)
	%	$$
	%	where each $e_t$ is a normal, independent and identically distributed random variable. The parameters are estimated with a standard linear regression, which yields
	%	$$
	%	\hat{\kappa}=-\frac{\log (\hat{b})}{\Delta t}, \quad \hat{\mu}=\frac{\hat{a}}{1-\hat{b}}, \quad \frac{\hat{\sigma}}{\sqrt{2 \hat{\kappa}}}=\sqrt{\frac{\hat{\sigma}_e^2}{1-\hat{b}^2}} .
	%	$$
	%	Note that this is only defined for $b<1$ which is equivalent to parameter restrictions that the OU process is mean-reverting. 
	
%	To account for regime shifts in financial markets, we further extend this framework by introducing regime-switching dynamics. In particular, we consider two alternative specifications based on the OU process: the Markov Regime-Switching Model (MRS) and the Hierarchical Dirichlet Process Hidden Markov Model (HDP-HMM). These allow for shifts in mean-reversion behaviour across different market states.
	
	To account for regime shifts in financial markets, we further extend this framework by introducing regime-switching dynamics. Based on the OU process, the methods such as Markov Regime-Switching Model (MRS) and Hierarchical Dirichlet Process Hidden Markov Model (HDP-HMM) allow for shifts in mean-reversion behaviour across different market states. Details of the regime-based OU process are provided in Appendix~\ref{sec:rs_ou_model}.

	\subsubsection{Arbitrage trading}
	An arbitrage allocation is defined as a mapping from $\mathbb{R}^p$ to $\mathbb{R}$ within a function class $\boldsymbol{W}$, which assigns a weight $w_{n, t-1}^\epsilon$ for the arbitrage portfolio $\epsilon_{n, t-1}$ in the investment strategy based solely on the arbitrage signal $z_{n, t-1}$ :
	\begin{equation}
		\boldsymbol{w}^\epsilon(\cdot): z_{n, t-1} \rightarrow w_{n, t-1}^\epsilon
	\end{equation}
	
	The residual allocation vector is denoted by $\boldsymbol{w_{t-1}^{\epsilon}}=\left(\begin{array}{lll}w^\epsilon_{1, t-1} & \cdots & w^\epsilon_{N, t-1}\end{array}\right)^\top$. We consider two flexible trading policies, extreme gradient boosting (XGB) and feedforward neural networks (FFN). 
	Both allocation functions are trained using a Sharpe-ratio-based objective.

	\paragraph{\textbf{Extreme Gradient Boosting (XGB)}}
	
%	XGBoost provides an nonlinear allocation rule based on gradient-boosted
%	decision trees:
%		\begin{equation}
%		\boldsymbol{w_{t-1}^{\epsilon \mid \mathrm{XGB}}}\left(z_{t-1}\right)=\boldsymbol{g^{\mathrm{XGB}}}\left(z_{t-1}\right)
%	\end{equation}
%	Gradient-boosted tree models are particularly useful in this setting because they
%	can be trained using custom differentiable objectives through first- and second-order
%	derivatives. In contrast to standard decision trees and random forests, XGBoost
%	allows us to optimise portfolio-level objectives such as the negative net Sharpe
%	ratio. It also provides an exact Tree SHAP implementation for local feature
%	attribution in tree ensembles. We therefore use XGBoost as the representative
%	gradient-boosted tree model. 
%%	To control overfitting, we use shallow trees and select
%%	the model complexity using validation performance.

	The extreme gradient boosting (XGB) provides a nonlinear allocation rule based on
	gradient-boosted decision trees and is designed for efficient and scalable
	learning \citep{ChenGuestrin2016}. The allocation function is represented as
	an additive ensemble of regression trees,
	\begin{equation}
		\boldsymbol{w}^{\epsilon \mid \mathrm{XGB}}(z_{n, t-1})
		=
		g_{\boldsymbol{\theta}}^{\mathrm{XGB}}(z_{n, t-1})
		=
		\sum_{b=1}^{B}
		\nu h_b(z_{n, t-1}),
	\end{equation}
	where $h_b$ denotes the $b$-th regression tree, $B$ is the number of
	boosting iterations, $\nu$ is the learning-rate (shrinkage) parameter, and
	$\boldsymbol{\theta}$ collects the parameters of the fitted tree ensemble.
	At each boosting iteration, an additional regression tree is added
	sequentially to improve the portfolio objective conditional on the current
	ensemble.
	
	A key advantage of XGB is its ability to accommodate custom
	differentiable objectives using first- and second-order derivative
	information. This allows the tree ensemble to be trained directly with
	respect to a portfolio-level, non-additive objective such as the negative net Sharpe ratio. The gradient of the portfolio loss with respect to the model outputs is used to construct
	the boosting updates.
	We therefore employ XGB as the representative gradient-boosted tree
	allocation model. Model complexity is controlled through the structure of
	the individual trees, including restrictions on the maximum tree depth and
	the minimum child weight.

	\paragraph{\textbf{Feedforward Neural Network (FFN)}}
	
	The feedforward neural network (FFN) provides another flexible allocation function by learning a nonlinear mapping from signals to allocation weights:
	\begin{equation}
		\boldsymbol{w^{\epsilon \mid \mathrm{FFN}}}\left(z_{n, t-1}\right)=	g_{\boldsymbol\theta}^{\mathrm{FFN}}\left(z_{n, t-1}\right)
	\end{equation}
	
%	Following \citet{GuijarroOrdonezPelgerZanotti2025}, we employ ReLU as the activation function. For a feedforward network with $L^{\mathrm{FFN}}$ layers, each hidden layer takes the output from the previous layer and transforms it into an output as
%	\begin{equation}
%		\begin{aligned}
%			x^{(l)} & =\operatorname{ReLU}\left(W^{(l-1) \top} x^{(l-1)}+w_0^{(l-1)}\right)=\operatorname{ReLU}\left(w_0^{(l-1)}+\sum_{k=1}^{K^{(l-1)}} w_k^{(l-1)} x_k^{(l-1)}\right) \\
%			y & =W^{\left(L^{\mathrm{FFN}}\right) \top} x^{\left(L^{\mathrm{FFN}}\right) }+w_0^{\left(L^{\mathrm{FFN}}\right)}
%		\end{aligned}
%	\end{equation}
%	with hidden layer outputs $x^{(l)}=\left(x_1^{(l)}, \ldots, x_{K^{(l)}}^{(l)}\right) \in \mathbb{R}^{K^{(l)}}$, parameters $W^{(l)}=\left(w_1^{(l)}, \ldots, w_{K^{(l)}}^{(l)}\right) \in$ $\mathbb{R}^{K^{(l)} \times K^{(l+1)}}$ for $l=1, \ldots, L^{\mathrm{FFN}}-1$ and $W^{\left(L^{\mathrm{FFN}}\right)} \in \mathbb{R}^{K^{(L^{\mathrm{FFN}})}}$, and where $\operatorname{ReLU}\left(x_k\right)=\max \left(x_k, 0\right)$. Here $x^{(0)} = z_{t-1}$ and $y = w_{t-1}^\epsilon$.

	Following \citet{GuijarroOrdonezPelgerZanotti2025}, we employ ReLU as the activation function. 
	Let $L^{\mathrm{FFN}}$ denote
	the number of hidden layers and let
	$\boldsymbol{x}_{n,t-1}^{(0)}=z_{n,t-1}$ denote the input
	feature vector for stock $n$ at time $t-1$. The hidden representations
	are recursively defined as
	\begin{equation}
		\boldsymbol{x}_{n,t-1}^{(\ell)}
		=
		\operatorname{ReLU} 
		\left(
		W^{(\ell)}
		\boldsymbol{x}_{n,t-1}^{(\ell-1)}
		+
		\boldsymbol{w}_{0}^{(\ell)}
		\right),
		\qquad
		\ell=1,\ldots,L^{\mathrm{FFN}},
	\end{equation}
	where
	$\boldsymbol{x}_{n,t-1}^{(\ell)}
	\in\mathbb{R}^{K^{(\ell)}}$
	is the output of hidden layer $\ell$,
	$W^{(\ell)}
	\in\mathbb{R}^{K^{(\ell)}\times K^{(\ell-1)}}$
	is the corresponding weight matrix, and
	$\boldsymbol{w}_{0}^{(\ell)}
	\in\mathbb{R}^{K^{(\ell)}}$
	is the bias vector. The ReLU activation function is applied elementwise
	and is defined as
	$\operatorname{ReLU}(x)=\max(x,0)$.
	The final hidden representation is mapped to the residual allocation
	weight through a linear output layer,
	\begin{equation}
		w_{n,t-1}^{\epsilon\mid\mathrm{FFN}}
		=
		\boldsymbol{w}^{(L^{\mathrm{FFN}}+1)\top}
		\boldsymbol{x}_{n,t-1}^{(L^{\mathrm{FFN}})}
		+
		w_{0}^{(L^{\mathrm{FFN}}+1)},
	\end{equation}
	where
	$\boldsymbol{w}^{(L^{\mathrm{FFN}}+1)}
	\in\mathbb{R}^{K^{(L^{\mathrm{FFN}})}}$
	and $w_{0}^{(L^{\mathrm{FFN}}+1)}\in\mathbb{R}$
	denote the weights and bias of the output layer, respectively.
	The parameter vector $\boldsymbol{\theta}$ collects all weights and
	biases of the network.
	
%	We mainly focus on Sharpe-ratio as the objective for training both XGBoost and FFN. 
	
%	(1) Sharpe-ratio maximisation

	We mainly focus on Sharpe ratio as the objective for training both XGB and FFN. 
	The arbitrage trading is to maximise the risk-adjusted returns, more specifically solving for
	\begin{equation}
		\quad \max _{\boldsymbol{w}^\epsilon \in \boldsymbol{W}} \frac{\mathbb{E}_{t-1}\left[{w_{t-1}^{R}}^{\top} R_{1:N, t} - c \tau_t \right]}{\sqrt{\operatorname{Var}_{t-1}\left[{w_{t-1}^{R}}^{\top} R_{1:N, t} - c \tau_t\right]}} \quad \text{with} \quad w_{t-1}^{R} = \frac{{w_{t-1}^{\epsilon}}^{\top} \Phi_{t-1}}{||{w_{t-1}^{\epsilon}}^{\top} \Phi_{t-1}||_1},
	\end{equation}
	where portfolio turnover at time $t$ be $\tau_t = \left\|
	\bm w_{t-1}^{R}
	-
	\bm w_{t-2}^{R}
	\right\|_1$ and $c$ is the proportional transaction-cost rate. When $c=0$, we obtain the gross performance metrics. Portfolio weights are normalised such that the sum of absolute stock weights equals one, i.e. $||w_{t-1}^R||_1 = 1$. This normalisation enforces a leverage constraint, as short positions are bounded by one.  
	
%	(2) Mean-variance optimisation
%	
%	Alternatively, the FFN can be trained to achieve the highest average return for a given level of variance risk via
%	\begin{equation}
%		\quad \max _{\boldsymbol{w}^\epsilon \in \boldsymbol{W}} \mathbb{E}_{t-1}\left[{w_{t-1}^{R}}^{\top} R_t \right]-\gamma \operatorname{Var}_{t-1}\left[{w_{t-1}^{R}}^{\top} R_t\right] \quad \text{with} \quad w_{t-1}^{R} = \frac{{w_{t-1}^{\epsilon}}^{\top} \Phi_{t-1}}{||{w_{t-1}^{\epsilon}}^{\top} \Phi_{t-1}||_1}
%	\end{equation}
%	for some risk-aversion parameter $\gamma$, which we set to $\gamma = 1$. 
	
%	Portfolio weights are normalised such that the sum of absolute stock weights equals one, i.e. $||w_{t-1}^R||_1 = 1$. This normalisation enforces a leverage constraint, as short positions are bounded by one.  

	%%%%%%%%%%%%%%%%%%%%%%%%%%%%%%%%%%%%%%%%%%%%%%%%%%%%%%%%%%%%
	\subsubsection{Portfolio performance evaluation}
	
	\paragraph{\textbf{Realised portfolio.}}
	
	In our setting, the initial capital is normalised to one dollar and is not reinvested. At time $t$,
	the portfolio consists of stock positions $\bm w_{t-1}^{R}$ and a residual cash
	position
	$
	1-\mathbf 1^\top \bm w_{t-1}^{R}.
	$
	Let the daily gross return of stock $n$ be
	$
	R_{n,t}^{\mathrm{gross}}
	=
	\frac{P_{n,t}-P_{n,t-1}}{P_{n,t-1}},
	$
	and let the corresponding excess return be
	$
	R_{n,t}
	=
	R_{n,t}^{\mathrm{gross}}-r_{f,t},
	$
	where $r_{f,t}$ is the risk-free rate.
	
	The portfolio gross profit and loss is
	$
	\mathrm{PnL}_{t}^{\mathrm{gross}}
	=
	\bm w_{t-1}^{R} {}^\top \bm R_{1:N, t}^{\mathrm{gross}}
	+
	\left(
	1-\mathbf 1^\top \bm w_{t-1}^{R}
	\right)r_{f,t}.
	$
	The corresponding portfolio excess return is
	$
%	R_{p,t}
%	=
	\bm w_{t-1}^{R} {}^\top
	\bm R_{1:N, t} = \mathrm{PnL}_{t}^{\mathrm{gross}}-r_{f,t},
	$
	where $\bm R_{1:N, t}$ denotes the vector of stock excess returns. 
%	The cumulative excess
%	return over the evaluation period is
%	$
%	\sum_{t=1}^{T}R_{p,t}.
%	$
%	
%	The annualised Sharpe ratio is defined as
%	$
%	\mathrm{SR}
%	=
%	\sqrt{252}
%	\frac{\mu}{\sigma},
%	$
%	with 
%	$
%	\mu=
%	\frac1T\sum_{t=1}^{T}R_{p,t}
%	$
%	and
%	$
%	\sigma=
%	\sqrt{
%		\frac1T
%		\sum_{t=1}^{T}
%		\left(R_{p,t}-\mu\right)^2
%	}.
%	$
	
	\paragraph{\textbf{Transaction costs and net performance.}}
	
	Market frictions are incorporated through proportional transaction costs. The cost
	incurred from rebalancing the portfolio from $\bm w_{t-2}^{R}$ to $\bm w_{t-1}^{R}$
	is
	$
	\operatorname{cost}_t
	=
	c
	\left\|
	\bm w_{t-1}^{R}
	-
	\bm w_{t-2}^{R}
	\right\|_1,
	$
	where $c$ is the proportional transaction-cost rate. When $c=0$, we could get the gross performance metrics. Let the portfolio turnover at time $t$ be $\tau_t = \left\|
	\bm w_{t-1}^{R}
	-
	\bm w_{t-2}^{R}
	\right\|_1$.
	
	The net portfolio profit and loss is
	$
	\mathrm{PnL}_{c, t}
	=
	\mathrm{PnL}_{0, t}
	-
	c \tau_t
	$ with $\mathrm{PnL}_{0, t} = \mathrm{PnL}_{t}^{\mathrm{gross}}$.
	The net portfolio excess return is therefore
	$
	R_{c,t}
	=
	R_{0,t}
	-
	c \tau_t
	=\bm w_{t-1}^{R} {}^\top
	\bm R_{1:N, t}-
	c \tau_t.
	$
	The cumulative net excess return over the evaluation period is
	$
	\sum_{t=1}^{T}
	R_{c,t}.
	$
	Finally, the annualised Sharpe ratio is defined as
	$
	\mathrm{SR}_c
	=
	\sqrt{252}
	\frac{\mu_c}{\sigma_c},
	$
	with daily mean return
	$
	\mu_c
	=
	\frac1T
	\sum_{t=1}^{T}
	R_{c,t}
	$
	and daily volatility
	$
	\sigma_c
	=
	\sqrt{
		\frac1T
		\sum_{t=1}^{T}
		\left(
		R_{c,t}
		-
		\mu_c
		\right)^2
	}.
	$
	%%%%%%%%%%%%%%%%%%%%%%%%%%%%%%%%%%%%%%%%%%%%%%%%%%%%%%%%%%%%

	% \subsection{Model-Level and Portfolio-Level Explainability}
	\subsection{Portfolio performance explainability}
	\label{sec:model_portfolio_explainability}

	This section presents a local first-order attribution-propagation framework for explaining machine-learning-based arbitrage portfolios. The framework characterizes the local sensitivities of sequential portfolio metrics with respect to feature exposures in the allocation function.
	
	The framework consists of two layers. The first attributes the residual allocation weights generated by the machine-learning allocation model to the input features. The second propagates these allocation-level attributions through the portfolio construction and performance evaluation pipeline to economically meaningful portfolio metrics, including daily excess return, mean return, volatility, turnover, gross Sharpe ratio, and net Sharpe ratio. By exploiting the differentiable structure of the portfolio-performance pipeline, the framework efficiently derives feature-level contribution time series for each portfolio metric. It therefore establishes a unified attribution chain from input signals to allocation decisions and ultimately to realised portfolio performance.
	
	The proposed framework offers three key advantages. First, it provides a unified attribution across the entire portfolio-performance pipeline, explicitly revealing how feature contributions propagate from allocation decisions to downstream portfolio metrics. Second, the propagated attributions exhibit strong empirical agreement with the corresponding estimated Shapley values, providing evidence that the proposed local first-order attribution captures the dominant attribution structure of finite-difference attribution. Third, the framework assigns an explicit economic interpretation to feature attribution by linking local feature contributions to feature exposures. For portfolio objectives whose attribution structures remain sufficiently stable over time, these local sensitivities can be used to guide post-training feature-exposure adjustment, potentially improving out-of-sample portfolio performance without retraining the allocation model.

	\subsubsection{Local first-order objective attribution}
	
	Let
	$
	\boldsymbol{\lambda}
	=
	(\lambda_1,\ldots,\lambda_p)^\top
	$
	denote the vector of feature-exposure coefficients. The baseline portfolio is defined by
	$
	\boldsymbol{\lambda}_0=\mathbf{1}_p.
	$
	The coefficients $\lambda_i$ are continuous variables in a neighborhood of $1$. In our setting, the features are cross-sectionally standardised, so $\boldsymbol{\lambda}$ controls the strength of the corresponding trading signals. Given the original feature matrix $
	\boldsymbol{z}
	=
	(z_1,\ldots,z_p)^\top
	$, the exposure-adjusted feature $z_i'$ is
	$
	z_i' = z_i\lambda_i,$ for $
	i=1,\ldots,p.
	$
	
	Let $g_{\boldsymbol\theta}$ be a trained allocation function whose parameters $\boldsymbol\theta$ are held fixed during attribution and feature-exposure calibration. The residual-weight vector is
	$
	\boldsymbol w_{t-1}^\epsilon(\boldsymbol\lambda)
	=
	g_{\boldsymbol\theta}
	\left(
	\boldsymbol\lambda\odot\boldsymbol z_{t-1}
	\right).
	$
	All subsequent portfolio quantities can therefore be represented as functions of the same exposure vector
	$
	Q(\boldsymbol\lambda),
	$
	where $Q$ may denote a component of the residual-weight vector, a tradable stock weight, turnover, a daily portfolio return, mean return, volatility, or the Sharpe ratio.

%	For a trained allocation function $g_{\theta}(\cdot)$ with input features $F$, the model output and the resulting sequential portfolio metrics under different feature exposures can be expressed as functions of $\boldsymbol{\lambda}$, denoted by $Q(\boldsymbol{\lambda})$. Here, $Q$ may represent residual weights, stock weights, portfolio turnover, daily returns, mean returns, volatility, or the Sharpe ratio.
	
	For a specific portfolio metric $Q$, a first-order expansion around the baseline $\boldsymbol{\lambda}_0=\mathbf{1}_p$ gives
	
	\begin{equation}
		Q(\boldsymbol{\lambda})
		=
		Q(\mathbf{1}_p)
		+
		\sum_{i=1}^{p}
		\phi_i(Q)(\lambda_i-1)
		+
		O\!\left(
		\|\boldsymbol{\lambda}-\mathbf{1}_p\|^2
		\right),
		\label{eq:first_order}
	\end{equation}
	
	where
	\[
	Q
	\in
	\left\{
	w^\epsilon_{n,t-1},
	w^R_{n,t-1},
	\tau_t,
	R_{c,t},
	\mu_c,
	\sigma_c,
	SR_c
	\right\},
	\]
	and $c$ denotes the proportional transaction-cost rate.
	
	Hence, $\phi_i(Q)$ is the local slope of the portfolio objective with respect to the exposure applied to feature $i$. We define the local objective attribution of feature $i$ as
	
	\begin{equation}
		\phi_i(Q)
		=
		\left.
		\frac{
			\partial Q(\boldsymbol{\lambda})
		}{
			\partial\lambda_i
		}
		\right|_{\boldsymbol{\lambda}=\mathbf{1}_p}.
		\label{eq:local_attr}
	\end{equation}
	
	The approximation error is determined by the second-order term
	$
	O\!\left(
	\|\boldsymbol{\lambda}-\mathbf{1}_p\|^2
	\right),
	$
	which becomes smaller as $\boldsymbol{\lambda}$ approaches the baseline $\boldsymbol{\lambda}_0=\mathbf{1}_p$.
	
	Equation \ref{eq:first_order} also provides the economic interpretation of the proposed attribution, explicitly connecting feature contribution $\phi_i(Q)$ and the corresponding exposure $\lambda_i$. Holding all other exposures fixed, a small increase in $\lambda_i$ raises $Q$ locally when $\phi_i(Q)>0$, whereas a small reduction in $\lambda_i$ raises $Q$ locally when $\phi_i(Q)<0$. More generally, for a small multivariate intervention $\Delta\boldsymbol\lambda$,
	$
	Q(\mathbf1_p+\Delta\boldsymbol\lambda)
	-
	Q(\mathbf1_p)
	=
	\boldsymbol\phi(Q)^\top
	\Delta\boldsymbol\lambda
	+
	O\!\left(
	\|\Delta\boldsymbol\lambda\|^2
	\right),
	$
	where
	$
	\boldsymbol\phi(Q)
	=
	\left(
	\phi_1(Q),\ldots,\phi_p(Q)
	\right)^\top.
	$
	Thus, any small adjustment satisfying
	$$
	\boldsymbol\phi(Q)^\top
	\Delta\boldsymbol\lambda>0
	$$
	is a local ascent direction for the target metric.
	
	\paragraph{\textbf{Attribution propagation through portfolio construction}}
%	As with Shapley-value estimation, evaluating sufficiently many exposure vectors $\boldsymbol{\lambda}$ to estimate every $\phi_i(Q)$ is computationally expensive. Instead, we efficiently estimate the attribution using the local derivative
%	
%	\[
%	\phi_i(Q)
%	=
%	\left.
%	\frac{
%		\partial Q(\boldsymbol{\lambda})
%	}{
%		\partial\lambda_i
%	}
%	\right|_{\boldsymbol{\lambda}=\mathbf{1}_p}.
%	\]
	
	Based on the model output attribution $\phi_{i}(w^\epsilon_{n,t-1})
	=
	\left.
	\left(
	\frac{
		\partial w^\epsilon_{n,t-1}(\boldsymbol{\lambda})
	}{
		\partial\lambda_i
	}
	\right)
	\right|_{\boldsymbol{\lambda}=\mathbf1_p}$, the local attributions of the sequential portfolio metrics are as follows.
	
%	\[
%	\phi_{i}(w^\epsilon_{n,t})
%	=
%	\left.
%	\left(
%	\frac{
%		\partial w^\epsilon_{n,t}(\boldsymbol{\lambda})
%	}{
%		\partial\lambda_i
%	}
%	\right)
%	\right|_{\boldsymbol{\lambda}=\mathbf1_p},
%	\]
	(1) Tradable stock allocation weights attribution:
	\begin{equation}
	\phi_{i}(w^R_{n,t-1})
	=
	\left.
	\left(
	\frac{
		\partial w^R_{n,t-1}(\boldsymbol{\lambda})
	}{
		\partial\lambda_i
	}
	\right)
	\right|_{\boldsymbol{\lambda}=\mathbf1_p}
	=
	\left.
	\left(
	\frac{
		\partial w^R_{n,t-1}(\boldsymbol{\lambda})
	}{
		\partial \bm w_{t-1}^{\epsilon}(\boldsymbol{\lambda})
	}
	\frac{
		\partial \bm w_{t-1}^{\epsilon}(\boldsymbol{\lambda})
	}{
		\partial\lambda_i
	}
	\right)
	\right|_{\boldsymbol{\lambda}=\mathbf1_p}.
	\end{equation}
	
	(2) Portfolio daily turnover and excess return attribution: For $Q \in \{\tau_t, R_{c,t}\}$,
	\begin{equation}
		\begin{aligned}
			\phi_{i}(Q)
			=
			\left.
			\left(
			\frac{
				\partial Q(\boldsymbol{\lambda})
			}{
				\partial\lambda_i
			}
			\right)
			\right|_{\boldsymbol{\lambda}=\mathbf1_p}
			=
			\sum_n
			\left.
			\left(
			\frac{
				\partial Q(\boldsymbol{\lambda})
			}{
				\partial \boldsymbol{w}^{R}_{n,t}(\boldsymbol{\lambda})
			}
			\frac{
				\partial \boldsymbol{w}^{R}_{n,t}(\boldsymbol{\lambda})
			}{
				\partial\lambda_i
			}
			\right)
			\right|_{\boldsymbol{\lambda}=\mathbf1_p} \triangleq \sum_n 	\phi_{i, n}(Q),
		\end{aligned}
	\end{equation}
	with $\boldsymbol{w}^{R}_{n,t}(\boldsymbol{\lambda})
	:=
	\bigl(
	w^{R}_{n,t-1}(\boldsymbol{\lambda}),
	\,w^{R}_{n,t-2}(\boldsymbol{\lambda})
	\bigr)^{\top}$.
	
	(3) Portfolio mean return, volatility and Sharpe ratio attribution: For $Q \in \{\mu_c, \sigma_c, SR_c\}$,
	\begin{equation}
		\begin{aligned}
			\phi_i(Q)
			&=
			\left.
			\frac{\partial Q(\boldsymbol{\lambda})}
			{\partial\lambda_i}
			\right|_{\boldsymbol{\lambda}=\mathbf1_p}
			=
			\sum_t
			\left.
			\left(
			\frac{\partial Q(\boldsymbol{\lambda})}
			{\partial R_{c,t}(\boldsymbol{\lambda})}
			\frac{\partial R_{c,t}(\boldsymbol{\lambda})}
			{\partial\lambda_i}
			\right)
			\right|_{\boldsymbol{\lambda}=\mathbf1_p} \triangleq \sum_t \phi_{i, t}(Q)
			\\
			&=
			\sum_t\sum_n
			\left.
			\left(
			\frac{\partial Q(\boldsymbol{\lambda})}
			{\partial R_{c,t}(\boldsymbol{\lambda})}
			\frac{\partial R_{c,t}(\boldsymbol{\lambda})}
			{\partial \boldsymbol{w}^{R}_{n,t}(\boldsymbol{\lambda})}
			\frac{\partial \boldsymbol{w}^{R}_{n,t}(\boldsymbol{\lambda})}
			{\partial\lambda_i}
			\right)
			\right|_{\boldsymbol{\lambda}=\mathbf1_p}\triangleq \sum_t\sum_n\phi_{i, n, t}(Q).
		\end{aligned}
	\end{equation}

	As long as the local attribution of the allocation-model output can be obtained, the attributions of downstream portfolio metrics can be propagated efficiently through the chain rule, provided that these metrics are differentiable functions of the model output. Therefore, the proposed local first-order attribution-propagation framework provides a unified attribution of sequential portfolio metrics, from allocation weights to the Sharpe ratio.

	\subsubsection{Relation to Shapley attribution}
	\label{sec:direct_shapley_validation}

	%	The proposed portfolio-level attribution framework is based on first-order propagation of model-level SHAP values. It is therefore not an exact Shapley decomposition of nonlinear portfolio objectives such as the Sharpe ratio. To assess the reliability of the approximation for key performance metrics, we also compute direct Shapley values using gross and net Sharpe ratios as the value functions.
	
	Direct Shapley values measure the realised change in the target portfolio objective when the information contained in a feature is added to a coalition of other features, averaged over all possible coalitions. The Shapley value can be decomposed into the main effect of each feature and the pairwise interaction effects between features. They therefore provide a model-agnostic benchmark for evaluating whether the propagated local first-order attributions identify economically meaningful feature effects. 
	%	In particular, we compare the sign, ranking, and magnitude of the propagated SHAP contributions with the corresponding direct Shapley values.
	
	Although direct Shapley values are theoretically appealing, they are computationally expensive because they require evaluating a large number of feature coalitions. They also provide only an aggregate attribution to the final objective and do not naturally reveal the intermediate economic channels through which a feature affects performance. For example, a direct Shapley value for the net Sharpe ratio does not indicate whether a positive contribution arises from higher gross return, lower volatility, lower turnover, or a combination of these channels. In contrast, the proposed propagation framework produces a complete attribution chain from allocation weights to turnover, return, volatility, and Sharpe-ratio components. It also yields local contribution time series at the stock-time level, which are not readily available from direct Shapley evaluation.
	
	We additionally compute Shapley interaction effects for portfolio objectives such as the Sharpe ratio. The interaction component measures the extent to which the joint contribution of two features differs from the sum of their individual contributions. If interaction effects are small relative to main effects, feature contributions are approximately separable. This provides support for analysing and adjusting feature exposures individually. For different target metrics, the interaction effects can differ even within the same feature universe.
	%	Conversely, large interaction effects indicate that feature-level interventions should be interpreted with caution because the economic contribution of one feature depends strongly on the presence of other features.

%	The first-order propagation framework provides an efficient attribution chain from model outputs to portfolio objectives. To assess the reliability of this approximation for key portfolio metrics, we also compute direct Shapley values for gross and net Sharpe ratios.

	\paragraph{\textbf{Shapley values}}
	\label{sec:attribution_relation_to_shapley}
	
	%	If we take $\boldsymbol{\lambda}=\mathbf{0}_M$, then
	%	
	%	\[
	%	Q(\mathbf{1}_M)
	%	\approx
	%	Q(\mathbf{0}_M)
	%	+
	%	\sum_{i=1}^{M}
	%	\phi_i(Q).
	%	\]
	%	
	%	Taking
	%	$
	%	\psi_0=Q(\mathbf{0}_M)
	%	$
	%	as the reference value, and assuming that the higher-order terms are negligible, we recover an approximation to the additive explanation of SHAP:
	%	
	%	\[
	%	Q(\mathbf{1}_M)
	%	\approx
	%	\psi_0
	%	+
	%	\sum_{i=1}^{M}
	%	\psi_i(Q),
	%	\]
	%	
	%	where $\psi_i(Q)$ denotes the Shapley value of feature $i$ for objective $Q$, a coalition-based feature-removal effect.

	The proposed attribution is defined in a continuous feature-exposure space. It is related but distinct from the coalition-based Shapley value \citep{Shapley1953}.
	
	For a target metric $Q$, define a coalition game
	$
	v_Q(S)
	=
	Q(\boldsymbol\lambda^S),
	$
	where
	\begin{equation}
	\lambda_i^S
	=
	\begin{cases}
		1, & i\in S,\\
		0, & i\notin S.
	\end{cases}
	\end{equation}
	A coalition $S$ represents the subset of features available to the model.
	The Shapley value of feature $i$ is
	\begin{equation}
		\psi_i(Q)
		=
		\sum_{S\subseteq\mathcal P\setminus\{i\}}
		\frac{|S|!(p-|S|-1)!}{p!}
		\left[
		v_Q(S\cup\{i\})-v_Q(S)
		\right].
		\label{eq:shapley}
	\end{equation}
	with 
	$
	\mathcal{P}=\{1,\ldots,p\}
	$
	denote the index set of the input features $\mathbf{z}$. The Shapley value $\psi_i(Q)$ is the average marginal contribution of feature $i$ across coalitions.
%	 i.e. $	\psi_i(Q)
%	=
%	\mathbb{E}_S \big[ v_Q(S\cup\{i\})-v_Q(S) \big].$
	
	If $Q(\mathbf0_p)$ is well defined, Shapley efficiency implies the additivity
	\begin{equation}
	Q(\mathbf1_p)
	=
	Q(\mathbf0_p)
	+
	\sum_{i=1}^{p}
	\psi_i(Q).
	\end{equation}
	In the present setting, $\mathbf0_p$ corresponds to replacing all cross-sectionally standardized features by their zero reference level. 
	
	Applying the local expansion in equation \ref{eq:first_order} at $\boldsymbol\lambda=\mathbf0_p$ gives
	$
	Q(\mathbf0_p)
	=
	Q(\mathbf1_p)
	-
	\sum_{i=1}^{p}
	\phi_i(Q)
	+
	O(p),
	$
	or, more informatively,
	$
	Q(\mathbf1_p)
	\approx
	Q(\mathbf0_p)
	+
	\sum_{i=1}^{p}
	\phi_i(Q).
	$
	This equation is an endpoint first-order extrapolation from $\mathbf1_p$ to $\mathbf0_p$. It resembles the exact Shapley decomposition, but the two are not generally identical. The local derivative
	$
	\phi_i(Q)
	=
	\partial_iQ(\mathbf1_p)
	$
	measures the marginal effect at the all-feature state, whereas
	$
	\psi_i(Q)
	$
	averages finite inclusion effects over all possible coalitions. Curvature and feature interactions can therefore produce differences between the two attributions.
	
	A more precise local connection can be obtained by defining a coalition game that switches exposures between $1-\delta$ and $1$. Let $\psi_i^{(\delta)}(Q)$ be its Shapley value. Under local differentiability,
	$
	\lim_{\delta\rightarrow0}
	\frac{\psi_i^{(\delta)}(Q)}
	{\delta}
	\approx
	\phi_i(Q).
	$
	Thus, the derivative attribution is the continuous local limit of a coalition-based feature-exposure attribution.

	Empirically, the estimated local first-order attribution
	$\widehat{\boldsymbol{\phi}}(Q)$
	and the corresponding estimated Shapley values
	$\widehat{\boldsymbol{\psi}}(Q)$ could be compared using correlation, sign consistency, rank agreement, and approximation error. Strong agreement provides empirical evidence that the local first-order attribution reliably captures the dominant feature-contribution structure of the finite-difference attribution, with relatively small interaction effects. We validate the propagated attribution of portfolio metrics against estimated Shapley values, with particular emphasis on the Sharpe ratio as the target metric.

	\paragraph{\textbf{Shapley interaction effects}}
	
%	We further examine pairwise feature interactions using the Shapley
%	interaction index. 
	For two distinct features $i,j\in\mathcal P$ and a
	coalition
	$
	S\subseteq\mathcal P\setminus\{i,j\},
	$
	the second-order marginal contribution of features $i$ and $j$ to target
	metric $Q$ is defined as
	\begin{equation}
		\delta_{ij}^{Q}(S)
		=
		v_Q(S\cup\{i,j\})
		-
		v_Q(S\cup\{i\})
		-
		v_Q(S\cup\{j\})
		+
		v_Q(S).
		\label{eq:second_order_marginal}
	\end{equation}
	The quantity $\delta_{ij}^{Q}(S)$ measures the incremental contribution
	generated by the joint inclusion of features $i$ and $j$, beyond the sum
	of their individual contributions conditional on coalition $S$.
	The Shapley interaction effect between features $i$ and $j$ is defined as
	\begin{equation}
		\psi_{ij}(Q)
		=
		\sum_{S\subseteq\mathcal P\setminus\{i,j\}}
		\frac{
			|S|!\,(p-|S|-2)!
		}{
			2(p-1)!
		}
		\delta_{ij}^{Q}(S),
		\qquad i\neq j.
		\label{eq:shapley_interaction}
	\end{equation}
%	Equivalently, $\psi_{ij}(Q)
%	=
%	\frac{1}{2}
%	\mathbb{E}_{S}
%	\left[
%		\delta_{ij}^{Q}(S)
%	\right].$
	The factor $1/2$ allocates the pairwise interaction equally between the
	two interacting features. Hence, $\psi_{ij}(Q)=\psi_{ji}(Q)$ represents
	the share of the pairwise interaction attributed to each feature.
	
	Given this definition, the Shapley value of feature $i$ can be decomposed
	as
	\begin{equation}
		\psi_i(Q)
		=
		\psi_i^{\mathrm{main}}(Q)
		+
		\sum_{j\neq i}
		\psi_{ij}(Q),
		\label{eq:shapley_decomposition}
	\end{equation}
	where $\psi_i^{\mathrm{main}}(Q)$
%	\begin{equation}
%		\psi_i^{\mathrm{main}}(Q)
%		=
%		\psi_i(Q)
%		-
%		\sum_{j\neq i}
%		\psi_{ij}(Q)
%		\label{eq:shapley_main_effect}
%	\end{equation}
	denotes the main contribution attributable to feature $i$ after removing its
	allocated pairwise interaction effects.

	\subsubsection{Model-level attribution}
	\label{sec:allocation_level_attribution}

	The proposed attribution-propagation framework is agnostic to the choice of local attribution estimator for the allocation model. For linear allocation models, the local attribution of the model output is given exactly by the product of the estimated model coefficient and the corresponding feature value. Details of the linear model are provided in Appendix~\ref{sec:linear_model}. For more complex allocation models, such as tree-based models and neural networks, the local attribution can be estimated using existing explainability methods. In this paper, following \citet{LundbergEtAl2020} and \citet{LundbergLee2017}, we employ efficient SHAP (SHapley Additive exPlanations) methods, including TreeSHAP for XGBoost (XGB) and DeepSHAP for feedforward neural networks (FFN), to estimate the local attribution of the allocation-model output. The estimated allocation-level attribution is then propagated through the differentiable portfolio-performance pipeline via the chain rule to obtain the attribution of downstream portfolio metrics.
	
%	The attribution propagation framework is agnostic to the choice of local attribution estimator for the allocation models. For linear allocation model, the local attribution of output for each feature is exactly the multiplication of corresponding estimated parameter and given feature values. For complex allocation models, such as tree-based models and neural networks, SHAP (SHapley Additive exPlanation) Values could be used as a unified measure of feature importance. In this paper, we employ efficient SHAP methods, including Tree SHAP for XGBoost and Deep SHAP for FFN, to estimate the local attribution of the allocation-model output, which is subsequently propagated through the portfolio-performance pipeline via the chain rule. 
%	The propagated attribution of portfolio metrics is then validated against its corresponding estimated Shapley values, with particular emphasis on the Sharpe ratio as the target metric.
	
	For asset $n$ at time $t-1$, the allocation model maps the signal vector $z_{n,t-1}\in\mathbb{R}^p$ to a residual arbitrage weight,
	$
		w_{n,t-1}^{\epsilon|\mathcal M}
		=
		g^{\mathcal M}_\theta(z_{n,t-1}),$ with $
		\mathcal M\in\{\mathrm{XGB},\mathrm{FFN}\}.
	$
	The local SHAP decomposition of the residual allocation output is
	\begin{equation}
		w_{n,t-1}^{\epsilon|\mathcal M}
		=
		\psi_{0}^{\mathcal M}(w^\epsilon_{n, t-1})
		+
		\sum_{i=1}^{p}
		\psi_{i}^{\mathcal M}(w^\epsilon_{n, t-1}),
	\end{equation}
	where $\psi_{i}^{\mathcal M}(w^\epsilon_{n, t-1})$ denotes the contribution of feature $i$ to the residual allocation decision under model $\mathcal M$. A positive value indicates that the information provided by feature $i$ increases the residual allocation weight, whereas a negative value indicates that it decreases the residual allocation weight.

%	In the additive explanation representation, the model output can be written as
%	\begin{equation}
%		w_{n,t-1}^{\epsilon}(z)
%		\approx
%		\phi_{0,n,t}
%		+
%		\sum_{i=1}^{M}
%		\phi_{i,n,t}(w^\epsilon)z_i,
%	\end{equation}
%	where $z_i\in\{0,1\}$ indicates whether feature $i$ is included in the coalition, a subset of features whose information is available to the model. The SHAP value $\phi_{i,n,t}(w^\epsilon)$ is the average marginal contribution of feature $i$ across coalitions.
	
	We use $\psi_{i}^{\mathcal M}(w^\epsilon_{n, t-1})$ to estimate $\phi_{i}^{\mathcal M}(w^\epsilon_{n, t-1})$. The global allocation-level contribution of feature $i$ can be aggregated along the cross-sectional dimension and time dimension:
	$
	\phi_i^{\mathcal M}(w^\epsilon)
	=
	\sum_{t=1}^{T}
	\phi_{i}^{\mathcal M}(w^\epsilon_{t-1})
	=
	\sum_{t=1}^{T}
	\sum_{n=1}^{N}
	\phi_{i}^{\mathcal M}(w^\epsilon_{n, t-1}).
	$
%	$
%		\phi_{i,n}(w^\epsilon)
%		=
%		\sum_{t=1}^{T}
%		\phi_{i,n,t}(w^\epsilon), 
%	$
%	$
%		\phi_{i,t}(w^\epsilon)
%		=
%		\sum_{n=1}^{N}
%		\phi_{i,n,t}(w^\epsilon),
%	$
%	and 
%	$
%		\phi_i(w^\epsilon)
%		=
%		\sum_{t=1}^{T}
%		\sum_{n=1}^{N}
%		\phi_{i}(w^\epsilon_{n, t-1}).
%	$
	Global importance based on absolute local attributions could be given by 
	$
		\mathrm{Imp}_i^{\mathcal M}(w^\epsilon)
		=
		\mathbb{E}_{n,t-1}
		\left[
		\left|
		\phi_{i}^{\mathcal M}(w^\epsilon_{n, t-1})
		\right|
		\right].
	$
	
	This allocation-level layer answers which features the model uses to determine long or short residual positions. It does not, by itself, determine whether those allocation decisions improve portfolio performance. This motivates the portfolio-level attribution layer described below.
	More generally, the framework also applies to models that predict stock returns when portfolio weights are a differentiable function of the predicted returns, allowing feature attributions to be propagated from return predictions through portfolio allocation to portfolio-level objectives.
	
	\subsubsection{Portfolio-level attribution}
	\label{sec:portfolio_level_attribution}
	
	Across different types of allocation models $\mathcal M$, the propagation from model-level to portfolio-level attribution follows the same framework, covering portfolio turnover, gross and net daily excess returns, mean returns, volatility, and Sharpe ratios.
	
	\paragraph{\textbf{Tradable stock allocation weights attribution.}}
	
	The residual allocation vector $\bm w_{t-1}^{\epsilon}$ is transformed into stock-level portfolio weights through the residual-to-stock mapping matrix $\Phi_{t-1}$. Define the unnormalised stock exposure vector
	$
	\tilde{\bm w}_{t-1}^{R} = 
	{\bm w_{t-1}^{\epsilon}}^{\top} \Phi_{t-1},
	$
	and the normalised realised stock portfolio weights
	$
	\bm w_{t-1}^{R}
	=
	\frac{
		\tilde{\bm w}_{t-1}^{R}
	}{
		\|\tilde{\bm w}_{t-1}^{R}\|_1
	}.
	$
	The $\ell_1$ normalisation imposes a leverage constraint.
	
	Because the mapping from residual weights to unnormalised stock exposures is linear, the allocation-level attribution propagates as
	$
	\bm \phi_i(\tilde{\bm w}_{t-1}^{R})
	=
	{\bm \phi_i(\bm w_{t-1}^{\epsilon})}^{\top} \Phi_{t-1},
	$
	where $\bm \phi_i(\bm w_{t-1}^{\epsilon})$ is the vector of feature-$i$ attributions to residual weights across assets.
	The normalisation step is nonlinear. We therefore use a first-order approximation. For stock $n$,
	\begin{equation}
		\phi_i(w_{n,t-1}^{R})
		=
		\frac{
			\phi_i(\tilde{w}^{R}_{n,t-1})\| \tilde{\bm w}^{R}_{t-1}\|_1
			-
			\tilde{w}^{R}_{n,t-1}
			\sum_{m=1}^{N}
			\operatorname{sign}(\tilde{w}^{R}_{m,t-1})
			\phi_i(\tilde{w}^{R}_{m,t-1})
		}{
			\|\tilde{\bm w}^{R}_{t-1}\|_1^2
		},
		\label{eq:weight_shap_normalization}
	\end{equation}
	where $ \tilde{\bm w}^{R}_{t-1}$ is from baseline portfolio under $\boldsymbol{\lambda}=\mathbf1_p$.

	\paragraph{\textbf{Portfolio turnover attribution.}}
	
	We incorporate transaction costs through proportional turnover costs. For stock $n$, turnover is
	$
	\tau_{n,t}
	=
	\left|
	w_{n,t-1}^{R}
	-
	w_{n,t-2}^{R}
	\right|,
	$
	and aggregate portfolio turnover is
	$
	\tau_t
	=
	\sum_{n=1}^{N}
	\tau_{n,t}
	=
	\left\|
	\bm w_{t-1}^{R}
	-
	\bm w_{t-2}^{R}
	\right\|_1.
	$
	The transaction cost at time $t$ is
	$
	\mathrm{TC}_t
	=
	c\tau_t,
	$
	where $c$ is the proportional transaction-cost rate. 
	%	In the baseline empirical specification, $c=0.0005$.
	
	%	\paragraph{Turnover attribution.}
	
	%	Using a first-order approximation to the absolute-value function, the feature contribution to stock-level turnover is
	%	\begin{equation}
		%		\phi_i(\tau_{n,t})
		%		=
		%		\operatorname{sign}
		%		\left(
		%		w_{n,t-1}^{R}
		%		-
		%		w_{n,t-2}^{R}
		%		\right)
		%		\left[
		%		\phi_i(w_{n,t-1}^{R})
		%		-
		%		\phi_i(w_{n,t-2}^{R})
		%		\right].
		%	\end{equation}
	The portfolio-level turnover attribution is
	\begin{equation}
		\phi_i(\tau_t)
		%		=
		%		\sum_{n=1}^{N}
		%		\phi_i(\tau_{n,t})
		= 
		\sum_{n=1}^{N}
		\operatorname{sign}
		\left(
		w_{n,t-1}^{R}
		-
		w_{n,t-2}^{R}
		\right)
		\left[
		\phi_i(w_{n,t-1}^{R})
		-
		\phi_i(w_{n,t-2}^{R})
		\right].
	\end{equation}
	Equivalently, $\phi_i(\tau_t)
	=
	\sum_{n=1}^{N}
	\phi_i(\tau_{n,t})$, 
	where $\phi_i(\tau_{n,t})$ denotes the stock-level turnover attribution. 
	The contribution to average turnover is
	$
	\phi_i(\bar{\tau})
	=
	\frac{1}{T}
	\sum_{t=1}^{T}
	\phi_i(\tau_t).
	$

	\paragraph{\textbf{Portfolio daily excess return attribution.}}
	
	Let $R_{n,t}$ denote the excess return of stock $n$ at time $t$. The realised portfolio gross excess return is
	$
	R_{0,t}
	=
	\sum_{n=1}^{N}
	w_{n,t-1}^{R}R_{n,t}.
	$
	Using linearity in portfolio weights, the feature contribution to daily portfolio gross excess return is
	\begin{equation}
		\phi_i(R_{0,t})
		=
		\sum_{n=1}^{N}
		\phi_i(w_{n,t-1}^{R})R_{n,t}.
	\end{equation}
	%	Equivalently, $\phi_i(R_{0,t})
	%	=
	%	\sum_{n=1}^{N}
	%	\phi_{i,n}(R_{0, t})$, 
	%	where $\phi_{i,n}(R_{0, t})$ denotes the stock-level attribution. 
	%	Equivalently, the stock-level contribution is
	%	$
	%		\phi_{i,n}(R_{0, t})
	%		=
	%		\phi_i(w_{n,t-1}^{R})R_{n,t},
	%	$ and $
	%		\phi_i(R_{0,t})
	%		=
	%		\sum_{n=1}^{N}
	%		\phi_{i,n}(R_{0, t}).
	%	$
	
	%	\paragraph{Daily net return attribution.}
	
	Assuming a transaction cost of $c$ per unit of turnover, the daily net excess return is
	$
	R_{c,t}
	=
	R_{0,t}
	-
	c\tau_t.
	$
	Therefore, the feature contribution to daily net excess return is
	\begin{equation}
		\phi_i(R_{c,t})
		=
		\phi_i(R_{0,t})
		-
		c\phi_i(\tau_t).
	\end{equation}
	%	Equivalently, $\phi_i(R_{c,t})
	%	=
	%	\sum_{n=1}^{N}
	%	\phi_{i,n}(R_{c, t})$,
	%	where 
	%	$
	%		\phi_{i,n}(R_{c, t})
	%		=
	%		\phi_{i,n}(R_{0, t})
	%		-
	%		c\phi_i(\tau_{n,t})
	%	$
	%	at the stock level.
	
	Equivalently, $\phi_i(R_{c,t})
	=
	\sum_{n=1}^{N}
	\phi_{i,n}(R_{c, t})$ with $c\ge0$,
	where 
	$
	\phi_{i,n}(R_{c, t})
	=
	\phi_{i,n}(R_{0, t})
	-
	c\phi_i(\tau_{n,t})
	$
	at the stock level.

	\paragraph{\textbf{Mean return attribution.}}
	
	Define the sample gross mean return over the evaluation window as
	$
	\mu_0
	=
	\frac{1}{T}
	\sum_{t=1}^{T}
	R_{0,t}.
	$
	The feature contribution to gross mean return is
	\begin{equation}
		\phi_i(\mu_0)
		=
		\frac{1}{T}
		\sum_{t=1}^{T}
		\phi_i(R_{0,t}).
	\end{equation}
	%	At the local stock-time level,
	%	$
	%		\phi_{i,n,t}(\mu_0)
	%		=
	%		\frac{1}{T}
	%		\phi_{i,n}(R_{0, t}).
	%	$
	%	We could write the cross-sectional and time-series contribution to portfolio mean return as
	%	$\phi_{i,n}(\mu)= \sum_{t=1}^T \phi_{i,n,t}(\mu) $ and $\phi_{i,t}(\mu)= \sum_{n=1}^N \phi_{i,n,t}(\mu) =  \frac{1}{T}\phi_{i}(R_{p,t})$.
	%	The annualised mean-return attribution is $252\,\phi_i(\mu)$.

	%	\paragraph{Net mean return attribution.}
	
	The average net return is
	$
	\mu_c
	=
	\frac{1}{T}
	\sum_{t=1}^{T}
	R_{c,t}
	=
	\mu_0
	-
	c\bar{\tau}.
	$
	Hence, the feature contribution to net mean return is
	\begin{equation}
		\phi_i(\mu_c)
		=
		\phi_i(\mu_0)
		-
		c\phi_i(\bar{\tau}).
	\end{equation}
	%	At the local level,
	%	$
	%		\phi_{i,n,t}(\mu_c)
	%		=
	%		\phi_{i,n,t}(\mu_0)
	%		-
	%		\frac{c}{T}
	%		\phi_i(\tau_{n,t}).
	%	$
	
	The portfolio net mean-return attribution of feature $i$ can be written as
	$
	\phi_i(\mu_c)
	\equiv
	\sum_{t=1}^{T}\phi_{i,t}(\mu_c)
	\equiv
	\sum_{t=1}^{T}\sum_{n=1}^{N}\phi_{i,n,t}(\mu_c)
	$ with $ c\geq 0,
	$
	where the stock-time-level attribution is
	$
	\phi_{i,n,t}(\mu_c)
	=
	\frac{1}{T}\phi_{i,n}(R_{c,t}).
	$
	%	and the corresponding time-series attribution is
	%	$
	%	\phi_{i,t}(\mu_c)
	%	=
	%	\frac{1}{T}\phi_i(R_{c,t}).
	%	$ 
	The annualised net mean-return attribution is $252\,\phi_i(\mu_c)$.
	
	We can clearly see that the feature contribution to the portfolio net mean return can be expressed as its contribution to the gross mean return, adjusted for the contribution of transaction costs through the average portfolio turnover over the trading period.
	
	%	Portfolio net mean return attribution of feature $i$ could be written as
	%	$
	%	\phi_i(\mu_c)
	%	\;\equiv\; \sum_{n=1}^N\phi_{i,n}(\mu_c) \;\equiv\; 
	%	\sum_{t=1}^T\phi_{i,t}(\mu_c) \;\equiv\; 
	%	\sum_{t=1}^T \sum_{n=1}^N \phi_{i,n,t}(\mu_c),
	%	$
	%	with time-series contribution
	%	$\phi_{i,t}(\mu_c)= \sum_{n=1}^N \phi_{i,n,t}(\mu_c) =  \frac{1}{T}\phi_{i}(R_{c,t}) = \frac{1}{T} \phi_{i}(R_{0,t})  - c\cdot \frac{1}{T} \phi_i(\tau_{t}) =\phi_{i,t}(\mu_c) - \frac{c}{T} \phi_i(\tau_{t})$.

	%	$\phi_{i,t}(\mu_c) =\phi_{i,t}(\mu_0) - \frac{c}{T} \phi_i(\tau_{t})$ and 
	%	$
	%		\phi_{i,n,t}(\mu_c)
	%		=
	%		\phi_{i,n,t}(\mu_0)
	%		-
	%		\frac{c}{T}
	%		\phi_i(\tau_{n,t}).
	%	$

	\paragraph{\textbf{Volatility attribution.}}
	
	%	Let
	%	$
	%	\sigma^2
	%	=
	%	\frac{1}{T}
	%	\sum_{t=1}^{T}
	%	(R_{p,t}-\mu)^2
	%	$
	%	and
	%	$
	%	\sigma=\sqrt{\sigma^2}.
	%	$
	%	Using a first-order delta-method approximation,
	%	\begin{equation}
		%		\phi_i(\sigma^2)
		%		\approx
		%		\frac{1}{T}
		%		\sum_{t=1}^{T}
		%		2(R_{p,t}-\mu)
		%		\left[
		%		\phi_i(R_{p,t})
		%		-
		%		\phi_i(\mu)
		%		\right].
		%	\end{equation}
	%	At the stock-time level,
	%	\begin{equation}
		%		\phi_{i,n,t}(\sigma^2)
		%		=
		%		\frac{2}{T}
		%		(R_{p,t}-\mu)
		%		\left[
		%		\phi_{i,n,t}(R_p)
		%		-
		%		\phi_{i,n}(\mu)
		%		\right].
		%	\end{equation}
	%	The volatility attribution is then
	%	\begin{equation}
		%		\phi_i(\sigma)
		%		\approx
		%		\frac{1}{2\sigma}
		%		\phi_i(\sigma^2),
		%	\end{equation}
	%	with local contribution
	%	\begin{equation}
		%		\phi_{i,n,t}(\sigma)
		%		=
		%		\frac{1}{2\sigma}
		%		\phi_{i,n,t}(\sigma^2).
		%	\end{equation}
	%	We could write the cross-sectional and time-series contribution to portfolio volatility as
	%	$\phi_{i,n}(\sigma)= \sum_{t=1}^T \phi_{i,n,t}(\sigma) $ and $\phi_{i,t}(\sigma)= \sum_{n=1}^N \phi_{i,n,t}(\sigma) =  \frac{1}{\sigma T} (R_{p,t}-\mu)\left(\phi_{i}(R_{p,t})-\phi_{i}(\mu)\right)$.
	%	The annualised volatility attribution is $\sqrt{252}\,\phi_i(\sigma)$.
	%	
	%	
	%	
	%	\paragraph{Net volatility attribution.}
	
	Define portfolio volatility
	$
	\sigma_c
	=
	\sqrt{
		\frac{1}{T}
		\sum_{t=1}^{T}
		\left(
		R_{c,t}
		-
		\mu_c
		\right)^2}.
	$
	The volatility attribution is
	\begin{equation}
		\phi_i\!\left(\sigma_c\right)
		=
		\frac{1}{T \sigma_c}
		\sum_{t=1}^{T}
		\left(
		R_{c,t}
		-
		\mu_c
		\right)
		\left[
		\phi_i(R_{c,t})
		-
		\phi_i(\mu_c)
		\right].
	\end{equation}
	
	The portfolio net volatility attribution of feature $i$ can be written as
	$
	\phi_i(\sigma_c)
	\equiv
	\sum_{t=1}^{T}\phi_{i,t}(\sigma_c)
	\equiv
	\sum_{t=1}^{T}\sum_{n=1}^{N}\phi_{i,n,t}(\sigma_c)
	$ with $ c\geq 0,
	$
	where the stock-time-level attribution is
	$
	\phi_{i,n,t}(\sigma_c)
	=
	\frac{1}{T \sigma_c}
	\left(
	R_{c,t}
	-
	\mu_c
	\right)
	\left[
	\phi_{i,n}(R_{c, t})
	-
	\phi_{i,n}(\mu_c)
	\right].
	$
	%	and the corresponding time-series attribution is
	%	$
	%	\phi_{i,t}(\sigma_c)=\frac{1}{T \sigma_c}(R_{c,t}-\mu_c)\left[\phi_{i}(R_{c,t})-\phi_{i}(\mu_c)\right].
	%	$ 
	The annualised net volatility attribution is
	$\sqrt{252}\,\phi_i(\sigma_c)$.
	
	%	Define
	%	$
	%		\sigma_c^2
	%		=
	%		\frac{1}{T}
	%		\sum_{t=1}^{T}
	%		\left(
	%		R_{c,t}
	%		-
	%		\mu_c
	%		\right)^2.
	%	$
	%	The delta-method attribution is
	%	\begin{equation}
		%		\phi_i\!\left(\sigma_c^2\right)
		%		\approx
		%		\frac{1}{T}
		%		\sum_{t=1}^{T}
		%		2
		%		\left(
		%		R_{c,t}
		%		-
		%		\mu_c
		%		\right)
		%		\left[
		%		\phi_i(R_{c,t})
		%		-
		%		\phi_i(\mu_c)
		%		\right].
		%	\end{equation}
	%	The corresponding volatility attribution is
	%	\begin{equation}
		%		\phi_i(\sigma_c)
		%		\approx
		%		\frac{1}{2\sigma_c}
		%		\phi_i\!\left(\sigma_c^2\right).
		%	\end{equation}
	%	At the local stock-time level,
	%	$
	%		\phi_{i,n,t}(\sigma_c)
	%		=
	%		\frac{1}{2\sigma_c}
	%		\frac{2}{T}
	%		\left(
	%		R_{c,t}
	%		-
	%		\mu_c
	%		\right)
	%		\left[
	%		\phi_{i,n}(R_{c, t})
	%		-
	%		\phi_{i,n}(\mu_c)
	%		\right].
	%	$
	%	The annualised net volatility attribution is
	%	$\sqrt{252}\,\phi_i(\sigma_c)$.
	
	Equivalently, since $R_{c,t}=R_{0,t}-c\tau_t$, net variance attribution can be decomposed as
	$
	\phi_i(\sigma_c^2)
	=
	\phi_i(\sigma_0^2)
	+
	c^2\phi_i(\sigma_{\tau}^2)
	-
	2c\,\phi_i(\mathrm{Cov}(R_{0,t},\tau_t)),
	$
	where $\sigma_{\tau}^2$ is the variance of portfolio turnover. This decomposition shows that transaction costs affect net return variance through both the variability of turnover and its covariance with gross portfolio returns.
	
	%	Portfolio net volatility of feature $i$ could be written as
	%	$
	%	\phi_i\!\left(\sigma_c\right)
	%	\;\equiv\; \sum_{n=1}^N\phi_{i,n}\left(\sigma_c\right) \;\equiv\; 
	%	\sum_{t=1}^T\phi_{i,t}\left(\sigma_c\right) \;\equiv\; 
	%	\sum_{t=1}^T \sum_{n=1}^N \phi_{i,n,t}\left(\sigma_c\right),
	%	$
	%	with time-series contribution $\phi_{i,t}(\sigma_c)= \sum_{n=1}^N \phi_{i,n,t}(\sigma_c) =  \frac{1}{\sigma_c T} (R_{c,t}-\mu_c)\left(\phi_{i}(R_{c,t})-\phi_{i}(\mu_c)\right)$.
	
	%	\paragraph{Gross Sharpe ratio attribution.}
	%	
	%	The annualised gross Sharpe ratio is
	%	$
	%	SR
	%	=
	%	\sqrt{252}
	%	\frac{\mu}{\sigma}.
	%	$
	%	Applying a first-order approximation gives
	%	\begin{equation}
		%		\phi_i(SR)
		%		\approx
		%		\sqrt{252}
		%		\frac{
			%			\phi_i(\mu)\sigma
			%			-
			%			\mu\phi_i(\sigma)
			%		}{
			%			\sigma^2
			%		}.
		%		\label{eq:gross_sr_attr}
		%	\end{equation}
	%	At the stock-time level,
	%	\begin{equation}
		%		\phi_{i,n,t}(SR)
		%		=
		%		\sqrt{252}
		%		\frac{
			%			\phi_{i,n,t}(\mu)\sigma
			%			-
			%			\mu\phi_{i,n,t}(\sigma)
			%		}{
			%			\sigma^2
			%		}.
		%	\end{equation}
	%	We could write the cross-sectional and time-series contribution to gross Sharpe ratio as
	%	$\phi_{i,n}(SR)= \sum_{t=1}^T \phi_{i,n,t}(SR) = \sqrt{252}\frac{\phi_{i,n}(\mu)\sigma-\mu\,\phi_{i,n}(\sigma)}{\sigma^2}$ and $\phi_{i,t}(SR)= \sum_{n=1}^N \phi_{i,n,t}(SR) =  \sqrt{252}\frac{\phi_{i,t}(\mu)\sigma-\mu\,\phi_{i,t}(\sigma)}{\sigma^2}$.
	%	
	%	This decomposition shows that a feature improves the Sharpe ratio when it increases mean return, reduces volatility, or both.
	
	%	\paragraph{Net Sharpe ratio attribution.}
	\paragraph{\textbf{Sharpe ratio attribution.}}
	
	The annualised Sharpe ratio is
	$
	SR_c
	=
	\sqrt{252}
	\frac{
		\mu_c
	}{
		\sigma_c
	}.
	$
	The first-order attribution is
	\begin{equation}
		\phi_i(SR_c)
		\approx
		\sqrt{252}
		\frac{
			\phi_i(\mu_c)\sigma_c
			-
			\mu_c\phi_i(\sigma_c)
		}{
			\sigma_c^2
		}.
		\label{eq:net_sr_attr}
	\end{equation}
	with $\mu_c$ and $\sigma_c$ given by baseline portfolio under $\boldsymbol{\lambda}=\mathbf1_p$.
	
	The annualised portfolio Sharpe-ratio attribution of feature $i$ can be
	$
	\phi_i(SR_c)
	\equiv
	\sum_{t=1}^{T}\phi_{i,t}(SR_c)
	\equiv
	\sum_{t=1}^{T}\sum_{n=1}^{N}\phi_{i,n,t}(SR_c)
	$ with $ c\geq 0.
	$
	At the local stock-time level,
	$
	\phi_{i,n,t}(SR_c)
	=
	\sqrt{252}
	\frac{
		\phi_{i,n,t}(\mu_c)\sigma_c
		-
		\mu_c\phi_{i,n,t}(\sigma_c)
	}{
		\sigma_c^2
	}.
	$
	The corresponding cross-sectional and time-series attributions can be obtained
	by aggregation as $\phi_{i,n}(SR_c)= \sum_{t=1}^T \phi_{i,n,t}(SR_c)$ and $\phi_{i,t}(SR_c)= \sum_{n=1}^N \phi_{i,n,t}(SR_c) $ respectively. 
	This attribution shows that a feature improves the Sharpe ratio when it increases mean return, reduces volatility, or both.
	
	%	We could get the cross-sectional and time-series contribution to Sharpe ratio as
	%	$\phi_{i,n}(SR_c)= \sum_{t=1}^T \phi_{i,n,t}(SR_c)$ and $\phi_{i,t}(SR_c)= \sum_{n=1}^N \phi_{i,n,t}(SR_c) $. This attribution shows that a feature improves the Sharpe ratio when it increases mean return, reduces volatility, or both.

	\paragraph{\textbf{Interpretation.}}
	
	The portfolio-level attribution layer addresses a different question from
	allocation-level attribution. While allocation-level attribution identifies
	the features that drive long or short residual positions, portfolio-level
	attribution identifies which features contribute to realised economic
	performance and through which channels. Specifically, the decomposition
	separates each feature's contribution to mean return, volatility, turnover, and gross and net Sharpe ratios. This distinction is important because a feature that strongly influences allocation weights does not necessarily improve risk-adjusted portfolio performance. A feature may also increase gross returns while reducing the net Sharpe ratio if it generates higher turnover or volatility.
	
	For a portfolio metric $Q$ with allocation model $\mathcal{M}$,
	the proposed framework provides feature attributions at both the
	stock-time and aggregated levels. The aggregate contribution of feature
	$i$ is obtained by summing its stock-time-level contributions:
	\begin{equation}
		\phi_i^{\mathcal{M}}(Q)
		=
		\sum_{t=1}^{T}
		\sum_{n=1}^{N}
		\phi_{i,n,t}^{\mathcal{M}}(Q).
	\end{equation}
	The framework therefore provides both cross-sectional and time-series
	decompositions of portfolio performance attribution. Provided that the allocation
	function admits a model-level attribution method, such as Tree SHAP for
	gradient-boosted trees or Deep SHAP for neural networks, the same
	first-order propagation mechanism can be applied to obtain portfolio-level
	economic attributions.

	\subsubsection{Statistical inference for local Sharpe attribution}
	
	The proposed local Sharpe attribution is estimated conditional on the fitted allocation model and the corresponding allocation-attribution rule. Under this conditional framework, uncertainty arises solely from the finite out-of-sample return path and, for net returns, the associated transaction-cost path.
	
	For portfolio objective $SR_c$, define the population moments
	\begin{equation}
		\mu_c
		=
		E[R_{c,t}],
		\quad
		\mu_{c,i}
		=
		E[\phi_i(R_{c,t})],
		\quad
		\gamma_c
		=
		E[(R_{c,t})^2],
		\quad
		\gamma_{c,i}
		=
		E[R_{c,t}\phi_i(R_{c,t})].
	\end{equation}
	
	Let
	$
		v_{c, i}
		=
		(\mu_c,\mu_{c,i},\gamma_c,\gamma_{c,i})^\top.
	$
	Then the local Sharpe-ratio attribution can be written as
	\begin{equation}
		\phi_i(SR_c) = f(v_{c, i})
		=
		\sqrt{252}
		\left[
		\frac{\mu_{c,i}}
		{\sqrt{\gamma_c-\mu_c^2}}
		-
		\frac{
			\mu_c
			(\gamma_{c,i}-\mu_c\mu_{c,i})
		}{
			(\gamma_c-\mu_c^2)^{3/2}
		}
		\right].
	\end{equation}
	
	Applying the multivariate Delta method yields the standard error of $\widehat{\phi}_i(SR_c)$
	
	\begin{equation}
			s
			\!\left(
			\widehat{\phi}_i(SR_c)
			\right)
			=
			\sqrt{
				\frac{
					\nabla f(\widehat{v}_{c, i})^\top
					\widehat{\Psi}_{c, i}
					\nabla f(\widehat{v}_{c, i})
				}{T}
			},
	\end{equation}
	
	where
	\(\widehat{\Psi}_{c, i}\)
	is the long-run covariance matrix of
	\((\widehat{\mu}_c,\widehat{\mu}_{c,i},\widehat{\gamma}_c,\widehat{\gamma}_{c,i})^\top\),
	accounting for serial dependence in the daily portfolio returns. 
%	The complete derivation, together with the construction of
%	\(\widehat{\Psi}_{c, i}\),
%	is provided in Appendix.
	
%	To conduct finite-sample inference, we employ a circular block bootstrap. Each bootstrap sample preserves the serial and cross-feature dependence by resampling common blocks of
%	
%	\[
%	(R_{c,t},
%	\phi_1(R_{c,t}),
%	\ldots,
%	\phi_p(R_{c,t})).
%	\]
%	
%	For each bootstrap replication, both the attribution estimator and its Delta-method standard error are recomputed. Confidence intervals are constructed using the studentized percentile-\(t\) bootstrap, and two-sided bootstrap \(p\)-values are reported for testing
%	
%	\[
%	H_0:\phi_i(SR_c)=0.
%	\]

	\paragraph{\textbf{Block long-run covariance estimator}}
	Following the block dependence structure of \citet{GotzeKunsch1996}, we estimate the long-run covariance matrix of the estimating moments using normalized block sums. Define the centered moment vector
	$
	V_{c,i,t}
	=
	\bigl(
	R_{c,t}-\widehat{\mu}_c,\;
	\phi_i(R_{c,t})-\widehat{\mu}_{c,i},\;
	(R_{c,t})^2-\widehat{\gamma}_c,\;
	R_{c,t}\phi_i(R_{c,t})-\widehat{\gamma}_{c,i}
	\bigr)^\top
	$ for $t = 1, ..., T$.
	
	For a block length \(b>1\), let the number of blocks be
	$
	\ell
	=
	\left\lfloor
	\frac{T}{b}
	\right\rfloor.
	$
	Within each block, define
	\begin{equation}
		\zeta_{c,i,j}
		=
		\frac{1}{\sqrt{b}}
		\sum_{t=(j-1)b+1}^{jb}
		V_{c,i,t},
		\qquad
		j=1,\ldots,\ell.
		\label{eq:block_sum}
	\end{equation}
	
%	Let
%	$
%	\bar{\zeta}_{c,i}
%	=
%	\frac{1}{\ell}
%	\sum_{j=1}^{\ell}
%	\zeta_{c,i,j}.
%	$
	The block long-run covariance estimator is
%	\begin{equation}
%		\widehat{\Psi}_{c,i}
%		=
%		\frac{1}{\ell}
%		\sum_{j=1}^{\ell}
%		\left(
%		\zeta_{c,i,j}
%		-
%		\bar{\zeta}_{c,i}
%		\right)
%		\left(
%		\zeta_{c,i,j}
%		-
%		\bar{\zeta}_{c,i}
%		\right)^\top.
%		\label{eq:block_cov}
%	\end{equation}
	\begin{equation}
	\widehat{\Psi}_{c,i}
	=
	\frac{1}{\ell}
	\sum_{j=1}^{\ell}
	\zeta_{c,i,j}
	\zeta_{c,i,j}^\top.
	\label{eq:block_cov}
	\end{equation}
	
	\paragraph{\textbf{Circular block bootstrap}}
%	To conduct finite-sample inference, we employ a circular block bootstrap. Each bootstrap sample preserves the serial and cross-feature dependence by resampling common blocks of
%	
%	\[
%	(R_{c,t},
%	\phi_1(R_{c,t}),
%	\ldots,
%	\phi_p(R_{c,t})).
%	\]
%	
%	For each bootstrap replication, both the attribution estimator and its Delta-method standard error are recomputed. Confidence intervals are constructed using the studentized percentile-\(t\) bootstrap, and two-sided bootstrap \(p\)-values are reported for testing
%	
%	\[
%	H_0:\phi_i(SR_c)=0.
%	\]
	
	To conduct finite-sample inference, we employ a circular block bootstrap. For joint inference, define
	$
		Y_{c,t}
		=
		\left(
		R_{c,t},
		\phi_1(R_{c,t}),
		\ldots,
		\phi_p(R_{c,t})
		\right)^\top,
		\label{eq:bootstrap_vector}
	$
	where \(p\) is the number of input features. For each bootstrap replication \(q=1,\ldots,B\), circular blocks of length \(b\) are sampled with replacement from
	\(\{Y_{c,t}\}_{t=1}^{T}\), concatenated, and truncated to length \(T\). All feature attributions are resampled using the same block indices to preserve both cross-feature and serial dependence.
	
	For each bootstrap sample, the moment estimates, attribution estimator, block covariance matrix, and Delta-method standard error are recomputed:
	$
		\widehat{\phi}_i^{*,q}(SR_c)
		=
		f\!\left(
		\widehat{v}_{c,i}^{*,q}
		\right)$ and $
		s
		\!\left(
		\widehat{\phi}_i^{*,q}(SR_c)
		\right)
		=
		\sqrt{
			\frac{
				\nabla f
				\!\left(
				\widehat{v}_{c,i}^{*,q}
				\right)^\top
				\widehat{\Psi}_{c,i}^{*,q}
				\nabla f
				\!\left(
				\widehat{v}_{c,i}^{*,q}
				\right)
			}{T}
		}.
		\label{eq:bootstrap_phi_se}
	$
	The centered studentized bootstrap root is
	$
		Z_{c,i}^{*, q}
		=
		\frac{
			\widehat{\phi}_i^{*,q}(SR_c)
			-
			\widehat{\phi}_i(SR_c)
		}{
			s
			\!\left(
			\widehat{\phi}_i^{*,q}(SR_c)
			\right)
		}.
		\label{eq:studentized_root}
	$
	Let \(q_{\tau}^{*}\) denote the empirical \(\tau\)-quantile of the studentized bootstrap roots. A two-sided percentile-\(t\) confidence interval for \(\phi_i(SR_c)\) is given by
	\[
		\left[
		\widehat{\phi}_i(SR_c)
		-
		q_{1-\alpha/2}^{*}
		s
		\!\left(
		\widehat{\phi}_i(SR_c)
		\right),
		\;
		\widehat{\phi}_i(SR_c)
		-
		q_{\alpha/2}^{*}
		s
		\!\left(
		\widehat{\phi}_i(SR_c)
		\right)
		\right].
		\label{eq:bootstrap_ci}
	\]
	The corresponding two-sided hypothesis
	$
	H_0:\;
	\phi_i(SR_c)=0
	$
	can then be tested using the studentized bootstrap distribution.

	\subsection{Explainability-driven feature exposure intervention}
	\label{sec:feature_exposure_intervention}
	
	The attribution framework can also be used to guide out-of-sample feature exposure interventions. Throughout the paper, feature intervention refers to changing the informational exposure of a feature. Since input features are standardised, setting a feature to zero corresponds to its average or signal-neutral state. The resulting intervention therefore measures the economic value of feature information relative to a neutral benchmark.
	
	Traditional Shapley values and SHAP explanations compare the effect of including versus excluding feature information. Feature exposure intervention generalises this idea by allowing feature information to be scaled continuously. A feature can be partially suppressed, left unchanged, or amplified before being passed into the trained allocation model. Importantly, the model parameters are not retrained. The intervention therefore isolates the portfolio effect of changing feature exposure while holding the learned allocation rule fixed.
	
%	The proposed intervention framework is based on two key considerations. 
%%	First, propagated feature contributions should be informative about the realised change in portfolio performance when feature exposures are adjusted, regardless of whether exposures are reduced or amplified. 
%	First, within the given feature universe, the main effects of individual features are expected to dominate their interaction effects when feature exposures are modified separately. Second, feature contributions should exhibit sufficient temporal persistence such that lagged contribution estimates remain informative for future trading periods. These properties are evaluated in our empirical study. Their validation provides supporting evidence for the effectiveness of explainability-driven feature exposure intervention while offering practical guidance and generalisable insights for its implementation.
	
	The proposed intervention framework is based on two key considerations. 
	%	First, propagated feature contributions should be informative about the realised change in portfolio performance when feature exposures are adjusted, regardless of whether exposures are reduced or amplified. 
	First, within the given feature universe, the intervention is most informative when
	the main effects of individual features are expected to dominate their interaction effects when feature exposures are modified separately. Second, feature contributions should exhibit sufficient temporal persistence such that lagged contribution estimates remain informative for future trading periods. These properties are evaluated in our empirical study. 
	% Their validation provides supporting evidence for the effectiveness of explainability-driven feature exposure intervention while offering practical guidance and generalisable insights for its implementation.
	Evidence in support of these properties helps explain the empirical effectiveness of the attribution-guided feature-exposure intervention and provides practical guidance for its implementation.

	All feature exposure multipliers are computed using only contribution information available before the beginning of the trading period in which they are applied, thereby avoiding look-ahead bias. The contribution signal may use only the sign of the estimated contribution, the rank of the contribution across features, or the full contribution magnitude. Exposure multipliers can be constructed either from the most recent estimation window or from recursively accumulated contribution information to improve temporal stability. Furthermore, the exposure rule can be model-specific or cross-model. In the model-specific case, FFN and XGBoost feature contributions generate separate exposure adjustments. In the cross-model case, feature contribution signals are averaged across models to extract economic information that is less sensitive to a particular model class. This design allows us to distinguish model-specific attribution patterns from more persistent economic feature effects. Stabilisation and cross-model averaging can serve as complementary regularisation mechanisms for the use of attribution information.

	\subsubsection{Feature-contribution signals}
	\label{sec:feature_contribution_signals}
	
	For each feature, we use its contribution to the target portfolio objective
	computed from information available prior to the corresponding out-of-sample
	trading period. Let
	$
	\phi_{i,y-1}^{\mathcal M_y}(Q)
	$
	denote the contribution of feature $i$ to target metric $Q$, computed over
	the validation period associated with the model $\mathcal M_y$ used in
	out-of-sample trading period $y$. In the empirical analysis, we focus
	primarily on the transaction-cost-adjusted Sharpe ratio $SR_c$ as the
	target metric.
	
	We consider three transformations of feature attributions for constructing
	feature-exposure signals.
	First, the sign transformation retains only the direction of the contribution:
	\begin{equation}
		T_{\mathrm{sign}}
		\left(
		\phi_{i,y-1}^{\mathcal M_y}(Q)
		\right)
		=
		\operatorname{sign}
		\left(
		\phi_{i,y-1}^{\mathcal M_y}(Q)
		\right).
	\end{equation}
	
	Second, the rank transformation combines the contribution direction with
	the cross-feature rank of its absolute magnitude:
	\begin{equation}
		T_{\mathrm{rank}}
		\left(
		\phi_{i,y-1}^{\mathcal M_y}(Q)
		\right)
		=
		\operatorname{sign}
		\left(
		\phi_{i,y-1}^{\mathcal M_y}(Q)
		\right)
		\operatorname{rank}
		\left(
		\left|
		\phi_{i,y-1}^{\mathcal M_y}(Q)
		\right|
		\right),
	\end{equation}
	where the rank is computed across features in ascending order of absolute
	attribution magnitude.
	
	Third, the magnitude transformation retains the full attribution:
	\begin{equation}
		T_{\mathrm{mag}}
		\left(
		\phi_{i,y-1}^{\mathcal M_y}(Q)
		\right)
		=
		\phi_{i,y-1}^{\mathcal M_y}(Q).
	\end{equation}
	
	Let
	$
	T_{i,y-1}^{\mathcal M_y}(Q)
	=
	T
	\left(
	\phi_{i,y-1}^{\mathcal M_y}(Q)
	\right)
	$
	denote the selected transformed attribution signal. We map this signal into
	a bounded feature-exposure multiplier according to
	\begin{equation}
		f_{\eta}
		\left(
		T_{i,y-1}^{\mathcal M_y}(Q)
		\right)
		=
		1
		+
		\eta
		\tanh
		\left(
		\frac{
			T_{i,y-1}^{\mathcal M_y}(Q)
		}{
			\operatorname{sd}_{j}
			\left(
			T_{j,y-1}^{\mathcal M_y}(Q)
			\right)
		}
		\right),
		\qquad
		\eta\in[0,1],
		\label{eq:feature_exposure_multiplier}
	\end{equation}
	where the standard deviation is computed across features. The resulting
	multiplier satisfies
	$
	f_{\eta}(\cdot)\in(1-\eta,1+\eta)
	$
	for $\eta>0$, while $f_0(\cdot)=1$ recovers the baseline exposure.
	
	The mapping is monotone and bounded. Positive attribution signals increase
	feature exposure, whereas negative signals reduce it. The parameter $\eta$
	controls the strength of the intervention and can be interpreted as the
	degree of confidence placed in the attribution information. Larger values
	place greater weight on the attribution signal, whereas smaller values retain
	feature exposures closer to those implied by the original trained model.

	\subsubsection{Unstabilised and stabilised exposure updates}
	\label{sec:lagged_recursive_exposure}
	
	We consider two out-of-sample exposure-update rules. The unstabilised rule
	uses only the attribution signal computed from the validation period
	immediately preceding out-of-sample trading period $y$:
	\begin{equation}
		\lambda_{i,y-1}^{\mathcal M_y}
		=
		f_{\eta}
		\left(
		T_{i,y-1}^{\mathcal M_y}(Q)
		\right).
		\label{eq:unstabilised_exposure}
	\end{equation}
	
	The stabilised rule recursively incorporates the previous feature exposure:
	\begin{equation}
		\lambda_{i,y-1}^{\mathcal M_y}
		=
		\lambda_{i,y-2}
		f_{\eta}
		\left(
		T_{i,y-1}^{\mathcal M_y}(Q)
		\right).
		\label{eq:stabilised_exposure}
	\end{equation}
	The initial exposure is set to
	$
	\lambda_{i,0}=1.
	$
	Both rules are free of look-ahead bias because
	$\phi_{i,y-1}^{\mathcal M_y}(Q)$ is computed using only information
	available prior to out-of-sample trading period $y$. Exposure coefficients
	may additionally be bounded within a prespecified interval,
	$
	\lambda_{\min}
	\leq
	\lambda_{i,y-1}
	\leq
	\lambda_{\max},
	$
	to prevent extreme feature scaling.
	
	The stabilised rule provides a form of temporal regularisation. The recursive
	update carries forward information embedded in previous exposure adjustments.
	This could reduce the sensitivity of the intervention to period-specific
	variation in the attribution signal.

	\subsubsection{Cross-model feature exposure}
	\label{sec:cross_model_feature_exposure}
	
	Cross-model averaging provides a second form of regularisation, operating
	across allocation models. To distinguish model-specific attribution
	information from components that are shared across model classes, we combine
	the exposure multipliers obtained from XGB and FFN.
	
	Let
	$
	\lambda_{i,y-1}^{\mathrm{XGB}}
	$
	and
	$
	\lambda_{i,y-1}^{\mathrm{FFN}}
	$
	denote the feature-exposure coefficients implied by the corresponding
	model-specific attribution signals. The cross-model exposure is defined as
	\begin{equation}
		\lambda_{i,y-1}^{\mathrm{CM}}
		=
		\frac{1}{2}
		\left(
		\lambda_{i,y-1}^{\mathrm{XGB}}
		+
		\lambda_{i,y-1}^{\mathrm{FFN}}
		\right).
		\label{eq:cross_model_exposure}
	\end{equation}
	The same cross-model exposure vector is subsequently applied to either
	trained allocation model. More generally,
	for a collection of allocation models $\mathfrak M$, the cross-model
	exposure can be written as
	$
	\lambda_{i,y-1}^{\mathrm{CM}}
	=
	\mathbb{E}_{\mathcal M\in\mathfrak M}
	\left[
	\lambda_{i,y-1}^{\mathcal M}
	\right].
	$
	
	Cross-model averaging regularises the intervention by attenuating
	model-specific variation in the attribution signal while retaining
	components that are common across allocation models. If feature
	contributions predominantly reflect model-specific noise, averaging across
	models should weaken the resulting intervention. Conversely, if the
	attributions contain economically meaningful information shared across
	model classes, the cross-model exposure should remain effective when
	applied to either allocation model.
	
	Stabilisation and cross-model averaging therefore provide complementary
	regularisation mechanisms. Stabilisation regularises feature exposures
	across trading periods, whereas cross-model averaging regularises them
	across allocation models. We evaluate both mechanisms separately and
	jointly in the out-of-sample feature-exposure experiments.

	\subsubsection{Feature-exposure adjustment without model retraining}
	
	Let
	$
	\boldsymbol{\lambda}_{y-1}
	=
	(\lambda_{1,y-1},\ldots,\lambda_{p,y-1})^\top
	$
	denote the resulting feature-exposure vector available prior to trading in
	out-of-sample period $y$, where
	$
	\boldsymbol{\lambda}_{y-1}=\mathbf{1}_p
	$
	corresponds to the baseline feature exposures. Depending on the
	intervention specification, $\boldsymbol{\lambda}_{y-1}$ is constructed
	from either model-specific or cross-model attribution information, with or
	without temporal stabilisation.
	
	The intervention directly uses the exposure coefficients
	$\boldsymbol{\lambda}$ introduced in the local first-order attribution
	framework. Importantly, the parameters $\boldsymbol{\theta}$ of the trained
	allocation model are held fixed throughout the intervention. Given the
	standardised feature matrix $\boldsymbol z_{t-1}$ used to construct the
	portfolio at trading time $t$ within out-of-sample period $y$, the
	exposure-adjusted features are defined as
	\begin{equation}
	\boldsymbol z_{t-1}'
	=
	\boldsymbol{\lambda}_{y-1}
	\odot
	\boldsymbol z_{t-1},
	\qquad t\in y,
	\end{equation}
	where the multiplication is applied feature-wise. Equivalently, for each
	feature $i$,
	$
	z_{i,t-1}'
	=
	\lambda_{i,y-1}z_{i,t-1},
	$
	with
	$
	i=1,\ldots,p
	$
	and $t\in y$.
	The exposure-adjusted features are then passed through the original trained
	allocation function,
	\begin{equation}
	\boldsymbol w_{t-1}^{\epsilon}
	(\boldsymbol{\lambda}_{y-1})
	=
	g_{\boldsymbol\theta}
	\left(
	\boldsymbol{\lambda}_{y-1}
	\odot
	\boldsymbol z_{t-1}
	\right),
	\qquad t\in y,
	\end{equation}
	with $\boldsymbol{\theta}$ unchanged. The resulting residual-weight vector
	is subsequently mapped into tradable stock weights and portfolio
	quantities using the same portfolio-construction procedure as for the
	baseline strategy. Hence, $\lambda_{i,y-1}>1$ increases the model's
	exposure to the economic information represented by feature $i$, whereas
	$\lambda_{i,y-1}<1$ reduces that exposure.
	
	This implementation establishes a direct link between the attribution and
	intervention stages. The local first-order attribution measures the
	sensitivity of portfolio objective $Q$ to the feature-exposure coefficients,
	while the intervention uses attribution information available prior to
	out-of-sample trading period $y$ to adjust those coefficients.
	Because the allocation parameters $\boldsymbol{\theta}$ remain unchanged,
	any difference between the baseline and intervention portfolios arises
	solely from changes in feature exposures. The intervention therefore
	provides a direct economic assessment of whether the attribution information
	can be translated into improved out-of-sample portfolio decisions.

	\section{Empirical Analysis}
	
%	In the empirical analysis section, we mainly include the arbitrage trading performance for the baseline portfolio by XGBoost (XGB) and feedforward neural network (FFN) as allocation function, portfolio metrics analysis for the baseline portfolios, the statistical significance of sharpe ratio attribution and the alignment of the local first-order attribution and estimated shapley values. Regarding to the feature exposure interventions, we provide the evidence of the cross-model similarity for portfolio metrics and attribution persistence across time. The economic significance of sharpe ratio improvement by using the feature exposure rules is given and the robustness for different transaction costs settings.
	
	In the empirical analysis, we first evaluate the arbitrage trading performance of the baseline portfolios constructed using extreme gradient boosting (XGB) and feedforward neural networks (FFN) as allocation functions. We then examine the portfolio-level metrics and their attributions, assess the statistical significance of the Sharpe-ratio attributions, and evaluate the extent to which the proposed local first-order attributions align with estimated Shapley values. For attribution-guided feature-exposure interventions, we provide evidence on the cross-model similarity of portfolio-level attributions and their persistence over time. We subsequently evaluate the economic significance of the resulting improvements in out-of-sample Sharpe ratios and assess the robustness of the feature-exposure strategies across alternative transaction-cost assumptions.

%	\subsection{Arbitrage trading performance and feature contribution analysis}
	
	\subsection{Arbitrage trading performance}
	\label{sec:baseline_portfolio}

%	For the empirical study, we collect daily equity data from October 2000 through December 2024. The sample consists of 346 U.S. stocks with large market capitalisation and high liquidity. Focusing on the most liquid securities helps mitigate trading frictions and market impact. Maintaining a common investment universe provides a consistent cross-sectional environment for evaluating the allocation
%	models and attribution-guided feature-exposure interventions.  
	
	For the empirical study, we use daily U.S. equity data from October 2000
	through December 2024. The sample consists of a universe of 346
	large-capitalisation and liquid U.S. stocks, which reduces concerns about trading frictions and market impact. 
%	All attribution-guided strategies are evaluated relative to their corresponding baseline portfolios within the same investment universe.
	Our empirical analysis focuses primarily on within-universe comparisons between the baseline portfolios and their corresponding attribution-guided strategies to assess the incremental economic value of portfolio-objective attributions.
%	allowing the analysis to isolate the incremental effect of the proposed feature-exposure interventions.
%	ensuring that performance differences are not driven by security selection.

	For the factor models, we use the Fama-French 5-factor specifications. Factor data are obtained from the Kenneth French Data Library. The 5-factor model includes the market, size, value, investment, and profitability factors. These factors are projected to our universe for constructing the real tradable portfolios. Stock factor loadings are estimated daily via rolling regressions of returns on the chosen factor set, using a 60-day window. The out-of-sample Residuals are then computed for each stock. This procedure yields daily residual series beginning in January 2001.  
	
%	Given these residuals, we generate trading signals in a rolling window framework. Specifically, a local lookback window of $L=30$ days is used to form signals, while models of signal extraction and allocation are estimated using a 5-year rolling window, with 4-year training window and 1-year validation window. For computational efficiency, models are re-estimated every 252 trading days using the most recent 5 years of data. Consequently, out-of-sample trading performance is evaluated from January 2006 through December 2024.  
	
	Given these residuals, we construct trading signals within a rolling-window framework. A local lookback window of $L=30$ trading days is used to construct the signals, while the signal-extraction and allocation models are estimated over a five-year rolling window comprising a four-year training period and a one-year validation period. For computational efficiency, the models are re-estimated every 252 trading days using the most recent five years of data. The resulting out-of-sample trading period spans January 2006 to December 2024.

	\paragraph{\textbf{Features in trading signals}}

	The allocation function uses an 18-dimensional trading-signal vector,
	$\bm{z}_{t-1}\in\mathbb{R}^{18}$, summarised in
	Table~\ref{tab:trading_signal_features}. The feature set combines information
	on the estimated regime-dependent Ornstein-Uhlenbeck dynamics, regime
	uncertainty and persistence, residual distributional characteristics, and
	market liquidity and tradability. Following \citet{YeoPapanicolaou2017}, we also
	include the goodness-of-fit measure $R^2$, as poor model fit may reduce the
	reliability of the estimated mean-reversion signals.
	
%	To allow the number of regimes to be inferred from the data, we employ a
%	Hierarchical Dirichlet Process Hidden Markov Model (HDP-HMM) for the
%	regime-dependent OU process. Details of its specification are
%	provided in Appendix~\ref{sec:HDPHMM}. The resulting features jointly
%	characterise the statistical properties and economic implementability of
%	the trading opportunities and serve as inputs to the allocation function.

		\begin{table}[htbp]
%		\begin{table}[H]
			\centering
			\caption{Trading signal features}
			\label{tab:trading_signal_features}
			\small
			\setlength{\tabcolsep}{4pt}
			\renewcommand{\arraystretch}{1.15}
			
			\resizebox{\textwidth}{!}{
				\begin{tabular}{
						>{\centering\arraybackslash}p{0.05\textwidth}
						>{\centering\arraybackslash}p{0.22\textwidth}
						>{\centering\arraybackslash}p{0.37\textwidth}
						>{\centering\arraybackslash}p{0.40\textwidth}
					}
					\toprule
					No. & Label & Feature & Definition \\
					\midrule
					
					% =====================================================
					% Group 1: OU Features
					% =====================================================
					\multicolumn{4}{c}{\textit{OU Features}} \\
					\midrule
					
					1
					& \texttt{kappa}
					& Mean-reversion speed
					& $\widehat{\kappa}_{s_{t-1}}$ \\
					
					2
					& \texttt{mu}
					& Long-run mean
					& $\widehat{\mu}_{s_{t-1}}$ \\
					
					3
					& \texttt{sigma\_OU}
					& OU volatility
					& $\widehat{\sigma}_{s_{t-1}}$ \\
					
					4
					& \texttt{cum\_res}
					& Residual process
					& $X^L_{t-1}$ \\
					
					5
					& \texttt{R2}
					& Goodness of fit
					& $R_{t-1}^{2}$ \\
					
					6
					& \texttt{prob\_signal}
					& Standardised trading signal
					& $\displaystyle
					\frac{
						X^L_{t-1}-\mu_{s_{t-1}}
					}{
						\sigma_{s_{t-1}}/
						\sqrt{2\kappa_{s_{t-1}}}
					}$ \\

					% =====================================================
					% Group 2: Regime-Based Features
					% =====================================================
					\midrule
					\multicolumn{4}{c}{\textit{Regime-Based Features}} \\
					\midrule
					
					7
					& \texttt{regime\_entropy}
					& Regime entropy
					& $\displaystyle
					-\sum_{s}
					p(s_{t-1}=s)
					\log p(s_{t-1}=s)$ \\
					
					8
					& \texttt{regime\_duration}
					& Expected regime duration
					& $\displaystyle
					\frac{1}
					{1-\pi_{s_{t-1},s_{t-1}}}$ \\
					
					9
					& \texttt{regime\_trans\_prob}
					& Maximum transition probability
					& $\displaystyle
					\max_{j=1,\ldots,r}
					\pi_{s_{t-1},j}$ \\

					% =====================================================
					% Group 3: Residual and Distributional Features
					% =====================================================
					\midrule
					\multicolumn{4}{c}{\textit{Residual and Distributional Features}} \\
					\midrule
					
					10
					& \texttt{pre\_res}
					& Realised residual
					& $\epsilon_{t-1}$ \\
					
					11
					& \texttt{pred\_res}
					& Predicted residual
					& $\widehat{\epsilon}_{t}$ \\
					
					12
					& \texttt{res\_std}
					& Residual volatility
					& $\operatorname{Std}
					(\epsilon_{t-L},\ldots,\epsilon_{t-1})$ \\
					
					13
					& \texttt{res\_skew}
					& Residual skewness
					& $\operatorname{Skew}
					(\epsilon_{t-L},\ldots,\epsilon_{t-1})$ \\
					
					14
					& \texttt{res\_kurt}
					& Residual kurtosis
					& $\operatorname{Kurtosis}
					(\epsilon_{t-L},\ldots,\epsilon_{t-1})$ \\
					
					15
					& \texttt{entropy\_pred\_res}
					& Entropy-residual interaction
					& $\displaystyle
					\frac{H_{t-1}}{\log r}
					\left|\widehat{\epsilon}_{t}\right|$ \\

					% =====================================================
					% Group 4: Market and Liquidity Features
					% =====================================================
					\midrule
					\multicolumn{4}{c}{\textit{Market and Liquidity Features}} \\
					\midrule
					
					16
					& \texttt{avg\_daily\_vol}
					& Average daily stock volume
					& $\displaystyle
					\frac{1}{L}
					\sum_{l=1}^{L}
					\operatorname{Volume}_{t-l}$ \\
					
					17
					& \texttt{log\_mkt\_cap}
					& Log market capitalisation
					& $\displaystyle
					\log\!\left(
					\text{Shares Outstanding}_{t-1}
					\times
					\text{Price}_{t-1}
					\right)$ \\
					
					18
					& \texttt{turnover}
					& Stock turnover
					& $\displaystyle
					\frac{
						\operatorname{Volume}_{t-1}
					}{
						\text{Shares Outstanding}_{t-1}
					}$ \\
					
					\bottomrule
				\end{tabular}
			}
			\vspace{2mm}
			
			\begin{minipage}{0.95\textwidth}
				\footnotesize
				\textit{Notes:}
				The table summarises the 18 trading-signal features used as inputs to
				the allocation function. The features are grouped into OU,
				regime-based, residual and distributional, and market and liquidity
				features according to their economic and statistical interpretation.
				The \textit{Label} column reports the feature labels used throughout
				the empirical analysis and attribution figures. Here, $s_{t-1}$
				denotes the prevailing regime at time $t-1$, $\pi_{ij}$ denotes the
				transition probability from regime $i$ to regime $j$, $r$ denotes
				the number of regimes, and $L$ denotes the lookback window.
			\end{minipage}
		\end{table}

	To capture mean reversion and regime switching more comprehensively, we consider additional regime-based features. Specifically, the entropy of the posterior distribution provides a compact measure of uncertainty, the expected duration reflects regime persistence, and the maximum transition probability identifies the most likely subsequent state. To allow the number of regimes to be inferred from the data, we employ a Hierarchical Dirichlet Process Hidden Markov Model (HDP-HMM) for the
	regime-dependent OU process. Details of its specification are provided in Appendix~\ref{sec:HDPHMM}. 
	
	We also consider the distributional properties of the residual series. Higher-order moments such as variance, skewness, and kurtosis capture heteroskedasticity and tail risk, which are particularly relevant for trading decisions. Furthermore, to account for the influence of regime confidence on arbitrage opportunities, we introduce an interaction term that combines normalised regime uncertainty with the magnitude of predicted residuals. This feature represents the intuition that large predicted deviations are more actionable when the regime assignment is reliable, given by low entropy.
	
	Alongside the statistical features from models, market microstructure information is non-negligible. Variables such as market capitalisation, recent average trading volume, and turnover capture liquidity and tradability, which are essential determinants of implementation feasibility and trading constraints.
	
%	These features jointly characterise market behaviour, regime dynamics, and model fit, and are subsequently used as inputs in the allocation function to learn portfolio weights.
	
	The resulting features jointly characterise the statistical properties and economic implementability of
	the trading opportunities and serve as inputs to the allocation function.

	\paragraph{\textbf{Trading performance of XGB and FFN}}
	\label{sec:trading_perf_xgb_ffn}
	
%	We compare a heavily regularized XGBoost and a FFN allocation model. Both models take the same 18 features as input and output residual weights. The models are trained with the objective of maximizing net Sharpe ratio. Transaction costs are set to 5 basis points per unit turnover. The FFN uses learning rate \(0.005\) and dropout \(0.3\). The XGBoost model uses \(\eta=0.001\), max depth \(=3\), minimum child weight \(=10{,}000\), and base score \(=0.01\).
	
	We consider extreme gradient boosting (XGB) and feedforward neural networks (FFN) as alternative allocation functions. Both models use the same 18 trading-signal features as inputs and generate residual portfolio weights as outputs. The FFN consists of four hidden layers with dimensions (32,16,8,4) and uses ReLU activation functions. XGB is implemented using shallow boosted trees. The allocation models are trained to maximise the net Sharpe ratio, incorporating transaction costs of 5 basis points per unit of portfolio turnover. Model hyperparameters are selected based on performance over the corresponding validation windows to control model complexity, including the learning rate, number of boosting rounds, maximum tree depth, and minimum child weight for XGB, and the learning rate and dropout rate for FFN.

		\begin{table}[htbp]
		\centering
		\caption{Baseline portfolio performance}
		\label{tab:baseline_performance}
		\small
		
		\resizebox{\textwidth}{!}{
			\begin{tabular}{cccccccc}
				\toprule
				Model &
				$SR^{gross}$ &
				$\mu^{gross}$ (\%) &
				$\sigma^{gross}$ (\%) &
				$SR^{net}$ &
				$\mu^{net}$ (\%) &
				$\sigma^{net}$ (\%) &
				Avg.\ Turnover (\%)  \\
				\midrule
				
				XGB &
				1.0207 &
				4.22 &
				4.13 &
				0.6175 &
				2.55 &
				4.13 &
				13.21  \\
				
				FFN &
				1.3051 &
				6.01 &
				4.61 &
				0.5750 &
				2.65 &
				4.61 &
				26.70\\
				\bottomrule
			\end{tabular}
		}
		
		\vspace{2mm}
		
		\begin{minipage}{0.95\textwidth}
			\footnotesize
			\textit{Notes:}
			The table reports the out-of-sample performance of the baseline
			portfolios. $SR$, $\mu$, and $\sigma$ denote the annualised Sharpe
			ratio, mean return, and volatility, respectively. Gross performance
			is evaluated before transaction costs, while net performance is
			evaluated after transaction costs of 5 basis points per unit of
			portfolio turnover. Avg.\ Turnover denotes the average daily portfolio
			turnover and is reported in percentage terms.
		\end{minipage}
	\end{table}
	
	\begin{figure}[htbp]
		\centering
		
		\begin{minipage}[t]{0.49\textwidth}
			\centering
			\includegraphics[width=\linewidth]
			{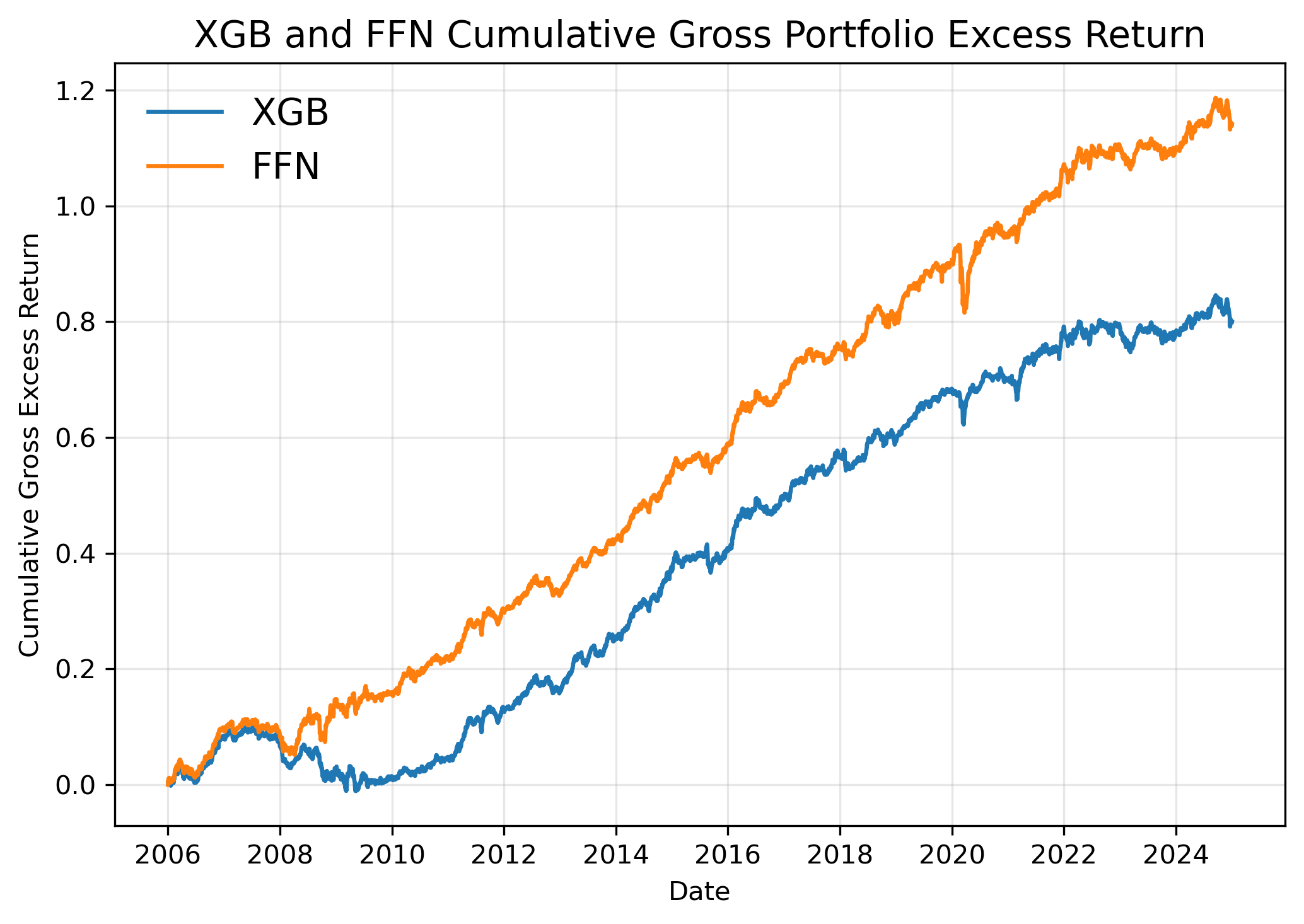}
			
			\small (a) Cumulative gross portfolio excess return
		\end{minipage}
		\hfill
		\begin{minipage}[t]{0.49\textwidth}
			\centering
			\includegraphics[width=\linewidth]
			{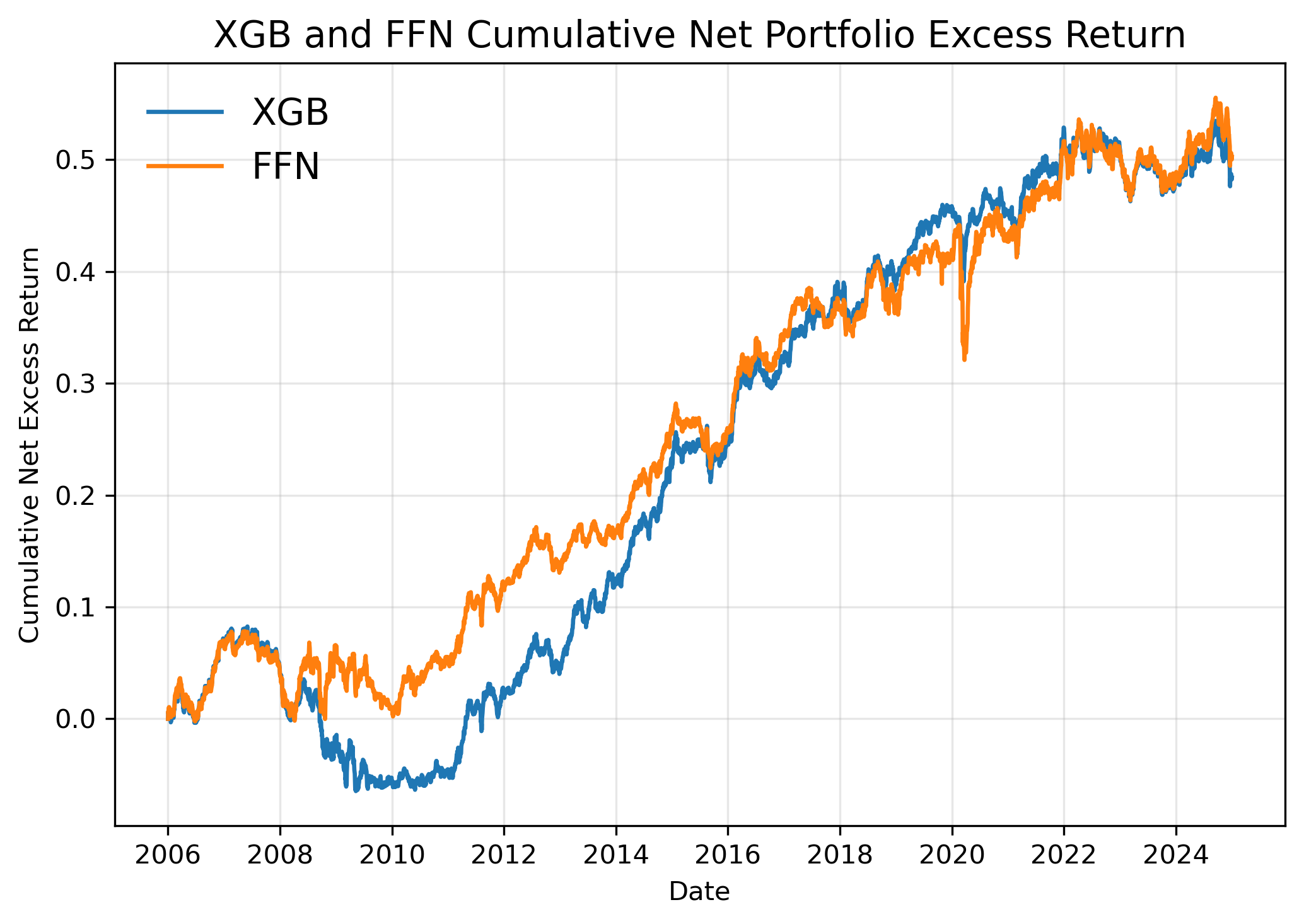}
			
			\small (b) Cumulative net portfolio excess return
		\end{minipage}
		
		%    	\caption{
			%    		Out-of-sample cumulative portfolio excess returns for the XGB and
			%    		FFN allocation models. Panel (a) reports cumulative gross excess
			%    		returns, while Panel (b) reports cumulative net excess returns after
			%    		proportional transaction costs of 5 basis points per unit of portfolio
			%    		turnover.
			%    	}
		\caption{
			Out-of-sample cumulative portfolio excess returns for the XGB and FFN allocation models. The left panel reports cumulative gross excess returns, while the right panel reports cumulative net excess returns after proportional transaction costs (c=5bp).
		}
		\label{fig:cumulative_return_xgb_ffn}
		
	\end{figure}

	\begin{figure}[htbp]
		\centering
		
		\begin{minipage}[t]{0.49\textwidth}
			\centering
			\includegraphics[width=\linewidth]
			{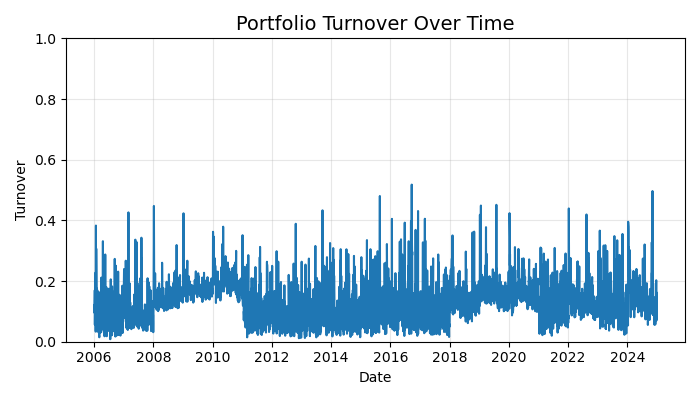}
			\small (a) XGB: Portfolio turnover
		\end{minipage}
		\hfill
		\begin{minipage}[t]{0.49\textwidth}
			\centering
			\includegraphics[width=\linewidth]
			{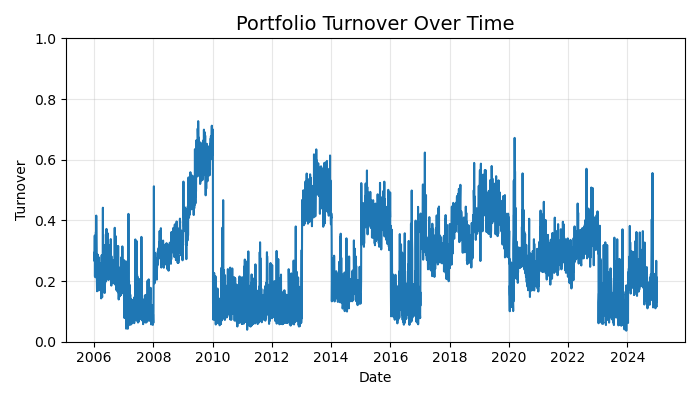}
			\small (b) FFN: Portfolio turnover
		\end{minipage}
		
		\caption{
			Out-of-sample portfolio turnover for the XGB and FFN allocation models.
		}
		
		\label{fig:turnover_xgb_ffn}
		
	\end{figure}

	Table~\ref{tab:baseline_performance} reports the resulting annualised gross and net portfolio performance statistics, together with average daily portfolio turnover. Figure~\ref{fig:cumulative_return_xgb_ffn} presents the cumulative gross and net portfolio excess returns, while Figure~\ref{fig:turnover_xgb_ffn} reports the daily portfolio turnover for the two allocation models. XGB exhibits substantially lower trading intensity than FFN, with average turnover of 13.21\%, compared with 26.70\% for FFN, consistent with the more heavily regularised structure of the tree-based allocation model. Although FFN achieves a higher gross Sharpe ratio than XGB (1.3051 versus 1.0207), its greater turnover leads to a larger deterioration in performance after transaction costs. Consequently, XGB achieves a slightly higher net Sharpe ratio than FFN (0.6175 versus 0.5750), highlighting the importance of trading intensity for the implementable performance of the allocation strategies.

	\subsection{Portfolio performance attribution analysis}
	\label{sec:feature_contribution_portfolio}
	
	We next examine how the learned features contribute not only to allocation weights, but also to portfolio-level economic metrics, including turnover, gross Sharpe ratio, and net Sharpe ratio. The goal is to understand whether the same feature signals that drive allocation decisions also explain portfolio performance, and whether such explanations are transferable across the two allocation models, XGBoost and FFN.
	
	\subsubsection{Feature contribution to allocation weights and portfolio turnover}
	\label{sec:feature_importance_weights}
	
	For each feature $i$, its contribution to portfolio metric $Q$ at time $t$ is obtained by aggregating the stock-level contributions,
	$
		\phi_{i,t}(Q)
		=
		\sum_{n=1}^{N}
		\phi_{i,n,t}(Q),
	$
	and the corresponding annualised feature contribution is defined as
	$
		\phi_i^{\mathrm{ann}}(Q)
		=
		\frac{252}{T}
		\sum_{t=1}^{T}
		\phi_{i,t}(Q),
%		=
%		\frac{252}{T}
%		\sum_{t=1}^{T}
%		\sum_{n=1}^{N}
%		\phi_{i,n,t}(Q),
	$
	where $T$ denotes the number of trading days in the evaluation period and $N$ denotes the number of stocks.

	\begin{figure}[htbp]
		\centering
		
		% ============================================================
		% Row 1: XGB
		% ============================================================
		\begin{minipage}[t]{0.49\linewidth}
			\centering
			\includegraphics[width=\linewidth]
			{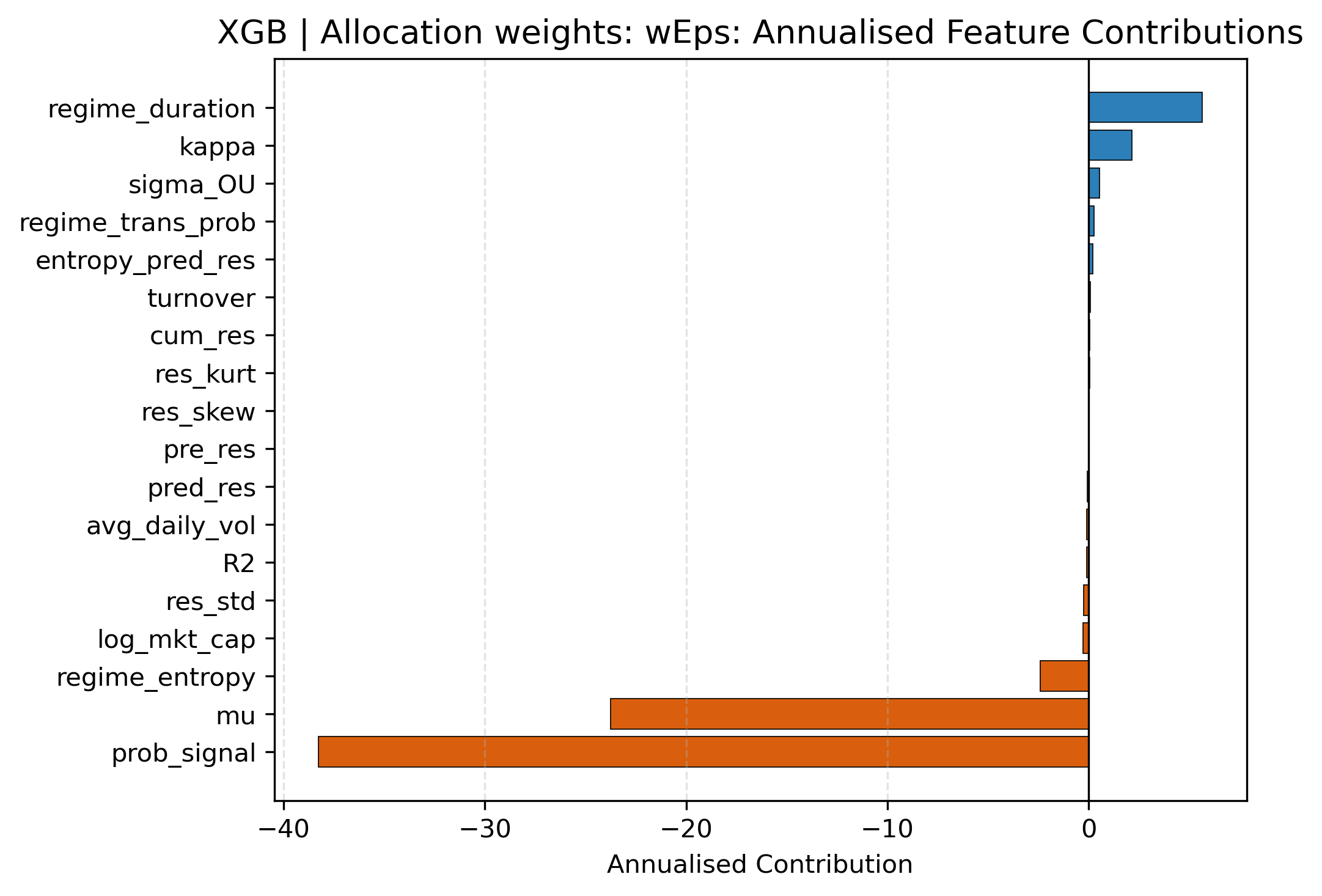}
			\small (a) XGB: Residual-weight attribution
			\label{fig:weights_contrib_xgb_weight}
		\end{minipage}
		\hfill
		\begin{minipage}[t]{0.49\linewidth}
			\centering
			\includegraphics[width=\linewidth]
			{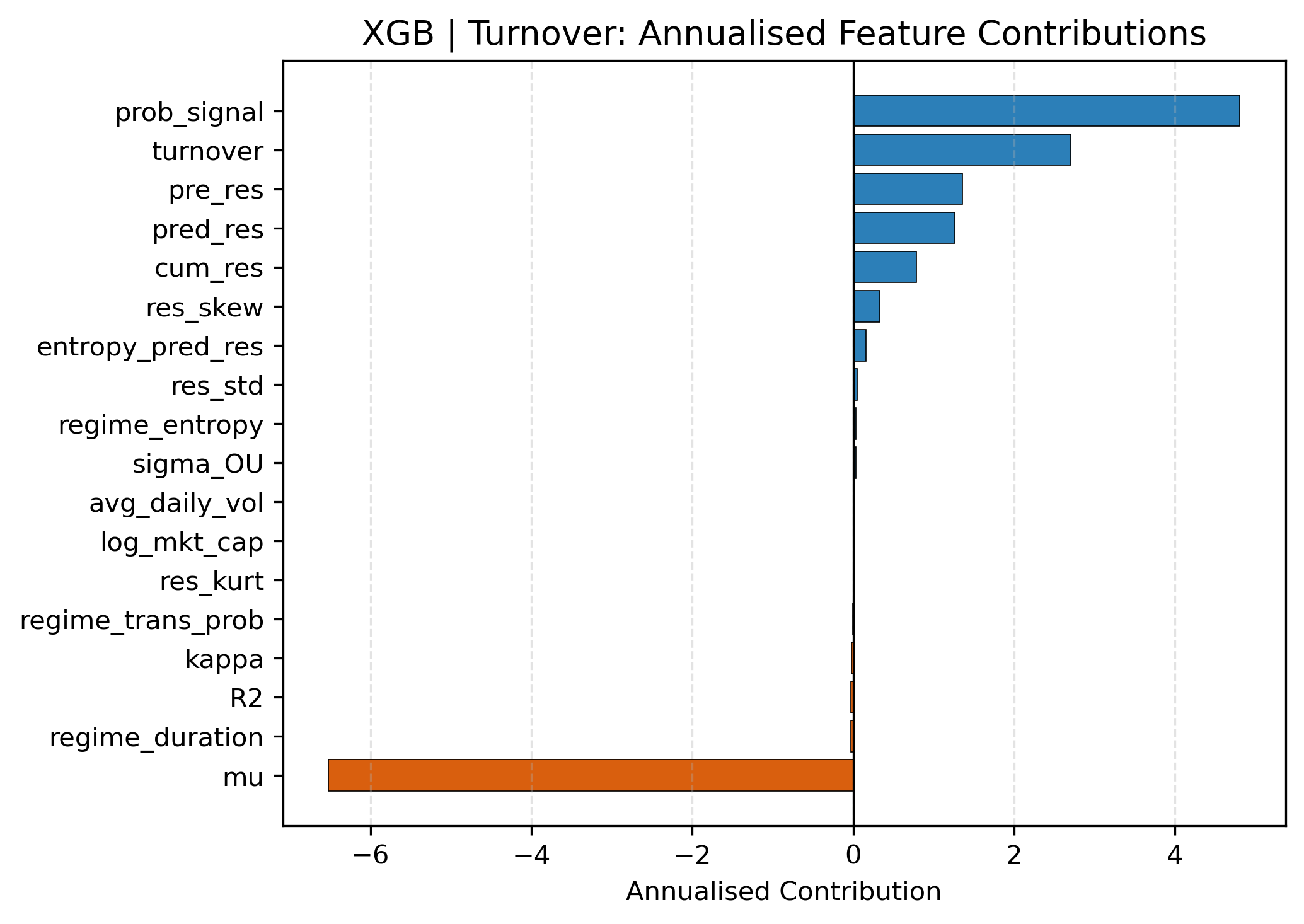}
			\small (b) XGB: Portfolio turnover attribution
			\label{fig:weights_contrib_xgb_turnover}
		\end{minipage}
		
		\vspace{0.5em}
		
		% ============================================================
		% Row 2: FFN
		% ============================================================
		\begin{minipage}[t]{0.49\linewidth}
			\centering
			\includegraphics[width=\linewidth]
			{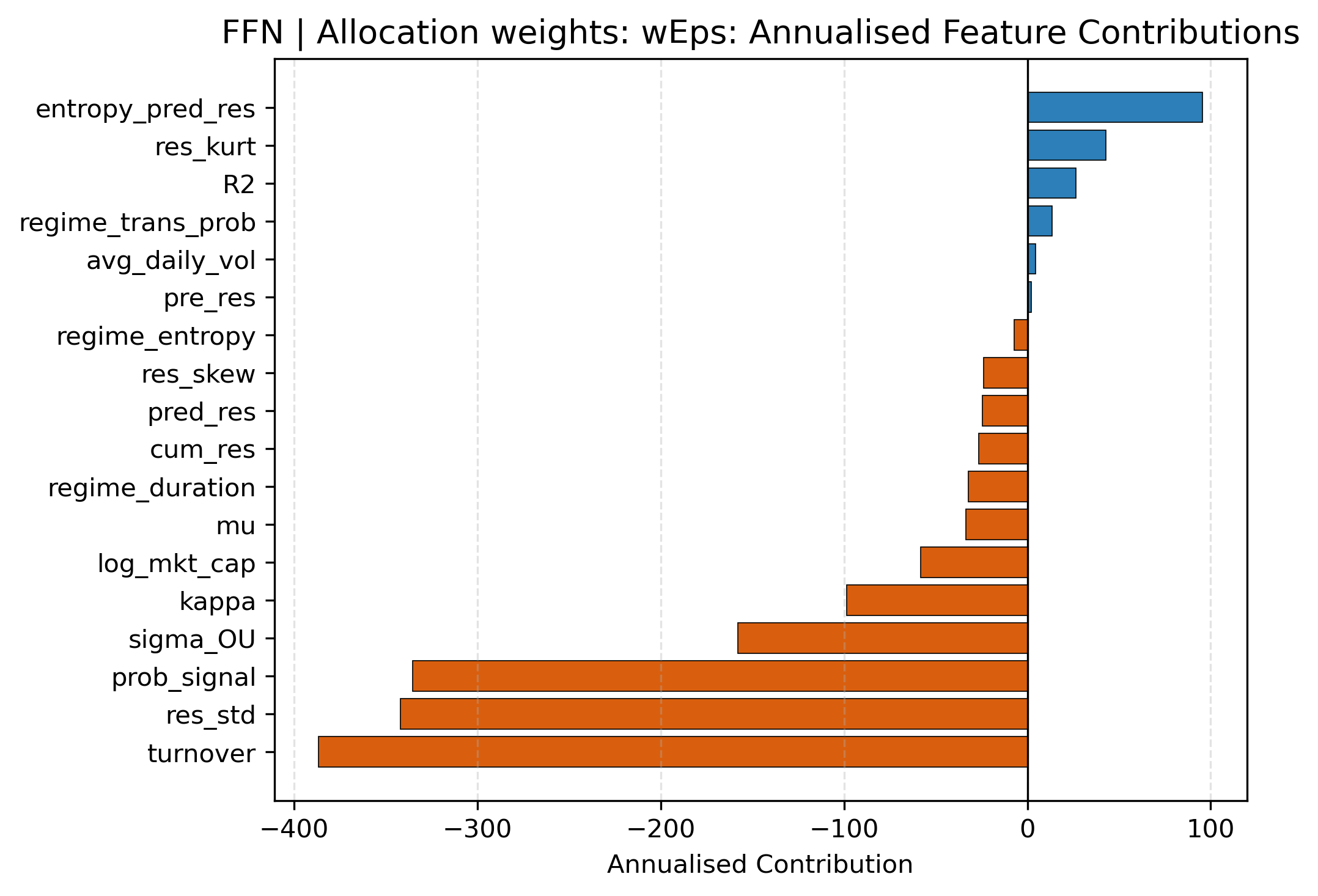}
			\small (c) FFN: Residual-weight attribution
			\label{fig:weights_contrib_ffn_weight}
		\end{minipage}
		\hfill
		\begin{minipage}[t]{0.49\linewidth}
			\centering
			\includegraphics[width=\linewidth]
			{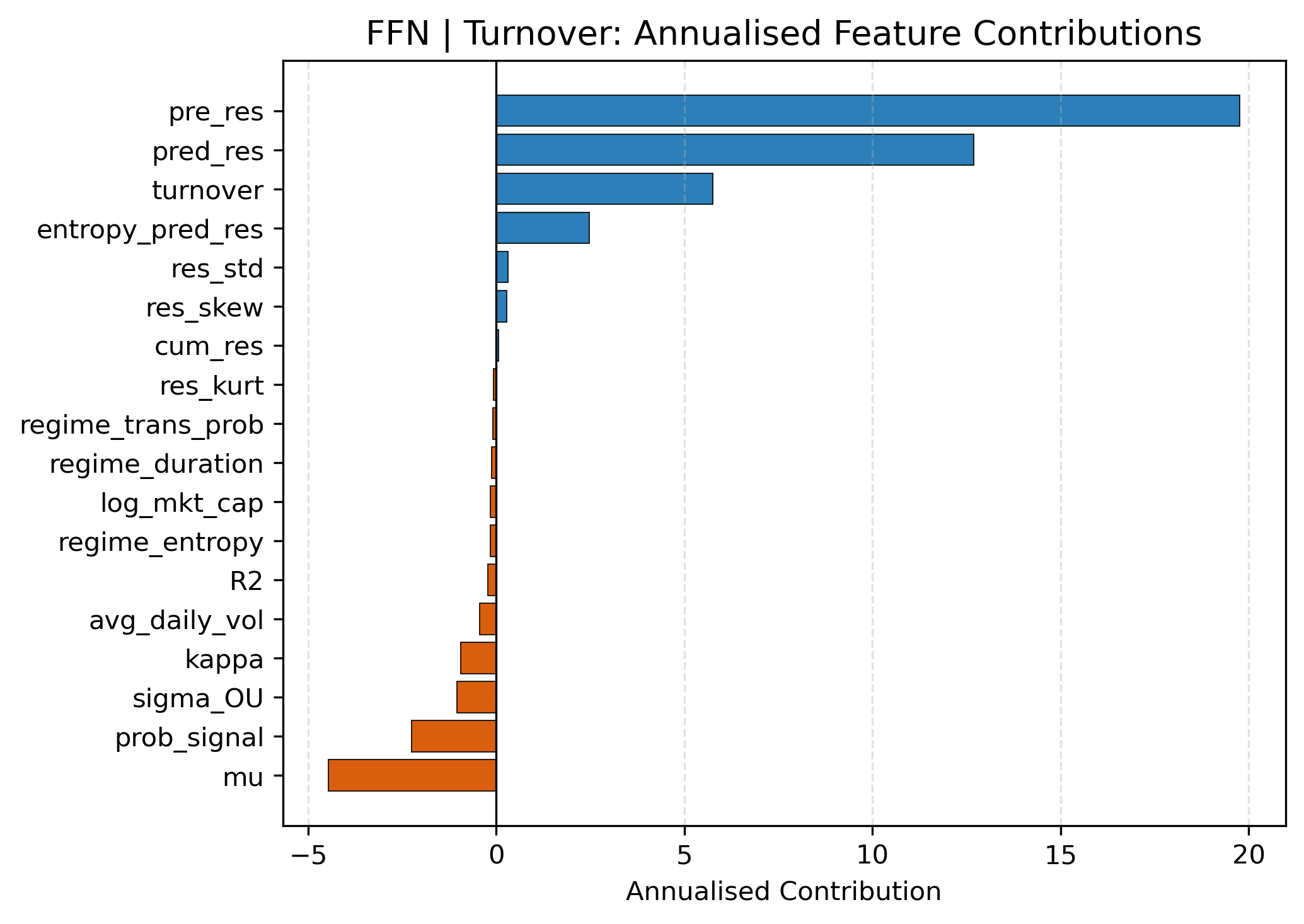}
			\small (d) FFN: Portfolio turnover attribution
			\label{fig:weights_contrib_ffn_turnover}
		\end{minipage}
		
		\caption{
			Annualised feature contributions to residual portfolio weights and
			portfolio turnover for XGB and FFN.
		}
		\label{fig:weights_turnover_contrib}
	\end{figure}
	
	Figure~\ref{fig:weights_turnover_contrib} reports the annualised feature contributions to the residual portfolio weights and portfolio turnover over the out-of-sample period from Jan 2006 to Dec 2024.
	The two models exhibit distinct attribution patterns. XGBoost concentrates its weight attribution mainly on \texttt{prob\_signal} and \texttt{mu}, suggesting a signal-confidence-driven allocation mechanism. By contrast, for FFN, the dominant contributors to the residual-weight component $w^\epsilon$ are \texttt{turnover}, \texttt{res\_std}, and \texttt{prob\_signal}, indicating that FFN places substantial emphasis on cost-related and volatility-related information when constructing portfolio weights. These features give negative contribution to the residual weights on average over the 19 years.

	The feature contribution to portfolio turnover is related to, but distinct from, the contribution to residual portfolio weights. Features that contribute substantially to the model output, such as \texttt{prob\_signal} and \texttt{mu}, also tend to exhibit large turnover contributions in both XGBoost and FFN. However, the relationship is not one-to-one. Several features with only modest contributions to residual weights generate disproportionately large turnover effects. For example, \texttt{pred\_res} and \texttt{pre\_res} exhibit relatively small contributions to residual weights but substantial contributions to portfolio turnover in both models. Conversely, some features that strongly influence residual weights have limited impact on portfolio turnover. Examples include \texttt{regime\_duration} and \texttt{kappa} in XGBoost, and \texttt{res\_kurt} and \texttt{res\_std} in FFN.
	
%	These results suggest that feature importance in the predictive model does not necessarily translate into feature importance for portfolio turnover. Since turnover directly determines transaction costs, the economic contribution of a feature depends not only on its effect on portfolio allocation but also on its impact on trading activity. This distinction is especially important when comparing gross and net portfolio performance, as the inclusion of transaction costs can significantly alter the economic outcomes.

	\paragraph{\textbf{Temporal dynamics of allocation weights and turnover attribution}}
	\label{sec:temporal_dynamics_selected_features}
	
	\begin{figure}[htbp]
		\centering
		
		\begin{minipage}[t]{0.49\linewidth}
			\centering
			\includegraphics[width=\linewidth]{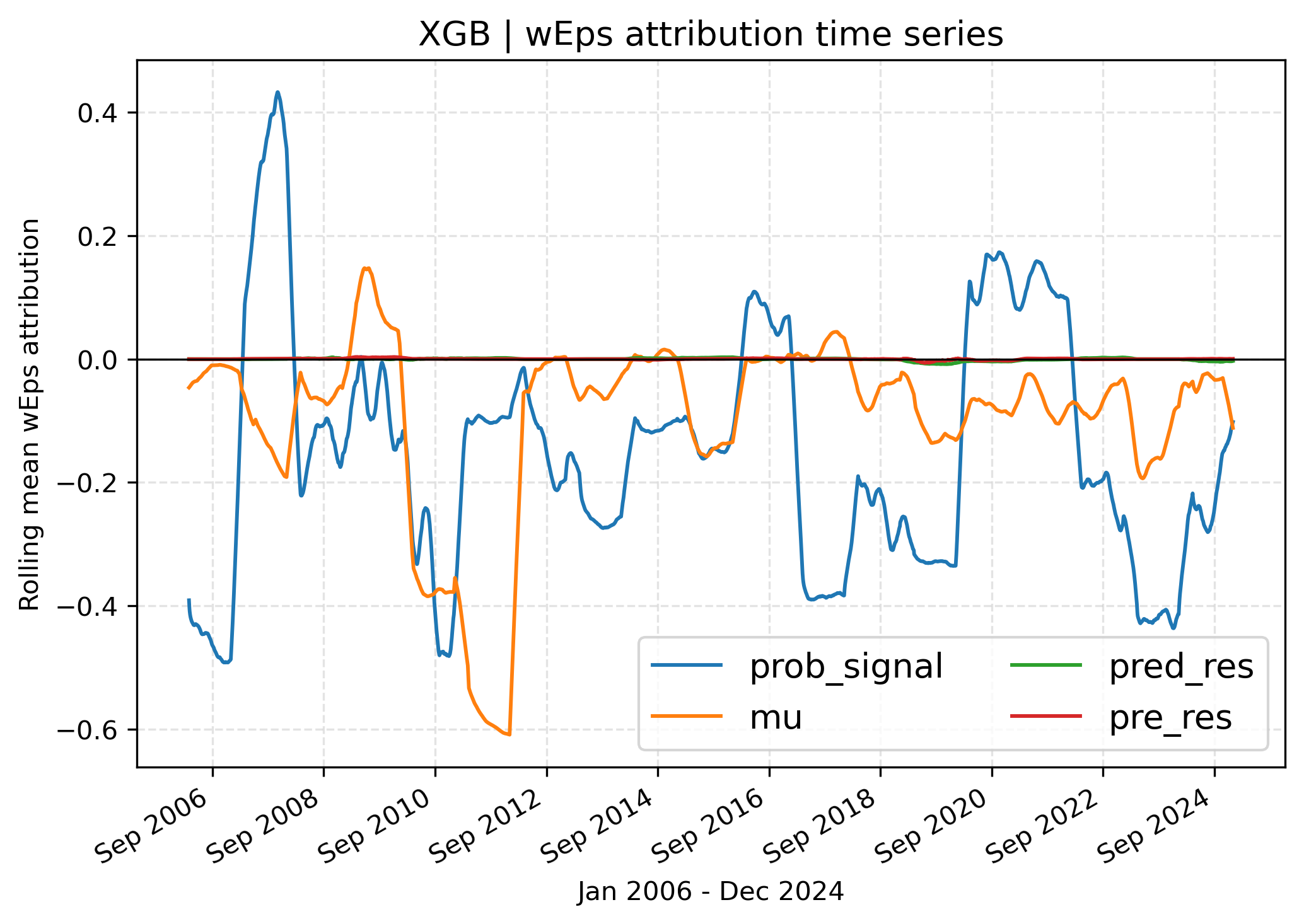}
			\small (a) XGB: $w^\epsilon$ attribution
			\label{fig:xgb_weps_selected}
		\end{minipage}
		\hfill
		\begin{minipage}[t]{0.49\linewidth}
			\centering
			\includegraphics[width=\linewidth]{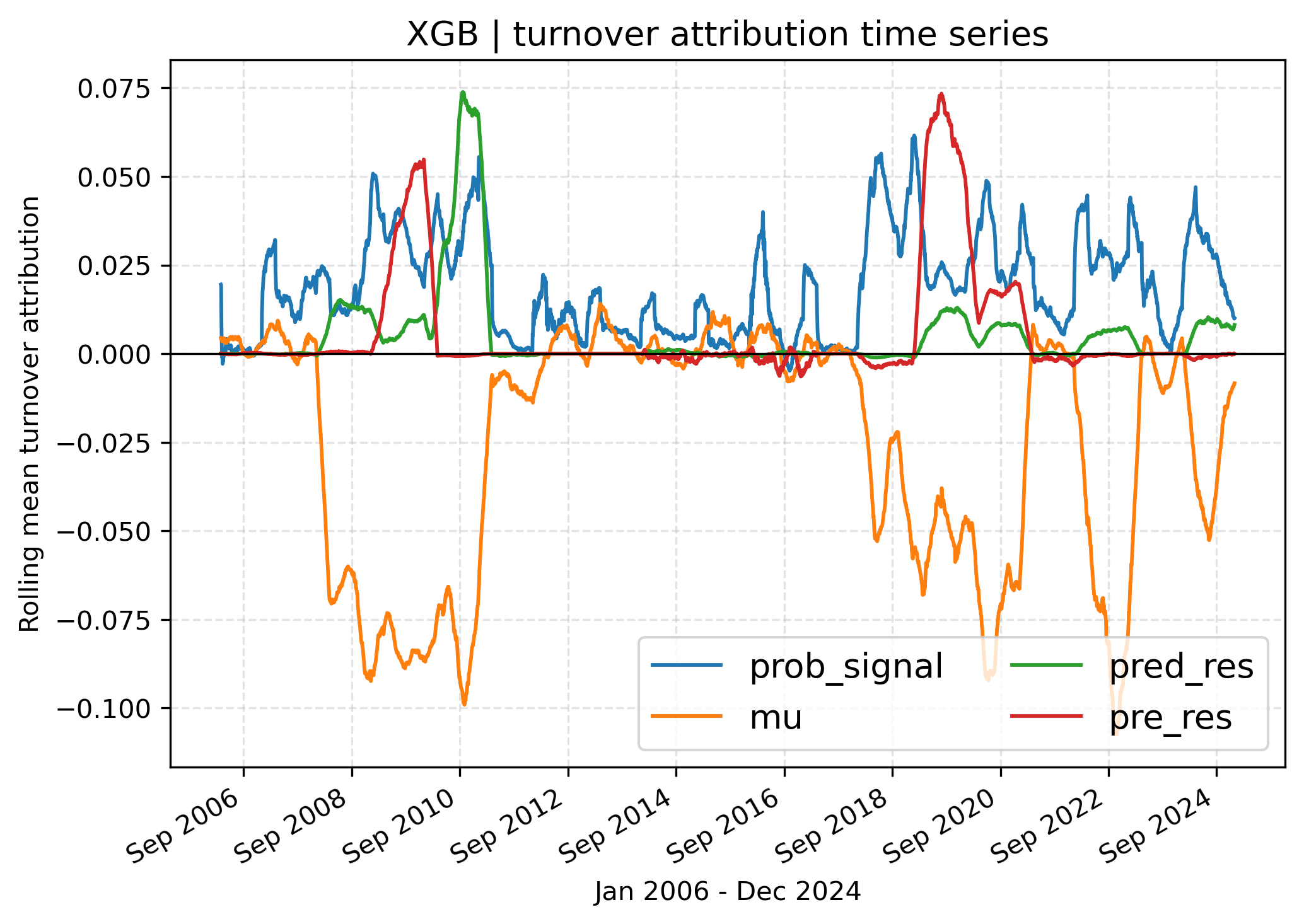}
			\small (b) XGB: portfolio turnover attribution
			\label{fig:xgb_turnover_selected}
		\end{minipage}
		
		\vspace{2mm}
		
		\begin{minipage}[t]{0.49\linewidth}
			\centering
			\includegraphics[width=\linewidth]{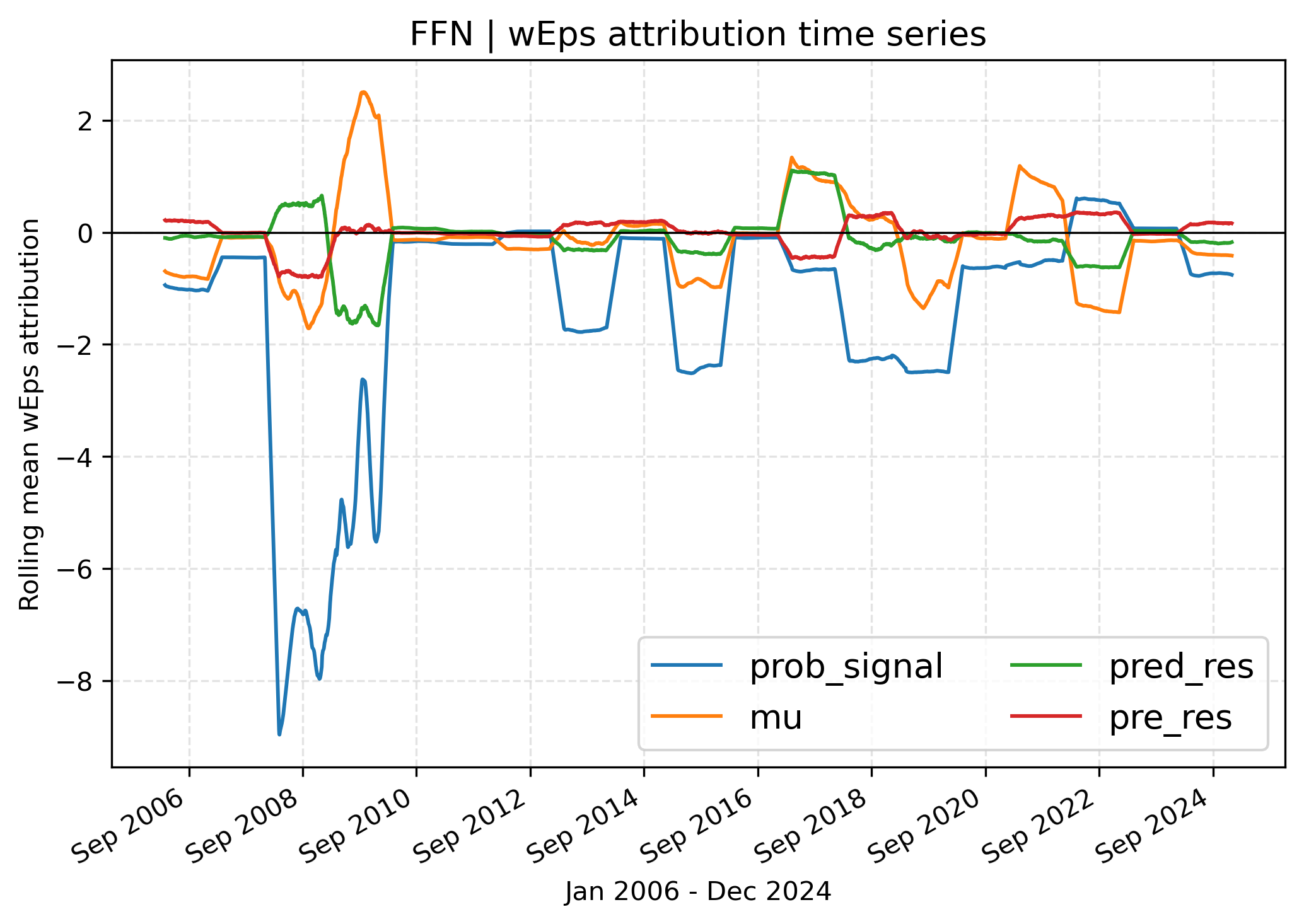}
			\small (c) FFN: $w^\epsilon$ attribution
			\label{fig:ffn_weps_selected}
		\end{minipage}
		\hfill
		\begin{minipage}[t]{0.49\linewidth}
			\centering
			\includegraphics[width=\linewidth]{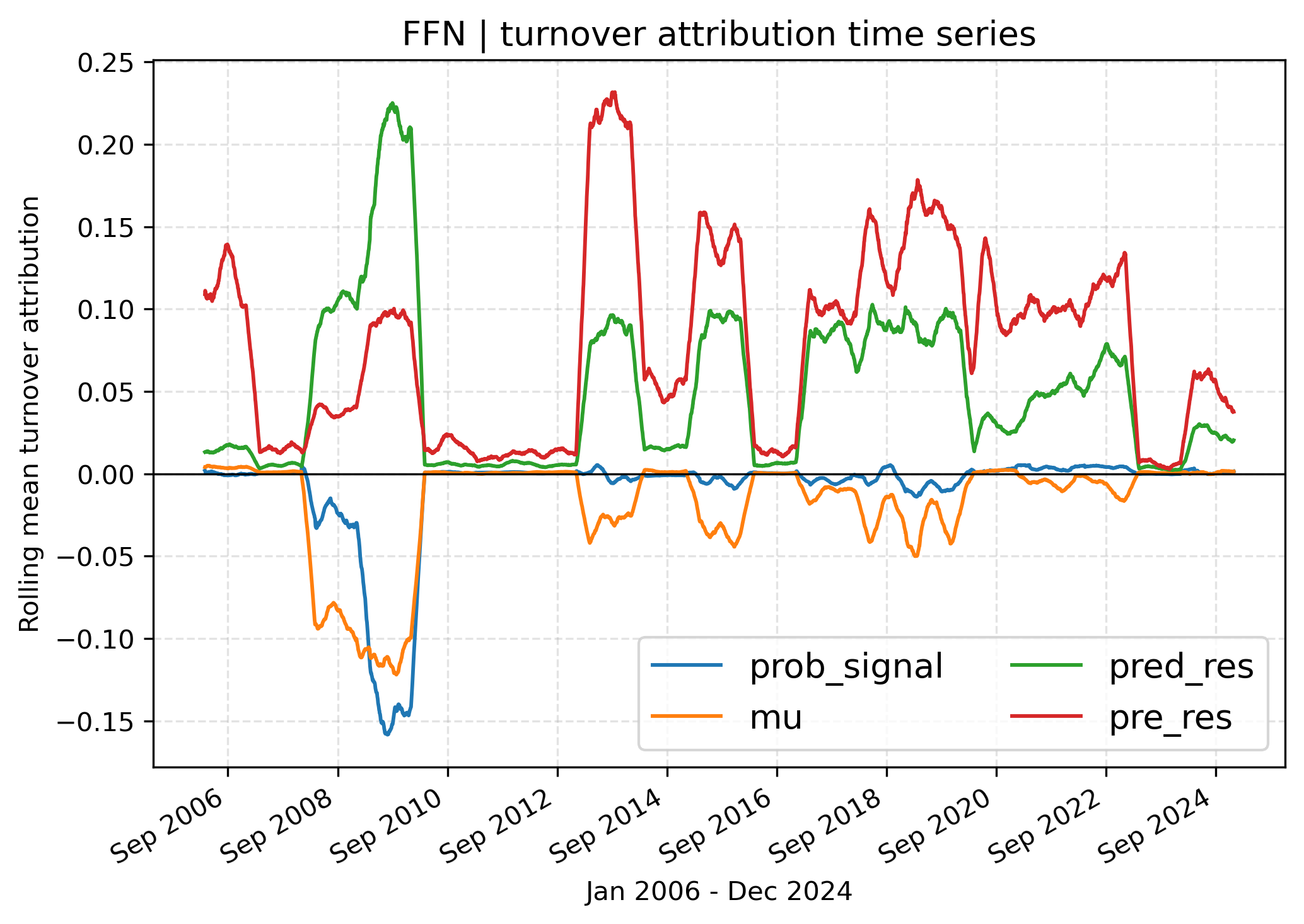}
			\small (d) FFN: portfolio turnover attribution
			\label{fig:ffn_turnover_selected}
		\end{minipage}
		
		\caption{63-day rolling mean residual-weight and portfolio turnover attribution time series for four representative features (\texttt{prob\_signal}, \texttt{mu}, \texttt{pred\_res}, and \texttt{pre\_res}) over 2006-2024. 
			%		FFN exhibits a stronger separation between allocation and trading dynamics, while XGBoost shows a smoother and more compact attribution pattern.
		}
		\label{fig:selected_feature_ts}
	\end{figure}
	
	To complement the annualised attribution results, Figure~\ref{fig:selected_feature_ts} reports the 63-day rolling mean residual-weight and portfolio turnover attributions over 2006-2024 for four representative features: \texttt{prob\_signal}, \texttt{mu}, \texttt{pred\_res}, and \texttt{pre\_res}. These features are selected because they repeatedly appear among the dominant contributors in the cumulative analysis and because they provide a compact view of the main economic channels underlying both portfolio construction and trading intensity. 
%	These features further shows that these dominant features are not only important in aggregate, but also display clear temporal variation.
	
	The results indicate that \texttt{prob\_signal} and \texttt{mu} are the primary contributors to residual portfolio weights across both models, exhibiting larger contributions than \texttt{pred\_res} and \texttt{pre\_res}. Conversely, \texttt{pred\_res} and \texttt{pre\_res} contribute more substantially to portfolio turnover dynamics over time. Notably, feature contributions to turnover become more pronounced during several stress periods.

	These results suggest that feature importance in the predictive model does not necessarily translate into feature importance for portfolio turnover. Since turnover directly determines transaction costs, the economic contribution of a feature depends not only on its effect on portfolio allocation but also on its impact on trading activity. This distinction is especially important when comparing gross and net portfolio performance, as the inclusion of transaction costs can significantly alter the economic outcomes.

	\subsubsection{Feature contribution to portfolio returns and Sharpe ratio}
	\label{sec:feature_contribution_metrics}

	\begin{figure}[htbp]
		\centering
		
		% ============================================================
		% Row 1: Mean return
		% ============================================================
		\begin{minipage}[t]{0.49\linewidth}
			\centering
			\includegraphics[width=\linewidth]
			{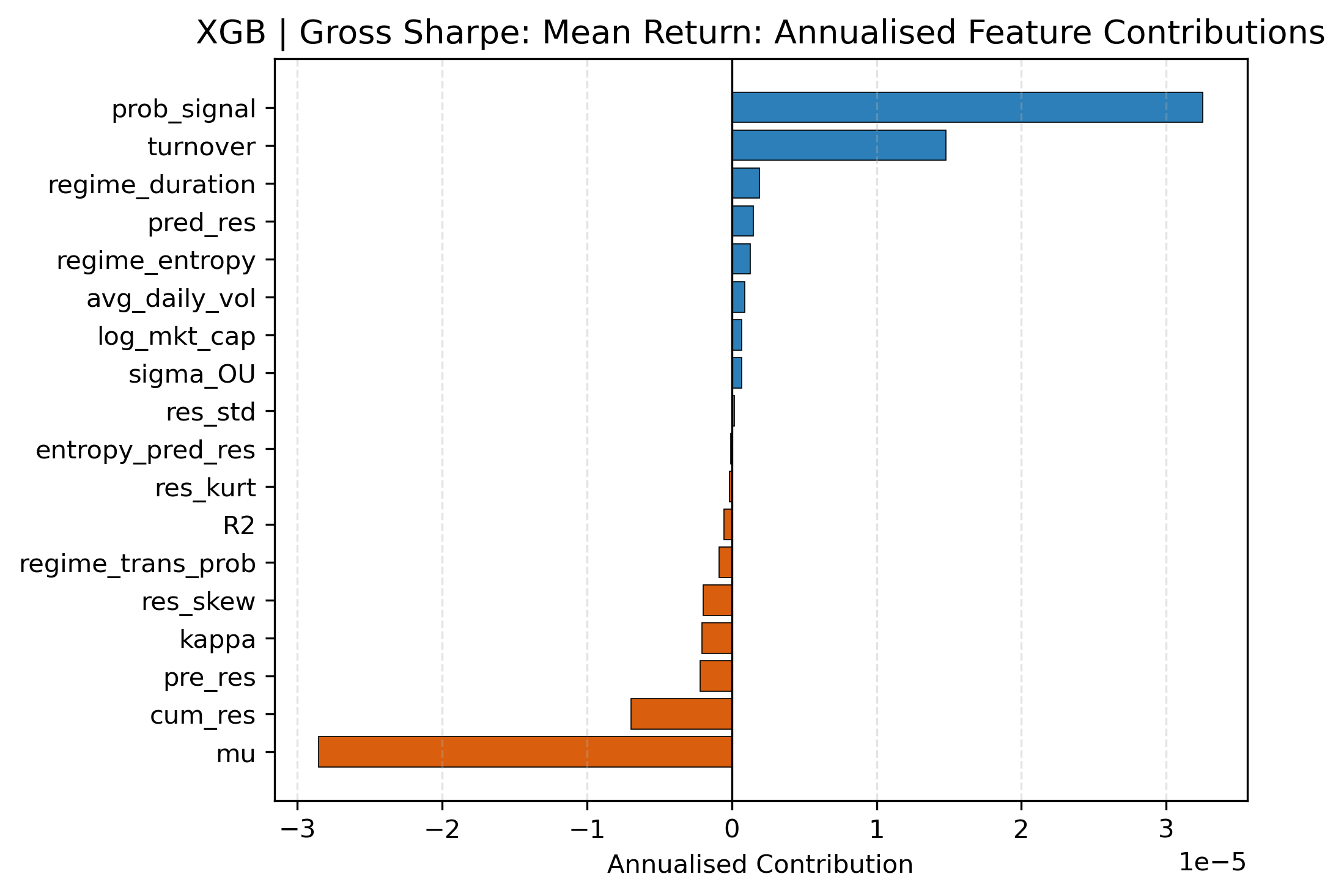}
			\small (a) XGB: Gross mean-return attribution
			\label{fig:xgb_gross_mu_contrib}
		\end{minipage}
		\hfill
		\begin{minipage}[t]{0.49\linewidth}
			\centering
			\includegraphics[width=\linewidth]
			{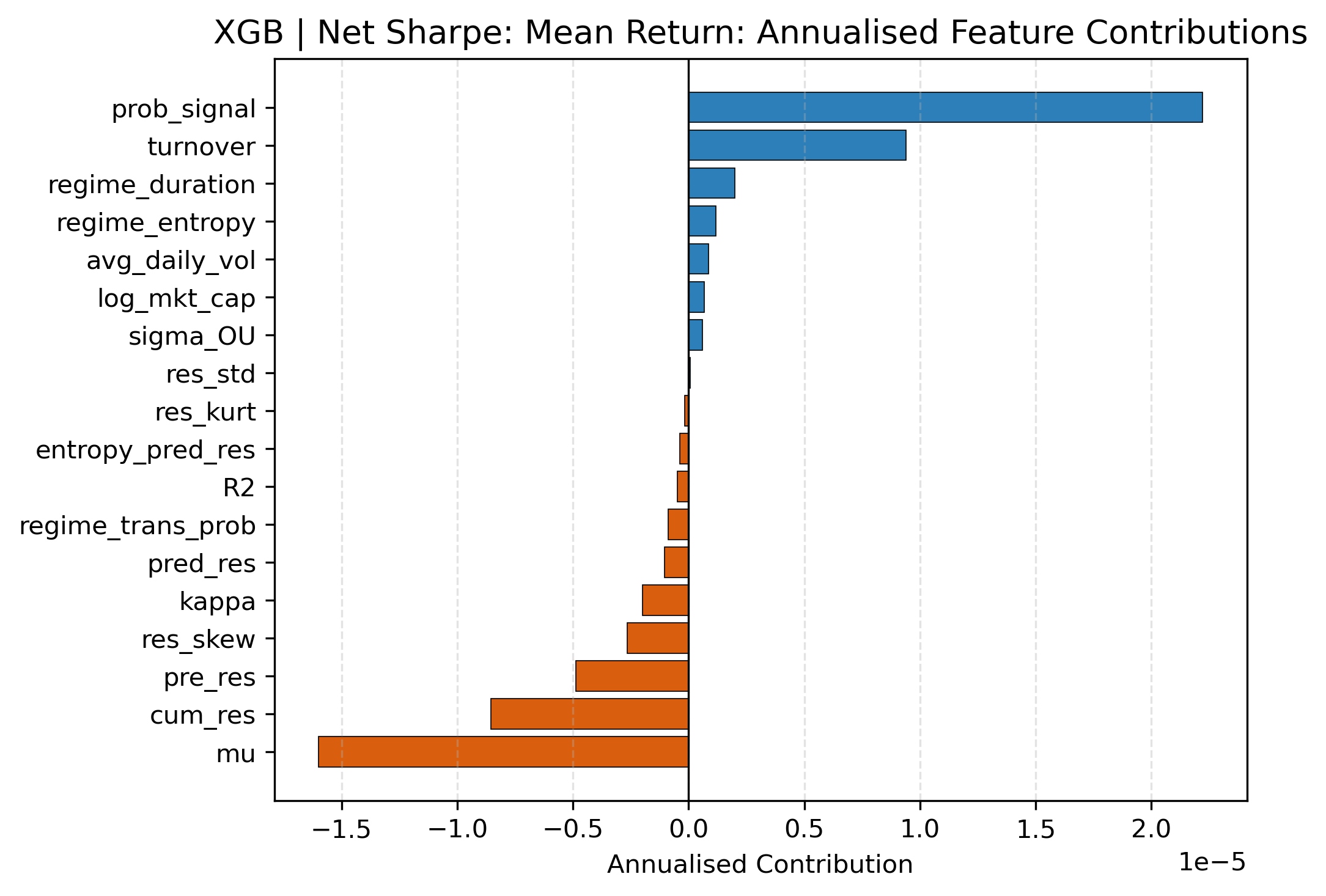}
			\small (b) XGB: Net mean-return attribution
			\label{fig:xgb_net_mu_contrib}
		\end{minipage}
		
		\vspace{0.5em}
		
		% ============================================================
		% Row 2: Volatility
		% ============================================================
		\begin{minipage}[t]{0.49\linewidth}
			\centering
			\includegraphics[width=\linewidth]
			{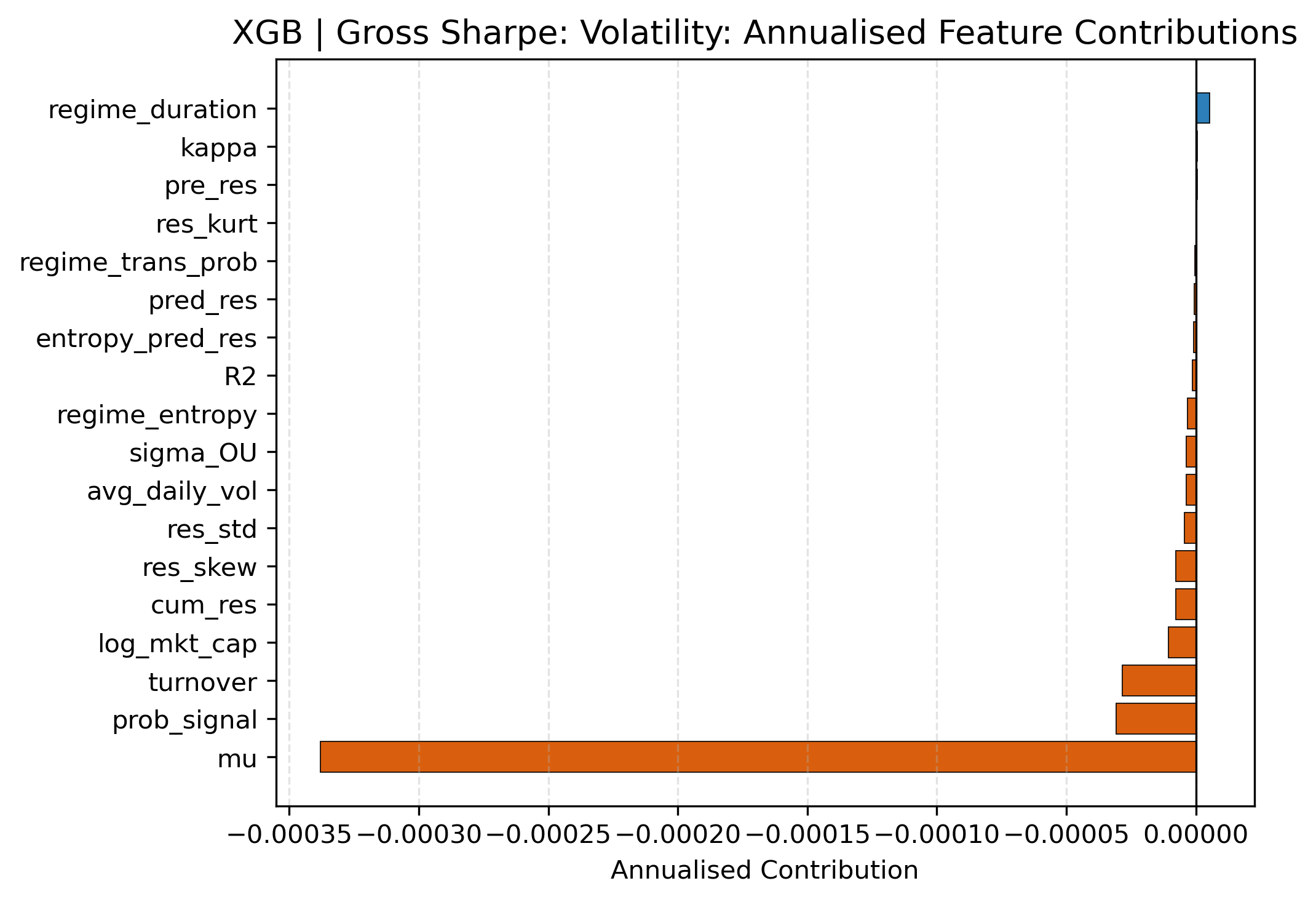}
			\small (c) XGB: Gross volatility attribution
			\label{fig:xgb_gross_sigma_contrib}
		\end{minipage}
		\hfill
		\begin{minipage}[t]{0.49\linewidth}
			\centering
			\includegraphics[width=\linewidth]
			{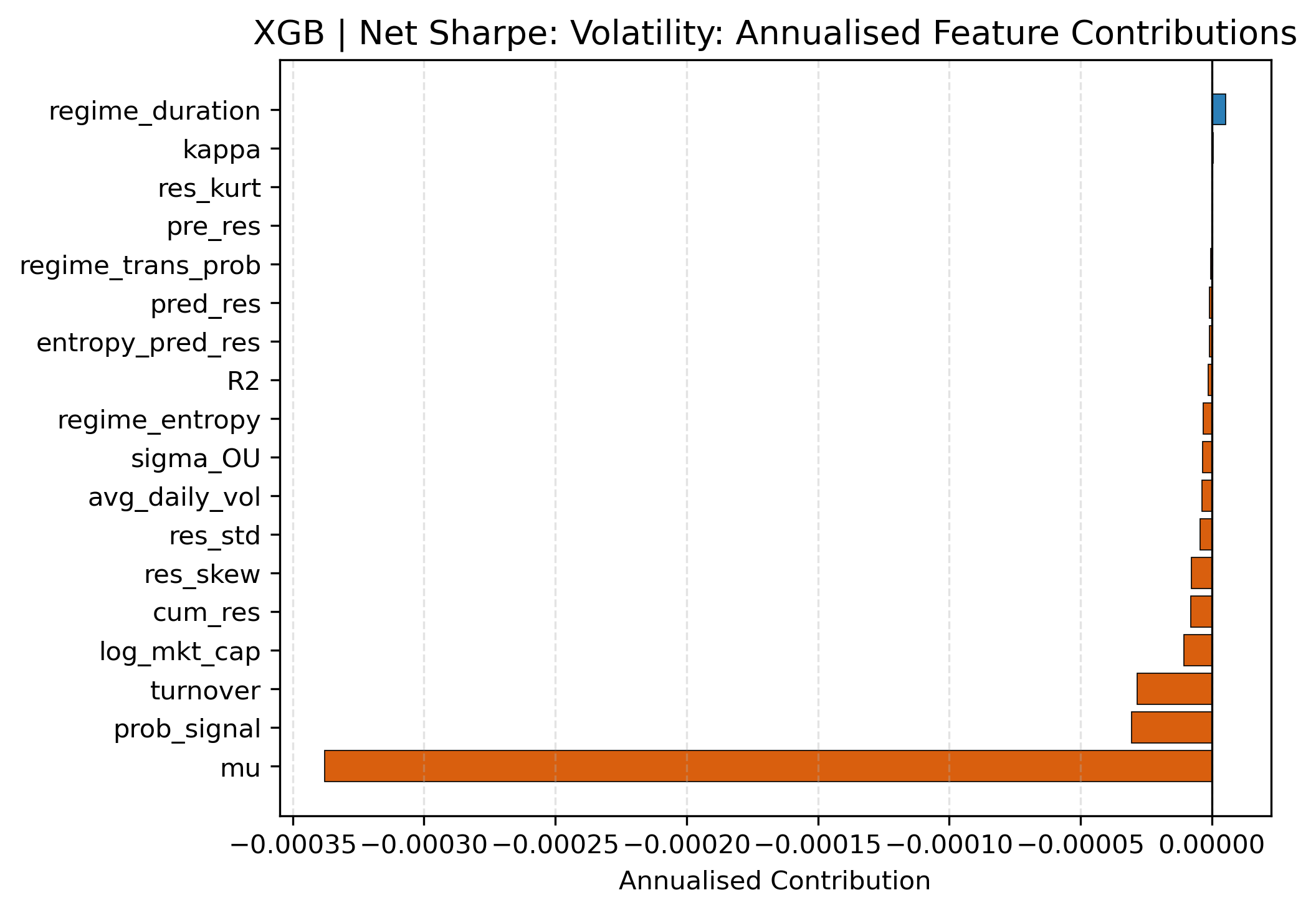}
			\small (d) XGB: Net volatility attribution
			\label{fig:xgb_net_sigma_contrib}
		\end{minipage}
		
		\vspace{0.5em}
		
		% ============================================================
		% Row 3: Sharpe ratio
		% ============================================================
		\begin{minipage}[t]{0.49\linewidth}
			\centering
			\includegraphics[width=\linewidth]
			{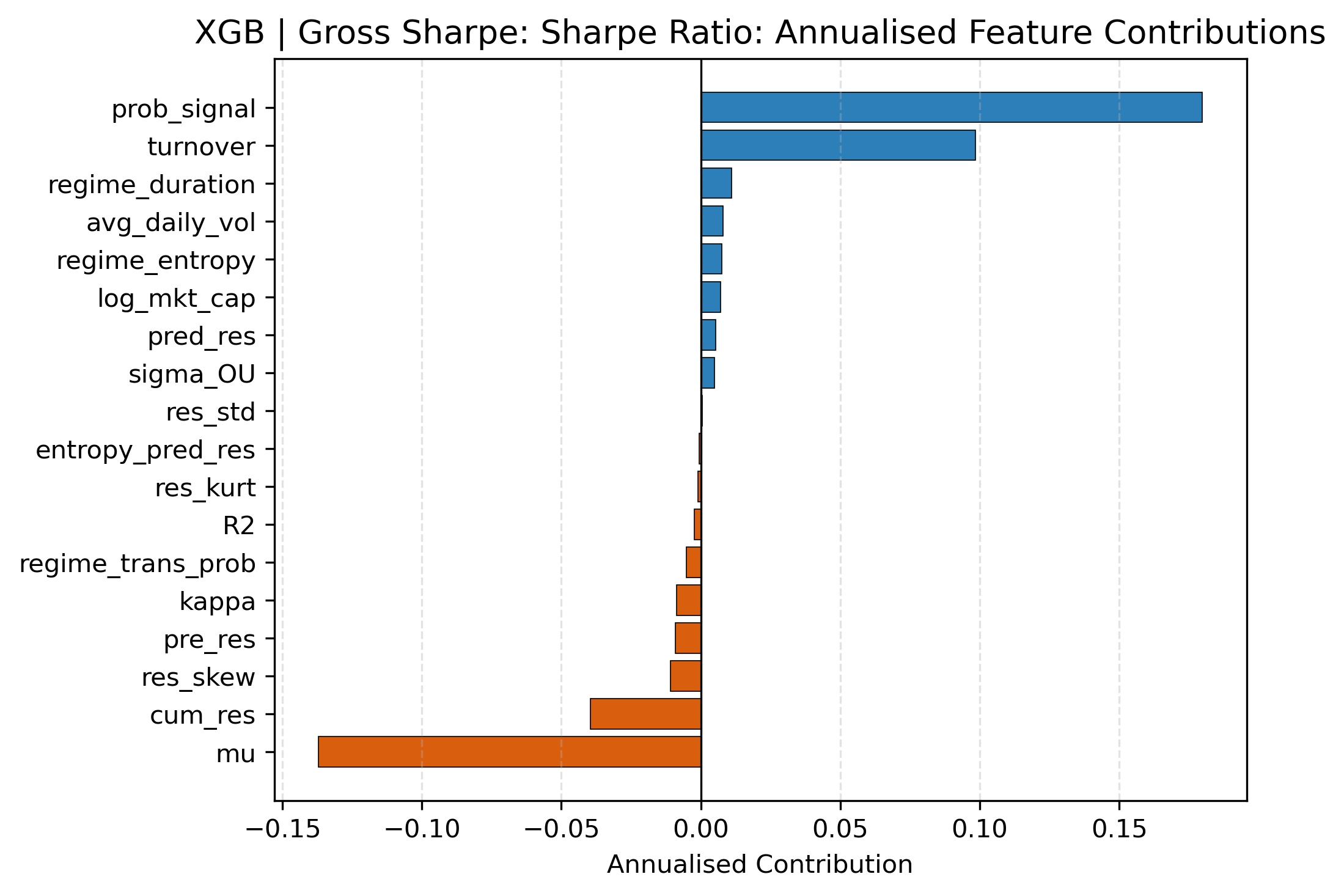}
			\small (e) XGB: Gross Sharpe-ratio attribution
			\label{fig:xgb_gross_sr_contrib}
		\end{minipage}
		\hfill
		\begin{minipage}[t]{0.49\linewidth}
			\centering
			\includegraphics[width=\linewidth]
			{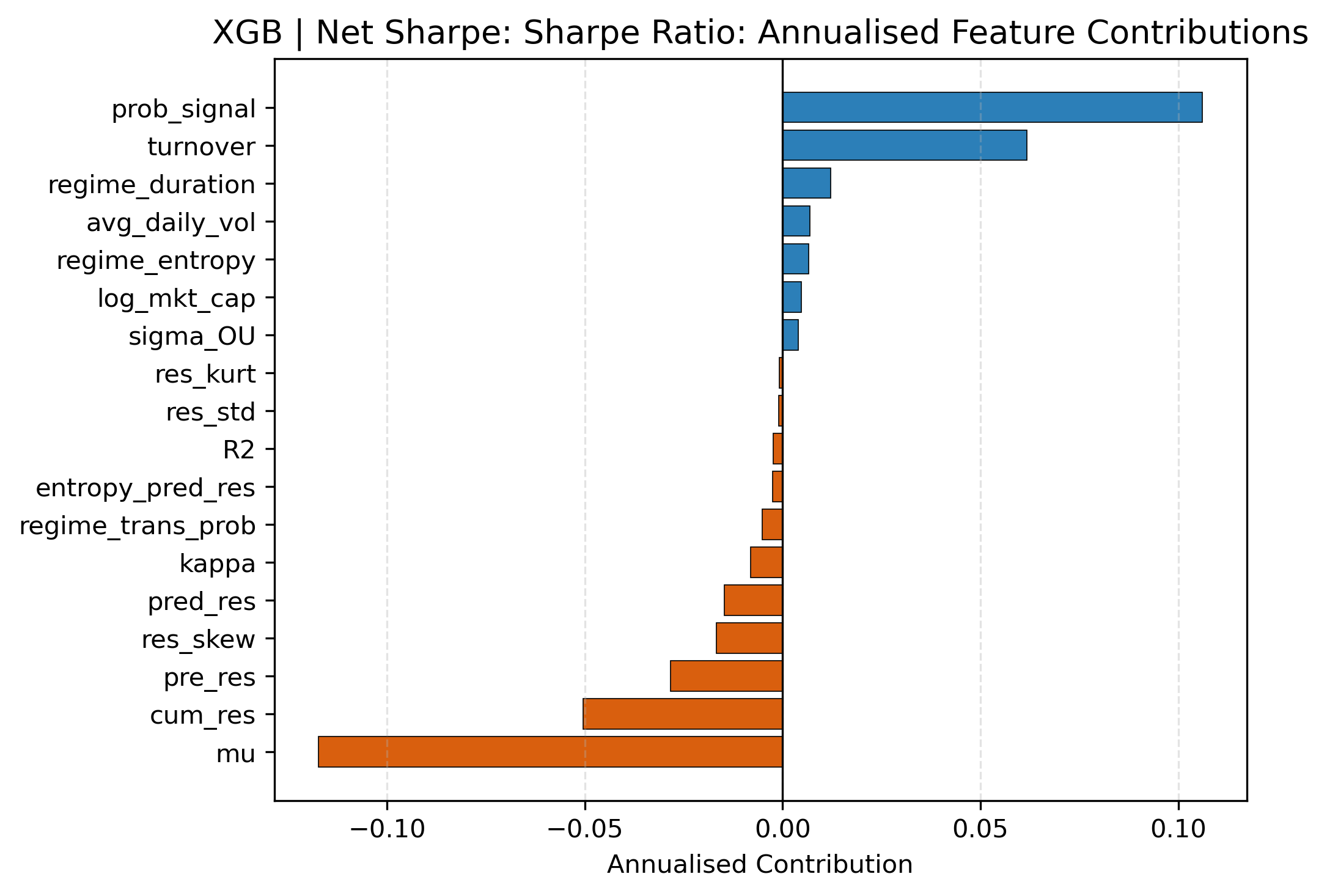}
			\small (f) XGB: Net Sharpe-ratio attribution
			\label{fig:xgb_net_sr_contrib}
		\end{minipage}
		
		\caption{
			Annualised feature contributions to XGB gross and net portfolio
			performance metrics. The left and right columns report gross and net
			attributions, respectively. From top to bottom, the rows report
			feature contributions to mean return, volatility, and the Sharpe ratio.
			Net performance incorporates transaction costs of 5 basis points per
			unit of portfolio turnover.
		}
		\label{fig:xgb_portfolio_metric_contrib}
		
	\end{figure}
	
	\begin{figure}[htbp]
		\centering
		
		% ============================================================
		% Row 1: Mean return
		% ============================================================
		\begin{minipage}[t]{0.49\linewidth}
			\centering
			\includegraphics[width=\linewidth]
			{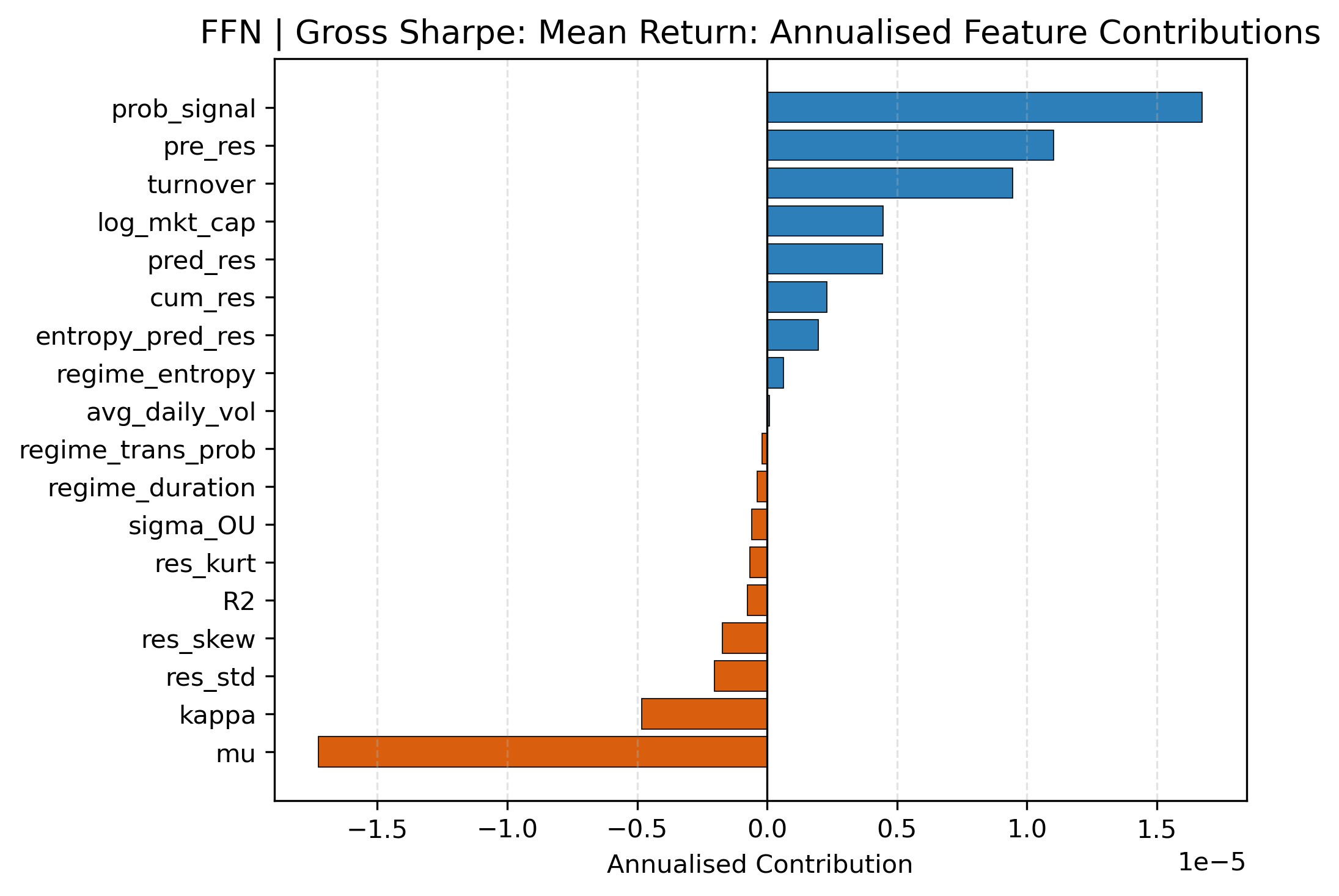}
			\small (a) FFN: Gross mean-return attribution
			\label{fig:ffn_gross_mu_contrib}
		\end{minipage}
		\hfill
		\begin{minipage}[t]{0.49\linewidth}
			\centering
			\includegraphics[width=\linewidth]
			{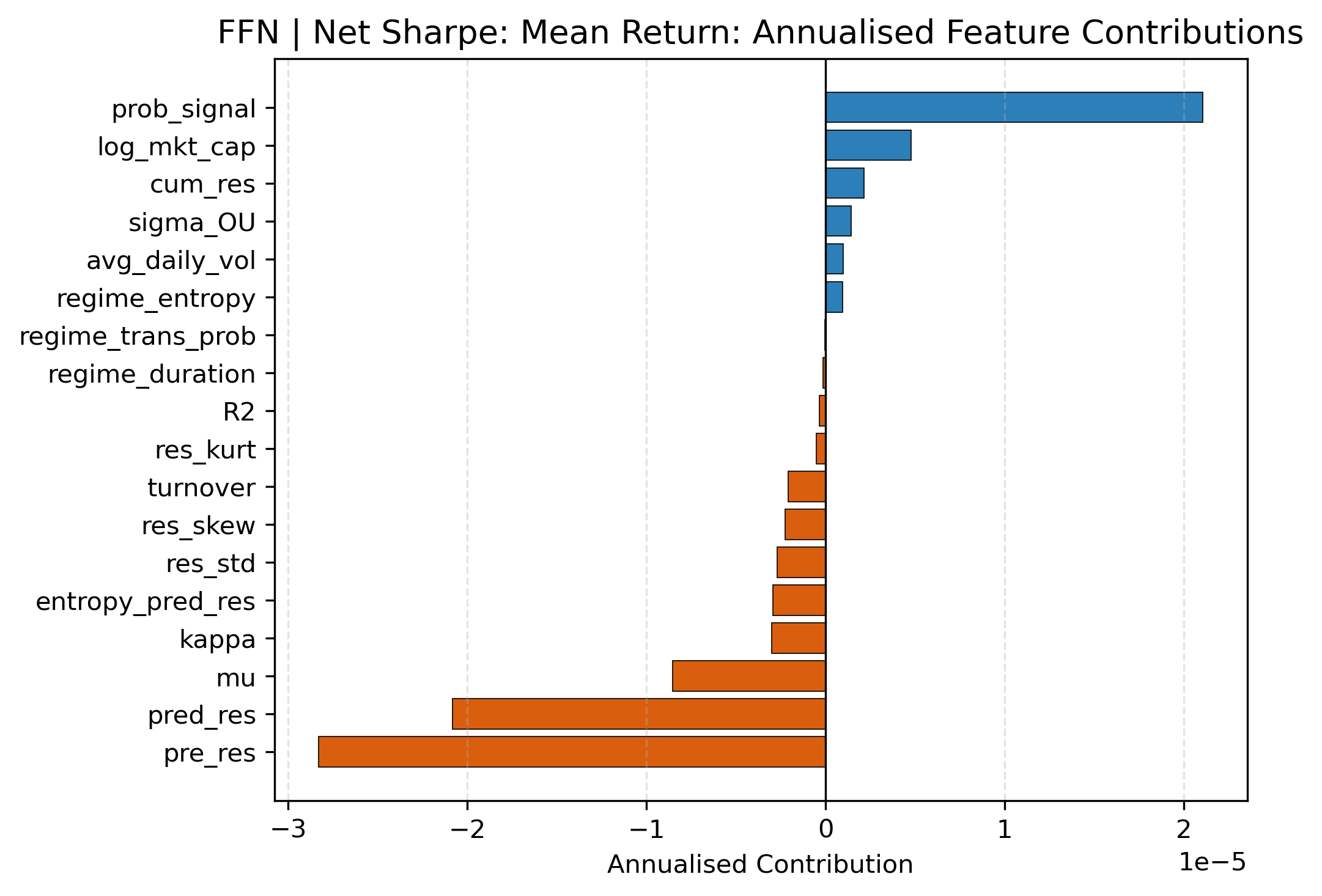}
			\small (b) FFN: Net mean-return attribution
			\label{fig:ffn_net_mu_contrib}
		\end{minipage}
		
		\vspace{0.5em}
		
		% ============================================================
		% Row 2: Volatility
		% ============================================================
		\begin{minipage}[t]{0.49\linewidth}
			\centering
			\includegraphics[width=\linewidth]
			{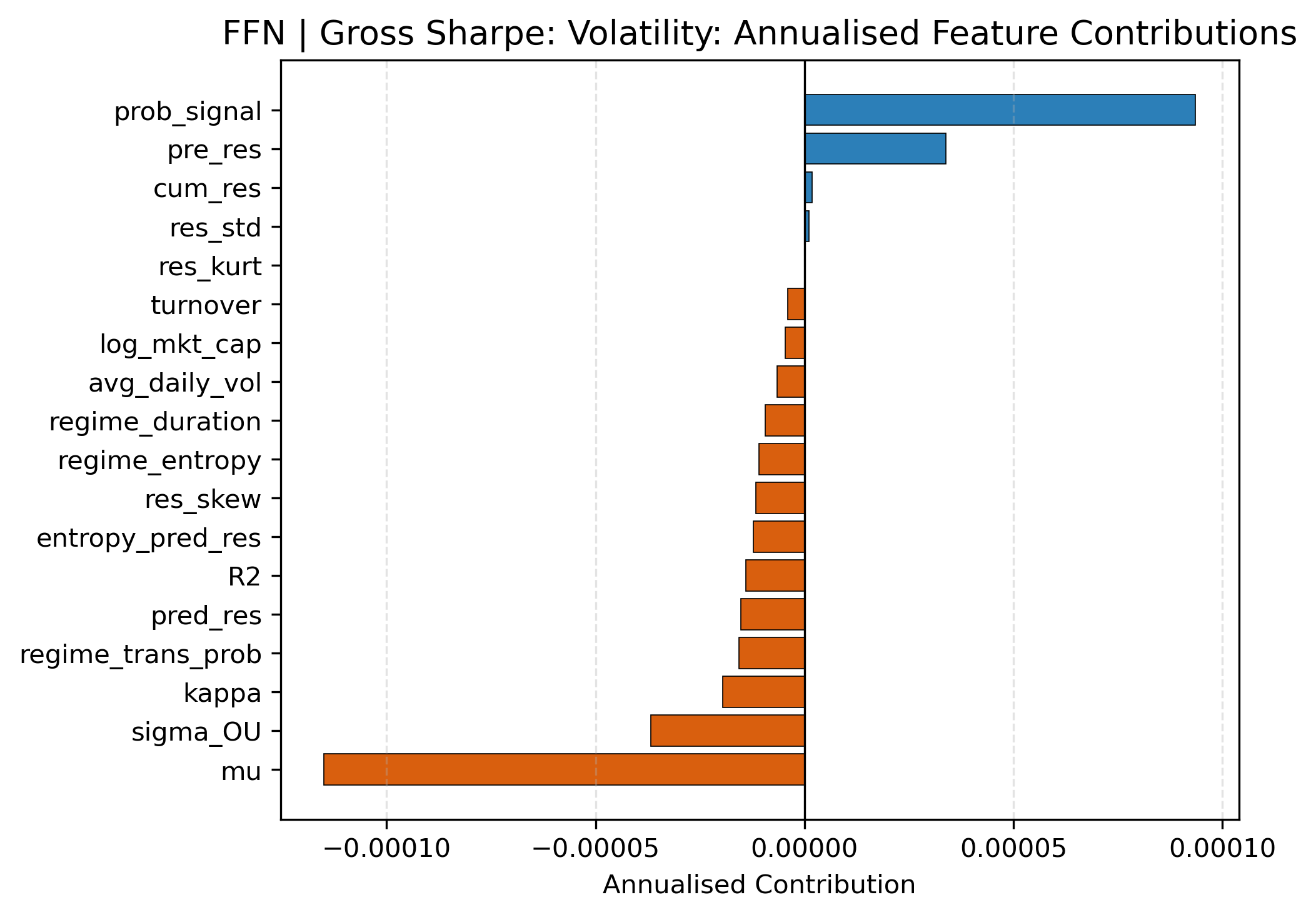}
			\small (c) FFN: Gross volatility attribution
			\label{fig:ffn_gross_sigma_contrib}
		\end{minipage}
		\hfill
		\begin{minipage}[t]{0.49\linewidth}
			\centering
			\includegraphics[width=\linewidth]
			{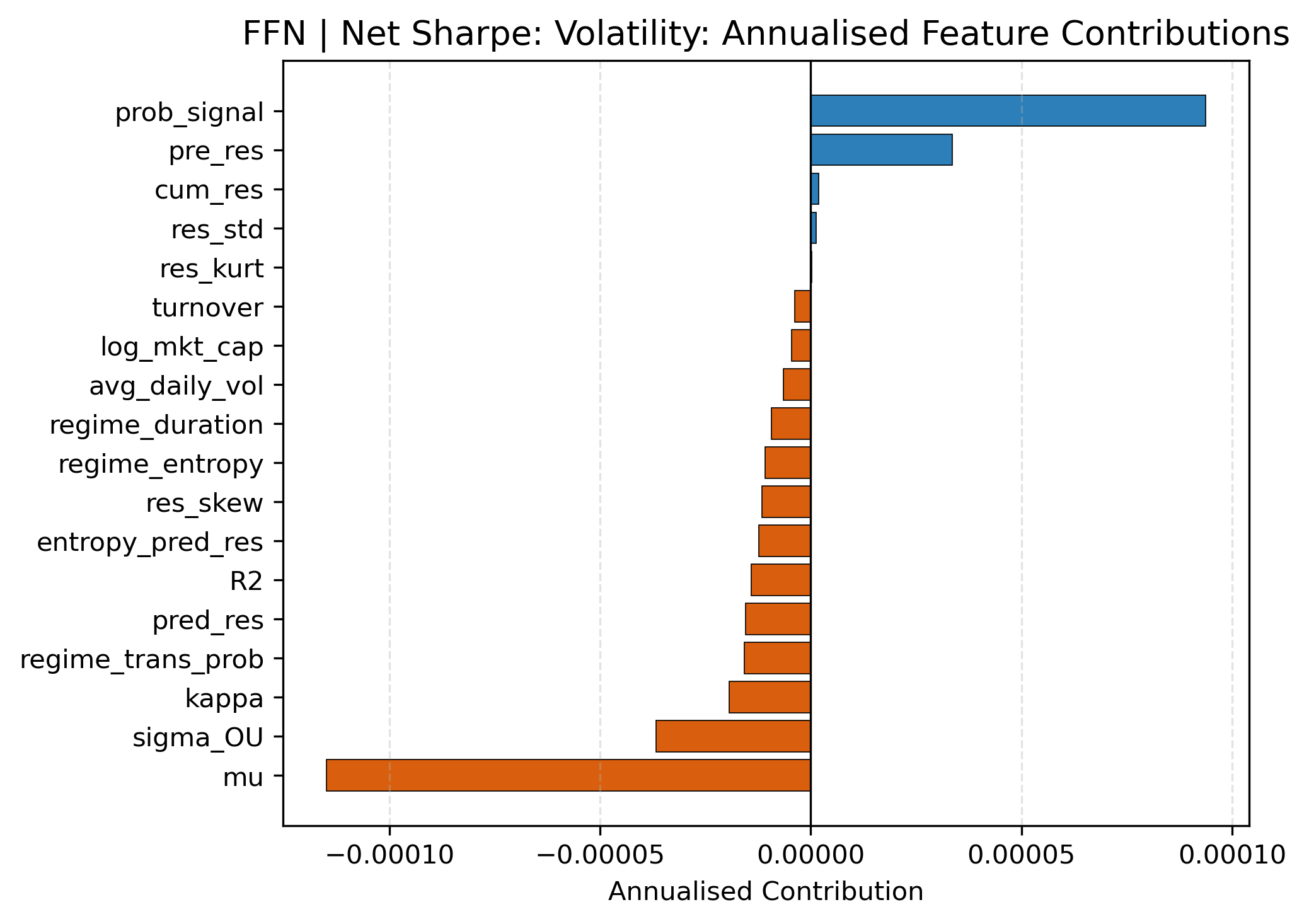}
			\small (d) FFN: Net volatility attribution
			\label{fig:ffn_net_sigma_contrib}
		\end{minipage}
		
		\vspace{0.5em}
		
		% ============================================================
		% Row 3: Sharpe ratio
		% ============================================================
		\begin{minipage}[t]{0.49\linewidth}
			\centering
			\includegraphics[width=\linewidth]
			{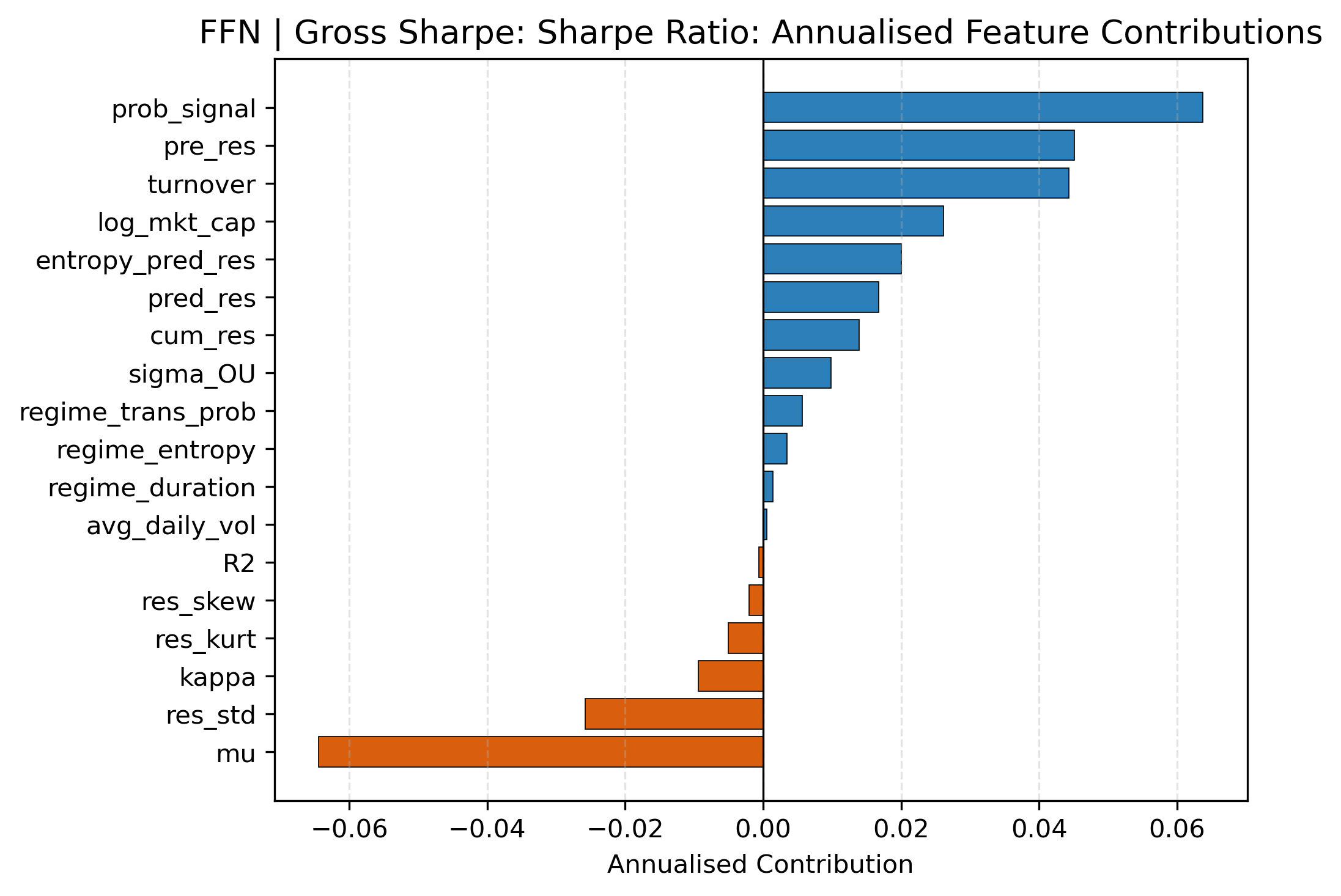}
			\small (e) FFN: Gross Sharpe-ratio attribution
			\label{fig:ffn_gross_sr_contrib}
		\end{minipage}
		\hfill
		\begin{minipage}[t]{0.49\linewidth}
			\centering
			\includegraphics[width=\linewidth]
			{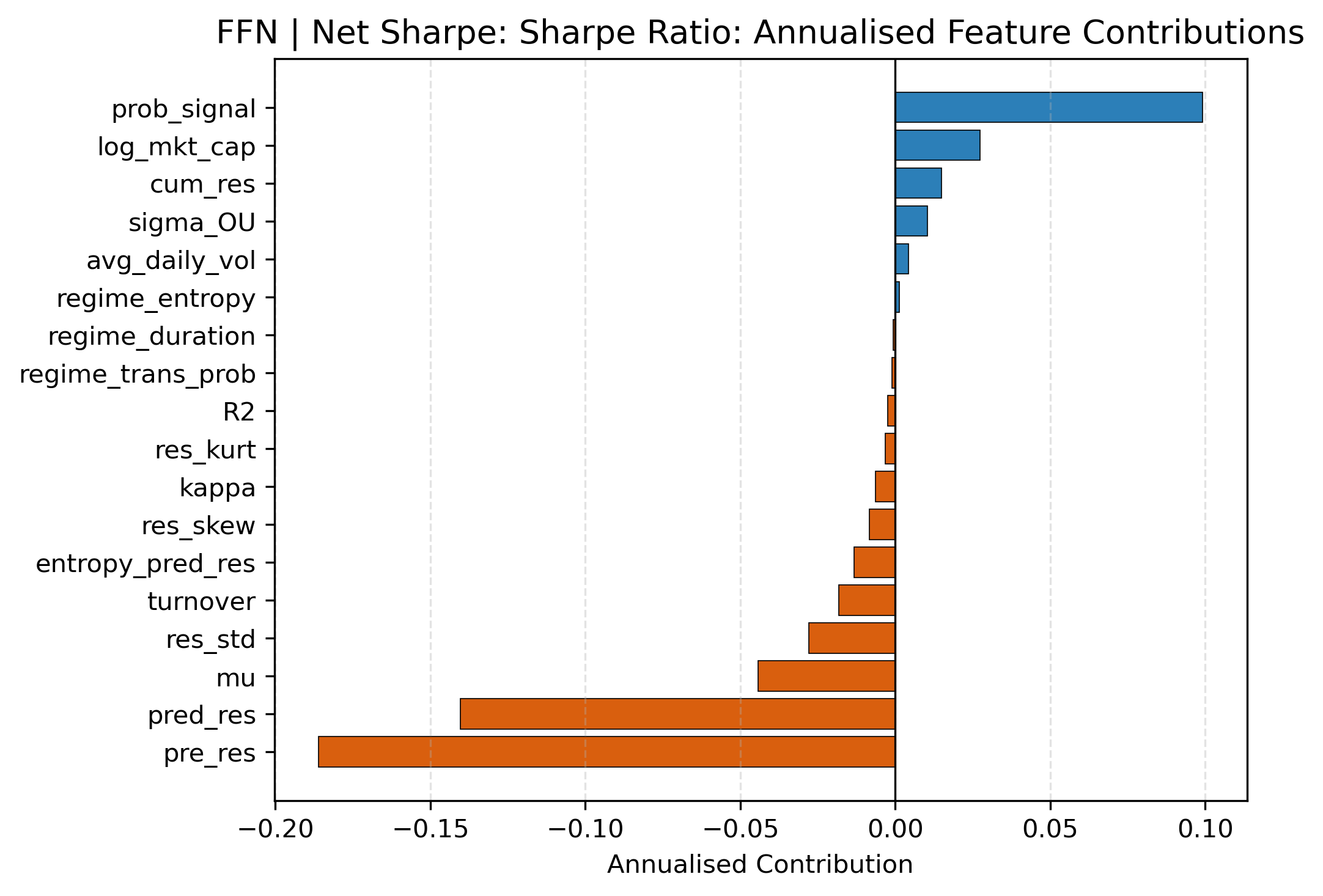}
			\small (f) FFN: Net Sharpe-ratio attribution
			\label{fig:ffn_net_sr_contrib}
		\end{minipage}
		
		\caption{
			Annualised feature contributions to FFN gross and net portfolio
			performance metrics. The left and right columns report gross and net
			attributions, respectively. From top to bottom, the rows report
			feature contributions to mean return, volatility, and the Sharpe ratio.
			Net performance incorporates transaction costs of 5 basis points per
			unit of portfolio turnover.
		}
		\label{fig:ffn_portfolio_metric_contrib}
		
	\end{figure}

	Figures~\ref{fig:xgb_portfolio_metric_contrib} and~\ref{fig:ffn_portfolio_metric_contrib} show the annualised feature contributions to portfolio gross and net performance metrics for XGB and FFN. The gross and net performance decompositions reveal a clear economic interpretation. For XGBoost, \texttt{mu} is the dominant negative contributor to both gross and net Sharpe, while \texttt{turnover} and \texttt{sigma\_OU} provide positive contributions, reflecting a lower-cost trading environment. For FFN, \texttt{pre\_res} and \texttt{pred\_res} contribute positively to gross Sharpe, but they become the most negative contributors to net Sharpe because their effect is accompanied by large turnover costs. In contrast, \texttt{prob\_signal} remains the strongest positive contributor to both gross and net Sharpe ratios.
	
	Moreover, the divergence between gross and net Sharpe ratio feature contributions is largely attributable to changes in mean-return contributions rather than volatility contributions, which remain relatively stable across the two specifications. This effect is especially evident for FFN, whose substantially higher turnover leads to a greater erosion of returns through transaction costs. Specifically, although 13 features contribute positively to the gross Sharpe ratio, only 6 features continue to contribute positively to the net Sharpe ratio. This result highlights the importance of evaluating feature contributions from a portfolio-performance perspective, as predictive importance alone does not necessarily translate into net economic value.

	\paragraph{\textbf{Gross-versus-net feature contribution dynamics}}
	\label{sec:gross_net_feature_dynamics}
	
\begin{figure}[htbp]
		\centering

		\begin{minipage}[t]{0.49\linewidth}
			\centering
			\includegraphics[width=\linewidth]{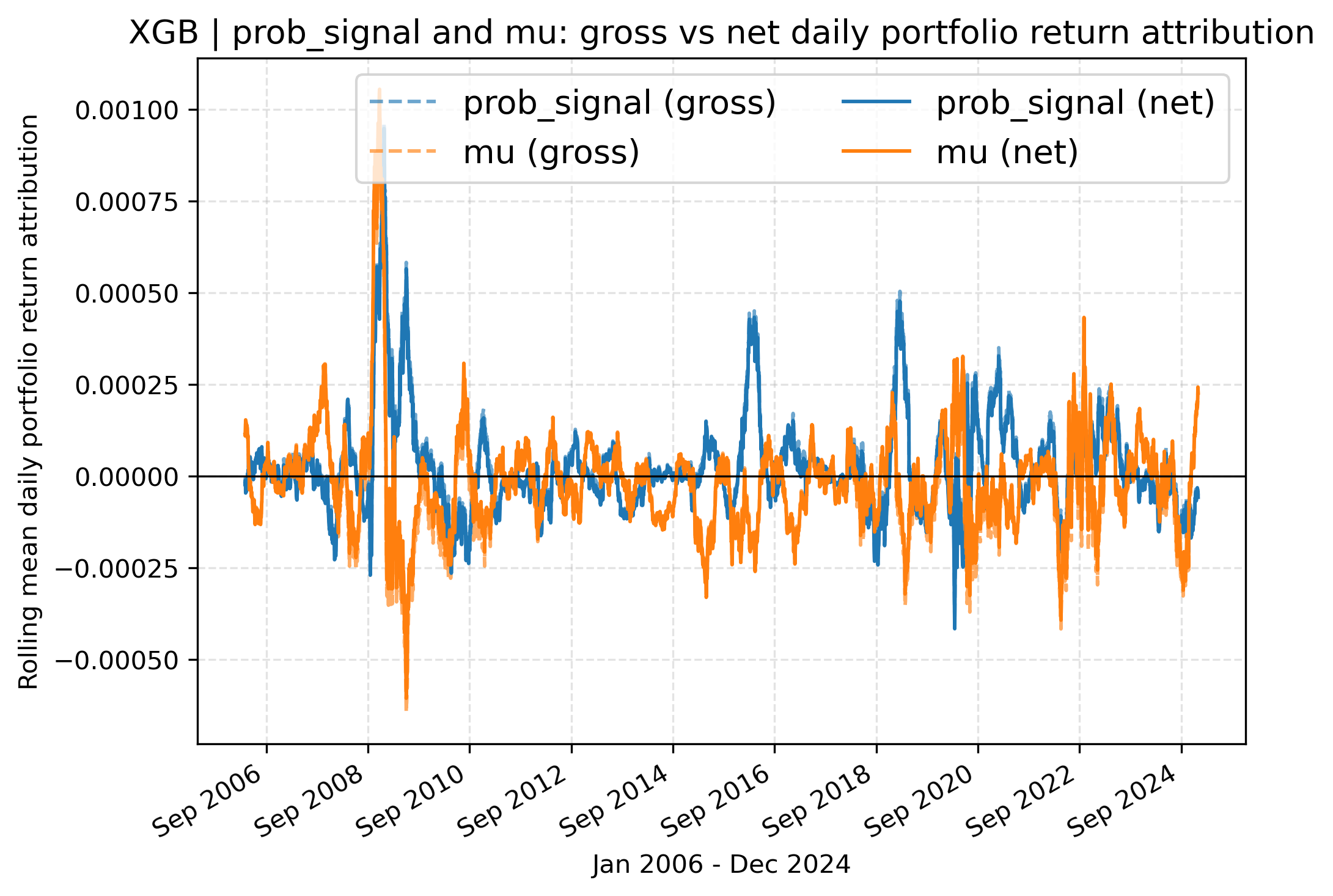}
			\small (a) XGB: \texttt{prob\_signal} and \texttt{mu}
			\label{fig:xgb_prob_mu_gross_net}
		\end{minipage}
		\hfill
		\begin{minipage}[t]{0.49\linewidth}
			\centering
			\includegraphics[width=\linewidth]{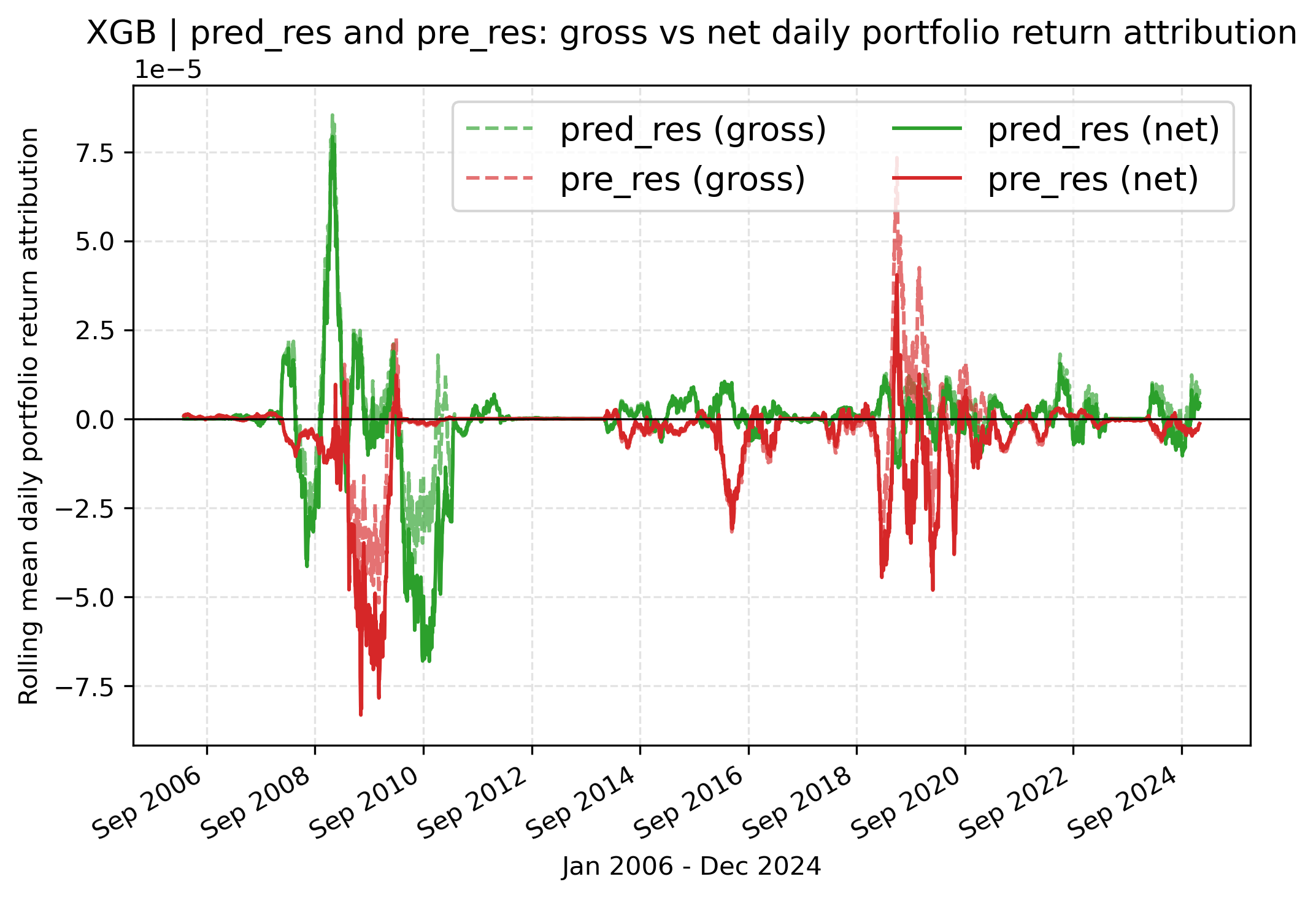}
			\small (b) XGB: \texttt{pred\_res} and \texttt{pre\_res}
			\label{fig:xgb_pred_pre_gross_net}
		\end{minipage}
		
		\vspace{2mm}

		\begin{minipage}[t]{0.49\linewidth}
			\centering
			\includegraphics[width=\linewidth]{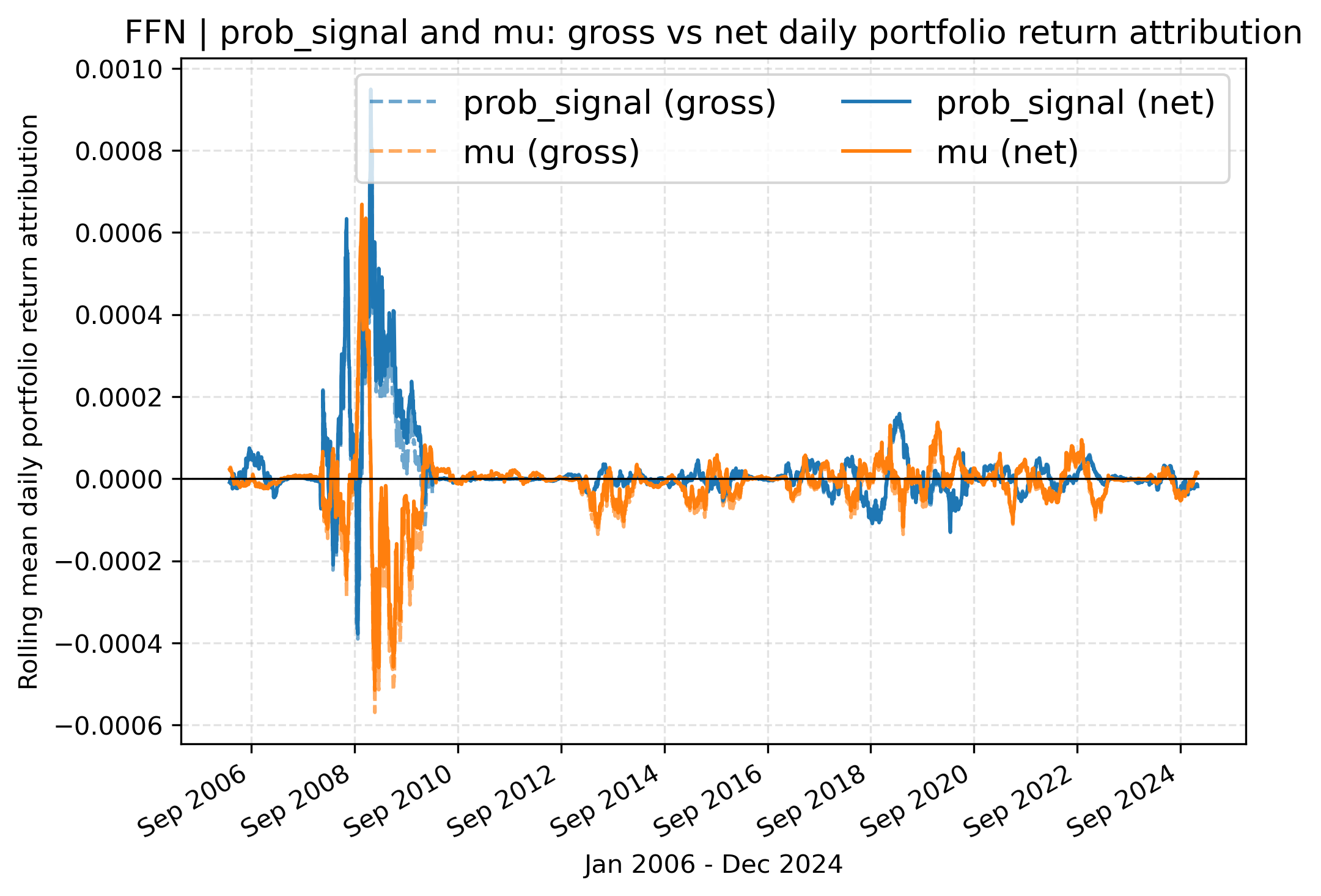}
			\small (c) FFN: \texttt{prob\_signal} and \texttt{mu}
			\label{fig:ffn_prob_mu_gross_net}
		\end{minipage}
		\hfill
		\begin{minipage}[t]{0.49\linewidth}
			\centering
			\includegraphics[width=\linewidth]{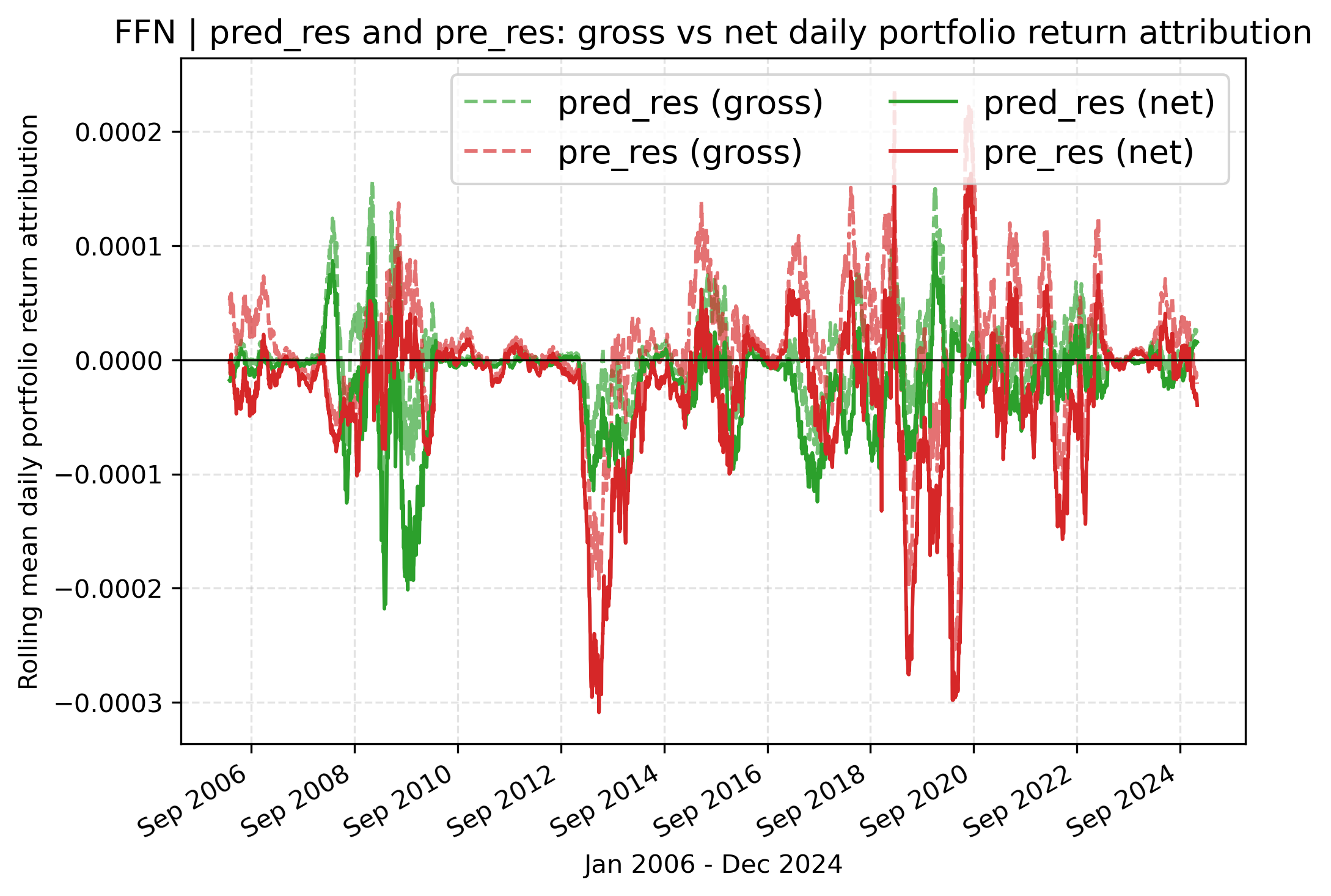}
			\small (d) FFN: \texttt{pred\_res} and \texttt{pre\_res}
			\label{fig:ffn_pred_pre_gross_net}
		\end{minipage}
		
		\caption{
			%			63-day rolling mean feature contributions to gross and net daily portfolio excess return components. Each panel overlays gross and net SHAP contribution paths for two representative features.
			63-day rolling mean gross and net daily portfolio excess return attribution time series for four representative features (\texttt{prob\_signal}, \texttt{mu}, \texttt{pred\_res}, and \texttt{pre\_res}) over 2006-2024. 
		}
		\label{fig:gross_net_dyn}
	\end{figure}

	Figure~\ref{fig:gross_net_dyn} compares the 63-day rolling feature contributions to gross and net daily portfolio excess return components over 2006-2024 for four representative features: \texttt{prob\_signal}, \texttt{mu}, \texttt{pred\_res}, and \texttt{pre\_res}. For both XGBoost and FFN, the contributions of \texttt{pre\_res} and \texttt{pred\_res} have strongly positive gross return attribution in several subperiods, but their net-return contributions are materially weaker and frequently turn negative after transaction costs are accounted for. This pattern indicates that these features generate economically meaningful gross performance but at the cost of substantially higher turnover. In contrast, \texttt{prob\_signal} and \texttt{mu} remains relatively stable between gross and net attribution, suggesting that signal-confidence information survives transaction-cost adjustment more effectively than residual-based features. Turnover is the key channel through which gross attribution is transformed into net attribution.
	
	\paragraph{\textbf{Statistical significance of Sharpe attribution}}
	
%	Table~\ref{tab:phi_significance}
%	summarises the statistical significance of the estimated local first-order Sharpe attributions for 18 features across the 19 out-of-sample years. The XGB and FFN specifications contain 278 and 342 available feature-year observations, respectively, since for XGBoost, not all the features are used by the trained model. The results indicate that the proposed local Sharpe attribution is not merely estimation noise. Across both allocation models, a nonnegligible number of feature-year pairs exhibit statistically significant positive or negative local effects on portfolio performance. On average, between 1.2 and 2.1 features are identified as having significant local Sharpe contributions in each out-of-sample year. Moreover, moving from gross to net Sharpe generally reduces the number of significantly positive attributions while increasing the number of significantly negative attributions, reflecting the additional impact of transaction costs on feature contributions.

		\begin{table}[htbp]
		\centering
		\caption{Significant local Sharpe attributions}
		\label{tab:phi_significance}
		\small
		
		\resizebox{\textwidth}{!}{
			%		\begin{tabular}{ccccccc}
				%			\toprule
				%			& \multicolumn{3}{c}{XGB}
				%			& \multicolumn{3}{c}{FFN} \\
				%			\cmidrule(lr){2-4}
				%			\cmidrule(lr){5-7}
				%			
				%			Metric
				%			& Positive & Negative & Total
				%			& Positive & Negative & Total \\
				%			\midrule
				%			
				%			Gross Sharpe ratio
				%			& 11 (4.0\%)
				%			& 14 (5.0\%)
				%			& 25 (9.0\%)
				%			& 19 (5.6\%)
				%			& 10 (2.9\%)
				%			& 29 (8.5\%) \\
				%			
				%			Net Sharpe ratio
				%			& 9 (3.2\%)
				%			& 15 (5.4\%)
				%			& 24 (8.6\%)
				%			& 14 (4.1\%)
				%			& 25 (7.3\%)
				%			& 39 (11.4\%) \\
				%			
				%			\bottomrule
				%		\end{tabular}
			
			\begin{tabular}{ccccccc}
				\toprule
				& \multicolumn{3}{c}{XGB}
				& \multicolumn{3}{c}{FFN} \\
				\cmidrule(lr){2-4}
				\cmidrule(lr){5-7}
				
				Metric
				& Positive & Negative & Total
				& Positive & Negative & Total \\
				\midrule
				
				Gross Sharpe ratio
				& 13 (4.7\%)
				& 15 (5.4\%)
				& 28 (10.1\%)
				& 16 (4.7\%)
				& 8 (2.3\%)
				& 24 (7.0\%) \\
				
				Net Sharpe ratio
				& 11 (4.0\%)
				& 15 (5.4\%)
				& 26 (9.4\%)
				& 10 (2.9\%)
				& 23 (6.7\%)
				& 33 (9.6\%) \\
				
				\bottomrule
			\end{tabular}
		}
		\vspace{2mm}
		
		\begin{minipage}{0.95\textwidth}
			\footnotesize
			\textit{Notes:}
			The table reports the number of feature-year pairs with significantly
			positive or negative local Sharpe-ratio attributions at the 5\% level,
			based on two-sided studentized circular block-bootstrap tests.
			The \textit{Total} column reports the combined number of significantly
			positive and negative feature-year pairs. Percentages relative to the
			total number of available feature-year observations are reported in
			parentheses. The XGB and FFN specifications contain 278 and 342
			available feature-year observations, respectively. Inference is based on 4,999 bootstrap replications using five-trading-day blocks to preserve short-run dependence in daily portfolio returns.
		\end{minipage}
	\end{table}
	
	Table~\ref{tab:phi_significance} summarises the statistical significance of the
	estimated local first-order Sharpe-ratio attributions for the 18 features across
	the 19 out-of-sample years. The XGB and FFN specifications contain 278 and 342
	available feature-year observations, respectively, as some features are not used
	by the fitted XGB models in particular estimation periods. The results provide
	evidence that the estimated local Sharpe attributions contain statistically
	meaningful variation rather than being driven solely by estimation noise. Across
	both allocation models, a non-negligible number of feature-year pairs exhibit
	significantly positive or negative contributions to the Sharpe ratio. On average,
	approximately 1.26 to 1.74 features per out-of-sample year have statistically
	significant local Sharpe contributions. Moreover, moving from gross to net Sharpe
	generally reduces the number of significantly positive attributions while
	increasing the number of significantly negative attributions, consistent with
	transaction costs altering the economic contribution of individual trading
	signals.

	\subsubsection{Alignment between the local first-order attributions and Shapley values}
	
%	The section is to show the reliability of our derived feature contribution to target metrics.
%	In this section, for each feature-year observation (278 for XGB and 342 for FFN), we compare the estimated shapley values with the corresponding main effects and our derived local first-order attributions for both gross and net Sharpe ratio. 
%	
%	The shapley value for each feature could be decomposed into its main effects and half of the interaction effects together with other features, i.e. $\psi_i(SR) = \psi_i^{main}(SR) + \frac{1}{2} \sum_{j: j\ne i} \psi_{i,j}(SR)$. If the main effects have a strong relationship with the shapley value, then the interaction effects are relatively small and the shapley value captures the key feature contribution directions for the finite differences over coalitions. In our feature universe, there are 18 main effects and 171 interaction effects to estimate. We use more than 500 coalitions to get the converged estimated shapley values and main effects. 
%	
%	According to section\ref{sec:attribution_relation_to_shapley}, the difference between real local first-order attribution and shapley value is mainly based on the higher-order effects. Thus, with small interaction effects, the estimated shapley value could be a method to show the reliability of the estimated local first-order attribution, indicating target metrics with respect to the feature exposures.

	This section assesses the reliability of the proposed local first-order
	attributions by comparing them with Shapley-based feature contributions. For
	each available feature-year observation, we
	compare the estimated Shapley value with the corresponding estimated main
	effect and the proposed local first-order attribution. The analysis is
	conducted separately for gross and net Sharpe ratios.
	
	The Shapley value of feature $i$ can be decomposed into its main effect and
	its allocated pairwise interaction effects, i.e.,
	$
	\psi_i(SR)
	=
	\psi_i^{\mathrm{main}}(SR)
	+
	\sum_{j\neq i}
	\psi_{ij}(SR).
	$
	Accordingly, a close alignment between the estimated main effects and the
	full Shapley values indicates that the interaction component does not
	substantially alter the feature-level contribution identified by the main
	effects. In particular, strong directional agreement suggests that the
	interaction terms generally refine the magnitude of the feature contributions
	rather than systematically reversing their signs. With 18 features, the
	decomposition involves 18 main effects and 153 distinct pairwise interaction
	effects. We estimate the Shapley values and main effects using more than 500
	sampled feature coalitions to obtain stable estimates.
	
	As discussed in Section~\ref{sec:attribution_relation_to_shapley}, the
	difference between the proposed local first-order attribution and the Shapley
	value arises from their treatment of nonlinear and higher-order effects. The
	local first-order attribution characterises the sensitivity of the target
	portfolio metric to feature exposure around the realised allocation strategy,
	whereas the Shapley value measures the average marginal contribution of a
	feature across feature coalitions. When interaction and higher-order effects
	do not materially alter the contribution direction, the Shapley estimates
	therefore provide a useful external benchmark for assessing whether our estimated local
	first-order attribution captures economically consistent feature-level
	information.

	\paragraph{\textbf{Estimated Sharpe ratio Shapley values and main effects}}
	\label{sec:main_interaction}

\begin{figure}[htbp]
		\centering
		
		% ============================================================
		% Row 1: Shapley vs Main Effects
		% ============================================================
		\begin{minipage}[t]{0.24\textwidth}
			\centering
			\includegraphics[width=\textwidth]
			{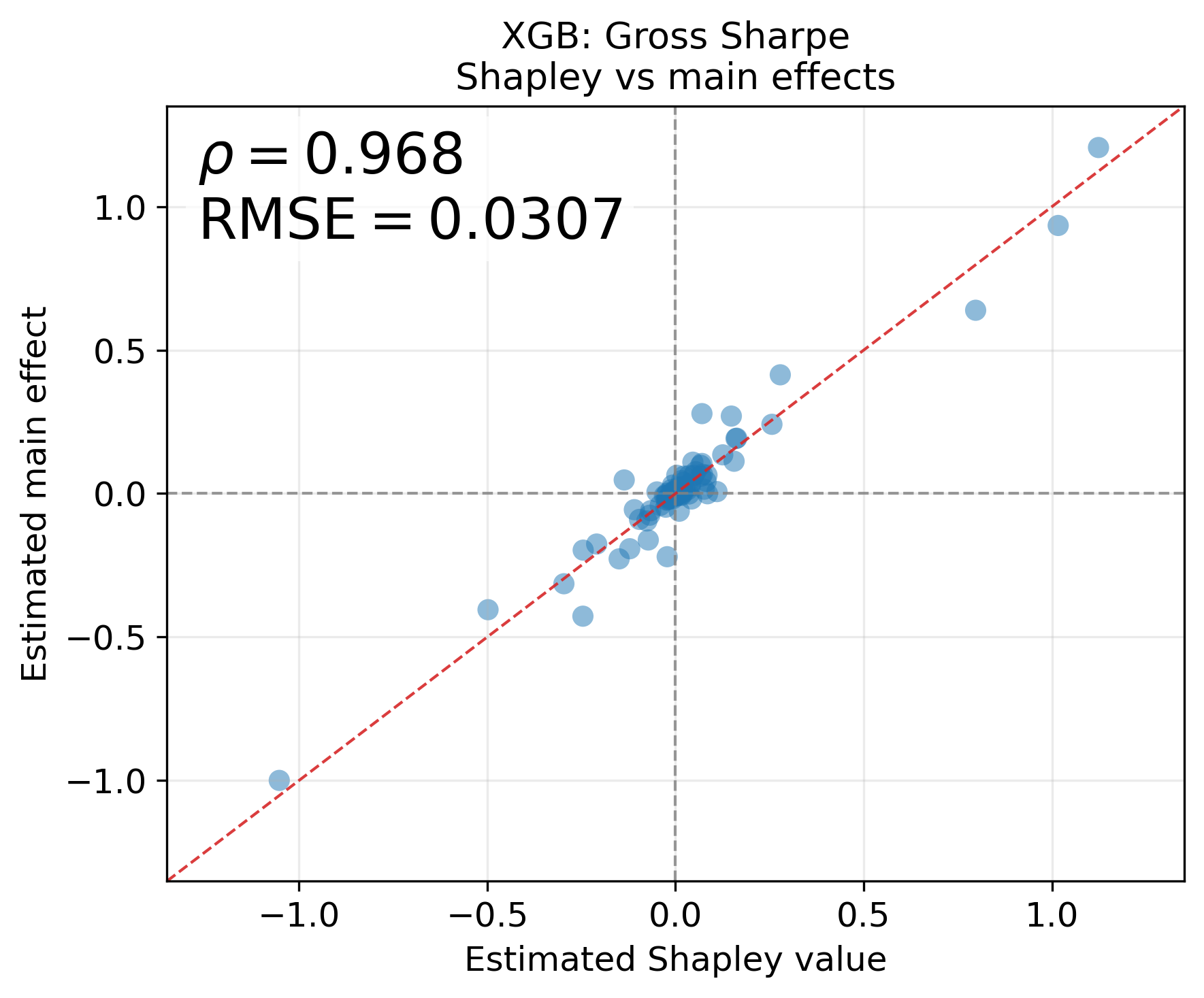}
			\small (a) XGB: Gross Sharpe
			\label{fig:xgb_gross_main_shapley}
		\end{minipage}
		\hfill
		\begin{minipage}[t]{0.24\textwidth}
			\centering
			\includegraphics[width=\textwidth]
			{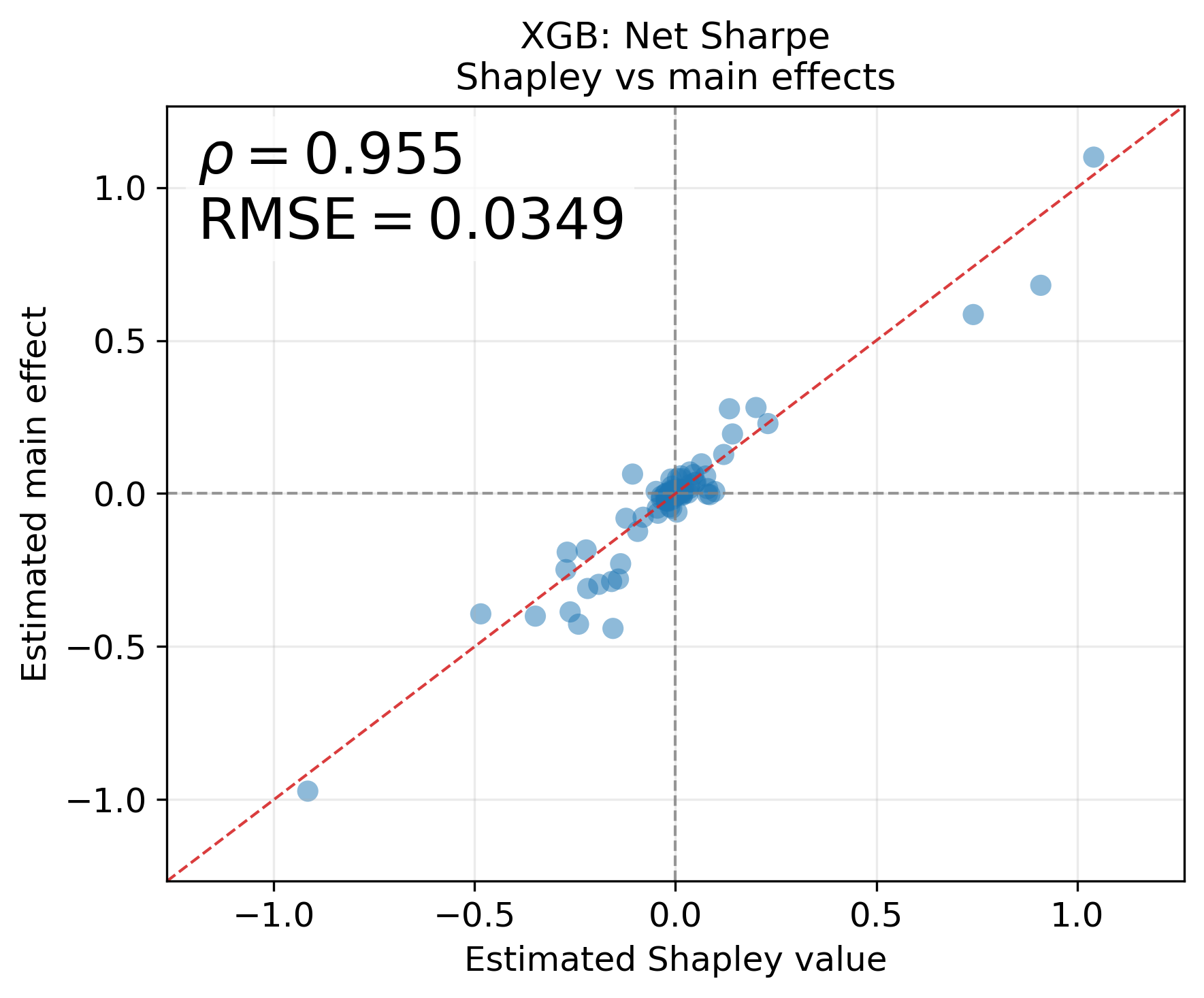}
			\small (b) XGB: Net Sharpe
			\label{fig:xgb_net_main_shapley}
		\end{minipage}
		\hfill
		\begin{minipage}[t]{0.24\textwidth}
			\centering
			\includegraphics[width=\textwidth]
			{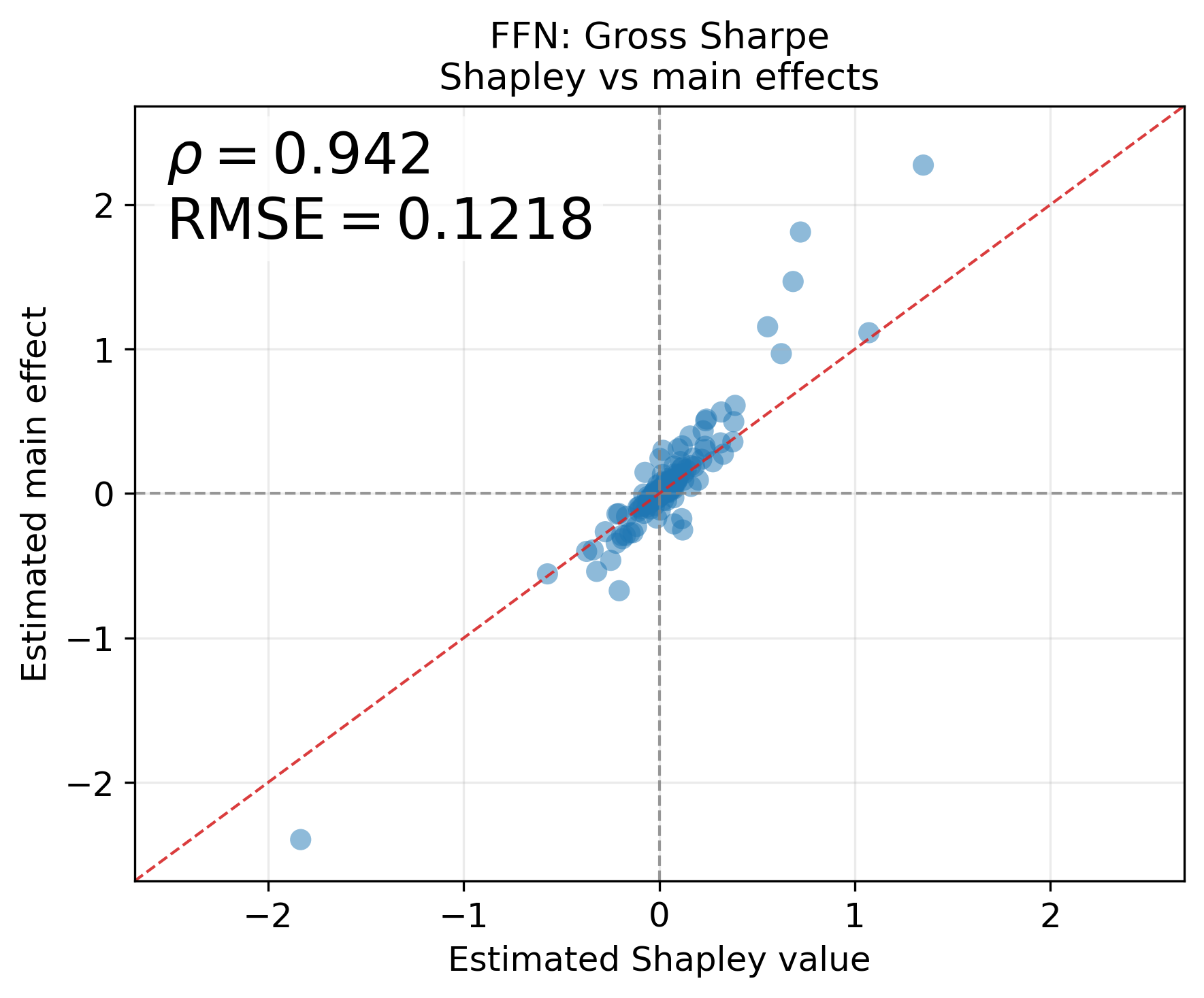}
			\small (c) FFN: Gross Sharpe
			\label{fig:ffn_gross_main_shapley}
		\end{minipage}
		\hfill
		\begin{minipage}[t]{0.24\textwidth}
			\centering
			\includegraphics[width=\textwidth]
			{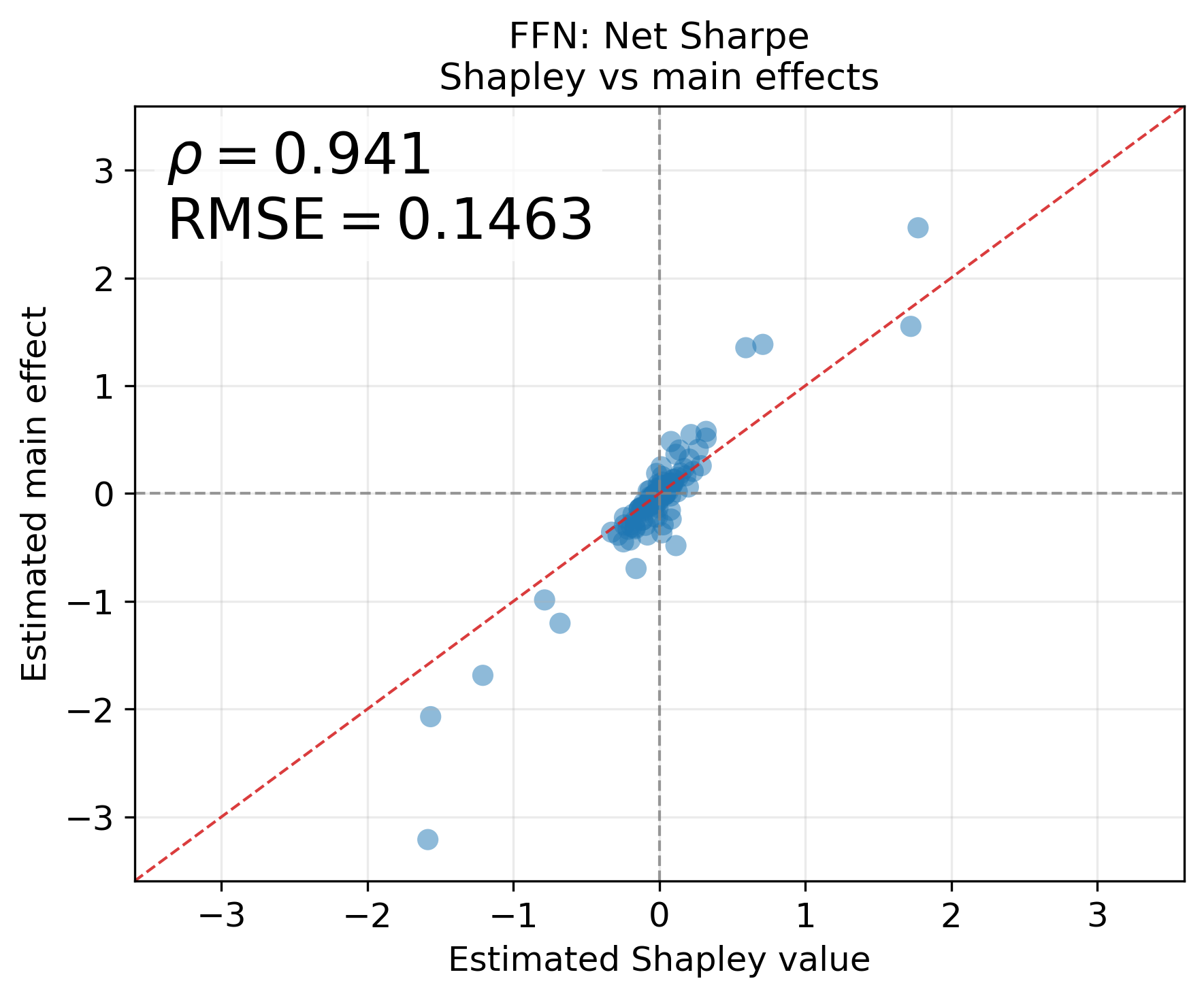}
			\small (d) FFN: Net Sharpe
			\label{fig:ffn_net_main_shapley}
		\end{minipage}
		
		%    	\caption*{\small Estimated Shapley values versus estimated main effects.}
		
		%    	\vspace{0.6em}
		%    	
		%    	% ============================================================
		%    	% Row 2: Shapley vs Local First-Order Attribution
		%    	% ============================================================
		%    	\begin{minipage}[t]{0.24\textwidth}
			%    		\centering
			%    		\includegraphics[width=\textwidth]
			%    		{figures/shapley/xgb_gross_SR_ours_vs_shapley.png}
			%    		\small (a) XGB: Gross Sharpe
			%    		\label{fig:xgb_gross_firstorder_shapley}
			%    	\end{minipage}
		%    	\hfill
		%    	\begin{minipage}[t]{0.24\textwidth}
			%    		\centering
			%    		\includegraphics[width=\textwidth]
			%    		{figures/shapley/xgb_net_SR_ours_vs_shapley.png}
			%    		\small (b) XGB: Net Sharpe
			%    		\label{fig:xgb_net_firstorder_shapley}
			%    	\end{minipage}
		%    	\hfill
		%    	\begin{minipage}[t]{0.24\textwidth}
			%    		\centering
			%    		\includegraphics[width=\textwidth]
			%    		{figures/shapley/ffn_gross_SR_ours_vs_shapley.png}
			%    		\small (c) FFN: Gross Sharpe
			%    		\label{fig:ffn_gross_firstorder_shapley}
			%    	\end{minipage}
		%    	\hfill
		%    	\begin{minipage}[t]{0.24\textwidth}
			%    		\centering
			%    		\includegraphics[width=\textwidth]
			%    		{figures/shapley/ffn_net_SR_ours_vs_shapley.png}
			%    		\small (d) FFN: Net Sharpe
			%    		\label{fig:ffn_net_firstorder_shapley}
			%    	\end{minipage}
		%    	
		%    	\caption*{\small Estimated Shapley values versus local first-order Sharpe attributions.}
		
		\caption{ 
			Comparison of estimated Shapley values with estimated main effects.
		}
		\label{fig:shapley_comparison_main}
	\end{figure}

	Figure~\ref{fig:shapley_comparison_main} compares the estimated Shapley values
	with the corresponding estimated main effects. Across the 278 available XGB
	and 342 FFN feature-year observations, the main effects exhibit a strong
	positive association with the Shapley values for both gross and net Sharpe
	ratios. The correlations exceed 0.94 in all four model-metric combinations.
	This strong alignment indicates that the main effects preserve most of the
	directional information contained in the full Shapley values. In particular,
	the interaction component does not appear to systematically reverse the signs
	of the estimated feature contributions. The interaction effects may
	nevertheless remain relevant for the magnitude of the contributions. For FFN,
	the estimated main effects tend to be larger in magnitude than the
	corresponding Shapley values, while retaining strong directional agreement.
	
	\paragraph{\textbf{Local first-order Sharpe ratio attributions and Shapley values}}
	
	\begin{figure}[htbp]
		\centering
		
		%    	% ============================================================
		%    	% Row 1: Shapley vs Main Effects
		%    	% ============================================================
		%    	\begin{minipage}[t]{0.24\textwidth}
			%    		\centering
			%    		\includegraphics[width=\textwidth]
			%    		{figures/shapley/xgb_gross_SR_main_vs_shapley.png}
			%    		\small (a) XGB: Gross Sharpe
			%    		\label{fig:xgb_gross_main_shapley}
			%    	\end{minipage}
		%    	\hfill
		%    	\begin{minipage}[t]{0.24\textwidth}
			%    		\centering
			%    		\includegraphics[width=\textwidth]
			%    		{figures/shapley/xgb_net_SR_main_vs_shapley.png}
			%    		\small (b) XGB: Net Sharpe
			%    		\label{fig:xgb_net_main_shapley}
			%    	\end{minipage}
		%    	\hfill
		%    	\begin{minipage}[t]{0.24\textwidth}
			%    		\centering
			%    		\includegraphics[width=\textwidth]
			%    		{figures/shapley/ffn_gross_SR_main_vs_shapley.png}
			%    		\small (c) FFN: Gross Sharpe
			%    		\label{fig:ffn_gross_main_shapley}
			%    	\end{minipage}
		%    	\hfill
		%    	\begin{minipage}[t]{0.24\textwidth}
			%    		\centering
			%    		\includegraphics[width=\textwidth]
			%    		{figures/shapley/ffn_net_SR_main_vs_shapley.png}
			%    		\small (d) FFN: Net Sharpe
			%    		\label{fig:ffn_net_main_shapley}
			%    	\end{minipage}
		%    	
		%    	\caption*{\small Estimated Shapley values versus estimated main effects.}
		%    	
		%    	\vspace{0.6em}
		
		% ============================================================
		% Row 2: Shapley vs Local First-Order Attribution
		% ============================================================
		\begin{minipage}[t]{0.24\textwidth}
			\centering
			\includegraphics[width=\textwidth]
			{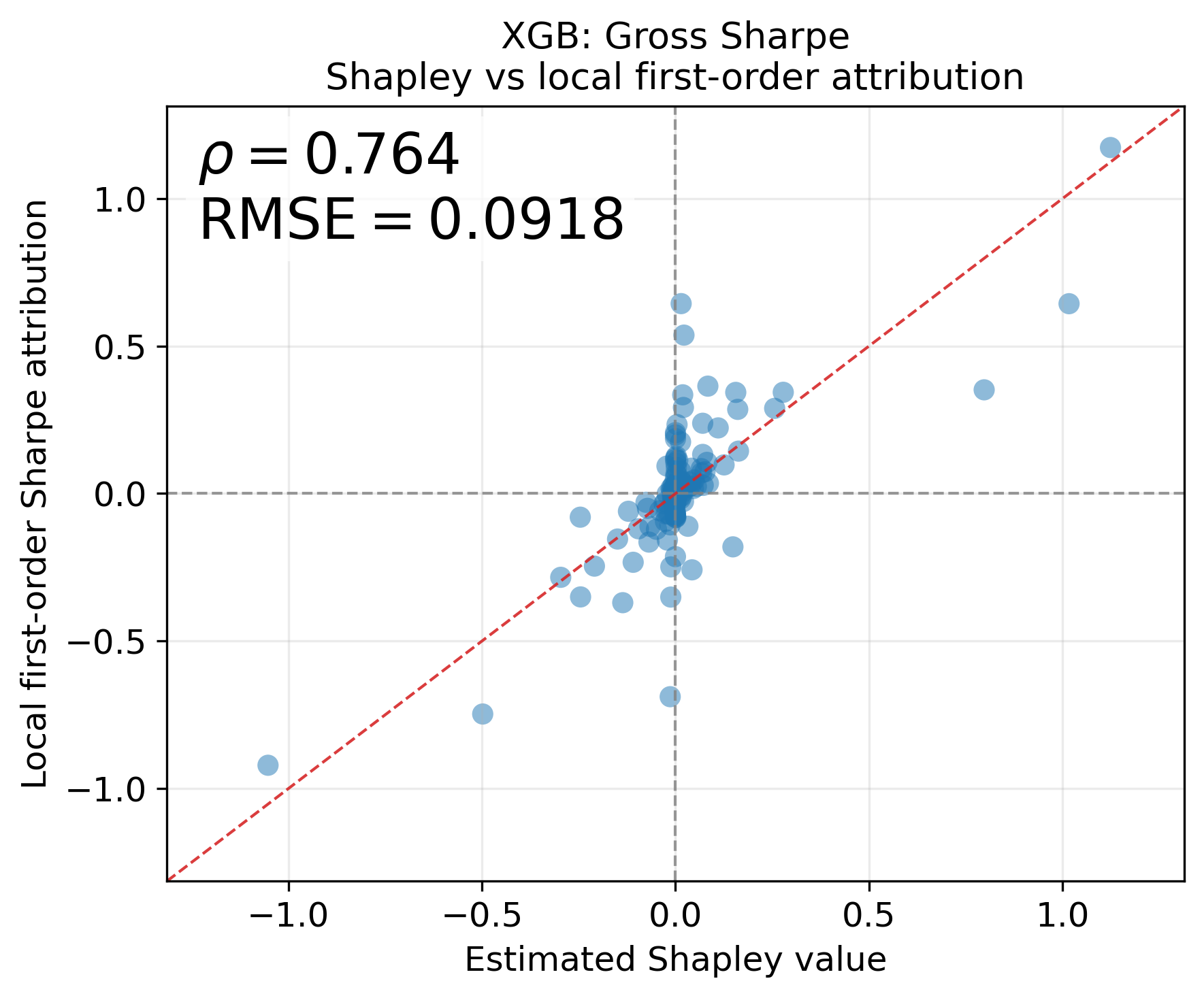}
			\small (a) XGB: Gross Sharpe
			\label{fig:xgb_gross_firstorder_shapley}
		\end{minipage}
		\hfill
		\begin{minipage}[t]{0.24\textwidth}
			\centering
			\includegraphics[width=\textwidth]
			{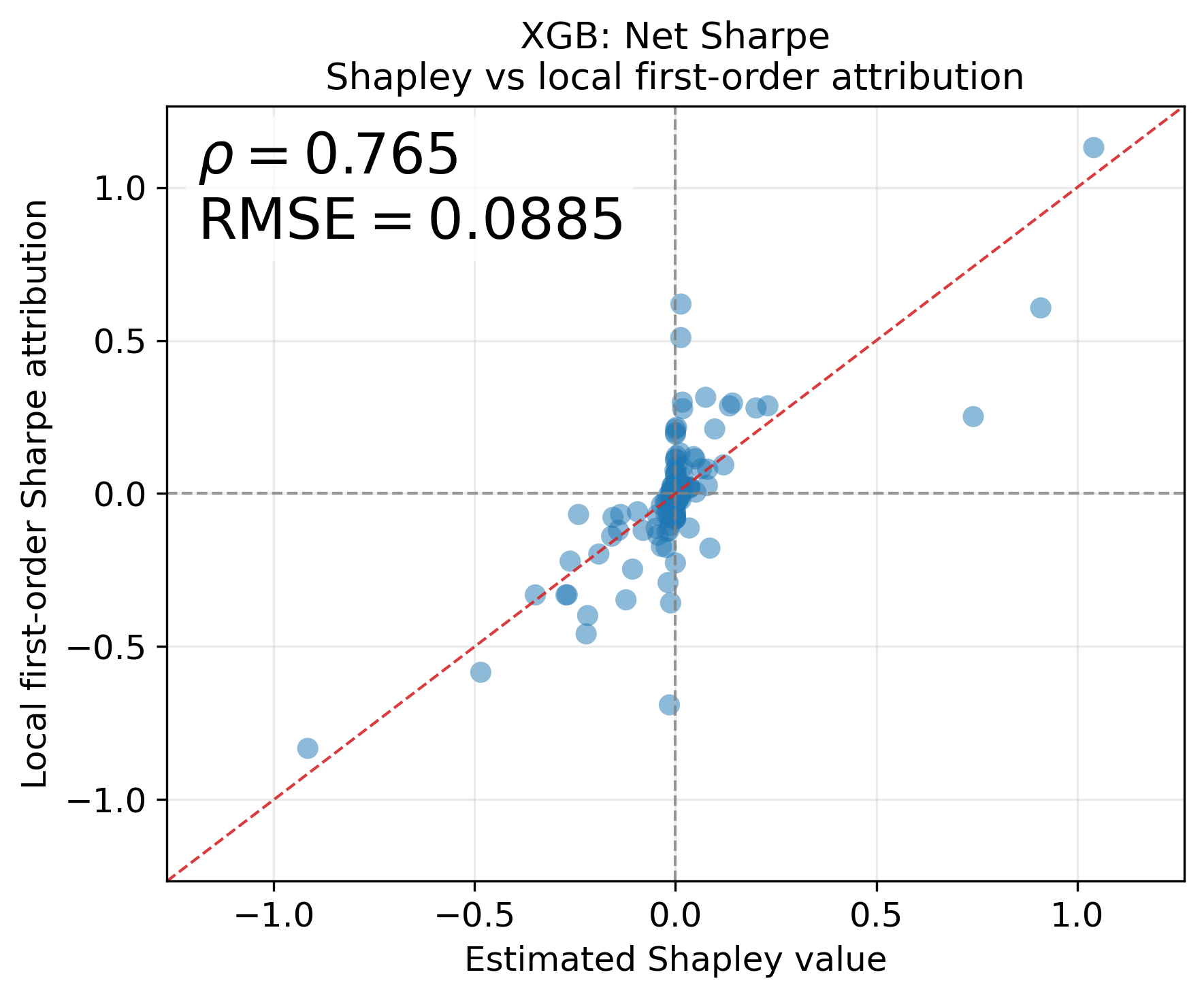}
			\small (b) XGB: Net Sharpe
			\label{fig:xgb_net_firstorder_shapley}
		\end{minipage}
		\hfill
		\begin{minipage}[t]{0.24\textwidth}
			\centering
			\includegraphics[width=\textwidth]
			{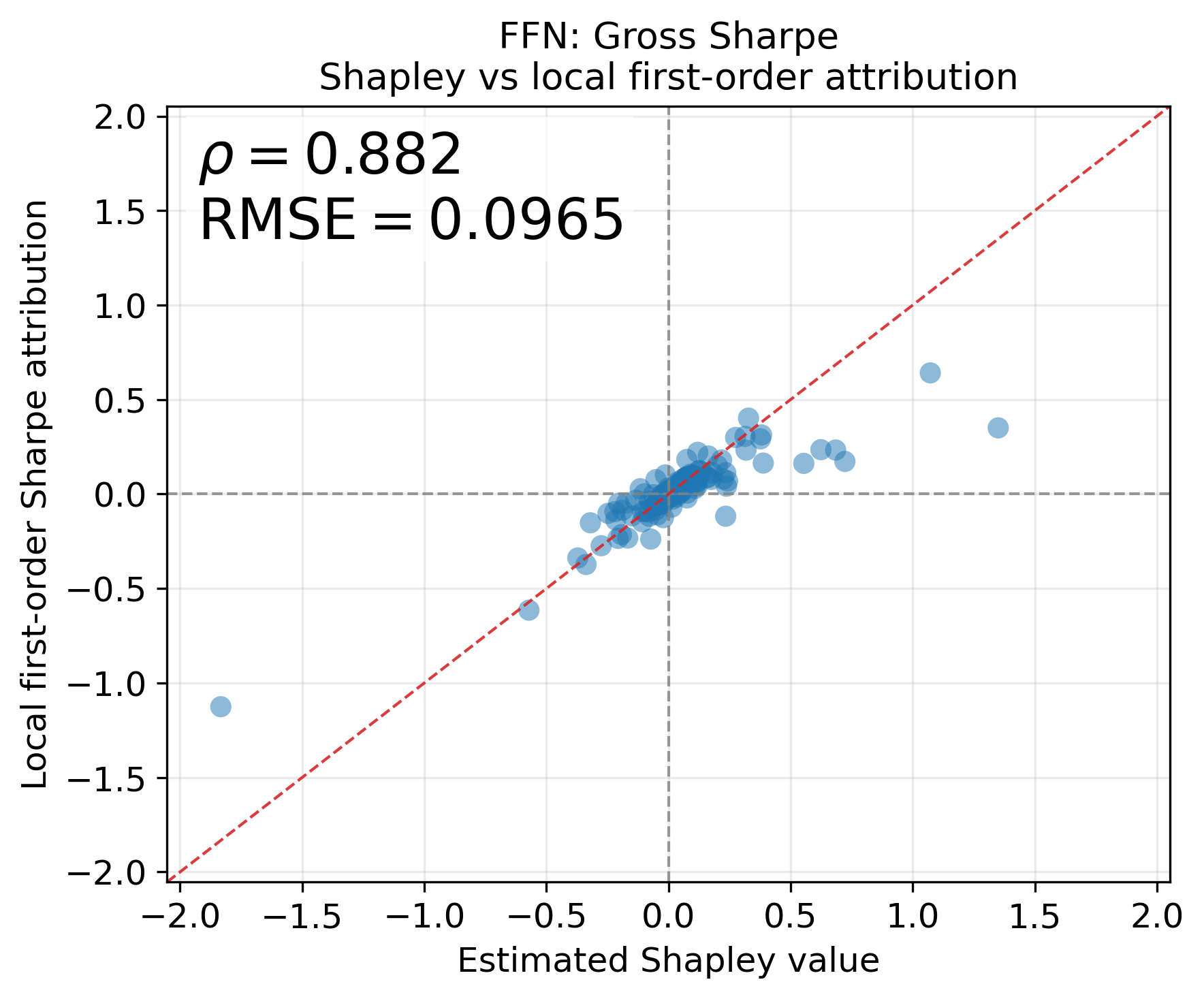}
			\small (c) FFN: Gross Sharpe
			\label{fig:ffn_gross_firstorder_shapley}
		\end{minipage}
		\hfill
		\begin{minipage}[t]{0.24\textwidth}
			\centering
			\includegraphics[width=\textwidth]
			{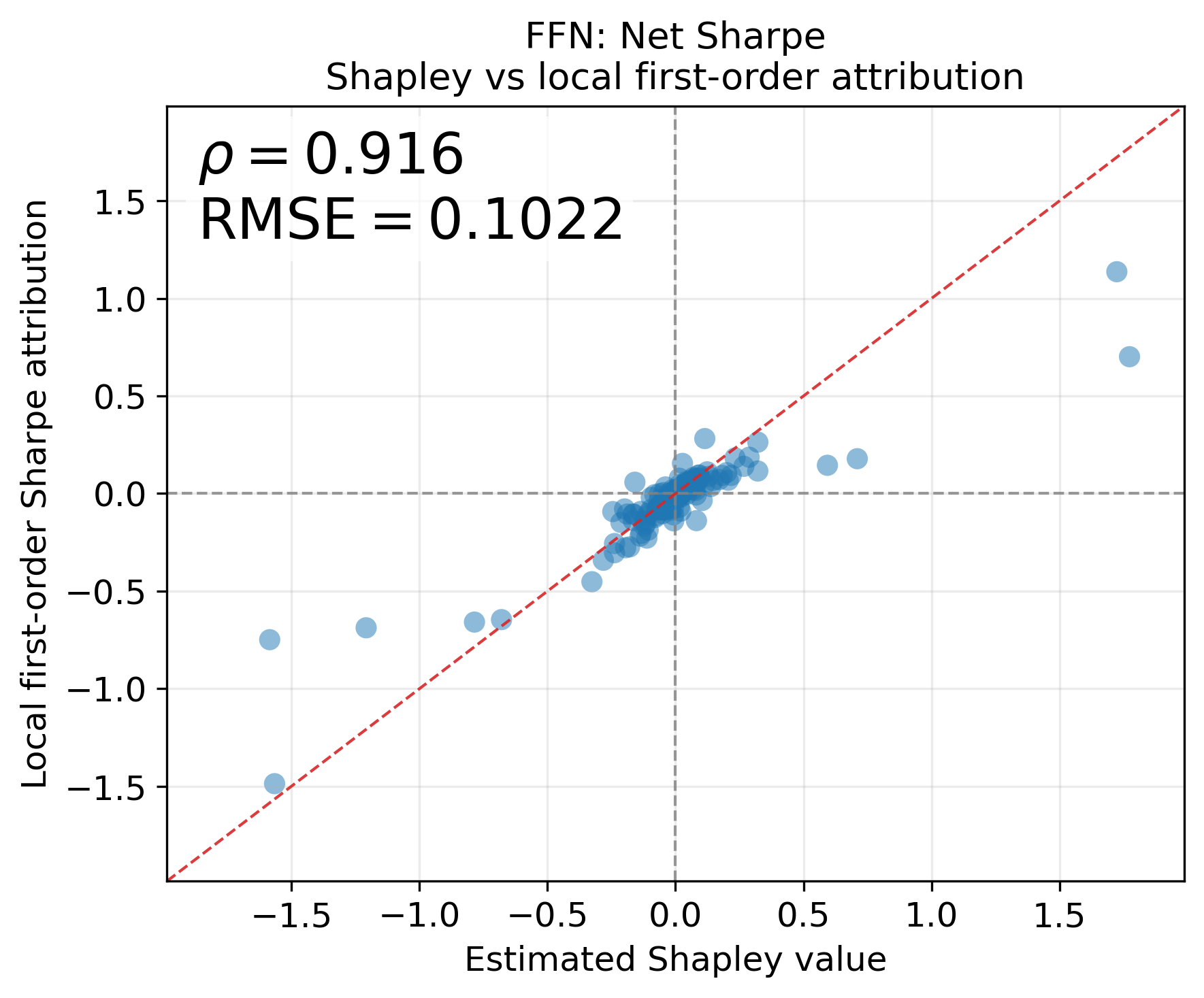}
			\small (d) FFN: Net Sharpe
			\label{fig:ffn_net_firstorder_shapley}
		\end{minipage}
		
		%    	\caption*{\small Estimated Shapley values versus local first-order Sharpe attributions.}
		
		\caption{
			Comparison of estimated Shapley values with local first-order attributions.
		}
		\label{fig:shapley_comparison_phi}
	\end{figure}

	Figure~\ref{fig:shapley_comparison_phi} compares the estimated Shapley values
	with the corresponding local first-order attributions. Across the 278 available
	XGB and 342 FFN feature-year observations, the proposed attributions exhibit
	a strong positive association with the Shapley values for both gross and net
	Sharpe ratios. The correlations are approximately 0.76 for XGB and 0.90 for
	FFN, with similar levels of alignment across the gross and net specifications.
	
	These results provide empirical support for the proposed local first-order
	attribution framework using independently estimated objective-level Shapley
	values as a benchmark. Together with the strong alignment between the Shapley
	values and their main effects, the results suggest that the local first-order attributions capture a substantial component of the feature-level contributions to portfolio Sharpe ratios. Although nonlinear and interaction effects may account for the remaining differences, the proposed attribution provides a computationally efficient measure of the local sensitivity of the Sharpe ratio with respect to feature exposures.

	\subsubsection{Attribution time-series correlation across features}
	
	To examine the dependence structure of feature contributions across different portfolio metrics, we compute pairwise Pearson correlations between the daily attribution time series,
	$
	\phi_{i,t}(Q),
	$
	for all feature pairs over the out-of-sample period from January 2006 to December 2024 ($4,781$ trading days).

		\begin{table}[htbp]
		\centering
		\caption{Within-model correlation of feature-contribution time series}
		\label{tab:within_model_corr}
		\small
		
		\begin{tabular}{ccccc}
			\toprule
			&
			\multicolumn{2}{c}{XGB}
			&
			\multicolumn{2}{c}{FFN}
			\\
			\cmidrule(lr){2-3}
			\cmidrule(lr){4-5}
			
			Metric
			& Mean $|\mathrm{corr}|$
			& Max $|\mathrm{corr}|$
			& Mean $|\mathrm{corr}|$
			& Max $|\mathrm{corr}|$
			\\
			\midrule
			
			Residual weight
			& 0.1416
			& 0.8586
			& 0.3851
			& 0.8482
			\\
			
			Turnover
			& 0.1105
			& 0.5432
			& 0.2691
			& 0.8154
			\\
			
			Gross Sharpe ratio
			& 0.0743
			& 0.3971
			& 0.0866
			& 0.5181
			\\
			
			Net Sharpe ratio
			& 0.0755
			& 0.4024
			& 0.0851
			& 0.5133
			\\
			
			\bottomrule
		\end{tabular}
		
		\vspace{2mm}
		
		\begin{minipage}{0.96\linewidth}
			\footnotesize
			\textit{Notes:}
			For each model, Pearson correlations are computed between all pairs
			of feature-contribution time series over the full out-of-sample period.
			The table summarises the off-diagonal entries of the resulting
			feature-correlation matrix across the 18 features.
			Mean $|\mathrm{corr}|$ denotes the average absolute pairwise
			correlation, while Max $|\mathrm{corr}|$ denotes the largest absolute
			pairwise correlation. Larger values indicate stronger dependence
			among feature-contribution dynamics within a model.
		\end{minipage}
		
	\end{table}
	
	\begin{figure}[htbp]
		\centering

		%%%%%%%%%%%%%%%%%%%% Row 2: XGB %%%%%%%%%%%%%%%%%%%%
		
		\begin{minipage}[b]{0.24\textwidth}
			\centering
			\includegraphics[width=\textwidth]{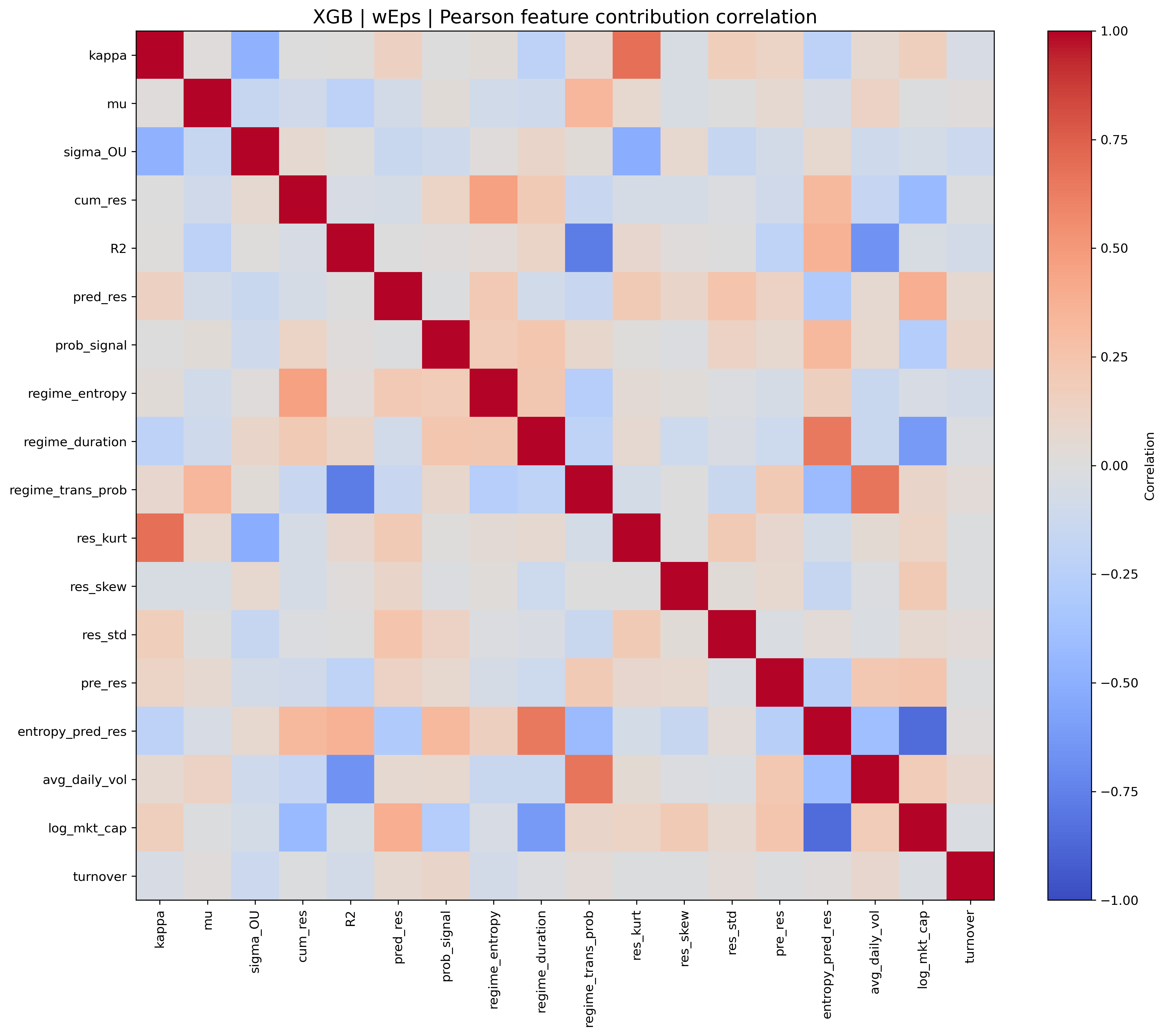}
			\small (a) XGB: $w^{\epsilon}$
		\end{minipage}
		\hfill
		\begin{minipage}[b]{0.24\textwidth}
			\centering
			\includegraphics[width=\textwidth]{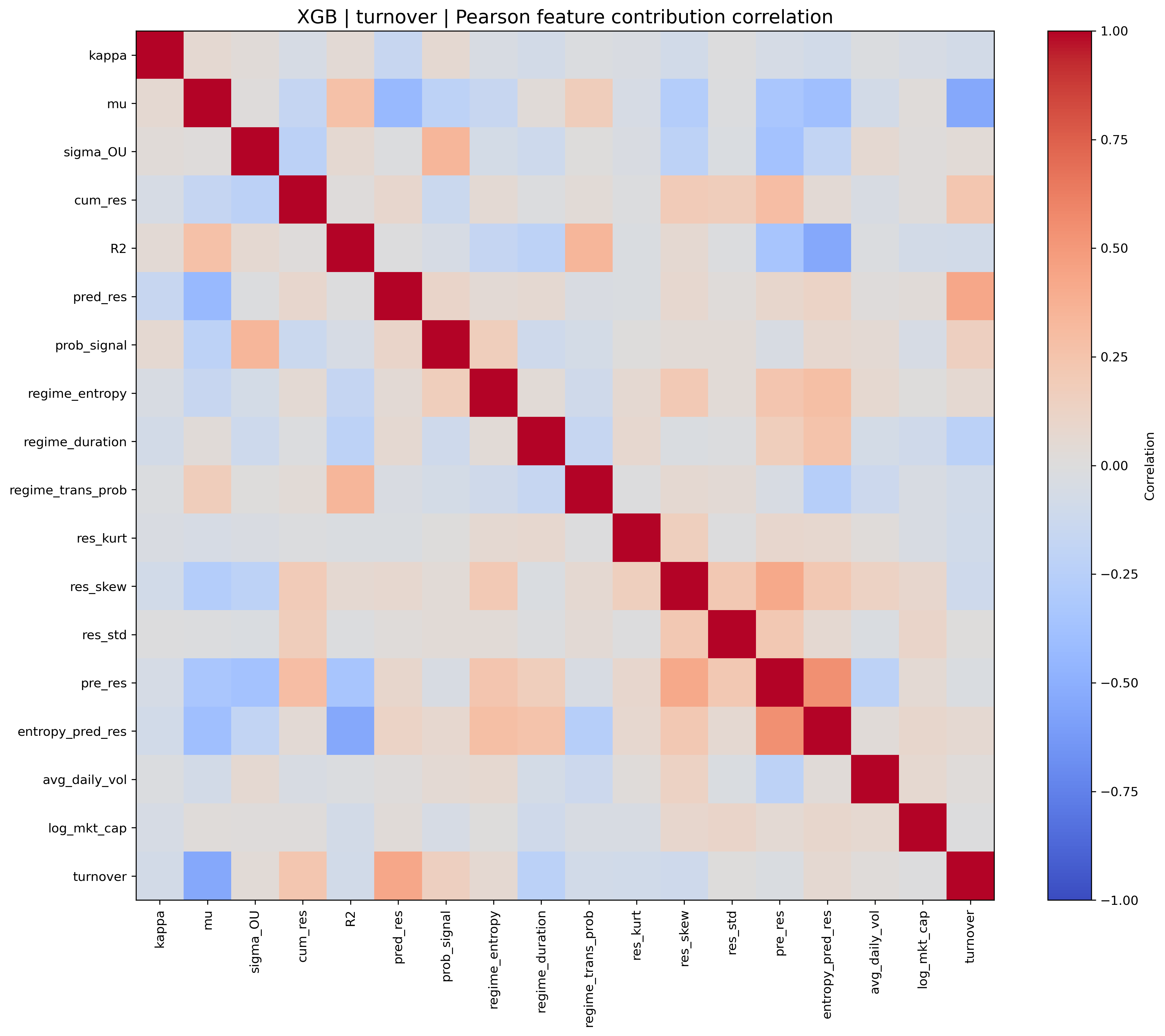}
			\small (b) XGB: Turnover
		\end{minipage}
		\hfill
		\begin{minipage}[b]{0.24\textwidth}
			\centering
			\includegraphics[width=\textwidth]{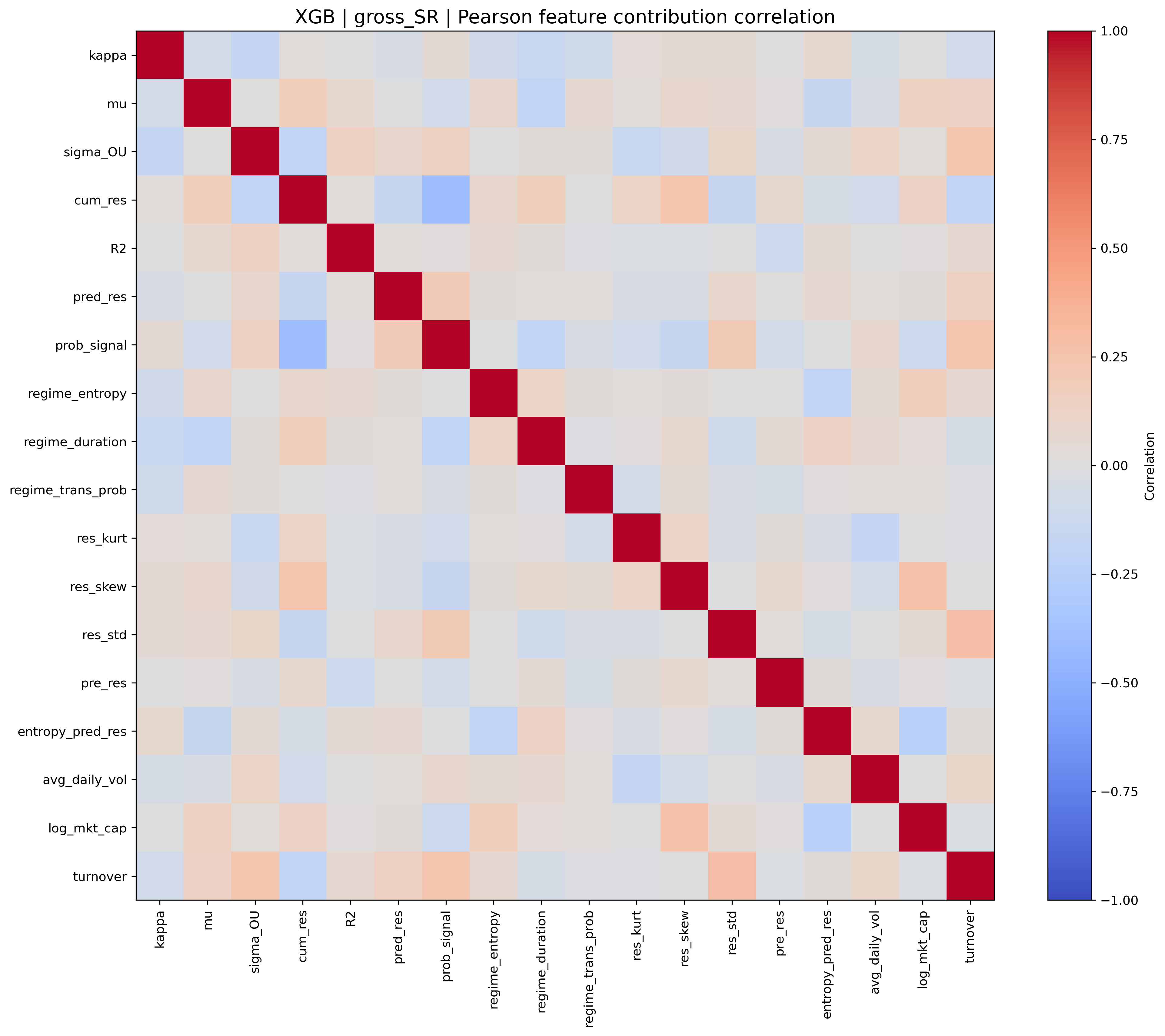}
			\small (c) XGB: Gross SR
		\end{minipage}
		\hfill
		\begin{minipage}[b]{0.24\textwidth}
			\centering
			\includegraphics[width=\textwidth]{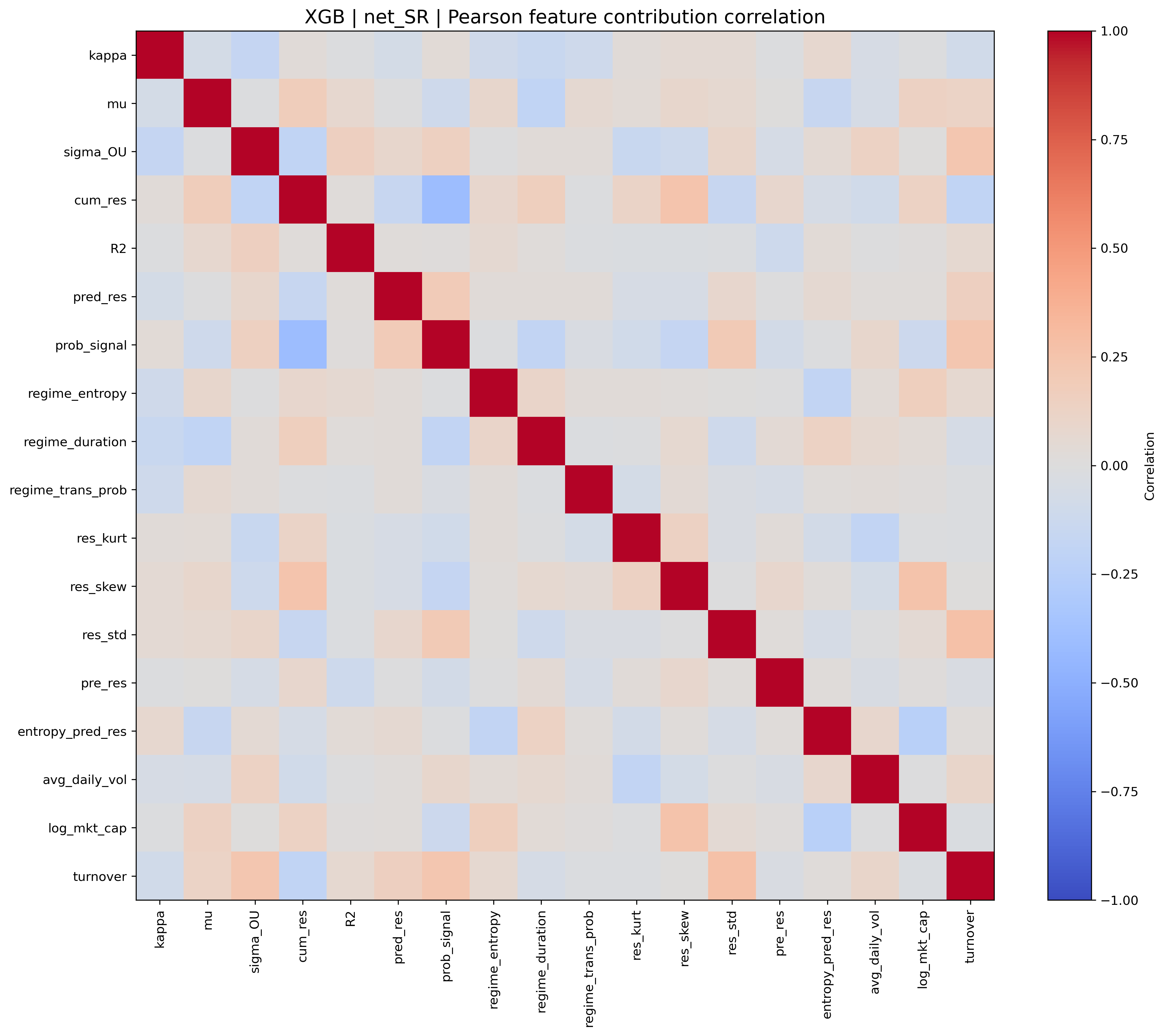}
			\small (d) XGB: Net SR
		\end{minipage}

		\vspace{0.3cm}
		
		%%%%%%%%%%%%%%%%%%%% Row 1: FFN %%%%%%%%%%%%%%%%%%%%
		
		\begin{minipage}[b]{0.24\textwidth}
			\centering
			\includegraphics[width=\textwidth]{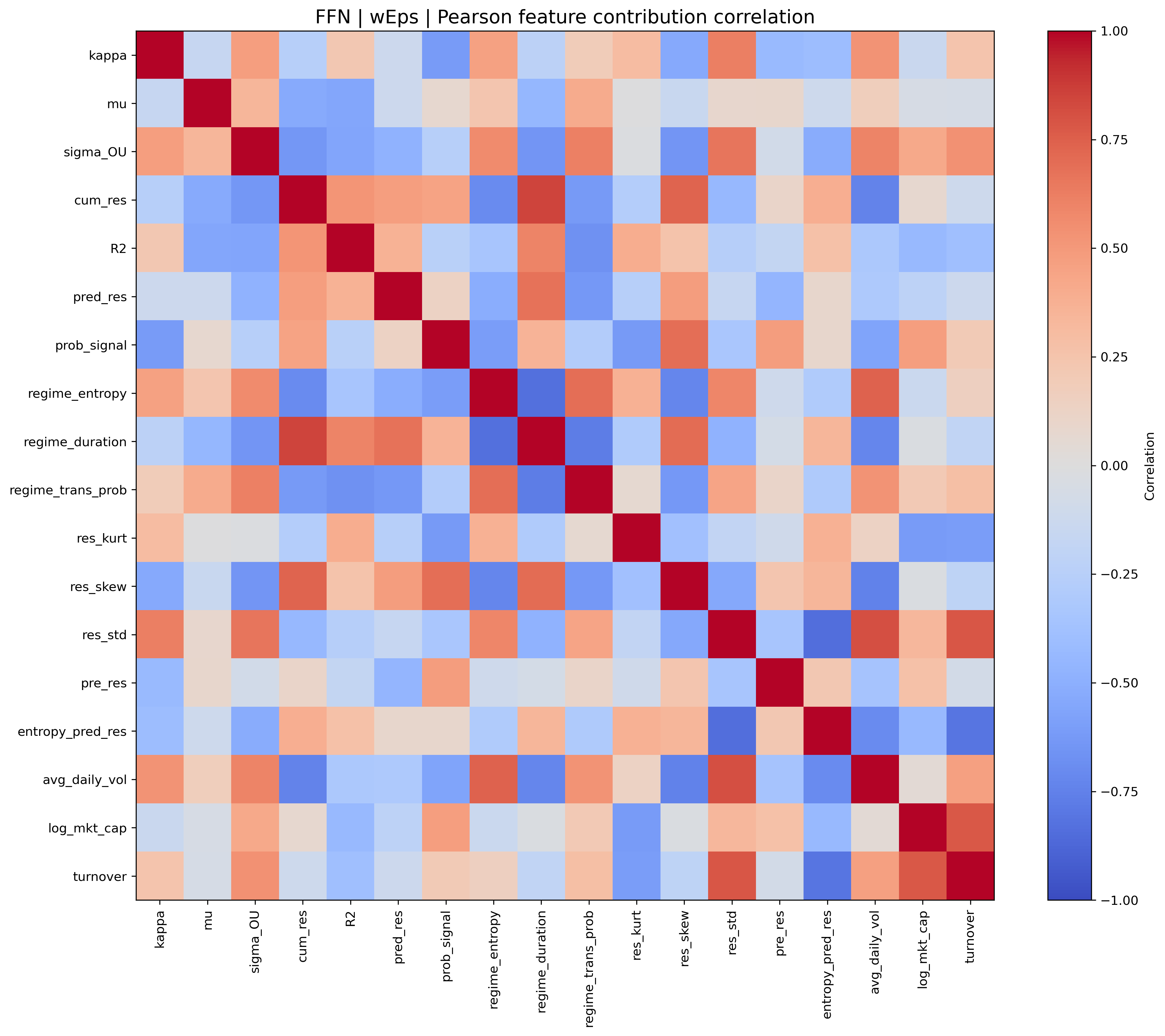}
			\small (e) FFN: $w^{\epsilon}$
		\end{minipage}
		\hfill
		\begin{minipage}[b]{0.24\textwidth}
			\centering
			\includegraphics[width=\textwidth]{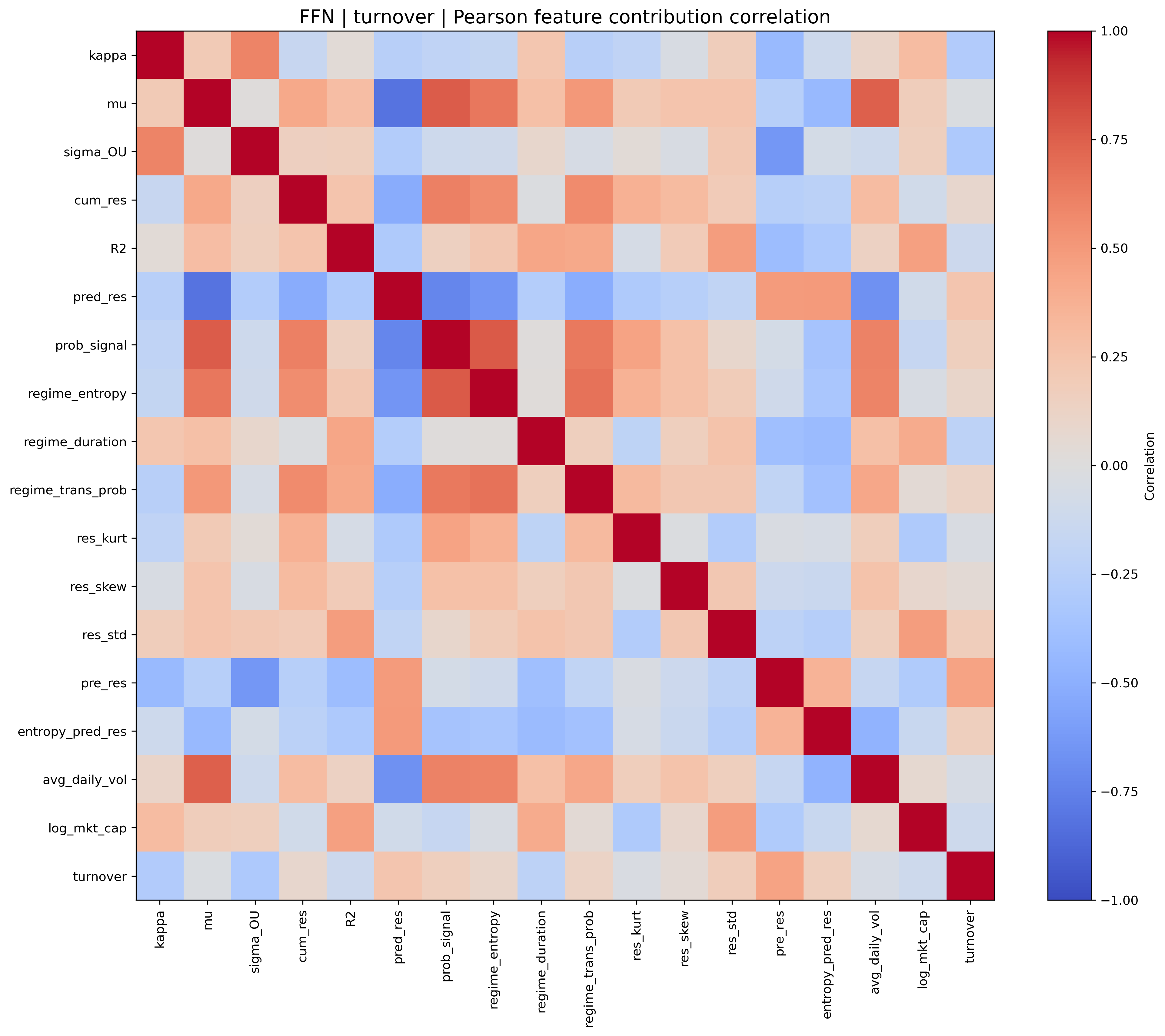}
			\small (f) FFN: Turnover
		\end{minipage}
		\hfill
		\begin{minipage}[b]{0.24\textwidth}
			\centering
			\includegraphics[width=\textwidth]{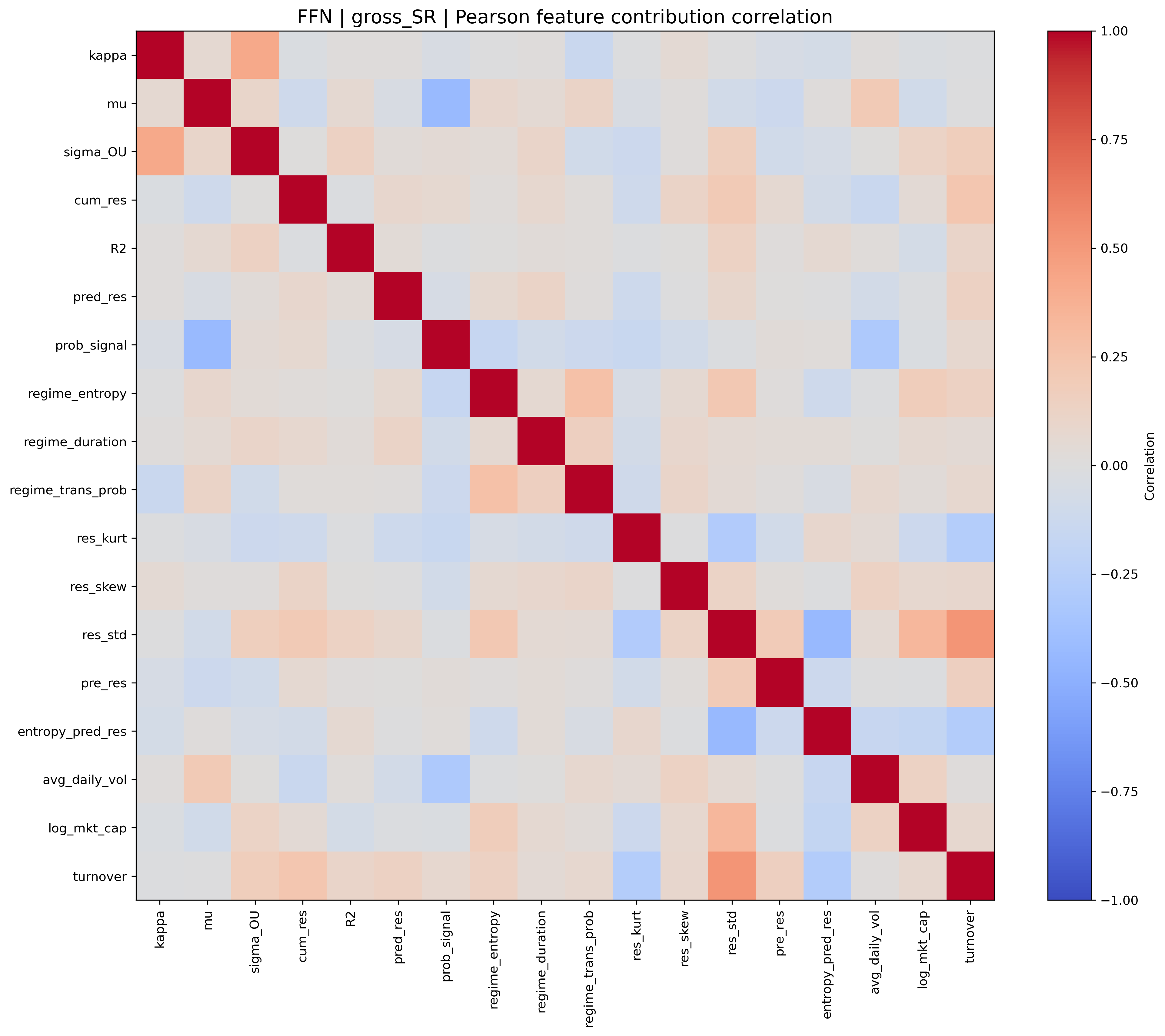}
			\small (g) FFN: Gross SR
		\end{minipage}
		\hfill
		\begin{minipage}[b]{0.24\textwidth}
			\centering
			\includegraphics[width=\textwidth]{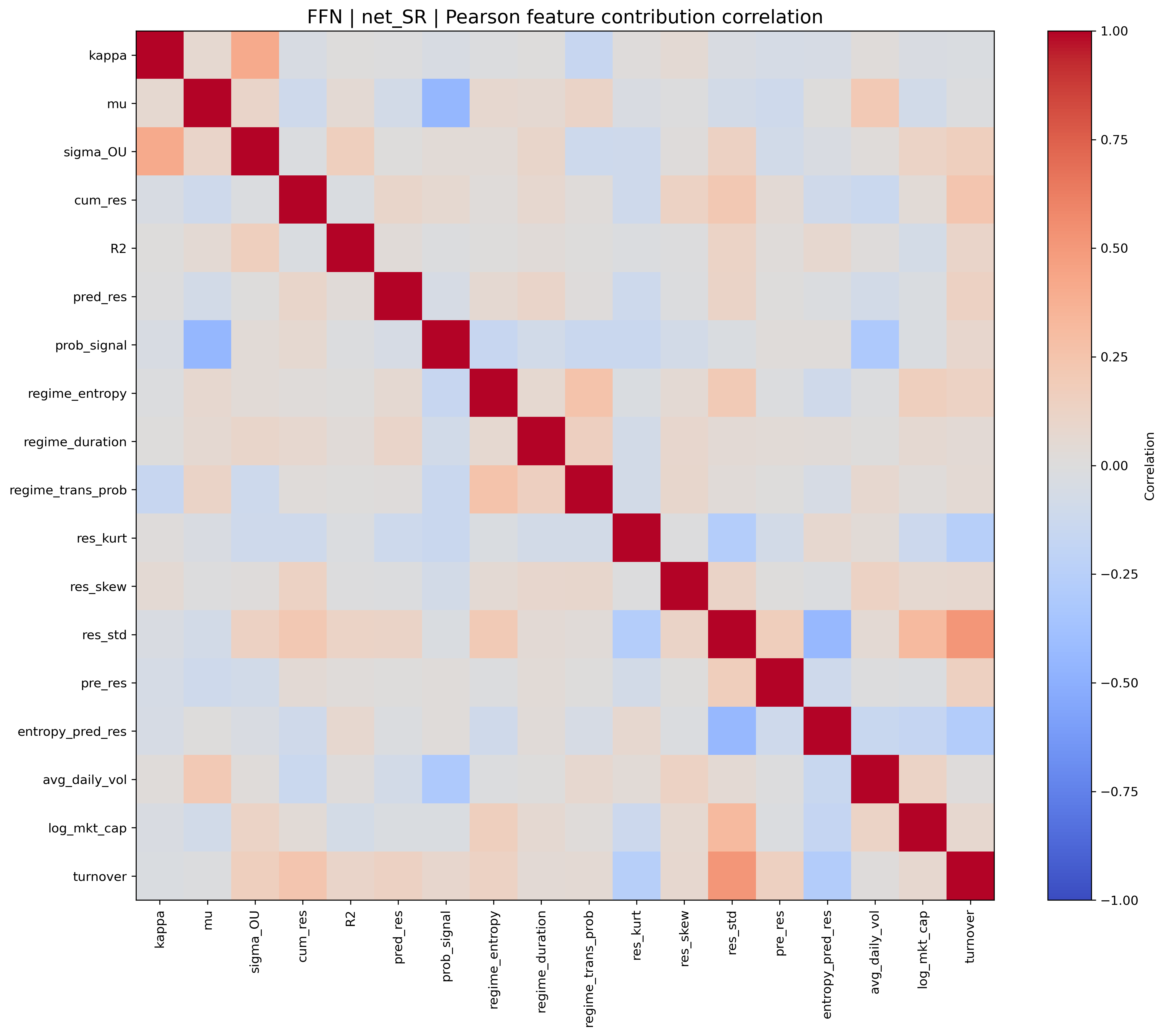}
			\small (h) FFN: Net SR
		\end{minipage}
		
		\caption{
			Within-model correlation matrices of daily feature-contribution time series for XGB and FFN.
			%			The first row reports pairwise Pearson correlations among the daily feature-contribution time series for the FFN model, while the second row reports the corresponding results for XGBoost. 
			Columns correspond to residual allocation weights ($w^{\epsilon}$), turnover, gross Sharpe ratio, and net Sharpe ratio, respectively. Correlations are computed over the out-of-sample period from January 2006 to December 2024. 
			%			Dependence is strongest at the allocation-weight level and becomes substantially weaker for portfolio-objective contributions, particularly for gross and net Sharpe ratios under FFN.
		}
		
		\label{fig:within_model_corr_heatmaps}
		
	\end{figure}

	Tables~\ref{tab:within_model_corr} report summary statistics of the off-diagonal entries of the feature-contribution correlation matrices for XGBoost and FFN, respectively. 
	Figures~\ref{fig:within_model_corr_heatmaps} present the full feature-contribution correlation matrices.
	We could observe that feature-contribution dependence is strongest at the allocation-weight level. For FFN, the mean absolute pairwise correlation decreases from $0.3851$ for residual-weight contributions to approximately $0.08$ for gross and net Sharpe-ratio contributions. This suggests that dependence among feature contributions is attenuated when contributions are propagated from model outputs to portfolio objectives.

	These results provide a partial explanation for why feature-level exposure interventions on portfolio metrics can remain effective in the presence of model-level interactions. For both XGB and FFN, the cross-feature correlations of attribution time series decline substantially as feature contributions are propagated from the allocation-model output, represented by residual portfolio weights, to portfolio-level Sharpe ratios. Although some features exhibit similar contribution dynamics at the allocation-weight level, their Sharpe-ratio attributions display considerably weaker linear dependence over time. This suggests that the propagation reduces the degree of common variation across feature contributions, resulting in a more separable attribution structure at the Sharpe-ratio level. Such separability provides supporting evidence for treating individual feature exposures as distinct intervention channels when implementing attribution-guided exposure adjustments.

	\subsubsection{Cross-model attribution similarity}
	\label{sec:cross-model_attribution_similarity}

	To evaluate whether different machine-learning models identify similar
	economic drivers, we compare the feature-attribution patterns generated by
	XGB and FFN. Let
	$
	\phi_{i,t}^{\mathcal M}(Q),
	$ with $
	\mathcal M\in\{\mathrm{XGB},\mathrm{FFN}\},
	$
	denote the attribution of feature $i$ at time $t$ under model $\mathcal M$
	for target metric $Q$, where $i=1,\ldots,p$ and $t=1,\ldots,T$.
	We examine cross-model similarity along two dimensions: the temporal
	dynamics of individual feature contributions and the cross-feature
	attribution structure at each point in time. Both dimensions are evaluated
	using Pearson correlation and sign consistency for portfolio metrics that
	are directly relevant to the economic interpretation of the allocation
	models, including residual weights, portfolio turnover, and gross and net
	Sharpe ratios.

	\paragraph{\textbf{Attribution time-series similarity.}}
	For each feature $i$, temporal agreement between the two models is measured
	by the Pearson correlation between its XGB and FFN attribution time series,
	\[
	\rho_i(Q)
	=
	\operatorname{corr}_{t}
	\left(
	\{\phi_{i,t}^{\mathrm{XGB}}(Q)\}_{t=1}^{T},
	\{\phi_{i,t}^{\mathrm{FFN}}(Q)\}_{t=1}^{T}
	\right).
	\]
	A larger value of $\rho_i(Q)$ indicates greater similarity in the temporal
	variation of feature $i$'s contribution across the two allocation models.
	Correlation does not necessarily imply directional agreement. We therefore
	also measure time-series sign consistency as
	\[
	SC_i(Q)
	=
	\frac{1}{|\mathcal T_i|}
	\sum_{t\in\mathcal T_i}
	\mathbf{1}
	\left[
	\operatorname{sign}
	\bigl(\phi_{i,t}^{\mathrm{XGB}}(Q)\bigr)
	=
	\operatorname{sign}
	\bigl(\phi_{i,t}^{\mathrm{FFN}}(Q)\bigr)
	\right],
	\]
	where $\mathbf{1}(\cdot)$ is the indicator function and $\mathcal T_i$
	denotes the set of time points for which both attribution values are
	nonzero for feature $i$. Thus, $SC_i(Q)\in[0,1]$ measures the proportion
	of time for which the two models assign the same contribution direction
	to feature $i$.

%	\paragraph{Attribution feature-structure similarity.}
%	To assess whether the two models identify a similar cross-feature
%	attribution structure, we compare their attribution vectors at each time
%	$t$. The feature-structure correlation is defined as
%	\[
%	\rho_t(Q)
%	=
%	\operatorname{corr}_{i}
%	\left(
%	\{\phi_{i,t}^{\mathrm{XGB}}(Q)\}_{i=1}^{p},
%	\{\phi_{i,t}^{\mathrm{FFN}}(Q)\}_{i=1}^{p}
%	\right).
%	\]
%	A larger value of $\rho_t(Q)$ indicates greater similarity in the
%	cross-feature attribution structure identified by the two models at time
%	$t$.
%	
%	The corresponding cross-feature sign consistency is defined as
%	\[
%	SC_t(Q)
%	=
%	\frac{1}{|\mathcal P_t|}
%	\sum_{i\in\mathcal P_t}
%	\mathbf{1}
%	\left[
%	\operatorname{sign}
%	\bigl(\phi_{i,t}^{\mathrm{XGB}}(Q)\bigr)
%	=
%	\operatorname{sign}
%	\bigl(\phi_{i,t}^{\mathrm{FFN}}(Q)\bigr)
%	\right],
%	\]
%	where $\mathcal P_t$ denotes the set of features for which both models
%	produce nonzero attributions at time $t$. Thus,
%	$SC_t(Q)\in[0,1]$ measures the proportion of available features for which
%	the two models assign the same contribution direction at time $t$.

	To examine whether cross-model agreement is concentrated among economically
	important features, we construct cumulative sets of the top-$K$ ranked
	features,
	$
	\mathcal P_K=\{(1),(2),\ldots,(K)\},
	$ with $ K=1,\ldots,p.
	$
	To obtain a model-neutral measure of feature importance, features are ranked
	separately for each target metric according to their average daily absolute
	attribution across the two models,
	\[
	I_i(Q)
	=
	\frac{1}{2}
	\left[
	\frac{1}{T}\sum_{t=1}^{T}
	\left|\phi_{i,t}^{\mathrm{FFN}}(Q)\right|
	+
	\frac{1}{T}\sum_{t=1}^{T}
	\left|\phi_{i,t}^{\mathrm{XGB}}(Q)\right|
	\right].
	\]
	Features are ordered such that
	$
	I_{(1)}(Q)\ge I_{(2)}(Q)\ge\cdots\ge I_{(p)}(Q),
	$
	where $(k)$ denotes the $k$-th most important feature for the corresponding
	target metric.
	
	\begin{figure}[htbp]
		\centering
		
		\begin{minipage}[t]{0.49\textwidth}
			\centering
			\includegraphics[width=\linewidth]
			{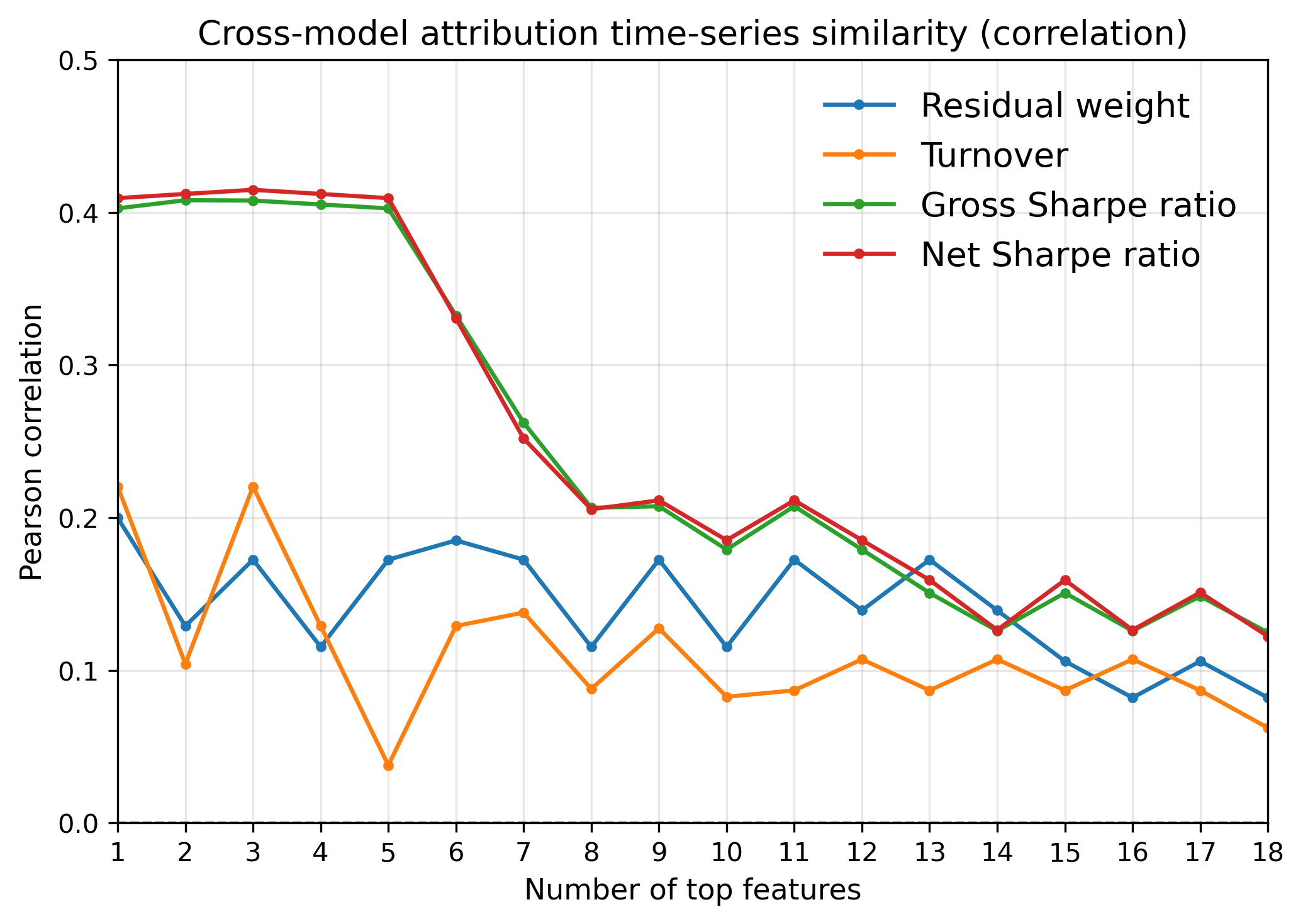}
			\small (a) Time-series similarity: Pearson correlation.
		\end{minipage}
		\hfill
		\begin{minipage}[t]{0.49\textwidth}
			\centering
			\includegraphics[width=\linewidth]
			{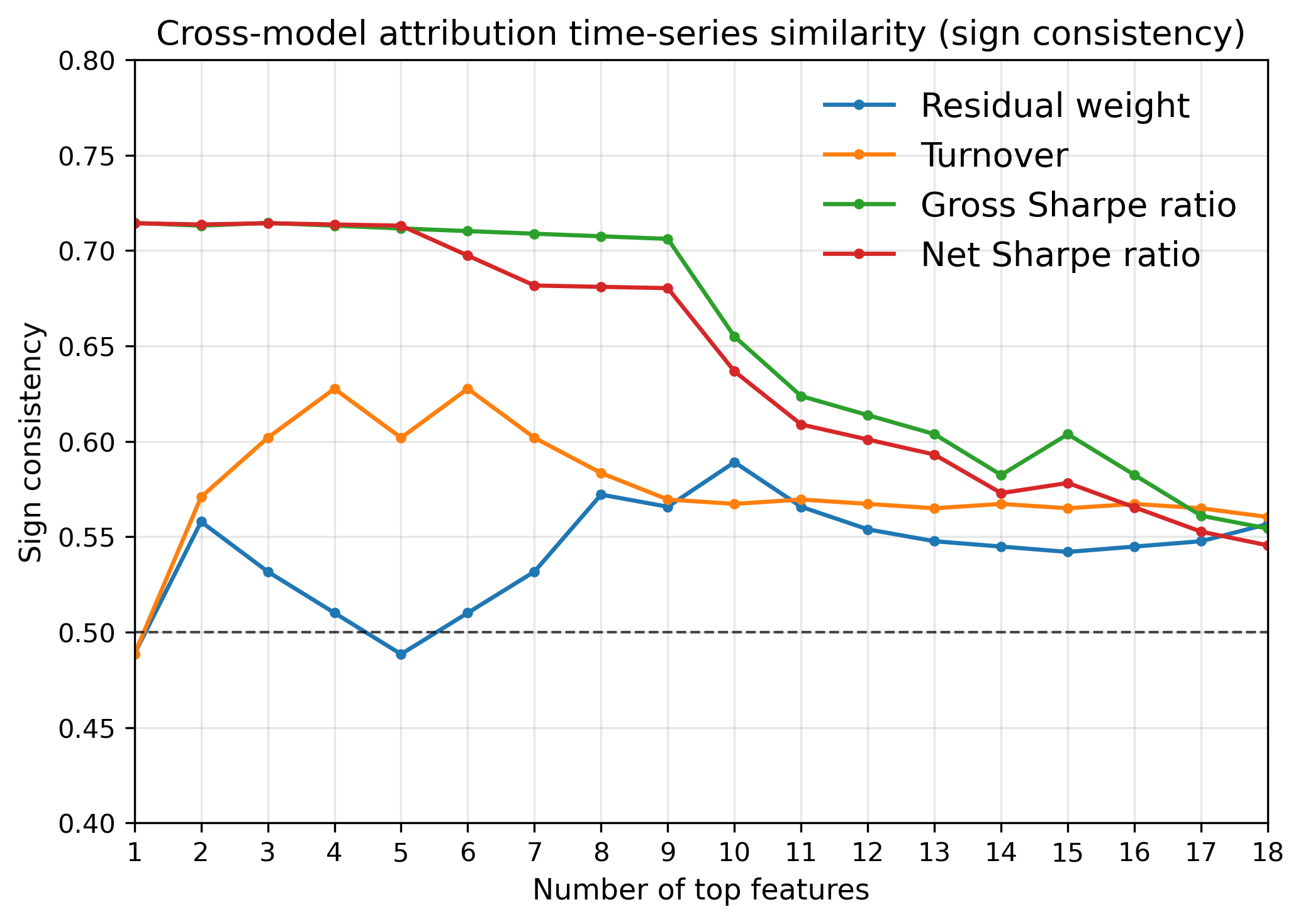}
			\small (b) Time-series similarity: sign consistency.
		\end{minipage}
		
		\caption{
			Cross-model similarity of attribution time series for top-ranked features.
			For each target metric $Q$, features are ranked by their average daily
			absolute attribution magnitude across the two models,
			$
			I_i(Q)
			=
			\frac{1}{2T}
			\sum_{t=1}^{T}
			\left(
			\left|\phi_{i,t}^{\mathrm{XGB}}(Q)\right|
			+
			\left|\phi_{i,t}^{\mathrm{FFN}}(Q)\right|
			\right).
			$
			For each $K=1,\ldots,18$, the top-$K$ set contains the $K$ features with
			the largest values of $I_i(Q)$. The left and right panels report,
			respectively, the median Pearson correlation and median sign consistency
			across the top-$K$ features. Each line corresponds to a different target
			metric: residual weights, portfolio turnover, gross Sharpe
			ratio, or net Sharpe ratio. }
		%			A decline as $K$ increases indicates that
		%			cross-model similarity is concentrated among features with larger
		%			attribution magnitudes.
		%			For each target metric, features are ranked according to the average absolute cumulative contribution across the two models,
		%			$
		%			I_f=
		%			\frac{
			%				\left|\sum_t\phi_{f,t}^{(\mathrm{FFN})}\right|
			%				+
			%				\left|\sum_t\phi_{f,t}^{(\mathrm{XGB})}\right|
			%			}{2}.
		%			$
		%			The left panel reports the mean (solid) and median (dashed) Pearson correlation of the top-$K$ ranked feature-contribution time series, while the right panel reports the corresponding sign consistency. Colours denote different portfolio objectives: residual weights ($w^{\epsilon}$), turnover, gross Sharpe ratio, and net Sharpe ratio. A declining trend indicates that cross-model agreement is concentrated among the most influential features.
		
		\label{fig:cross_model_topk_similarity}
	\end{figure}
	
	Figure~\ref{fig:cross_model_topk_similarity} reports cross-model similarity
	for residual weights, portfolio turnover, and gross and net Sharpe ratios
	as progressively lower-ranked features are included. Both correlation and
	sign consistency generally decline with $K$, indicating that cross-model
	agreement is concentrated among features with larger attribution
	magnitudes. This pattern is particularly pronounced for turnover and
	Sharpe-ratio attributions, for which the most influential features exhibit
	substantially stronger cross-model alignment than lower-ranked features.

	These results suggest that, although XGB and FFN generate distinct
	model-specific attribution patterns, the contributions of economically
	important features contain a common time-varying component across the two
	allocation models. The alignment is particularly evident for influential
	Sharpe-ratio features, with correlations of approximately 0.4 and sign-consistency rates of around 70\%, whereas lower-ranked features exhibit more
	model-specific dynamics. This common component provides supporting evidence
	for combining attribution information across models in the subsequent
	cross-model exposure intervention.

%	\paragraph{Cross-model similarity of feature-contribution structures}

%	\paragraph{Attribution feature-structure similarity}
%	For each feature, feature structure agreement between the two models is quantified using the Pearson correlation of their attribution structure,
%	$$
%	\rho_t (Q)= \operatorname{corr}\left(\{\phi_{i,t}^{\mathrm{XGB}}(Q)\}_{i=1}^{p},\{\phi_{i,t}^{\mathrm{FFN}}(Q)\}_{i=1}^{p}\right).
%	$$
%	A larger value of $\rho_t (Q)$ indicates that the two models havesimilar attribute structure to the corresponding economic feature.
%	
%	%	\paragraph{Sign consistency.}
%	Additionally, sign consistency is given by
%	$$
%	SC_t(Q) = \frac{1}{|\mathcal P_t|} \sum_{i\in\mathcal P_t} \mathbf 1\!\left( \operatorname{sign}(\phi_{i,t}^{\mathrm{XGB}}(Q)) = \operatorname{sign}(\phi_{f,t}^{\mathrm{FFN}}(Q)) \right),
%	$$
%	where $\mathbf 1(\cdot)$ is the indicator function and $\mathcal P_t$ denotes the set of observations for which both attribution values are non-zero at time $t$. The statistic $SC_t\in[0,1]$ measures the proportion of features that the two models assign the same contribution direction at time $t$.

	\paragraph{\textbf{Attribution feature-structure similarity.}}
	To assess whether the two models identify a similar cross-feature
	attribution structure, we compare their attribution vectors at each time
	$t$. The feature-structure correlation is defined as
	\[
	\rho_t(Q)
	=
	\operatorname{corr}_{i}
	\left(
	\{\phi_{i,t}^{\mathrm{XGB}}(Q)\}_{i=1}^{p},
	\{\phi_{i,t}^{\mathrm{FFN}}(Q)\}_{i=1}^{p}
	\right).
	\]
	A larger value of $\rho_t(Q)$ indicates greater similarity in the
	cross-feature attribution structure identified by the two models at time
	$t$.
	The corresponding cross-feature sign consistency is defined as
	\[
	SC_t(Q)
	=
	\frac{1}{|\mathcal P_t|}
	\sum_{i\in\mathcal P_t}
	\mathbf{1}
	\left[
	\operatorname{sign}
	\bigl(\phi_{i,t}^{\mathrm{XGB}}(Q)\bigr)
	=
	\operatorname{sign}
	\bigl(\phi_{i,t}^{\mathrm{FFN}}(Q)\bigr)
	\right],
	\]
	where $\mathcal P_t$ denotes the set of features for which both models
	produce nonzero attributions at time $t$. Thus,
	$SC_t(Q)\in[0,1]$ measures the proportion of available features for which
	the two models assign the same contribution direction at time $t$.

\begin{figure}[htbp]
		\centering
		
		\begin{minipage}[t]{0.49\textwidth}
			\centering
			\includegraphics[width=\linewidth]
			{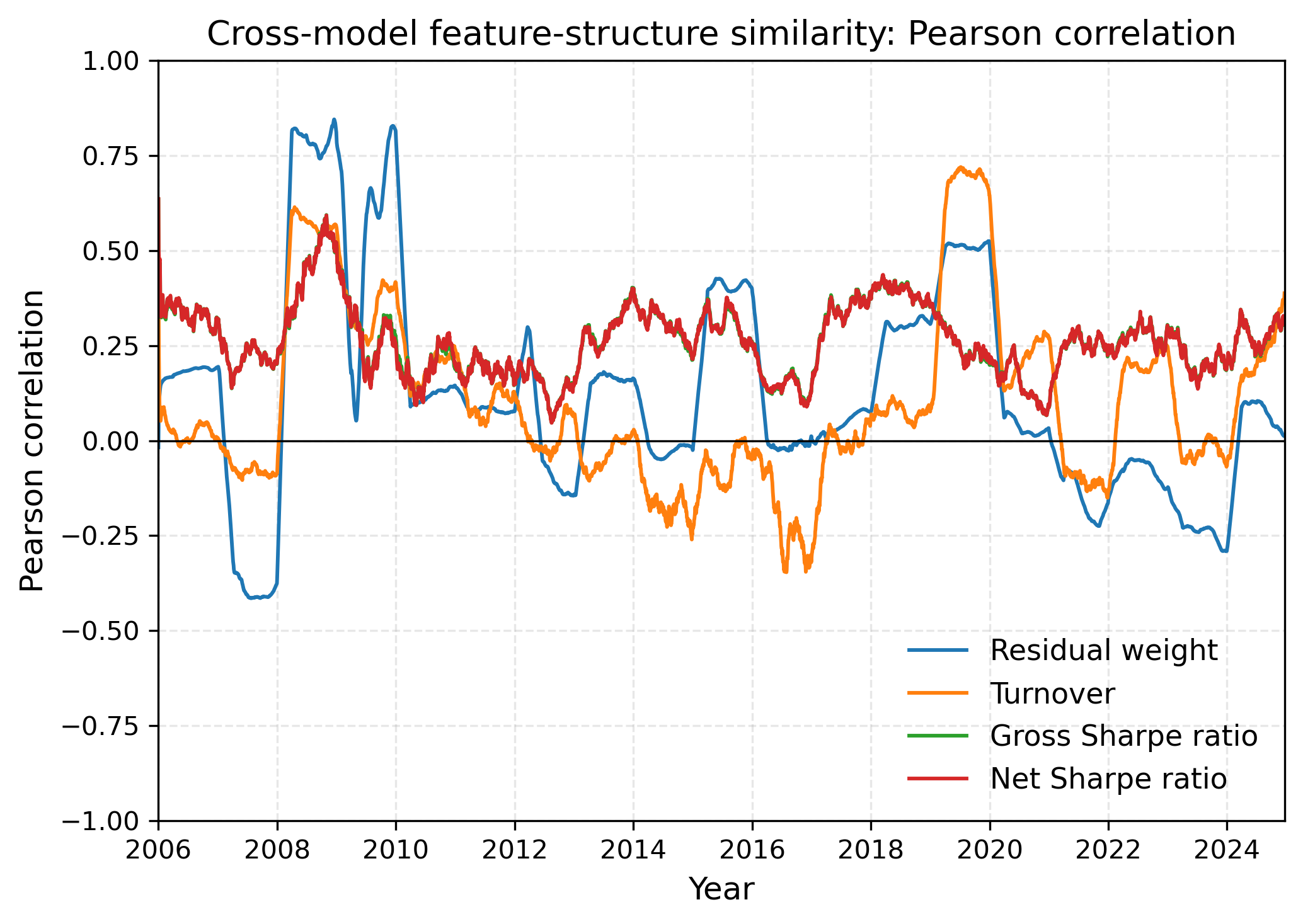}
			\small (a) Feature-structure correlation.
		\end{minipage}
		\hfill
		\begin{minipage}[t]{0.49\textwidth}
			\centering
			\includegraphics[width=\linewidth]
			{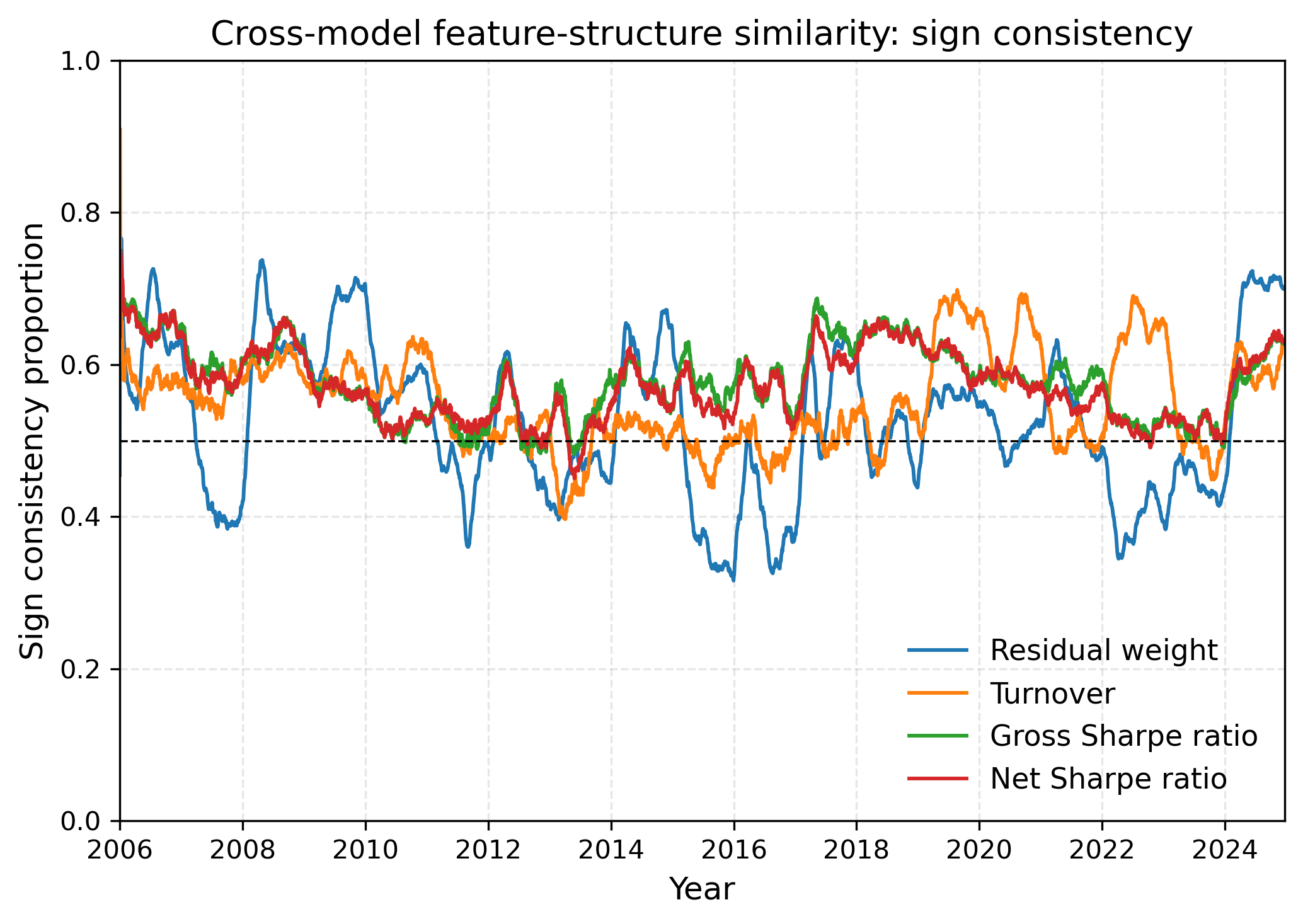}
			\small (b) Feature-structure sign consistency.
		\end{minipage}
		
		\caption{Evolution of cross-model feature-structure similarity between XGB and FFN. Each curve is a 63-day moving average computed from the daily similarity between the 18-dimensional feature-contribution vectors of the two models. The left panel reports Pearson correlations, while the right panel reports sign consistency proportions.}
		\label{fig:cross_model_feature_structure_similarity}
	\end{figure}

	Figure~\ref{fig:cross_model_feature_structure_similarity} shows evolution of cross-model feature-structure similarity between XGB and FFN. The feature-contribution structure exhibits increasing similarity across the two allocation models as the attribution propagates from the residual weights (model outputs) to portfolio performance metrics, particularly the gross and net Sharpe ratios. The similarity of residual-weight and turnover attributions is particularly pronounced during periods of market stress, such as the 2008 global financial crisis and the 2020 COVID-19 market disruption. In contrast, the feature-contribution structures for gross and net Sharpe ratios exhibit broader positive cross-model alignment over time, with mean correlations of approximately 0.26. The two models also assign the same contribution direction to approximately 58\% of features on average. This structural alignment suggests that the two allocation models capture a common component in their Sharpe-ratio attributions, providing empirical support for the use of cross-model averaged feature scaling.
	
	Given the yearly retraining design of the trading strategies, we also examine cross-model similarity in yearly feature-attribution structures. For model
	$\mathcal M$, the yearly attribution of feature $i$ to target metric $Q$ is
	defined as
	$
	\phi_{i,y}^{\mathcal M}(Q)
	=
	\sum_{t\in y}\phi_{i,t}^{\mathcal M}(Q),
	$
	yielding a cross-feature attribution vector for each of the 19
	out-of-sample years. We report the mean and median yearly
	cross-model correlation and sign consistency. The mean captures the
	average degree of agreement across years, whereas the median characterises
	a typical year and is less sensitive to unusually high or low similarity
	in individual years.
	
		\begin{table}[htbp]
		\centering
		\caption{Cross-model similarity of yearly feature-attribution structures}
		\label{tab:yearly_cross_model_similarity}
		\small
		\setlength{\tabcolsep}{4pt}
		
		\begin{tabular}{ccccccc}
			\toprule
			&
			\multicolumn{3}{c}{Correlation}
			&
			\multicolumn{3}{c}{Sign Consistency} \\
			\cmidrule(lr){2-4}
			\cmidrule(lr){5-7}
			
			Metric
			& Mean
			& Median
			& $\Pr(\rho>0)$
			& Mean
			& Median
			& $\Pr(SC>50\%)$ \\
			
			\midrule
			
			Residual weight
			& 0.1274
			& 0.0773$^{*}$
			& 63.16\%
			& 52.90\%
			& 54.55\%
			& 63.16\% \\
			
			Turnover
			& 0.0484
			& -0.0667
			& 47.37\%
			& 52.98\%
			& 54.55\%
			& 57.89\% \\
			
			Gross Sharpe ratio
			& 0.1808
			& 0.1519$^{**}$
			& 63.16\%
			& 56.62\%
			& 54.55\%$^{**}$
			& 57.89\% \\
			
			Net Sharpe ratio
			& 0.1747
			& 0.1899$^{**}$
			& 68.42\%
			& 53.67\%
			& 55.56\%
			& 57.89\% \\
			
			\bottomrule
		\end{tabular}
		
		\vspace{2mm}
		
		\begin{minipage}{0.95\textwidth}
			\footnotesize
			\textit{Notes:}
			For each of the 19 out-of-sample years, daily feature attributions are
			aggregated within the year to form cross-feature attribution vectors for
			XGB and FFN. Correlation denotes the Pearson correlation between the two
			attribution vectors within each year. Sign consistency denotes the
			proportion of features with the same nonzero attribution sign across the
			two models. $\Pr(\rho>0)$ is the proportion of years with positive
			cross-model correlation, while $\Pr(SC>50\%)$ is the proportion of years
			with sign consistency exceeding 50\%. Significance stars attached to the
			medians are based on one-sided Wilcoxon signed-rank tests against zero
			for correlation and 50\% for sign consistency.
			$^{***}$, $^{**}$, and $^{*}$ denote significance at the 1\%, 5\%, and
			10\% levels, respectively.
		\end{minipage}
	\end{table}
	
	In table~\ref{tab:yearly_cross_model_similarity}, these results provide evidence of a common cross-model component in yearly Sharpe-ratio attribution structures, despite relatively weak similarity in the attribution structures at the model-output level. 
	The median yearly correlations are 0.1519 for gross Sharpe
	and 0.1899 for net Sharpe, both of which are significantly positive at the
	5\% level based on one-sided Wilcoxon signed-rank tests. Positive
	cross-model correlations are observed in 63.16\% and 68.42\% of the 19
	out-of-sample years for gross and net Sharpe ratios, respectively. 
%	Gross
%	Sharpe attributions also exhibit significant directional agreement, with
%	median sign consistency of 54.55\%, whereas the corresponding evidence is
%	weaker for net Sharpe. Overall, these results provide evidence of a common
%	cross-model component in yearly Sharpe-ratio attribution structures,
%	supporting the use of attribution information across the two allocation
%	models.
	Under the same market conditions, different models may produce distinct turnover attributions, while their feature contributions to gross and net Sharpe ratios exhibit greater cross-model similarity.

	\subsection{Portfolio performance by feature exposure intervention}

	In this section, we implement feature-exposure interventions using only information available prior to each trading period. We first examine the persistence of feature attributions across time and then evaluate the economic significance of the resulting improvements in portfolio Sharpe ratios. Finally, we assess transaction-cost sensitivity by constructing net Sharpe-ratio attributions under alternative transaction-cost rates and evaluating portfolio performance under different out-of-sample transaction-cost assumptions.
	
	\subsubsection{Attribution persistence across time}
	\label{sec:attribution_persistence}

	We examine whether feature-attribution persistence varies across portfolio
	objectives. Consistent with the yearly retraining design, the feature
	contributions computed over the validation period provide the ex ante
	attribution signal for the subsequent out-of-sample trading period. Let
	$\mathcal M_y$ denote the model used for the $y$-th out-of-sample trading
	period, where $y=1,\ldots,19$ and $\mathcal M\in\{\mathrm{XGB},\mathrm{FFN}\}$. For each model specification, we compare the
	validation-period feature contributions,
	$
	\phi_{i,y-1}^{\mathcal M_y}(Q),
	$
	with the realised feature contributions in the subsequent out-of-sample
	trading period,
	$
	\phi_{i,y}^{\mathcal M_y}(Q),
	$
	for three portfolio targets: turnover, gross Sharpe ratio, and net Sharpe
	ratio.
	
	To assess the persistence of the cross-feature attribution structure, we
	compute, for each out-of-sample trading period $y=1,\ldots,19$, the Pearson
	correlation between the validation and subsequent out-of-sample attribution
	vectors:
	\[
	\rho_y^{\mathcal M_y}(Q)
	=
	\operatorname{corr}_{i}
	\left(
	\{\phi_{i,y-1}^{\mathcal M_y}(Q)\}_{i=1}^{p},
	\{\phi_{i,y}^{\mathcal M_y}(Q)\}_{i=1}^{p}
	\right).
	\]
	A larger value of $\rho_y^{\mathcal M_y}(Q)$ indicates greater persistence
	in the cross-feature attribution structure under model $\mathcal M_y$.
	The corresponding sign consistency is defined as
	\[
	SC_y^{\mathcal M_y}(Q)
	=
	\frac{1}{|\mathcal P_y^{\mathcal M_y}|}
	\sum_{i\in\mathcal P_y^{\mathcal M_y}}
	\mathbf{1}
	\left[
	\operatorname{sign}
	\bigl(\phi_{i,y-1}^{\mathcal M_y}(Q)\bigr)
	=
	\operatorname{sign}
	\bigl(\phi_{i,y}^{\mathcal M_y}(Q)\bigr)
	\right],
	\]
	where $\mathcal P_y^{\mathcal M_y}$ denotes the set of features for which
	the validation and out-of-sample attributions are both nonzero. Thus,
	$SC_y^{\mathcal M_y}(Q)\in[0,1]$ measures the proportion of available
	features whose contribution direction persists from the validation period
	to the subsequent out-of-sample trading period.

    \begin{table}[htbp]
    	\centering
    	\caption{Attribution persistence from validation to out-of-sample trading periods}
    	\label{tab:persistence_turnover_SR}
    	\small
    	\setlength{\tabcolsep}{5pt}
    	
    	\begin{tabular}{ccccccc}
    		\toprule
    		&
    		\multicolumn{3}{c}{Correlation}
    		&
    		\multicolumn{3}{c}{Sign Consistency} \\
    		\cmidrule(lr){2-4}
    		\cmidrule(lr){5-7}
    		
    		Model
    		& Mean
    		& Median
    		& $\Pr(\rho>0)$
    		& Mean
    		& Median
    		& $\Pr(SC>50\%)$ \\
    		
    		\midrule
    		
    		% ============================================================
    		% Panel A: Portfolio Turnover
    		% ============================================================
    		
    		\multicolumn{7}{c}{
    			\textbf{Panel A: Target = Portfolio Turnover}
    		} \\
    		\midrule
    		
    		XGB
    		& 0.8865
    		& 0.9363$^{***}$
    		& 100.00\%
    		& 83.87\%
    		& 81.25\%$^{***}$
    		& 89.47\% \\
    		
    		FFN
    		& 0.9897
    		& 0.9947$^{***}$
    		& 100.00\%
    		& 88.30\%
    		& 88.89\%$^{***}$
    		& 100.00\% \\
    		
    		\midrule
    		
    		% ============================================================
    		% Panel B: Gross Sharpe Ratio
    		% ============================================================
    		
    		\multicolumn{7}{c}{
    			\textbf{Panel B: Target = Gross Sharpe Ratio}
    		} \\
    		\midrule
    		
    		XGB
    		& 0.2379
    		& 0.3889$^{**}$
    		& 68.42\%
    		& 54.63\%
    		& 55.56\%
    		& 57.89\% \\
    		
    		FFN
    		& 0.0569
    		& 0.0705
    		& 63.16\%
    		& 50.58\%
    		& 44.44\%
    		& 36.84\% \\
    		
    		\midrule
    		
    		% ============================================================
    		% Panel C: Net Sharpe Ratio
    		% ============================================================
    		
    		\multicolumn{7}{c}{
    			\textbf{Panel C: Target = Net Sharpe Ratio}
    		} \\
    		\midrule
    		
    		XGB
    		& 0.2423
    		& 0.3111$^{**}$
    		& 68.42\%
    		& 56.23\%
    		& 55.56\%
    		& 52.63\% \\
    		
    		FFN
    		& 0.2453
    		& 0.2190$^{***}$
    		& 68.42\%
    		& 52.63\%
    		& 55.56\%
    		& 57.89\% \\
    		
    		\bottomrule
    	\end{tabular}
    	
    	\vspace{2mm}
    	
    	\begin{minipage}{0.95\textwidth}
    		\footnotesize
    		\textit{Notes:}
    		For each trading period, the cross-feature Pearson correlation between
    		the feature-attribution vectors over the validation period and
    		the subsequent out-of-sample period is computed. Sign consistency is
    		defined as the proportion of available features with the same nonzero
    		attribution sign across the two periods. The table reports summary
    		statistics across the 19 validation-to-OOS comparisons.
    		$\Pr(\rho>0)$ denotes the proportion of trading periods with positive
    		correlation, while $\Pr(SC>50\%)$ denotes the proportion with sign
    		consistency exceeding 50\%. Significance stars attached to the medians
    		are based on one-sided Wilcoxon signed-rank tests against zero for
    		correlation and 50\% for sign consistency.
    		$^{***}$, $^{**}$, and $^{*}$ denote significance at the 1\%, 5\%,
    		and 10\% levels, respectively.
    		
    		%    		Significance stars are based on one-sided Wilcoxon signed-rank tests of whether the distribution of the correlation statistics is shifted above zero, or whether the distribution of sign-consistency statistics is shifted above 50\%, respectively.
    	\end{minipage}
    	
    \end{table}
    
%    Table~\ref{tab:persistence_turnover_SR} reports validation-to-OOS persistence in
%    feature-attribution structures across the 19 trading periods. Persistence
%    varies substantially across portfolio objectives. Turnover attributions
%    are highly persistent for both allocation models, with the median correlations
%    are 0.936 for XGB and 0.995 for FFN, and the median sign-consistency rates
%    are 81.25\% and 88.89\%, respectively. Both measures are statistically
%    significant at the 1\% level. This strong persistence is consistent with
%    turnover being closely linked to the allocation decisions generated by the
%    models and therefore less directly exposed to short-term variation in
%    realised returns.
%    
%    Sharpe-ratio attributions exhibit weaker but economically relevant
%    persistence. For net Sharpe, the median validation-to-OOS correlations are
%    0.311 for XGB and 0.219 for FFN, both significantly positive at the 5\%
%    level or better. Gross Sharpe persistence is less consistent across models, with
%    the median correlation is significantly positive for XGB (0.389) but small
%    and insignificant for FFN (0.071). The median sign-consistency rates for net Sharpe attributions are approximately 55\% for both models. These results indicate that persistence is stronger in the overall cross-feature attribution structure than in feature-level directional agreement, as small attribution values may change sign around zero across periods. 

	Table~\ref{tab:persistence_turnover_SR} reports validation-to-OOS persistence in
	feature-attribution structures across the 19 trading periods. Persistence
	varies substantially across portfolio objectives. Turnover attributions are highly persistent for both allocation models, 
	with median correlations of 0.936 for XGB and 0.995 for FFN, and median sign-consistency rates of 81.25\% and 88.89\%, respectively. Both measures are statistically
	significant at the 1\% level. This strong persistence is consistent with
	turnover being closely linked to the allocation decisions generated by the
	models and therefore less directly exposed to short-term variation in
	realised returns.
	
	Sharpe-ratio attributions exhibit weaker but statistically detectable and economically relevant persistence. 
	For net Sharpe, the median validation-to-OOS correlations are
	0.311 for XGB and 0.219 for FFN, both significantly positive at the 5\%
	level or better. Gross Sharpe persistence is less consistent across models, with
	the median correlation  significantly positive for XGB (0.389) but small
	and insignificant for FFN (0.071). The median sign-consistency rates for net Sharpe attributions are approximately 55\% for both models. These results indicate that persistence is stronger in the overall cross-feature attribution structure than in feature-level directional agreement, as small attribution values may change sign around zero across periods.

    The more consistent persistence of net Sharpe attributions across the two models is consistent with the inclusion of transaction-cost effects, given the substantially stronger persistence observed in turnover attributions. In contrast, gross-return-related feature contributions may be more sensitive to changes in market conditions. Overall, the results provide supporting evidence that validation-period attributions contain economically useful information about subsequent out-of-sample attribution structures, particularly for turnover and net Sharpe ratio, motivating their use as inputs to the feature-exposure intervention rule.

	\paragraph{\textbf{Persistence over longer aggregation horizons.}}
	
	The preceding analysis evaluates validation-to-OOS attribution persistence
	at the annual horizon. We next examine whether the
	persistence of economically relevant feature contributions extends to
	longer horizons by aggregating annual attributions over adjacent,
	non-overlapping validation and out-of-sample windows. This analysis allows
	for the possibility that attribution patterns evolve gradually over time and become more apparent when short-term variation is aggregated.
	
	Because the allocation models are retrained annually, each feature
	attribution is associated with the corresponding model vintage. For an
	$H$-year aggregation horizon, let
	$\Phi_{i,y}^{\mathrm{Val},(H)}(Q)$ and
	$\Phi_{i,y}^{\mathrm{OOS},(H)}(Q)$ denote the cumulative validation and
	out-of-sample contributions, respectively, for feature $i$ and target
	metric $Q$. For terminal out-of-sample period $y$, these are defined as
	$
	\Phi_{i,y}^{\mathrm{Val},(H)}(Q)
	=
	\sum_{h=0}^{H-1}
	\phi_{i,\,y-(2H-1)+h}^{
		\mathcal M_{\,y-(2H-2)+h}
	}(Q),
	$
	and
	$
	\Phi_{i,y}^{\mathrm{OOS},(H)}(Q)
	=
	\sum_{h=0}^{H-1}
	\phi_{i,\,y-(H-1)+h}^{
		\mathcal M_{\,y-(H-1)+h}
	}(Q).
	$
	Thus, the validation window covers periods
	$y-(2H-1),\ldots,y-H$, while the immediately subsequent out-of-sample
	window covers periods $y-H+1,\ldots,y$. The two windows are adjacent,
	non-overlapping, and contain the same number of annual observations.
	
	The annual persistence analysis reported above corresponds to the special
	case $H=1$. We extend the aggregation horizon to $H=8$ to examine whether
	the validation-to-future attribution structure persists over longer
	horizons. With 19 annual out-of-sample trading periods, the number of
	available rolling validation-OOS window pairs at horizon $H$ is
	$
	N_H = 21-2H,
	$
	yielding $19,17,15,13,11,9,7,$ and $5$ comparisons for
	$H=1,\ldots,8$, respectively.
	For $H=1$, the cumulative contributions reduce to the annual
	validation-OOS comparison,
	$
	\Phi_{i,y}^{\mathrm{Val},(1)}(Q)
	=
	\phi_{i,y-1}^{\mathcal M_y}(Q)
	$ and $
	\Phi_{i,y}^{\mathrm{OOS},(1)}(Q)
	=
	\phi_{i,y}^{\mathcal M_y}(Q),
	$ with $
	y=1,\ldots,19.
	$
	At the two-year aggregation horizon ($H=2$), the corresponding cumulative
	contributions are
	$
	\Phi_{i,y}^{\mathrm{Val},(2)}(Q)
	=
	\phi_{i,y-3}^{\mathcal M_{y-2}}(Q)
	+
	\phi_{i,y-2}^{\mathcal M_{y-1}}(Q)
	$
	and
	$
	\Phi_{i,y}^{\mathrm{OOS},(2)}(Q)
	=
	\phi_{i,y-1}^{\mathcal M_{y-1}}(Q)
	+
	\phi_{i,y}^{\mathcal M_y}(Q),
	$ with $
	y=3,\ldots,19.
	$
	Thus, the two-year construction compares cumulative feature contributions
	over two consecutive validation periods with those realised over the
	immediately following two out-of-sample trading periods.

	For each horizon $H$, persistence is evaluated by the cross-feature
	Pearson correlation
	$
	\rho_{y}^{(H)}(Q)
	=
	\operatorname{corr}_{i}
	\left(
	\left\{
	\Phi_{i,y}^{\mathrm{Val},(H)}(Q)
	\right\}_{i=1}^{p},
	\left\{
	\Phi_{i,y}^{\mathrm{OOS},(H)}(Q)
	\right\}_{i=1}^{p}
	\right),
	$
	together with the corresponding sign consistency
	
	$
	SC_{y}^{(H)}(Q)
	=
	\frac{1}{
		|\mathcal P_{y}^{(H)}|
	}
	\sum_{
		i\in\mathcal P_{y}^{(H)}
	}
	\mathbf{1}
	\left[
	\operatorname{sign}
	\left(
	\Phi_{i,y}^{\mathrm{Val},(H)}(Q)
	\right)
	=
	\operatorname{sign}
	\left(
	\Phi_{i,y}^{\mathrm{OOS},(H)}(Q)
	\right)
	\right],
	$
	where $\mathcal P_{y}^{(H)}$ denotes the set of features with nonzero
	cumulative contributions in both windows.

	\begin{figure}[htbp]
		\centering
		
		%%%%%%%%%%%%%%%%%%%%%%%%%%%%%%%%%%%%%%%%%%%%%%%%%%%%%%%%%%%
		% Row 1
		%%%%%%%%%%%%%%%%%%%%%%%%%%%%%%%%%%%%%%%%%%%%%%%%%%%%%%%%%%%
		
		\begin{minipage}[t]{0.49\textwidth}
			\centering
			\includegraphics[width=\linewidth]
			{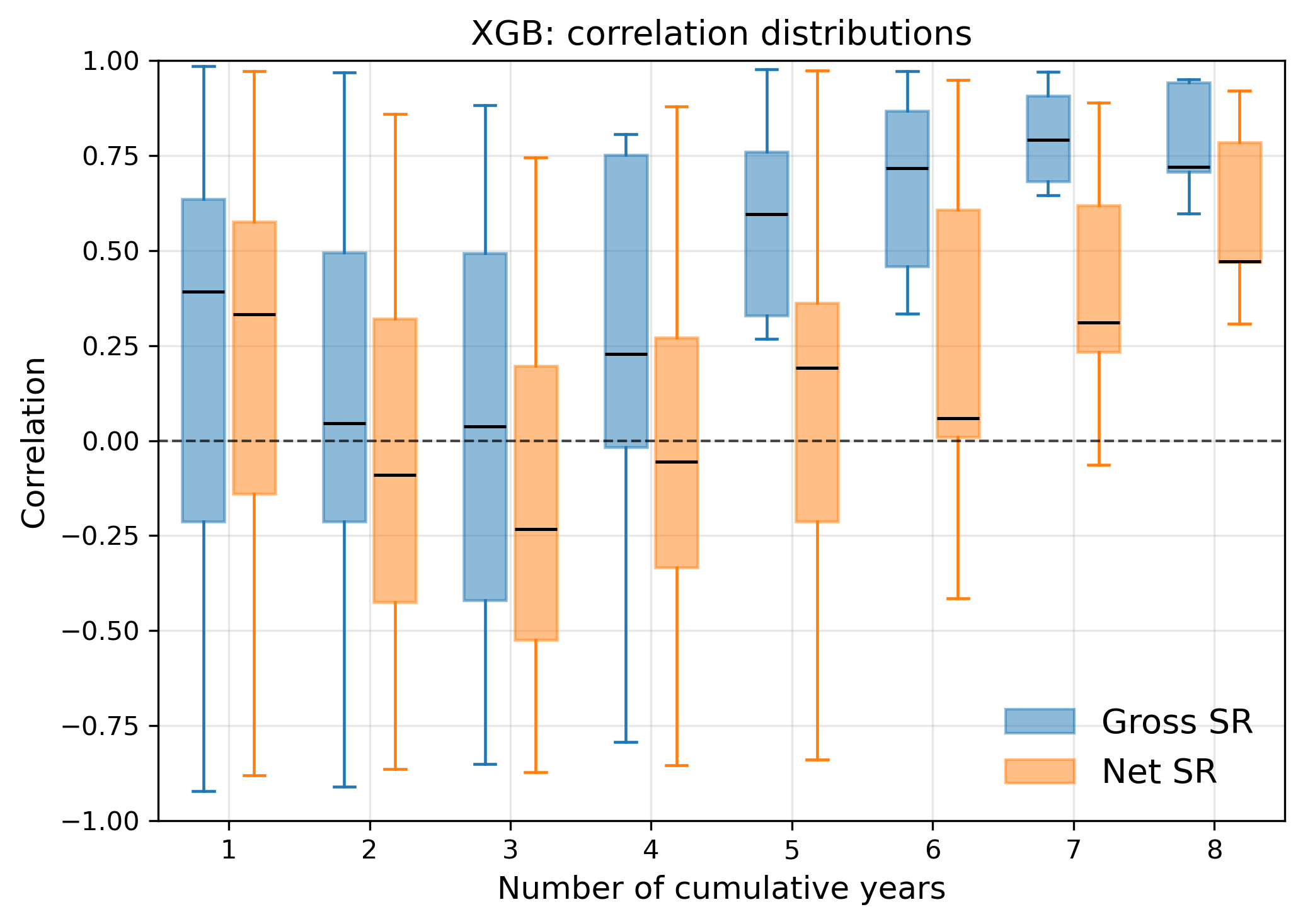}
			\small (a) XGB: Correlation
		\end{minipage}
		\hfill
		\begin{minipage}[t]{0.49\textwidth}
			\centering
			\includegraphics[width=\linewidth]
			{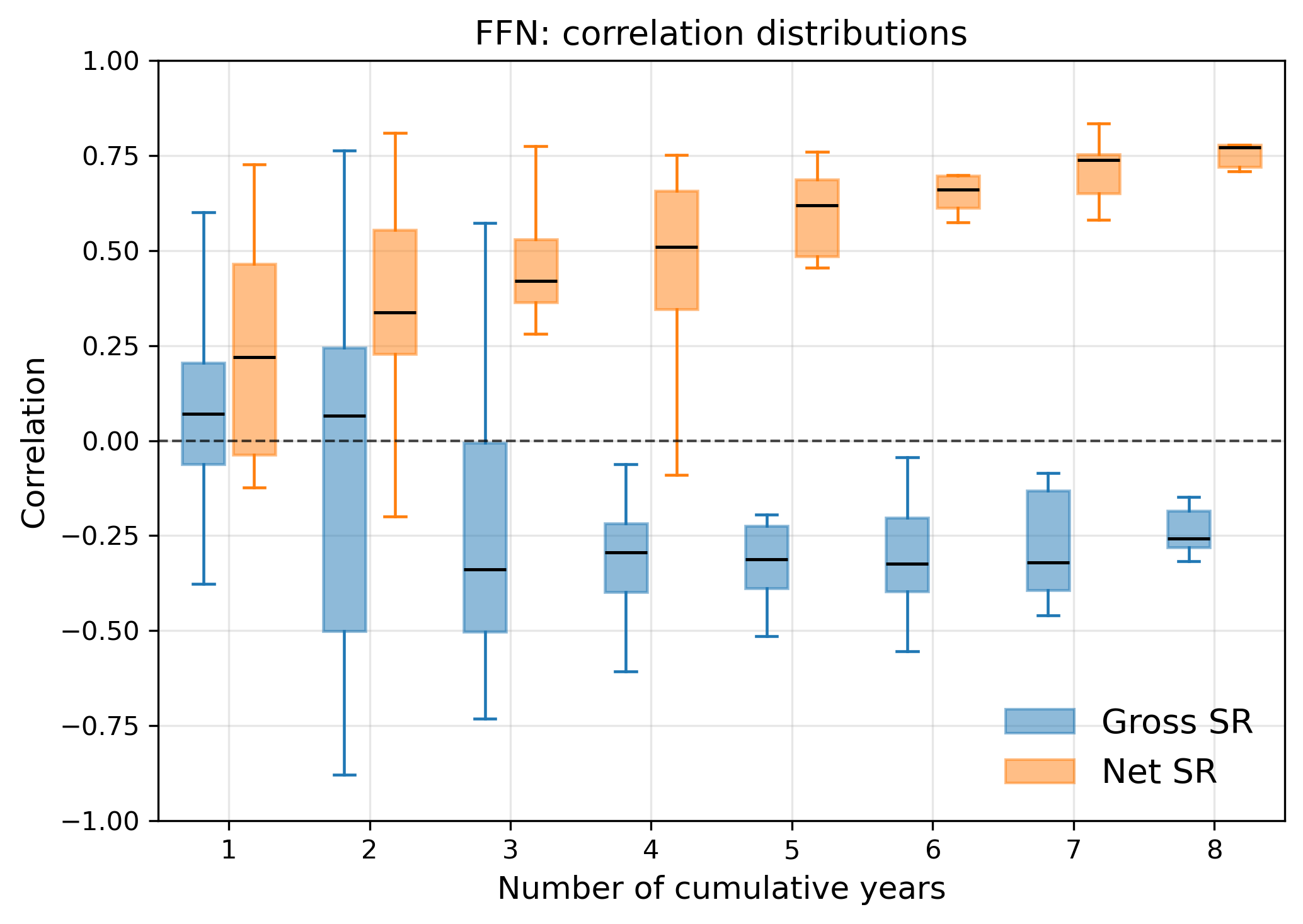}
			\small (b) FFN: Correlation
		\end{minipage}
		
		\vspace{2mm}
		
		%%%%%%%%%%%%%%%%%%%%%%%%%%%%%%%%%%%%%%%%%%%%%%%%%%%%%%%%%%%
		% Row 2
		%%%%%%%%%%%%%%%%%%%%%%%%%%%%%%%%%%%%%%%%%%%%%%%%%%%%%%%%%%%
		
		\begin{minipage}[t]{0.49\textwidth}
			\centering
			\includegraphics[width=\linewidth]
			{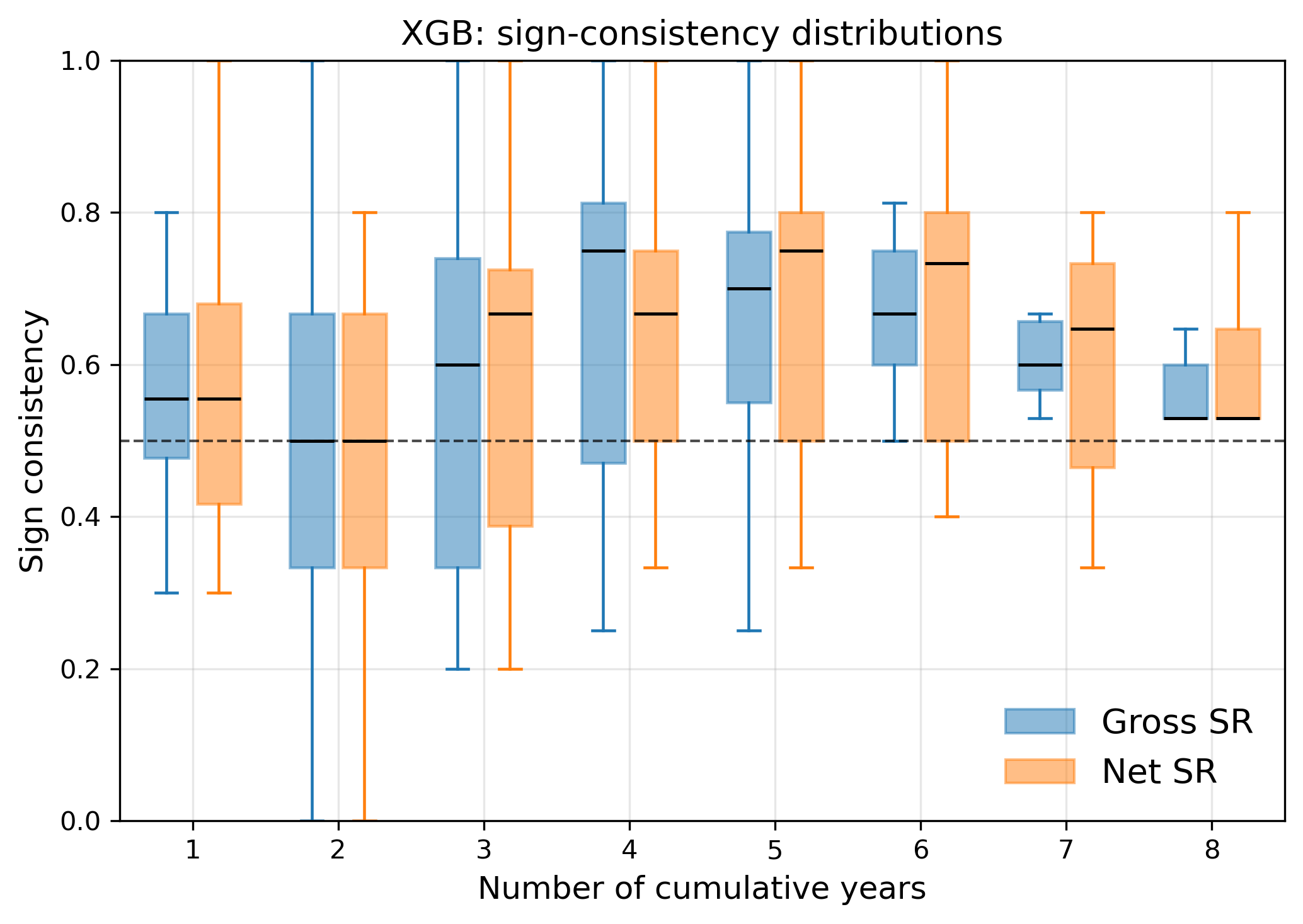}
			\small (c) XGB: Sign consistency
		\end{minipage}
		\hfill
		\begin{minipage}[t]{0.49\textwidth}
			\centering
			\includegraphics[width=\linewidth]
			{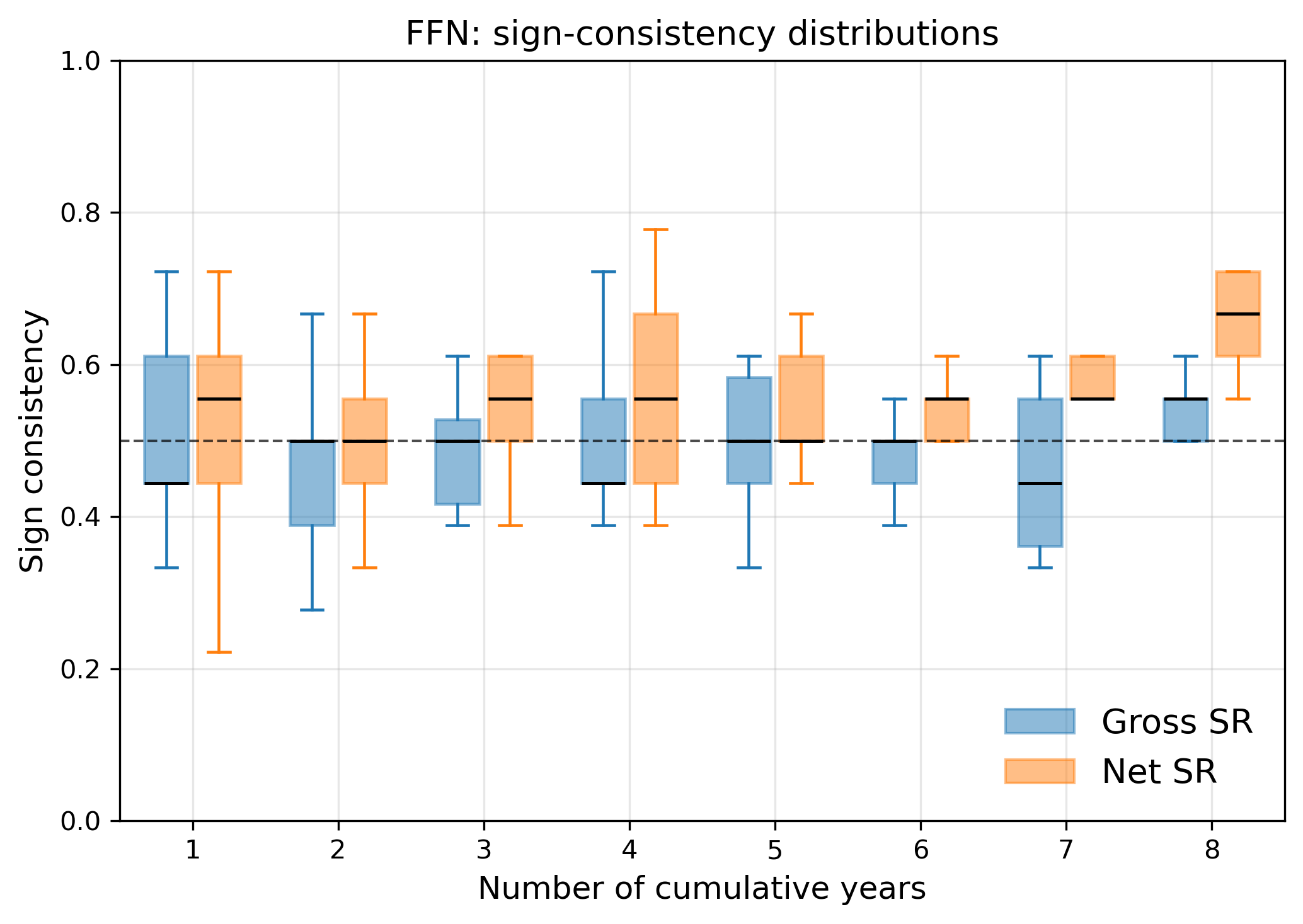}
			\small (d) FFN: Sign consistency
		\end{minipage}
		
		\caption{
			Validation-to-future consistency of cumulative feature contributions.
			For each aggregation horizon $H=1,\ldots,8$, feature contributions are
			aggregated over $H$ consecutive validation periods and compared with those
			aggregated over the immediately following $H$ out-of-sample trading periods.
			The comparison window is then rolled forward by one period, yielding
			$N_H=21-2H$ validation-OOS window pairs. 
%			Accordingly, there are 19
%			comparisons for $H=1$, 17 for $H=2$, and progressively fewer comparisons
%			as the aggregation horizon increases, with 5 remaining at $H=8$.
			Boxplots summarise the distributions of the cross-feature Pearson
			correlations $\rho_y^{(H)}(Q)$ (top row) and sign-consistency measures
			$SC_y^{(H)}(Q)$ (bottom row) across the rolling window pairs. At each
			horizon, gross and net Sharpe-ratio attributions are reported separately.
			The central line denotes the median, the lower and upper edges of each box
			denote the first and third quartiles, respectively, and the whiskers extend
			to the most extreme observations within 1.5 times the interquartile range.
		}
		\label{fig:cumulative_year_similarity_8y}
	\end{figure}

	Figure~\ref{fig:cumulative_year_similarity_8y} shows the validation-to-future consistency of cumulative feature contributions over aggregation horizons of up to eight years. Although the number of available rolling-window comparisons decreases as the horizon increases, longer aggregation horizons generally indicate stronger persistence in net Sharpe-ratio attribution structures for both XGB and FFN. Across multi-year horizons, the majority of correlations for cumulative net Sharpe attributions are positive, while sign consistency exceeds 50\% in the majority of rolling-window comparisons at several longer horizons, particularly beyond five years.

	The results also highlight differences between the two allocation models. For FFN, turnover attributions exhibit particularly strong persistence across trading periods. Given the relatively high turnover of the FFN portfolio, this persistent turnover component provides a plausible channel through which net Sharpe-ratio attributions retain information across periods. Consistent with this interpretation, cumulative net Sharpe attributions exhibit substantially stronger cross-period alignment than their gross Sharpe counterparts, for which persistence weakens and correlations become predominantly negative at longer horizons.
	
	For XGBoost, portfolio turnover is substantially lower, suggesting a smaller role for transaction-cost effects in the persistence of net Sharpe attributions. Instead, both gross and net Sharpe-ratio attribution structures generally exhibit stronger cross-period alignment as the aggregation horizon increases. Taken together, these results suggest that, although the relative roles of gross-performance and turnover channels differ between XGB and FFN, temporal aggregation reveals a persistent component in net Sharpe-ratio feature contributions under both allocation models.

	\subsubsection{Trading performance with feature-exposure interventions}
	
	The preceding analysis provides evidence that validation-period net
	Sharpe-ratio attributions contain economically relevant information about
	the subsequent out-of-sample attribution structure in section~\ref{sec:attribution_persistence} and that Sharpe-ratio attributions exhibit a common component across XGB and FFN in section~\ref{sec:cross-model_attribution_similarity}. We now examine whether these attribution relationships can be translated into improvements
	in realised out-of-sample portfolio performance. Following the intervention
	framework in Section~\ref{sec:feature_exposure_intervention}, we use net
	Sharpe-ratio attributions estimated prior to each trading period, based on
	a transaction-cost rate of $c=0.0005$, to determine feature exposures for
	the subsequent out-of-sample period.
	
	For each allocation model, we consider four attribution-guided exposure
	rules: (i) model-specific attributions without stabilisation;
	(ii) stabilised model-specific attributions; (iii) cross-model averaged
	attributions without stabilisation; and (iv) stabilised cross-model averaged
	attributions. We use the magnitude transformation, which preserves both the sign and magnitude of the attribution, and set the intervention-strength parameter to $\eta=0.5$, with feature exposures constrained to $0\leq\lambda\leq3$.
	All exposure adjustments are therefore constructed using
	information available prior to the corresponding out-of-sample trading
	period.

\begin{figure}[htbp]
		\centering
		
		\begin{minipage}[t]{0.49\textwidth}
			\centering
			\includegraphics[width=\linewidth]
			{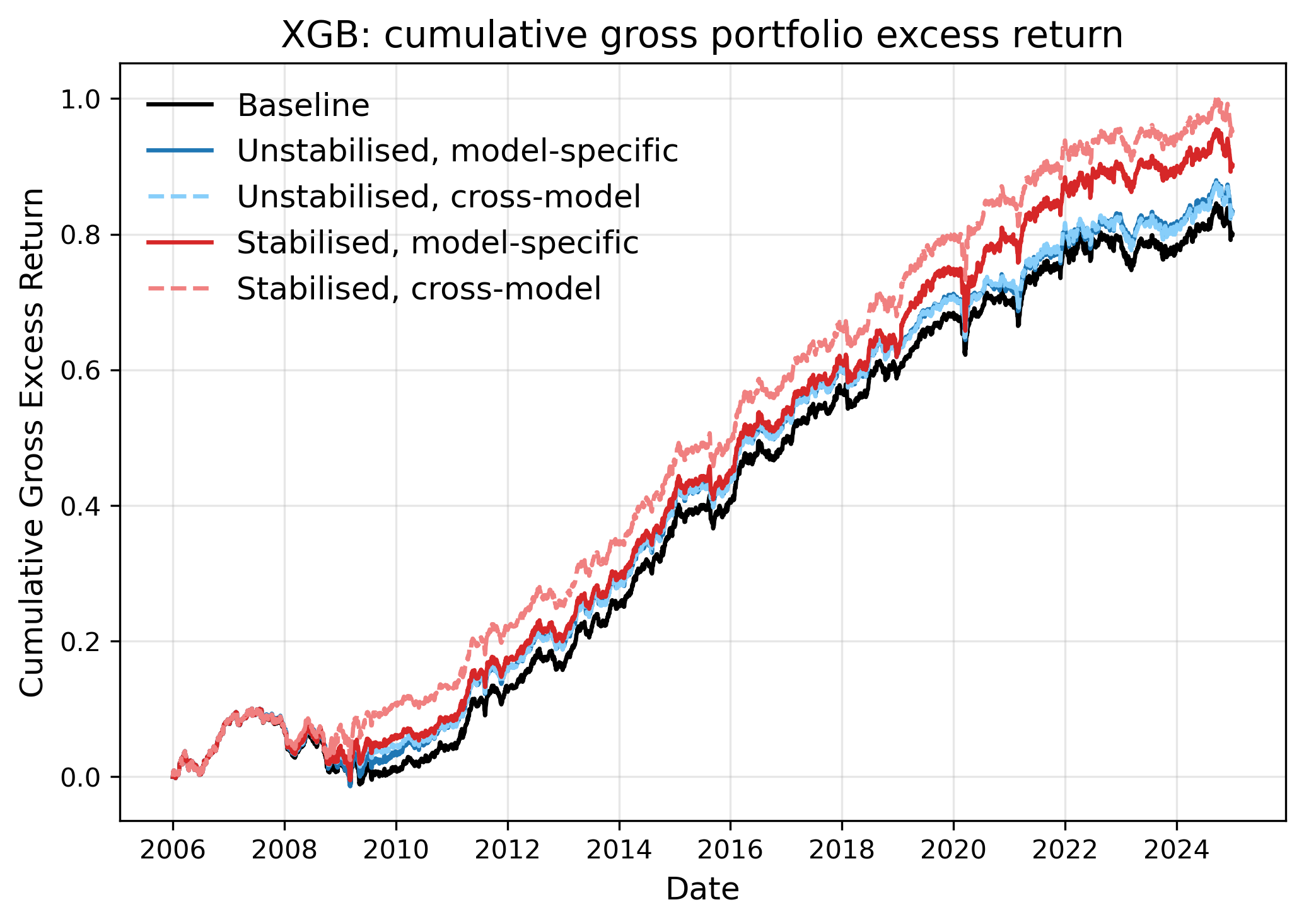}
			\small (a) XGB: gross return
		\end{minipage}
		\hfill
		\begin{minipage}[t]{0.49\textwidth}
			\centering
			\includegraphics[width=\linewidth]
			{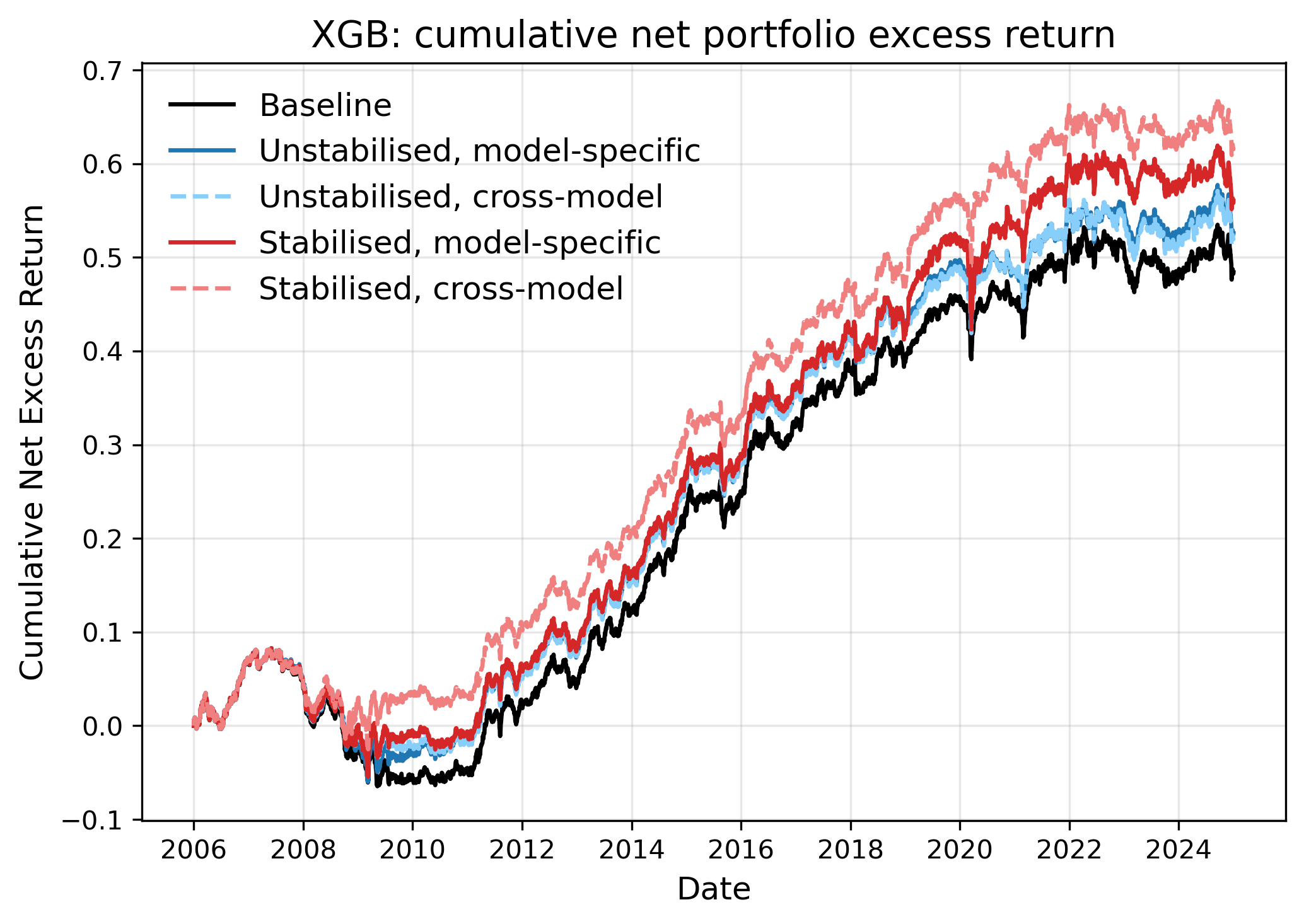}
			\small (b) XGB: net return
		\end{minipage}
		
		\vspace{2mm}
		
		\begin{minipage}[t]{0.49\textwidth}
			\centering
			\includegraphics[width=\linewidth]
			{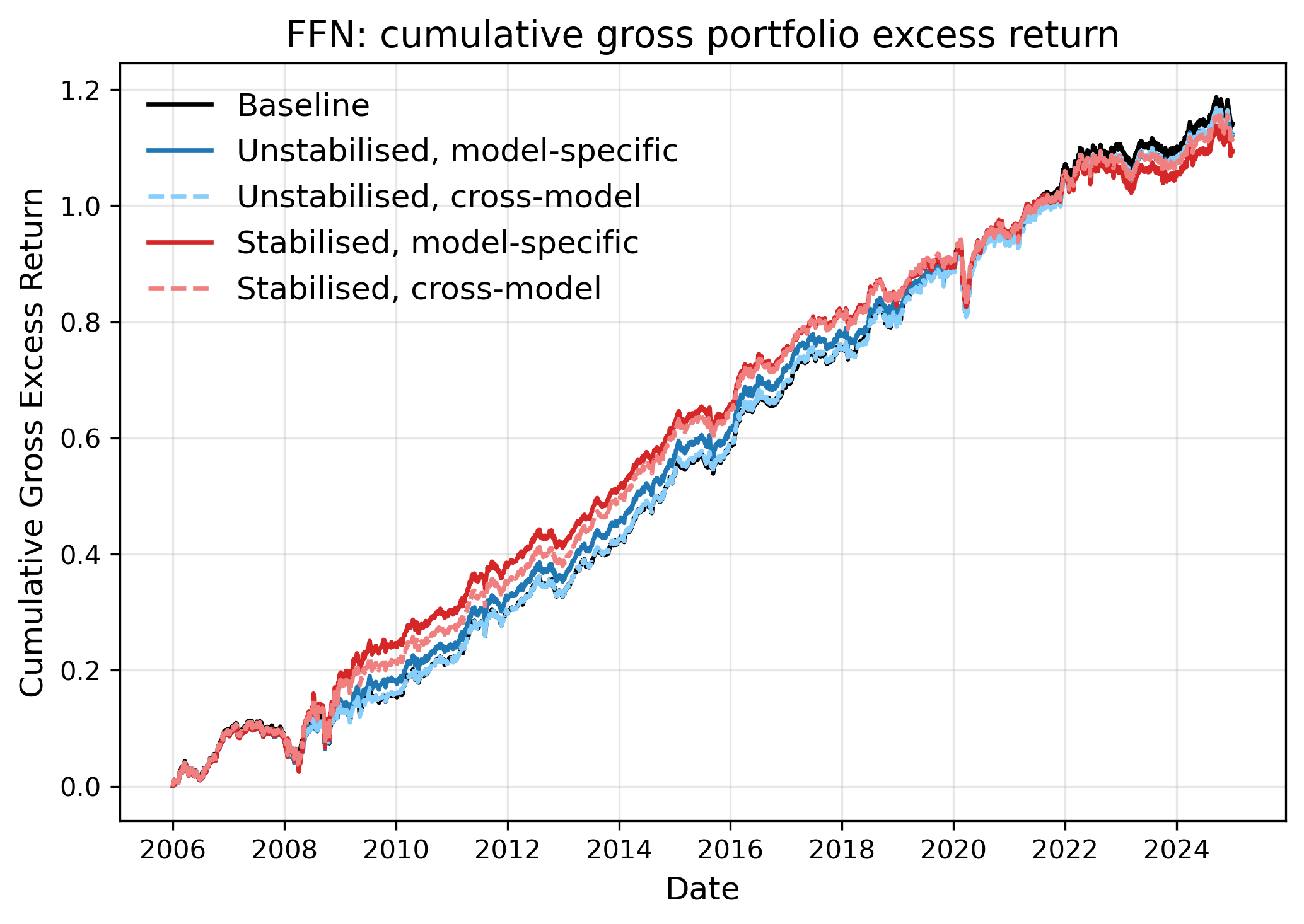}
			\small (c) FFN: gross return
		\end{minipage}
		\hfill
		\begin{minipage}[t]{0.49\textwidth}
			\centering
			\includegraphics[width=\linewidth]
			{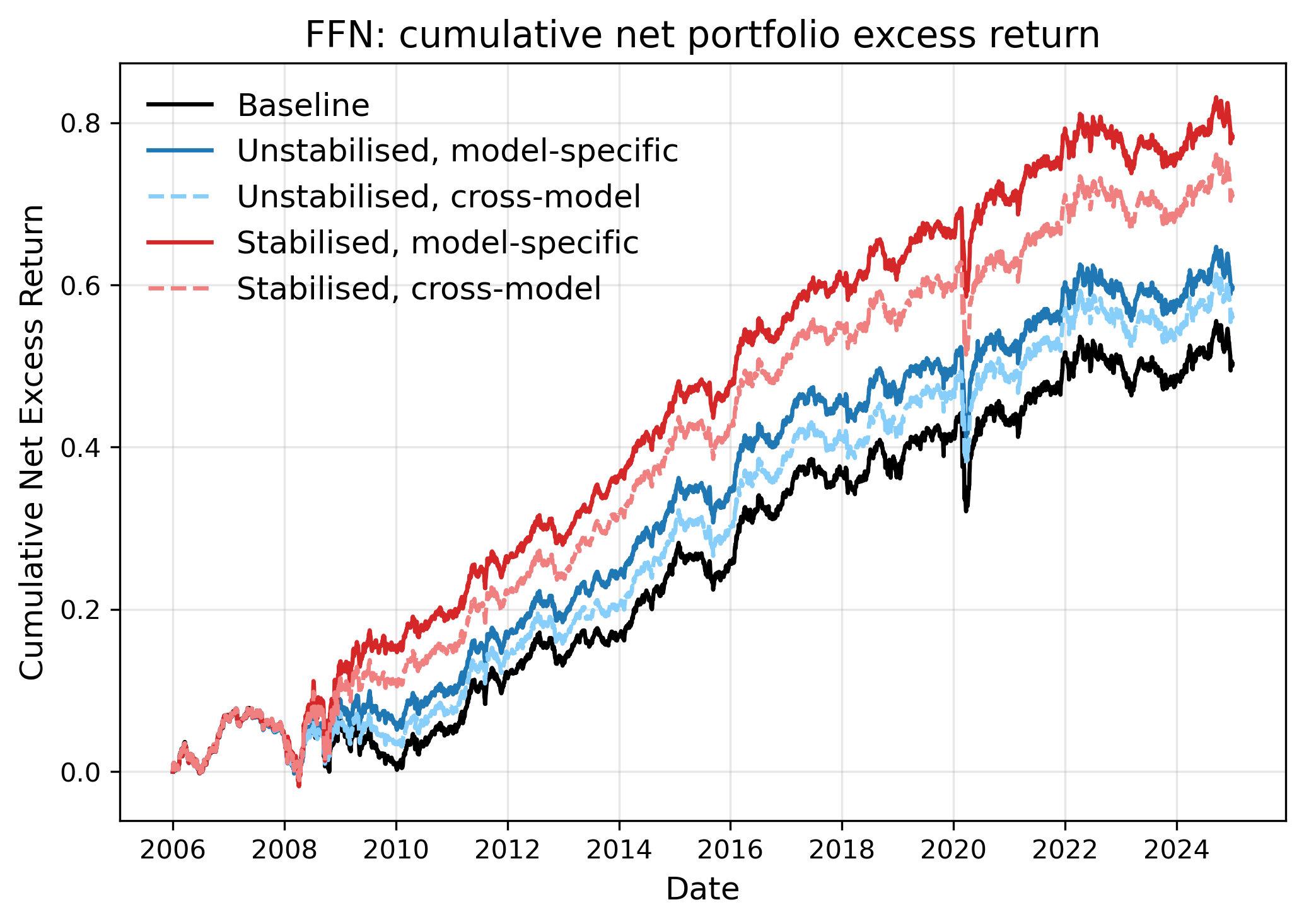}
			\small (d) FFN: net return
		\end{minipage}
		
		\caption{Cumulative gross and net portfolio excess returns with feature exposure interventions for XGB and FFN.}
		\label{fig:feature_scaling_xgb_ffn_gross_net_cumret}
	\end{figure}

\begin{figure}[htbp]
		\centering
		
		%%%%%%%%%%%%%%%%%%%%%%%%%%%%%%%%%%%%%%%%%%%%%%%%%%%%%%%%%%%%%%
		% Row 1: XGB
		%%%%%%%%%%%%%%%%%%%%%%%%%%%%%%%%%%%%%%%%%%%%%%%%%%%%%%%%%%%%%%
		
		\begin{minipage}[t]{0.24\textwidth}
			\centering
			\includegraphics[width=\linewidth]
			{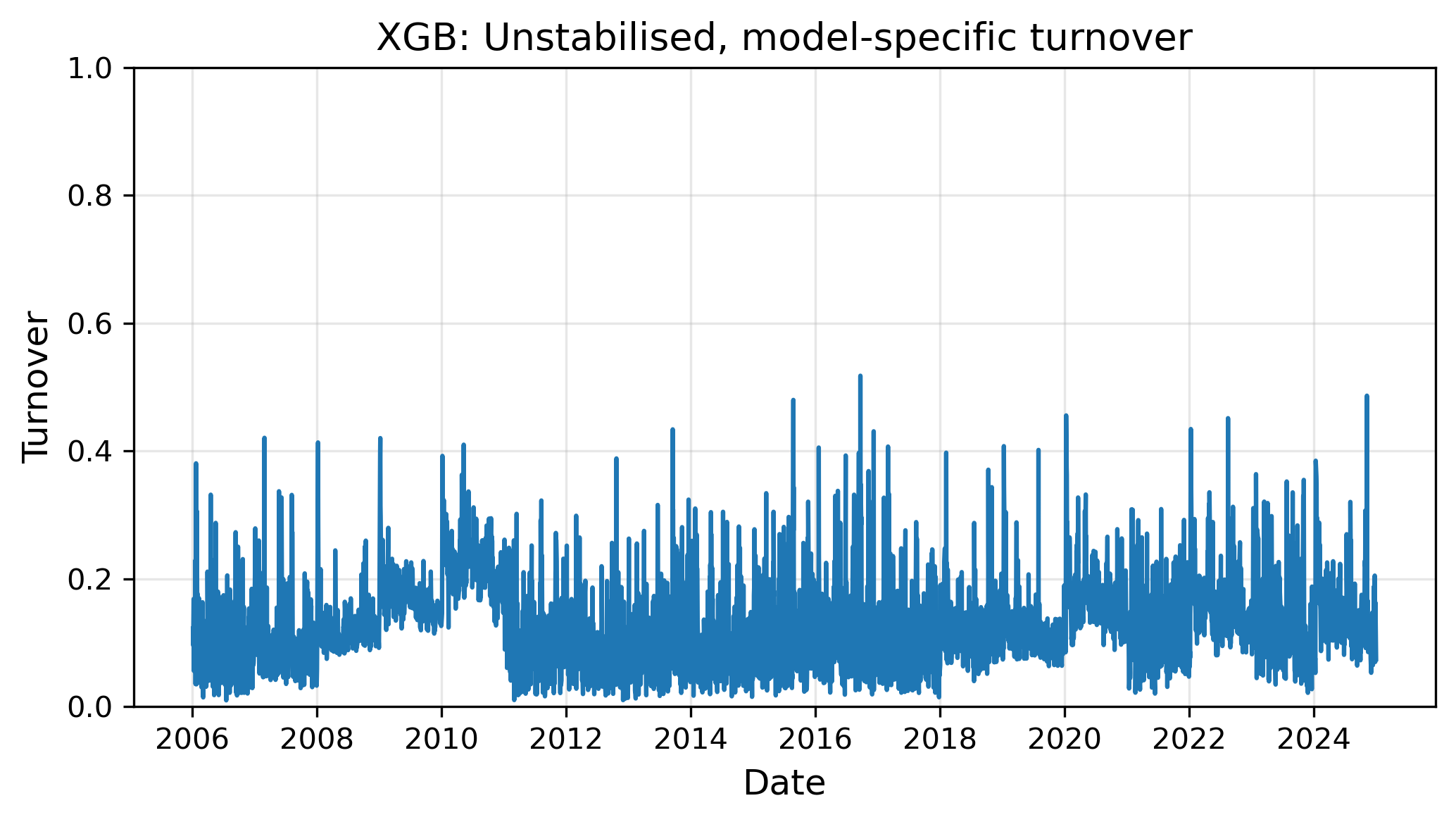}
			\small \centering Unstabilised, model-specific
		\end{minipage}
		\hfill
		\begin{minipage}[t]{0.24\textwidth}
			\centering
			\includegraphics[width=\linewidth]
			{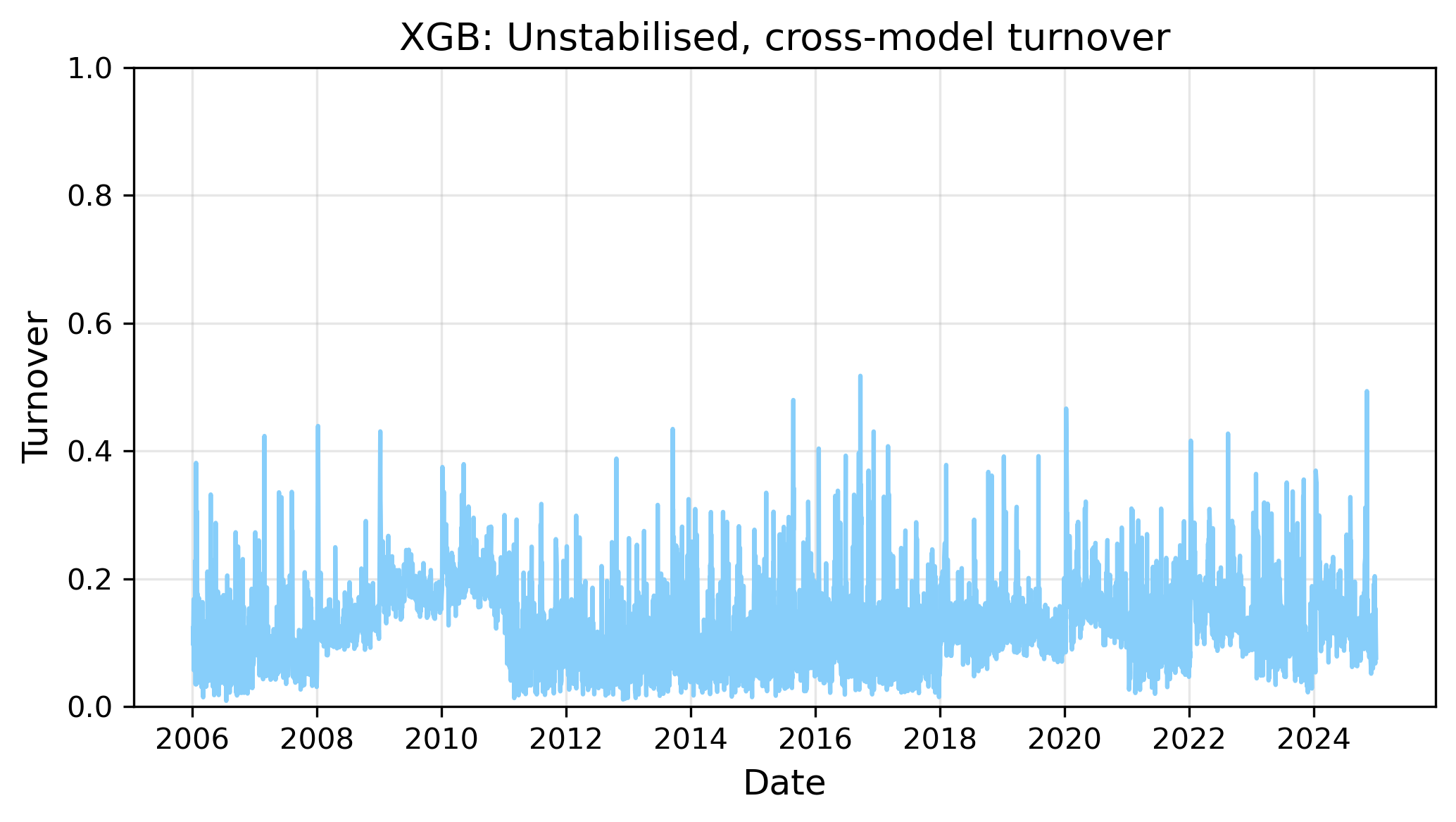}
			\small \centering Unstabilised, cross-model
		\end{minipage}
		\hfill
		\begin{minipage}[t]{0.24\textwidth}
			\centering
			\includegraphics[width=\linewidth]
			{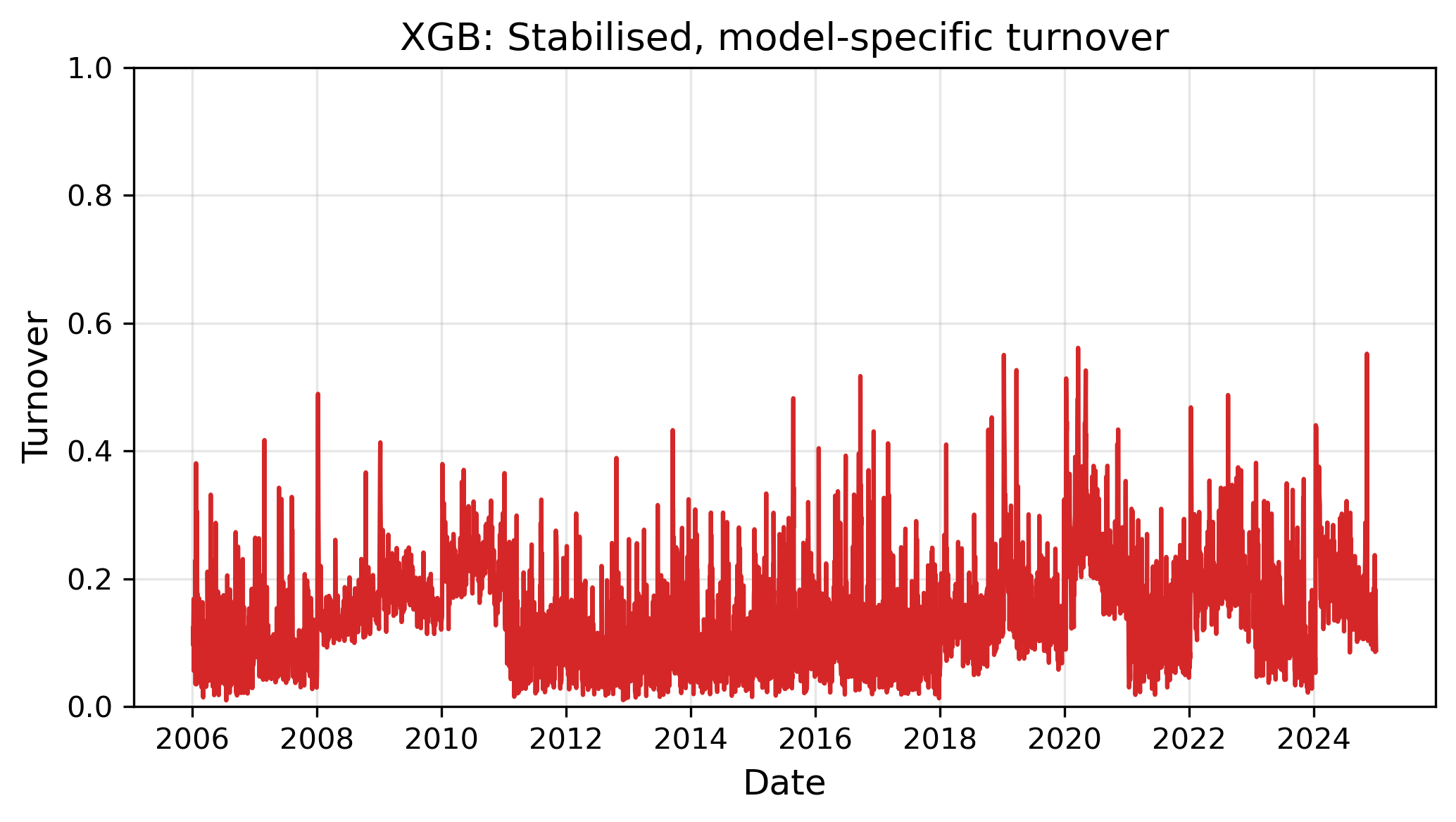}
			\small \centering Stabilised, model-specific
		\end{minipage}
		\hfill
		\begin{minipage}[t]{0.24\textwidth}
			\centering
			\includegraphics[width=\linewidth]
			{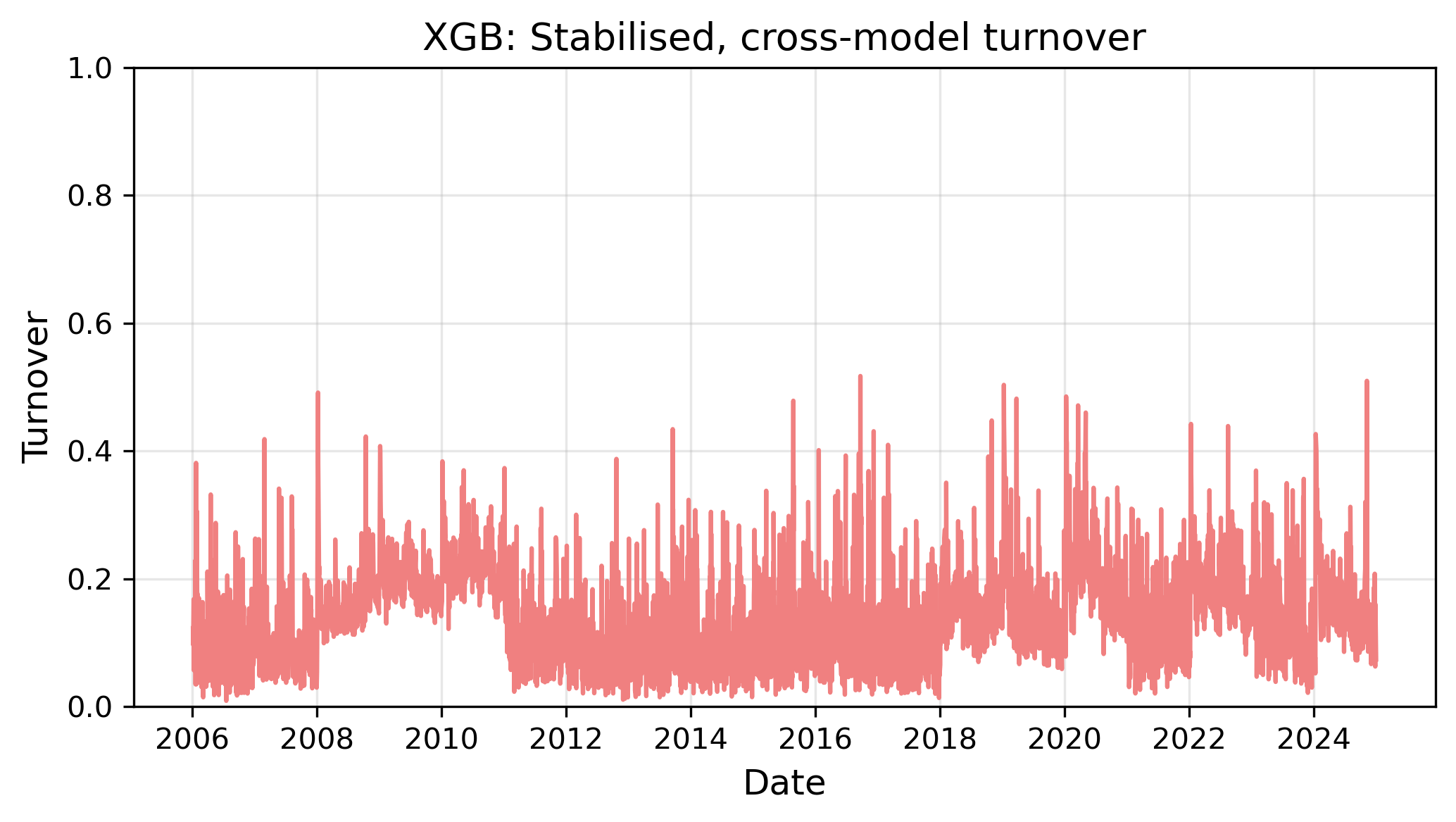}
			\small \centering Stabilised, cross-model
		\end{minipage}
		
		\vspace{1mm}
		
		{\small \textbf{(a) XGB}}
		
		\vspace{2mm}
		
		%%%%%%%%%%%%%%%%%%%%%%%%%%%%%%%%%%%%%%%%%%%%%%%%%%%%%%%%%%%%%%
		% Row 2: FFN
		%%%%%%%%%%%%%%%%%%%%%%%%%%%%%%%%%%%%%%%%%%%%%%%%%%%%%%%%%%%%%%
		
		\begin{minipage}[t]{0.24\textwidth}
			\centering
			\includegraphics[width=\linewidth]
			{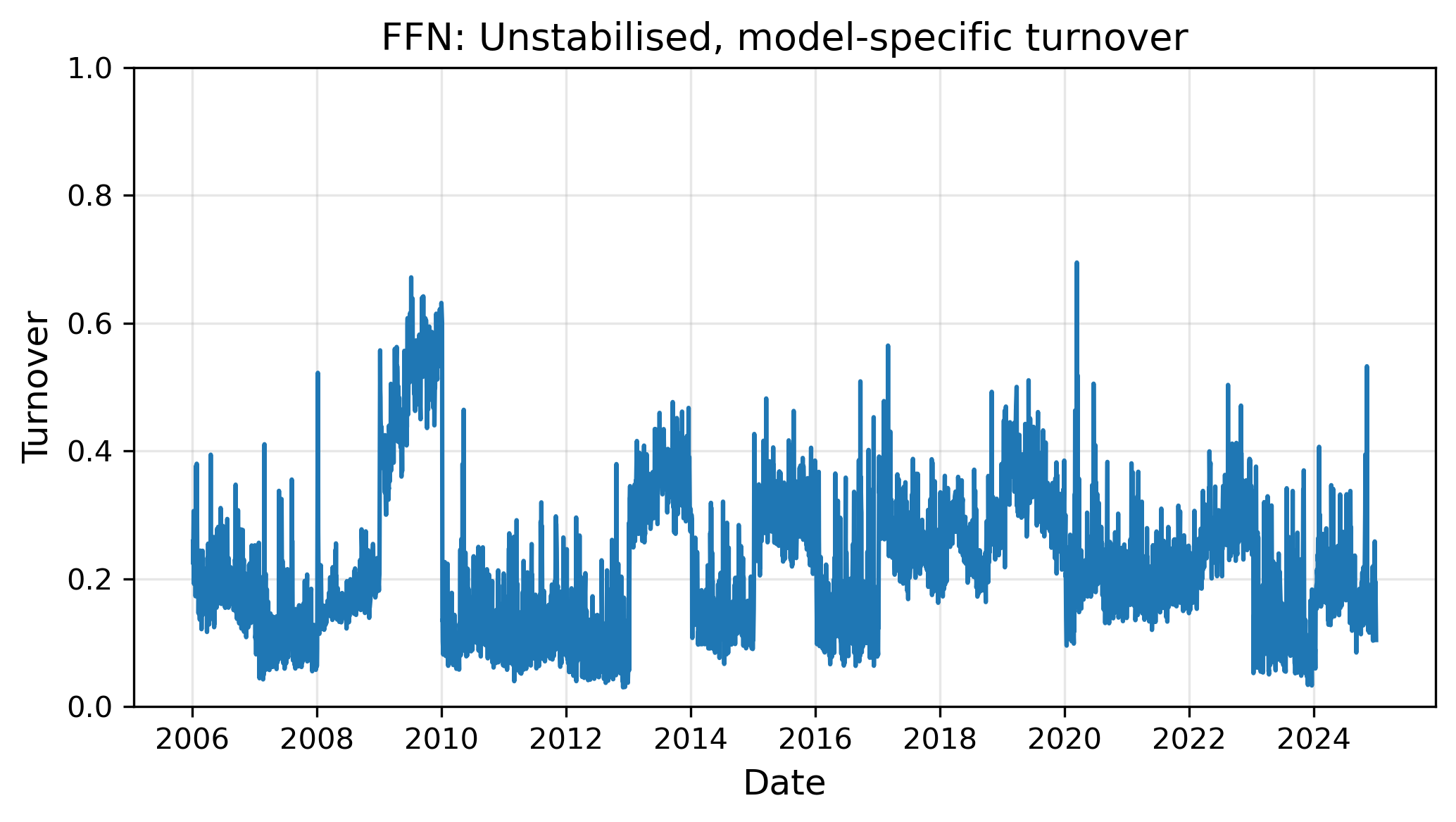}
			\small \centering Unstabilised, model-specific
		\end{minipage}
		\hfill
		\begin{minipage}[t]{0.24\textwidth}
			\centering
			\includegraphics[width=\linewidth]
			{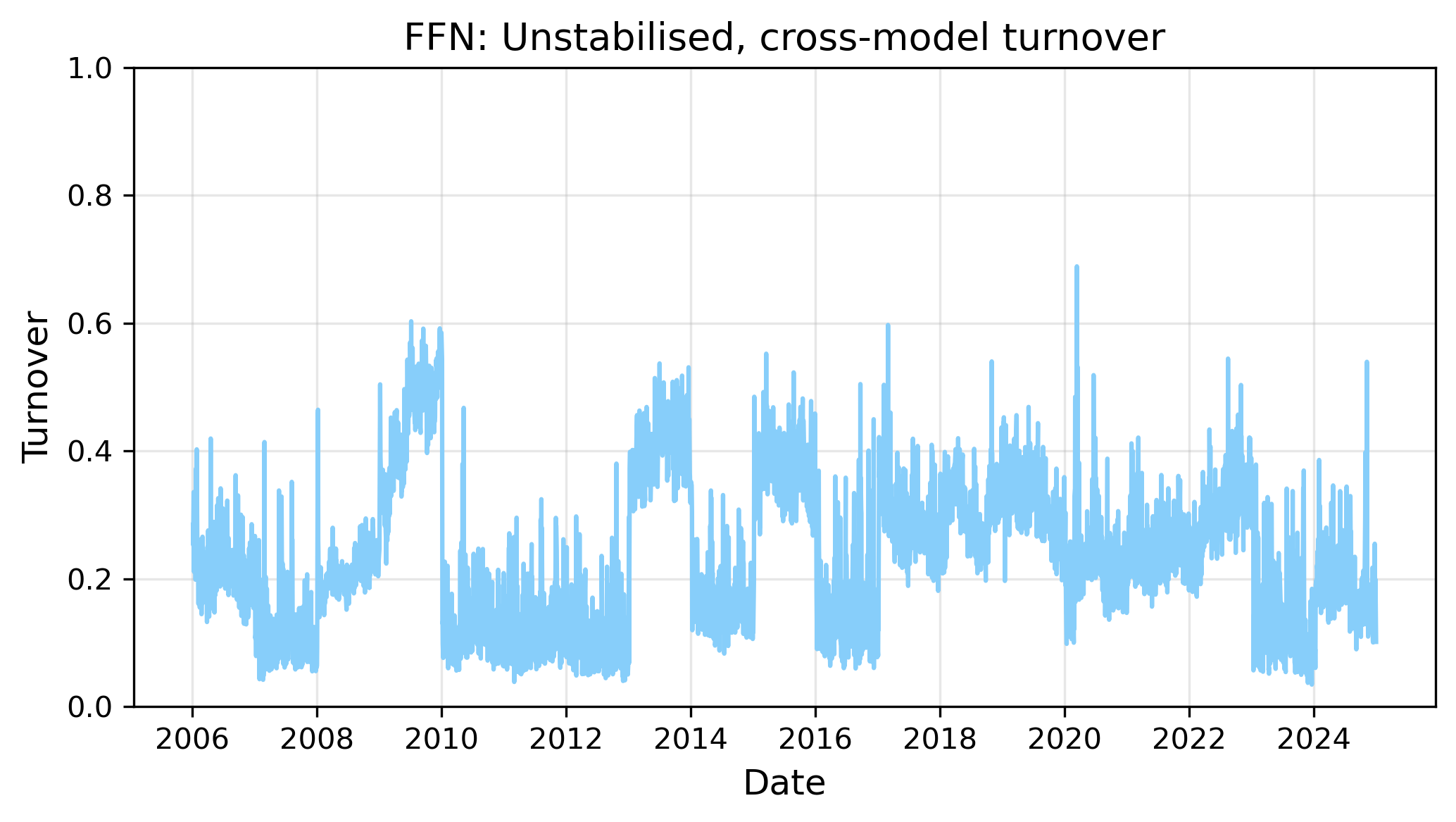}
			\small \centering Unstabilised, cross-model
		\end{minipage}
		\hfill
		\begin{minipage}[t]{0.24\textwidth}
			\centering
			\includegraphics[width=\linewidth]
			{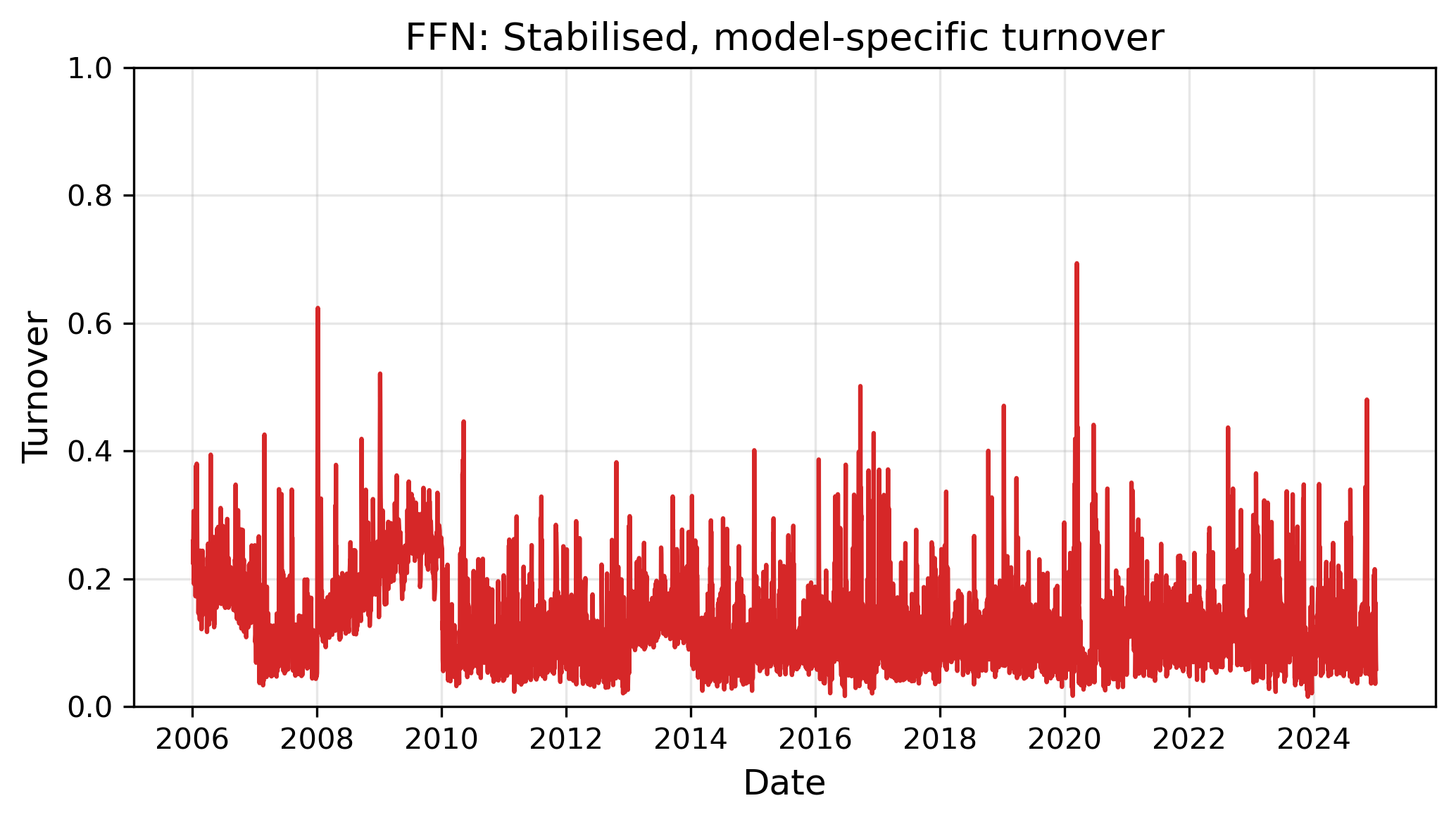}
			\small \centering Stabilised, model-specific
		\end{minipage}
		\hfill
		\begin{minipage}[t]{0.24\textwidth}
			\centering
			\includegraphics[width=\linewidth]
			{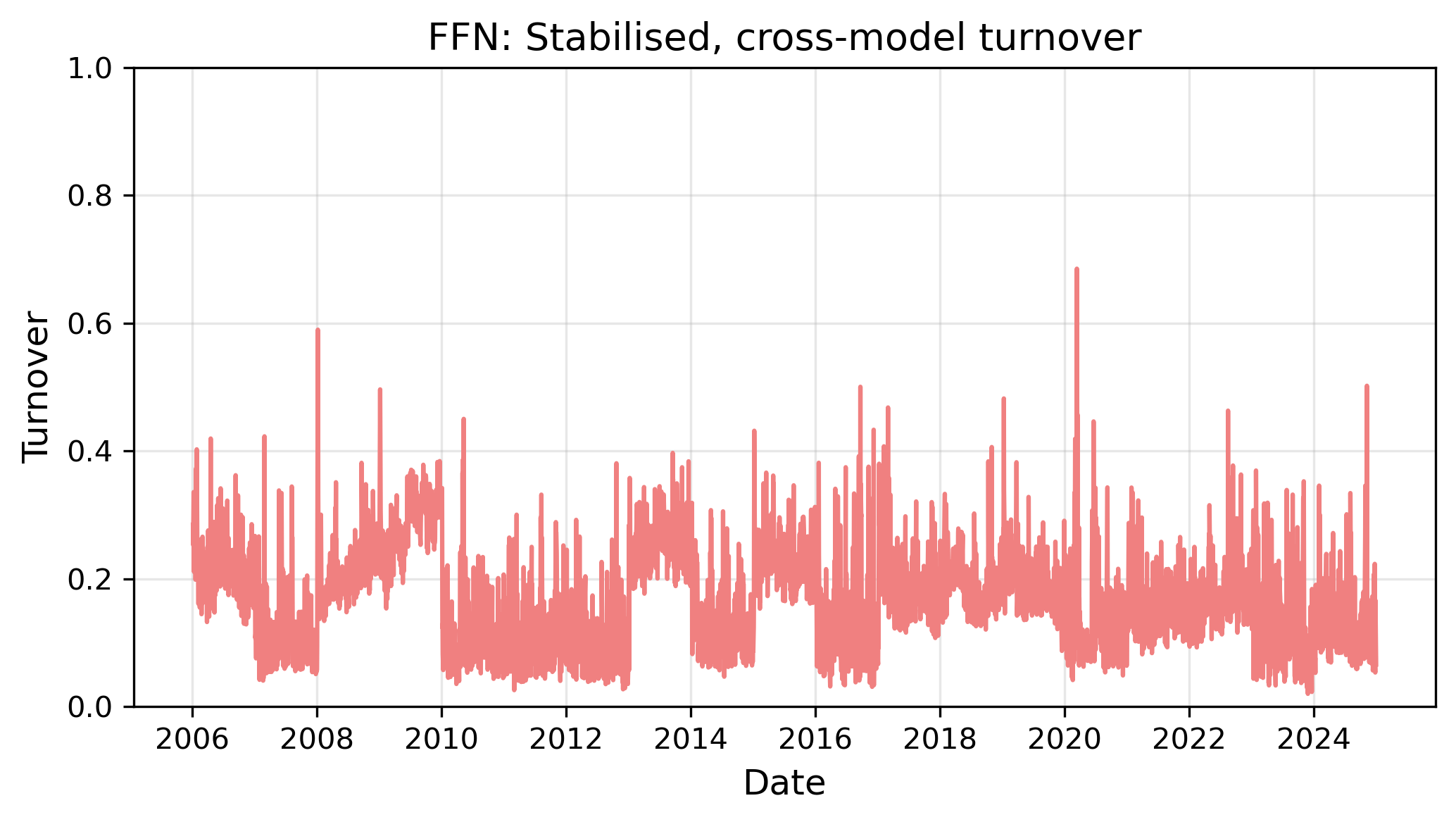}
			\small \centering Stabilised, cross-model
		\end{minipage}
		
		\vspace{1mm}
		
		{\small \textbf{(b) FFN}}
		
		\caption{Portfolio turnover dynamics with feature exposure interventions for XGB and FFN. The four columns compare unstabilised versus stabilised feature exposures and model-specific versus cross-model feature contributions.
		}
		\label{fig:feature_scaling_turnover_mag}
		
	\end{figure}

	Table~\ref{tab:feature_scaling_performance_all} reports the resulting
	portfolio performance. The attribution-guided interventions improve the net
	Sharpe ratio relative to the corresponding baseline portfolio under all
	eight model-rule combinations, with statistically significant improvements
	in seven of the eight cases.
	Figures~\ref{fig:feature_scaling_xgb_ffn_gross_net_cumret} and
	\ref{fig:feature_scaling_turnover_mag} provide the corresponding time-series
	evidence. Figure~\ref{fig:feature_scaling_xgb_ffn_gross_net_cumret} reports
	cumulative gross and net excess returns under the four intervention rules,
	while Figure~\ref{fig:feature_scaling_turnover_mag} reports the associated
	turnover dynamics.

	\begin{table}[htbp]
		\centering
		\caption{Portfolio performance under attribution-guided feature-exposure interventions}
		\small
		
		\resizebox{\textwidth}{!}{
				\begin{tabular}{ccccccccccl}
				\toprule
				Model &
				Update Rule &
				$SR^{gross}$ &
				$\mu^{gross}$ (\%) &
				$\sigma^{gross}$ (\%) &
				$SR^{net}$ &
				$\mu^{net}$ (\%) &
				$\sigma^{net}$ (\%) &
				Avg.\ Turnover (\%) &
				$\Delta SR^{gross}$ &
				$\Delta SR^{net}$ \\
				\midrule
				
				\multicolumn{11}{c}{\textbf{Panel A: Model-specific}}\\
				\midrule
				
				\multirow{2}{*}{\textbf{XGB}}
				& Unstabilised
				& 1.0612
				& 4.39
				& 4.14
				& 0.6695
				& 2.77
				& 4.14
				& 12.85
				& 0.0405
				& $0.0520^{*}$ \\
				
				& Stabilised
				& 1.0684
				& 4.75
				& 4.45
				& 0.6638
				& 2.95
				& 4.45
				& 14.27
				& 0.0477
				& 0.0463 \\
				
				\midrule
				
				\multirow{2}{*}{\textbf{FFN}}
				& Unstabilised
				& 1.2878
				& 5.91
				& 4.59
				& 0.6846
				& 3.14
				& 4.59
				& 21.98
				& -0.0173
				& $0.1096^{***}$ \\
				
				& Stabilised
				& 1.1531
				& 5.76
				& 5.00
				& 0.8250
				& 4.12
				& 5.00
				& 13.02
				& -0.1520
				& $0.2500^{***}$ \\
				
				\midrule
				\multicolumn{11}{c}{\textbf{Panel B: Cross-model}}\\
				\midrule
				
				\multirow{2}{*}{\textbf{XGB}}
				& Unstabilised
				& 1.0645
				& 4.38
				& 4.10
				& 0.6670
				& 2.74
				& 4.10
				& 12.96
				& 0.0438
				& $0.0495^{**}$ \\
				
				& Stabilised
				& 1.1678
				& 5.02
				& 4.30
				& 0.7559
				& 3.25
				& 4.30
				& 14.06
				& 0.1471
				& $0.1384^{**}$ \\
				
				\midrule
				
				\multirow{2}{*}{\textbf{FFN}}
				& Unstabilised
				& 1.3156
				& 5.92
				& 4.50
				& 0.6581
				& 2.96
				& 4.50
				& 23.48
				& 0.0105
				& $0.0831^{***}$ \\
				
				& Stabilised
				& 1.2224
				& 5.88
				& 4.81
				& 0.7800
				& 3.75
				& 4.81
				& 16.89
				& -0.0827
				& $0.2050^{***}$ \\
				
				\bottomrule
			\end{tabular}
		}
		
		\vspace{2mm}
		
		\begin{minipage}{0.95\textwidth}
			\footnotesize
			\textit{Notes:}
			The table reports out-of-sample portfolio performance under
			model-specific and cross-model attribution-guided feature-exposure
			interventions. Panel A uses attribution information from the
			corresponding allocation model, while Panel B uses cross-model averaged
			attribution information. The unstabilised rules use the attribution
			signal from the preceding validation period directly, whereas the
			stabilised rules recursively update the signal.
			$\Delta SR^{gross}$ and $\Delta SR^{net}$ denote changes in gross and
			net Sharpe ratios relative to the corresponding baseline portfolios
			reported in Section~\ref{sec:baseline_portfolio}. Statistical
			significance reported for $\Delta SR^{net}$ is based on the one-sided
			Ledoit-Wolf Sharpe-ratio improvement test.
			Inference is based on 4,999 bootstrap replications using five-trading-day blocks to preserve short-run dependence in daily portfolio returns.
			$^{***}$, $^{**}$, and $^{*}$ denote significance at the 1\%, 5\%,
			and 10\% levels, respectively.
%			The table reports portfolio performance under model-specific and
%			cross-model attribution-guided feature-exposure interventions.
%			The final column reports the statistical significance of the improvement
%			in out-of-sample net Sharpe ratio relative to the corresponding baseline
%			portfolio reported in section~\ref{sec:baseline_portfolio}. Statistical inference is based on the one-sided Ledoit-Wolf
%			Sharpe-ratio improvement test.
%			$^{***}$, $^{**}$, and $^{*}$ denote significance at the 1\%, 5\%,
%			and 10\% levels, respectively.
		\end{minipage}
		
%        \caption{Portfolio performance using model-specific and cross-model magnitude-based feature contributions. The final column reports the statistical significance of the improvement in out-of-sample net Sharpe ratio relative to the baseline portfolio based on the one-sided Ledoit-Wolf Sharpe-ratio improvement test ($^{***}p<0.01$, $^{**}p<0.05$, $^{*}p<0.10$).}
		\label{tab:feature_scaling_performance_all}
	\end{table}
	
	Under the model-specific unstabilised rule, the net Sharpe ratio of XGB
	increases from 0.6175 to 0.6695, an improvement of 0.0520 that is
	significant at the 10\% level. For FFN, the corresponding net Sharpe ratio
	increases from 0.5750 to 0.6846, with an improvement of 0.1096 that is
	significant at the 1\% level. These results indicate that even the
	unregularised model-specific attribution signal contains economically useful
	information for subsequent feature-exposure decisions.
	
	The effects of stabilisation and cross-model averaging differ across the
	two allocation models. For XGB, cross-model information is particularly
	useful. The cross-model intervention increases the net Sharpe ratio from
	0.6175 to 0.6670 without stabilisation and to 0.7559 with stabilisation,
	corresponding to improvements of 0.0495 and 0.1384, respectively, both
	significant at the 5\% level. The improvement under the stabilised
	cross-model rule is accompanied by an increase in gross Sharpe ratio from
	1.0207 to 1.1678, while average daily turnover remains close to the
	baseline level. These results suggest that XGB
	benefits more from combining attribution information across models than
	from temporal stabilisation alone.
	
	For FFN, temporal stabilisation plays a more prominent role. Under the
	stabilised model-specific rule, the net Sharpe ratio increases from 0.5750
	to 0.8250, while average daily turnover falls from 26.70\% to 13.02\%.
	Under the stabilised cross-model rule, the net Sharpe ratio reaches 0.7800
	and turnover declines to 16.89\%. Both net Sharpe improvements are
	significant at the 1\% level. Notably, these improvements occur despite
	lower gross Sharpe ratios than the FFN baseline, indicating that the gains
	are primarily associated with lower trading intensity and transaction
	costs.

	The results therefore reveal different transmission channels across the two
	model families. For XGB, whose baseline portfolio already exhibits relatively low turnover, attribution-guided feature scaling primarily improves gross risk-adjusted performance, particularly when cross-model information is combined with stabilisation. Cross-model averaging allows XGB to incorporate complementary attribution information from FFN, which tends to capture a broader set of potential arbitrage opportunities. This additional information may contribute to the observed improvement in the gross Sharpe ratio. For FFN, which has substantially
	higher baseline turnover, the main economic benefit of stabilisation arises
	through reduced trading intensity and the associated transaction costs.
	Despite these differences, attribution-guided feature-exposure adjustments
	increase net Sharpe ratios across both allocation models.

%	Figures~\ref{fig:feature_scaling_xgb_ffn_gross_net_cumret} and
%	\ref{fig:feature_scaling_turnover_mag} provide the corresponding time-series
%	evidence. Figure~\ref{fig:feature_scaling_xgb_ffn_gross_net_cumret} reports
%	cumulative gross and net excess returns under the four intervention rules,
%	while Figure~\ref{fig:feature_scaling_turnover_mag} reports the associated
%	turnover dynamics.
	
	Overall, the out-of-sample results provide direct economic evidence that
	feature attributions estimated prior to the trading period can be used to
	guide subsequent feature-exposure decisions. Stabilisation and cross-model
	averaging provide complementary forms of regularisation, although their
	relative benefits depend on the allocation model. The statistically
	significant improvements in net Sharpe ratio provide the primary economic
	validation of the attribution-guided intervention framework.

	\begin{table}[htbp]
		\centering
		%		\caption{Significance of arbitrage alphas}
		\caption{Factor-adjusted alphas of attribution-guided portfolios}
		\label{tab:portfolio_alpha}
		\small
		\setlength{\tabcolsep}{5pt}
		
		\resizebox{\textwidth}{!}{
			\begin{tabular}{ccclcclc}
				\toprule
				
				%			& &
				%			\multicolumn{3}{c}{Gross} &
				%			\multicolumn{3}{c}{Net} \\
				%			\cmidrule(lr){3-5}
				%			\cmidrule(lr){6-8}
				%			
				%			Model
				%			& Update Rule
				%			& $\alpha$ (\%)
				%			& $t_{\alpha}$
				%			& $R^2$ (\%)
				%			& $\alpha$ (\%)
				%			& $t_{\alpha}$
				%			& $R^2$ (\%) \\
				
				Model
				& Update Rule
				& $\alpha^{gross}$ (\%)
				& $t_{\alpha}^{gross}$
				& $R^{2,gross}$ (\%)
				& $\alpha^{net}$ (\%)
				& $t_{\alpha}^{net}$
				& $R^{2,net}$ (\%) \\

				\midrule
				
				% =====================================================
				% Baseline
				% =====================================================
				
				XGB
				& Baseline
				& 2.85
				& 3.19$^{***}$
				& 16.28
				& 1.19
				& 1.32
				& 16.26 \\
				
				FFN
				& Baseline
				& 4.48
				& 4.44$^{***}$
				& 17.49
				& 1.12
				& 1.11
				& 17.47 \\
				
				% =====================================================
				% Panel A: Model-specific
				% =====================================================
				
				\midrule
				\multicolumn{8}{c}{\textbf{Panel A: Model-specific}} \\
				\midrule
				
				\multirow{2}{*}{XGB}
				& Unstabilised
				& 3.02
				& 3.35$^{***}$
				& 16.18
				& 1.40
				& 1.55
				& 16.17 \\
				
				& Stabilised
				& 3.23
				& 3.33$^{***}$
				& 18.12
				& 1.43
				& 1.47
				& 18.10 \\
				
				\midrule
				
				\multirow{2}{*}{FFN}
				& Unstabilised
				& 4.42
				& 4.35$^{***}$
				& 16.85
				& 1.65
				& 1.62
				& 16.82 \\
				
				& Stabilised
				& 4.24
				& 3.83$^{***}$
				& 14.30
				& 2.60
				& 2.35$^{**}$
				& 14.30 \\
				
				% =====================================================
				% Panel B: Cross-model
				% =====================================================
				
				\midrule
				\multicolumn{8}{c}{\textbf{Panel B: Cross-model}} \\
				\midrule
				
				\multirow{2}{*}{XGB}
				& Unstabilised
				& 3.02
				& 3.39$^{***}$
				& 16.24
				& 1.39
				& 1.55
				& 16.22 \\
				
				& Stabilised
				& 3.62
				& 3.87$^{***}$
				& 17.27
				& 1.85
				& 1.97$^{**}$
				& 17.27 \\
				
				\midrule
				
				\multirow{2}{*}{FFN}
				& Unstabilised
				& 4.40
				& 4.43$^{***}$
				& 18.04
				& 1.44
				& 1.45
				& 18.02 \\
				
				& Stabilised
				& 4.36
				& 4.11$^{***}$
				& 15.57
				& 2.23
				& 2.10$^{**}$
				& 15.57 \\
				
				\bottomrule
			\end{tabular}
		}
		\vspace{2mm}
		
		\begin{minipage}{0.98\textwidth}
			\footnotesize
			%			\textit{Notes:}
			%			The table reports factor-adjusted performance for the baseline and
			%			attribution-guided feature-exposure strategies. The baseline portfolios
			%			are reported first as common benchmarks. Panel A reports model-specific
			%			feature-exposure adjustments, while Panel B reports cross-model
			%			feature-exposure adjustments. Gross performance is evaluated at zero
			%			transaction costs, while net performance is evaluated at a
			%			transaction-cost rate of 5 basis points per transaction. Alpha is
			%			estimated from a daily time-series regression of portfolio excess
			%			returns on the Fama-French five factors and is annualised.
			%			$t_{\alpha}$ denotes the Newey-West $t$-statistic for the null
			%			hypothesis $H_0:\alpha=0$, and $R^2$ of regression is reported in percentage terms.
			%			Statistical inference is two-sided. $^{***}$, $^{**}$, and $^{*}$
			%			denote significance at the 1\%, 5\%, and 10\% levels, respectively.
			\textit{Notes:}
			The table reports Fama-French five-factor alphas for the baseline and
			attribution-guided arbitrage portfolios. The baseline portfolios are reported first as
			benchmarks. Panel A reports model-specific feature-exposure interventions,
			while Panel B reports cross-model interventions. Net returns incorporate a transaction
			cost of 5 basis points per transaction. Alpha is estimated from daily
			time-series regressions of portfolio excess returns on the Fama-French
			five factors and is annualised. $t_{\alpha}$ denotes the Newey-West
			$t$-statistic for $H_0:\alpha=0$, and $R^2$ of regression is reported. Statistical inference is two-sided. $^{***}$, $^{**}$, and $^{*}$
			denote significance at the 1\%, 5\%, and 10\% levels, respectively.
		\end{minipage}
	\end{table}

	Table~\ref{tab:portfolio_alpha} reports the Fama-French five-factor
	alphas of the baseline and attribution-guided arbitrage portfolios.
	All gross alphas are positive and statistically significant at the 1\%
	level, indicating that the gross returns generated by the arbitrage
	portfolios are not fully explained by their exposures to the Fama-French
	five factors. In contrast, the net alphas of both baseline portfolios are
	positive but statistically insignificant after accounting for transaction
	costs.
	
	The feature-exposure interventions improve the factor-adjusted portfolios performance, with the strongest evidence arising under the
	stabilised specifications. For XGB, the stabilised cross-model intervention
	produces an annualised net alpha of 1.85\%, which is statistically
	significant at the 5\% level, compared with an insignificant baseline net
	alpha of 1.19\%. For FFN, the stabilised model-specific and cross-model
	interventions generate annualised net alphas of 2.60\% and 2.23\%,
	respectively, both significant at the 5\% level. 
	The results therefore indicate that the economic gains from feature-exposure
	interventions are not solely attributable to conventional Fama-French
	factor exposures. In addition to improving
	the targeted net Sharpe ratio, the attribution-guided interventions can
	generate statistically significant positive alphas after transaction costs.

	\subsubsection{Transaction cost sensitivity}
	\label{sec:tc_sensitivity}
	
	Since the proposed feature attribution is constructed with respect to the net Sharpe ratio, the learnt feature contributions may depend on the assumed transaction cost. We therefore examine the robustness of the proposed framework under different market friction assumptions.
	
	\begin{figure}[htbp]
		\centering
		
		% ==================== XGB: Unstabilised ====================
		\begin{minipage}[t]{0.46\textwidth}
			\centering
			\includegraphics[
			width=\linewidth,
			height=0.2\textheight,
			keepaspectratio
			]{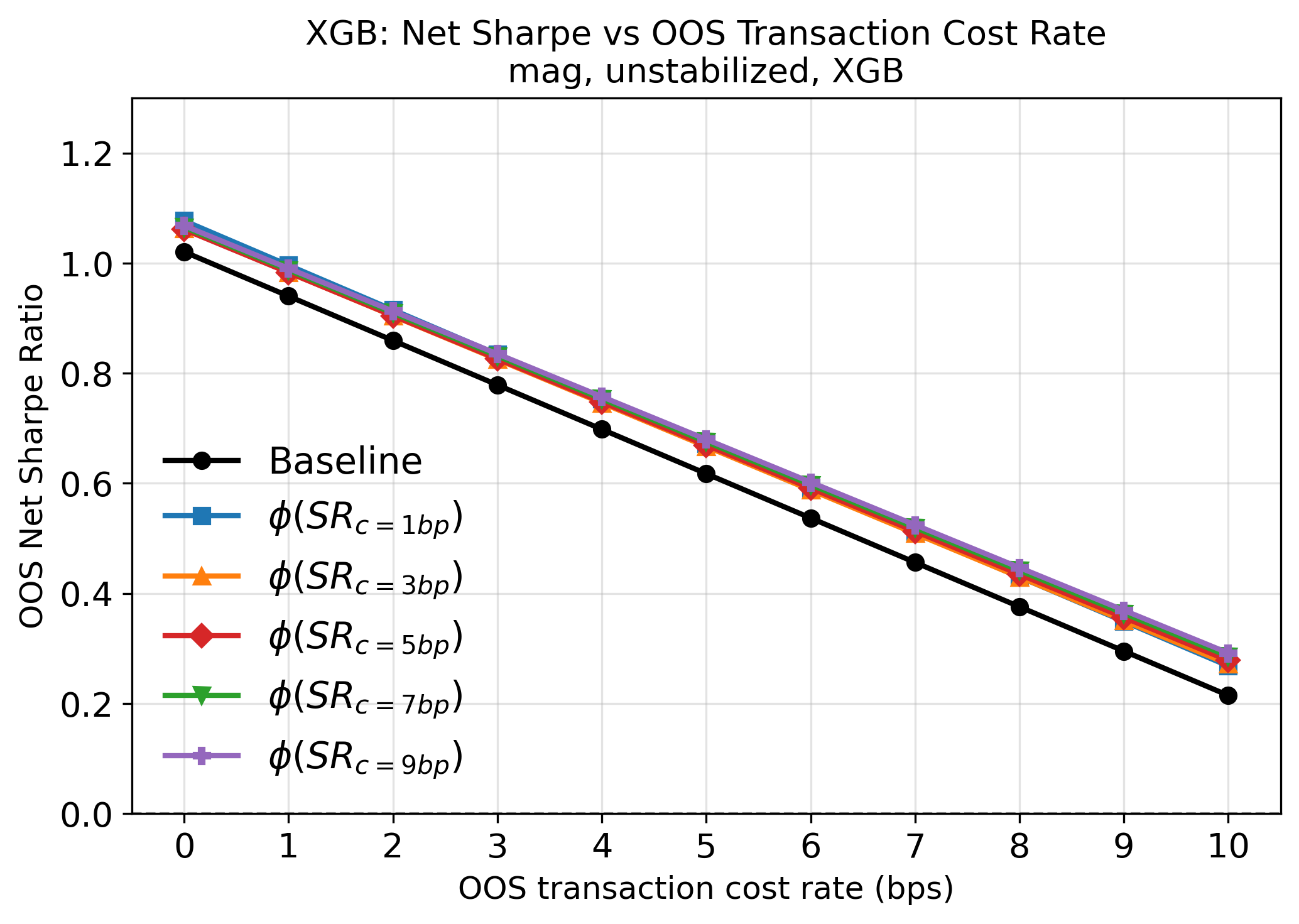}
			\small (a) XGB: Unstabilised, model-specific
			\label{fig:tc_xgb_ms_unstabilised}
		\end{minipage}
		\hfill
		\begin{minipage}[t]{0.46\textwidth}
			\centering
			\includegraphics[
			width=\linewidth,
			height=0.2\textheight,
			keepaspectratio
			]{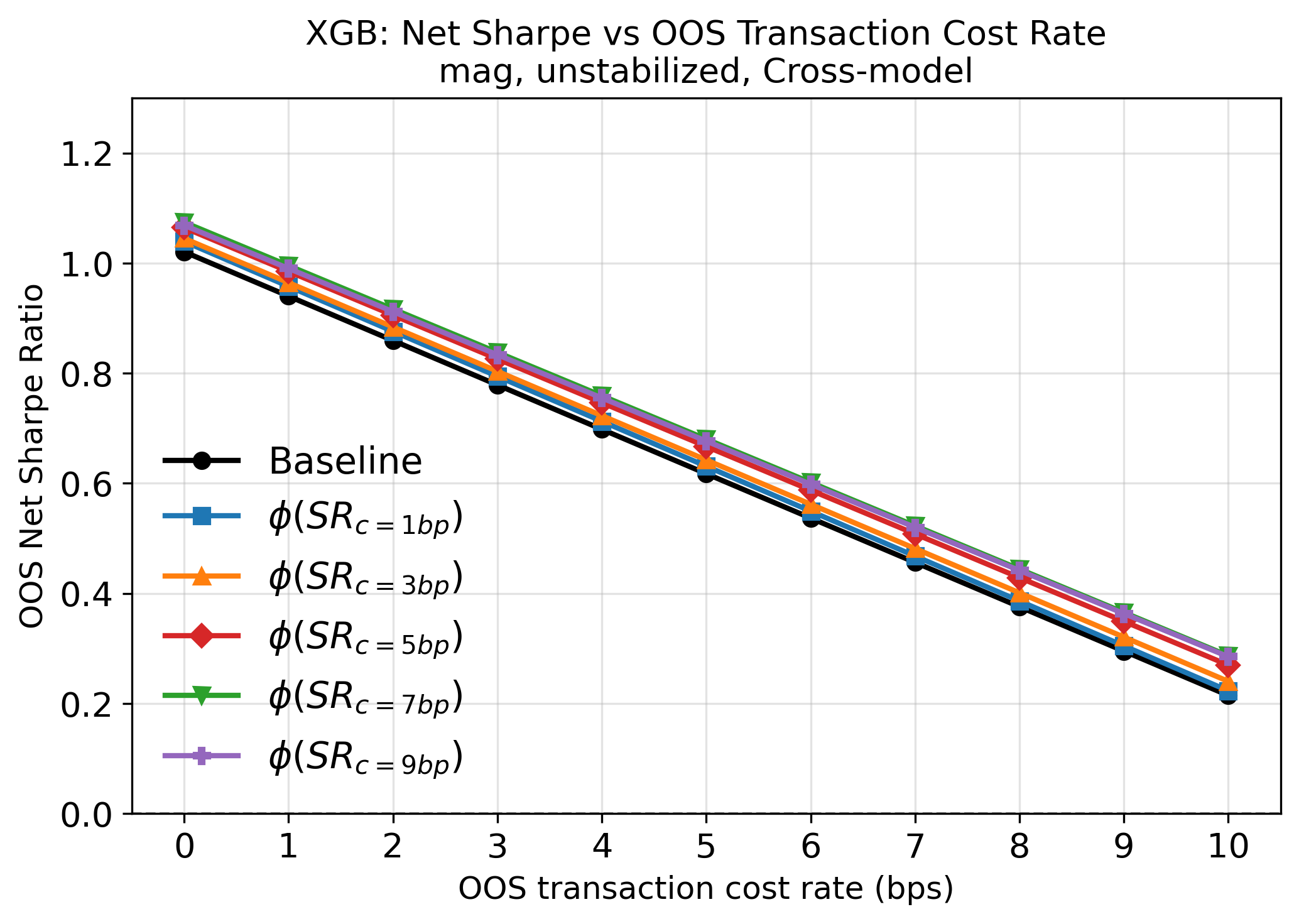}
			\small (b) XGB: Unstabilised, cross-model
			\label{fig:tc_xgb_cross_unstabilised}
		\end{minipage}
		
		\vspace{0.15em}
		
		% ==================== XGB: Stabilised ====================
		\begin{minipage}[t]{0.46\textwidth}
			\centering
			\includegraphics[
			width=\linewidth,
			height=0.2\textheight,
			keepaspectratio
			]{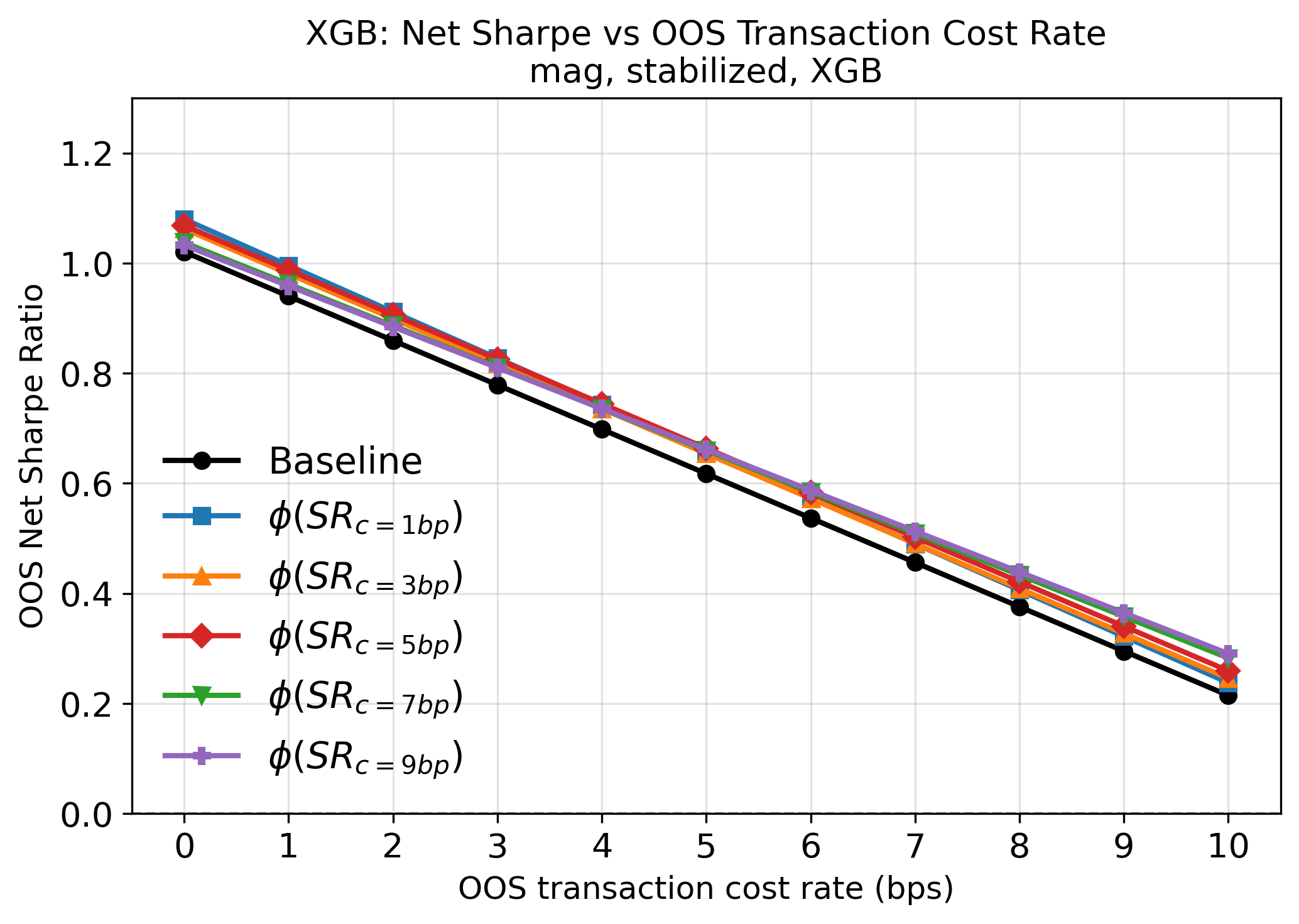}
			\small (c) XGB: Stabilised, model-specific
			\label{fig:tc_xgb_ms_stabilised}
		\end{minipage}
		\hfill
		\begin{minipage}[t]{0.46\textwidth}
			\centering
			\includegraphics[
			width=\linewidth,
			height=0.2\textheight,
			keepaspectratio
			]{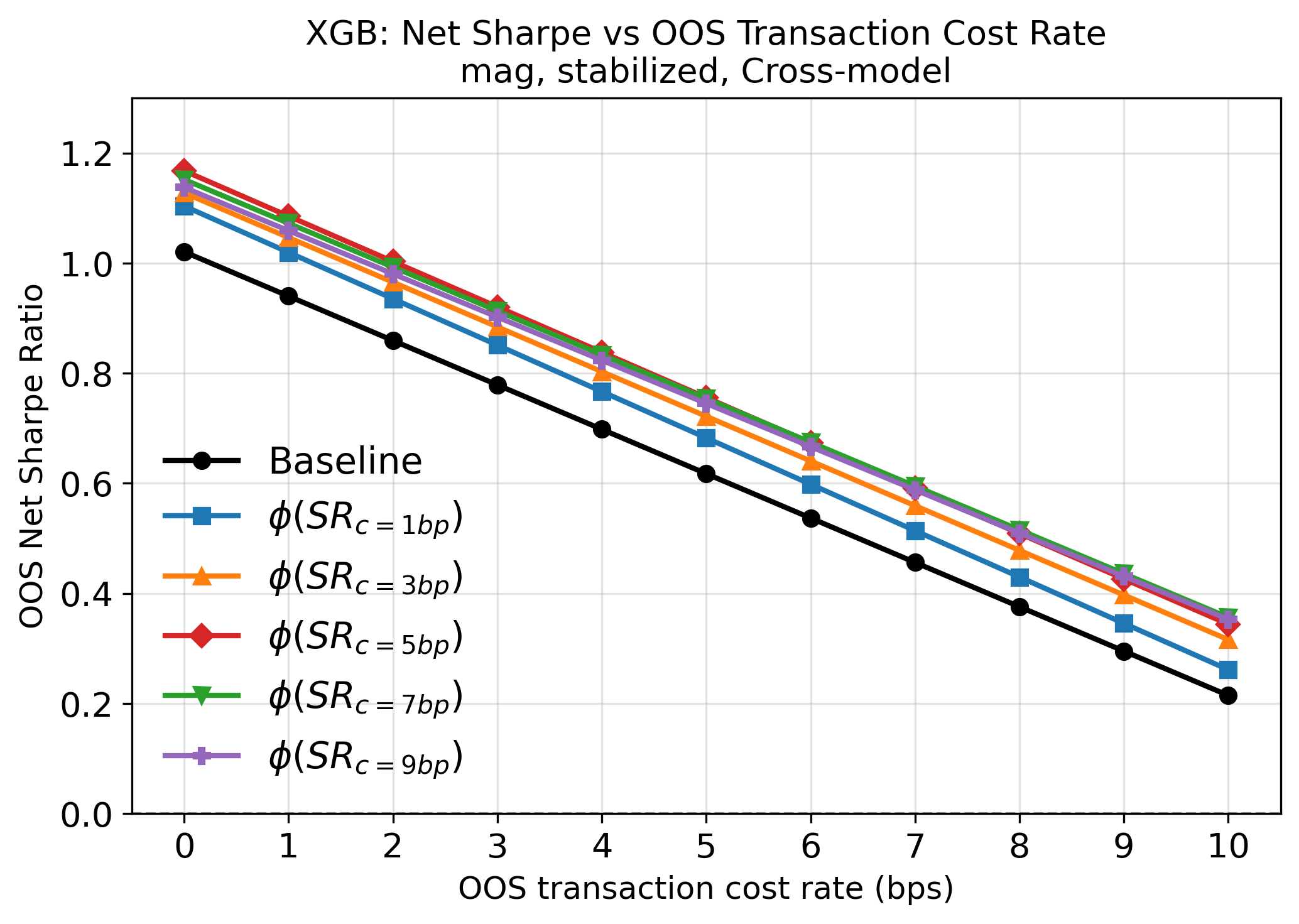}
			\small (d) XGB: Stabilised, cross-model
			\label{fig:tc_xgb_cross_stabilised}
		\end{minipage}
		
		\vspace{0.15em}
		
		% ==================== FFN: Unstabilised ====================
		\begin{minipage}[t]{0.46\textwidth}
			\centering
			\includegraphics[
			width=\linewidth,
			height=0.2\textheight,
			keepaspectratio
			]{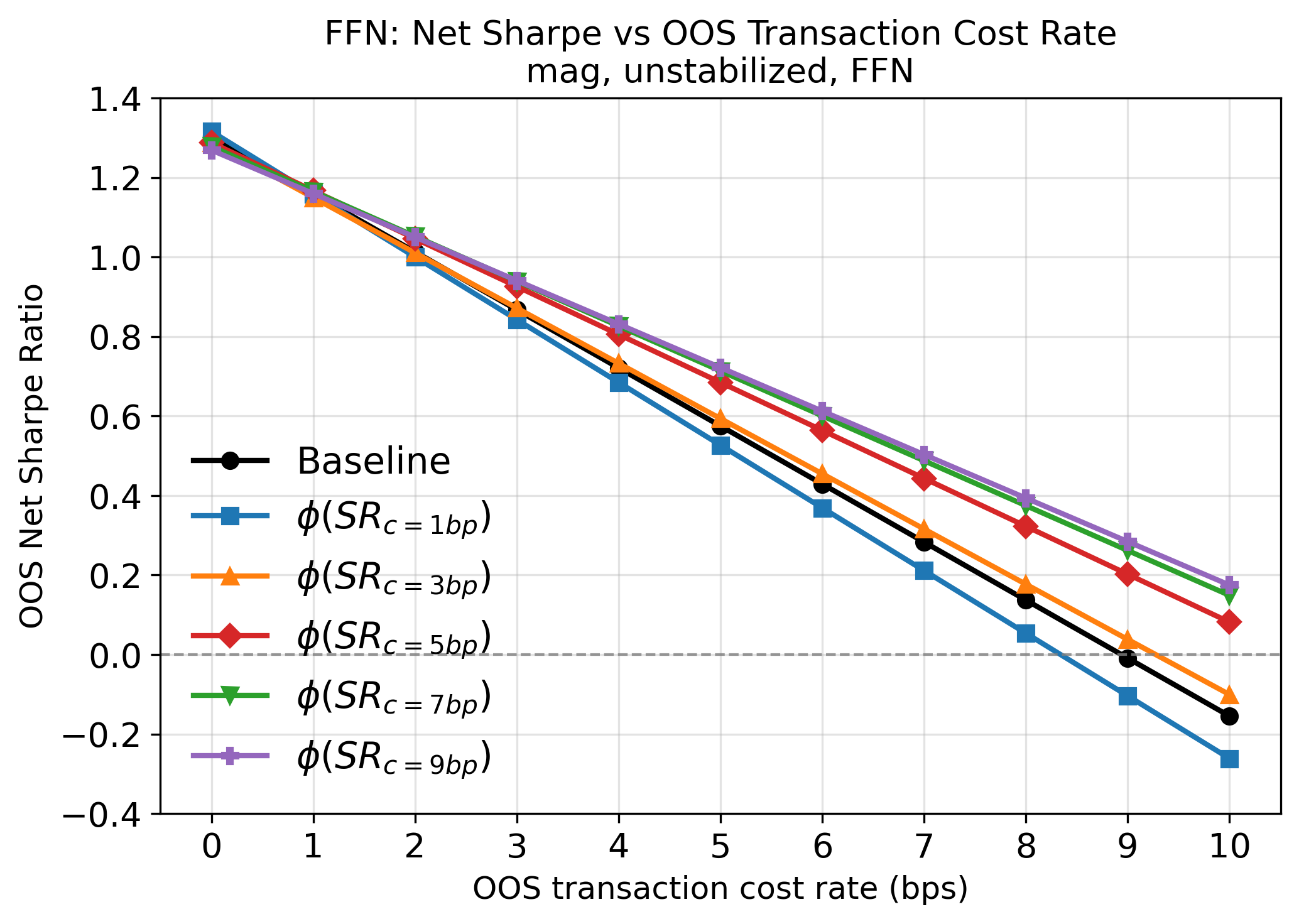}
			\small (e) FFN: Unstabilised, model-specific
			\label{fig:tc_ffn_ms_unstabilised}
		\end{minipage}
		\hfill
		\begin{minipage}[t]{0.46\textwidth}
			\centering
			\includegraphics[
			width=\linewidth,
			height=0.2\textheight,
			keepaspectratio
			]{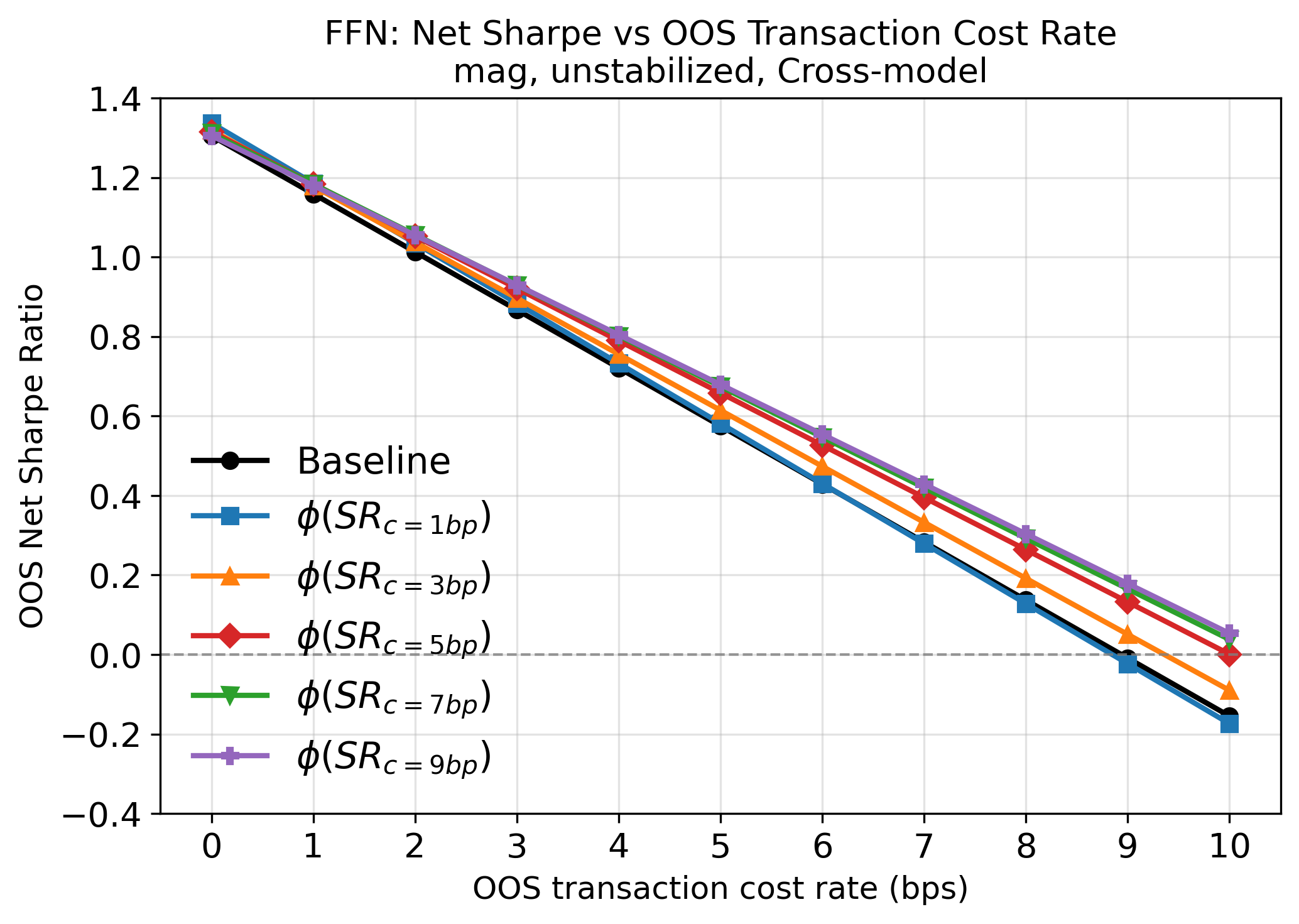}
			\small (f) FFN: Unstabilised, cross-model
			\label{fig:tc_ffn_cross_unstabilised}
		\end{minipage}
		
		\vspace{0.15em}
		
		% ==================== FFN: Stabilised ====================
		\begin{minipage}[t]{0.46\textwidth}
			\centering
			\includegraphics[
			width=\linewidth,
			height=0.2\textheight,
			keepaspectratio
			]{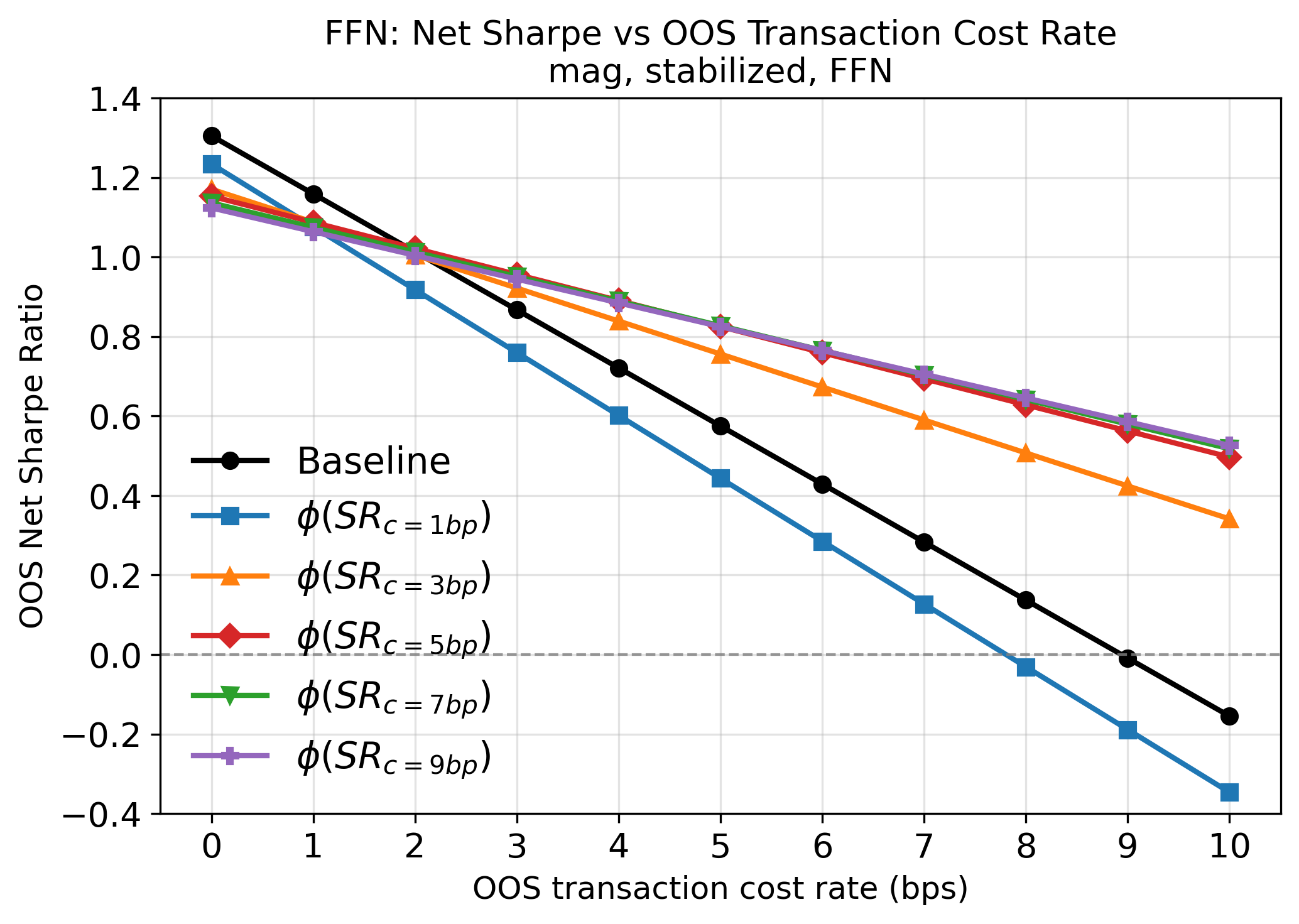}
			\small (g) FFN: Stabilised, model-specific
			\label{fig:tc_ffn_ms_stabilised}
		\end{minipage}
		\hfill
		\begin{minipage}[t]{0.46\textwidth}
			\centering
			\includegraphics[
			width=\linewidth,
			height=0.2\textheight,
			keepaspectratio
			]{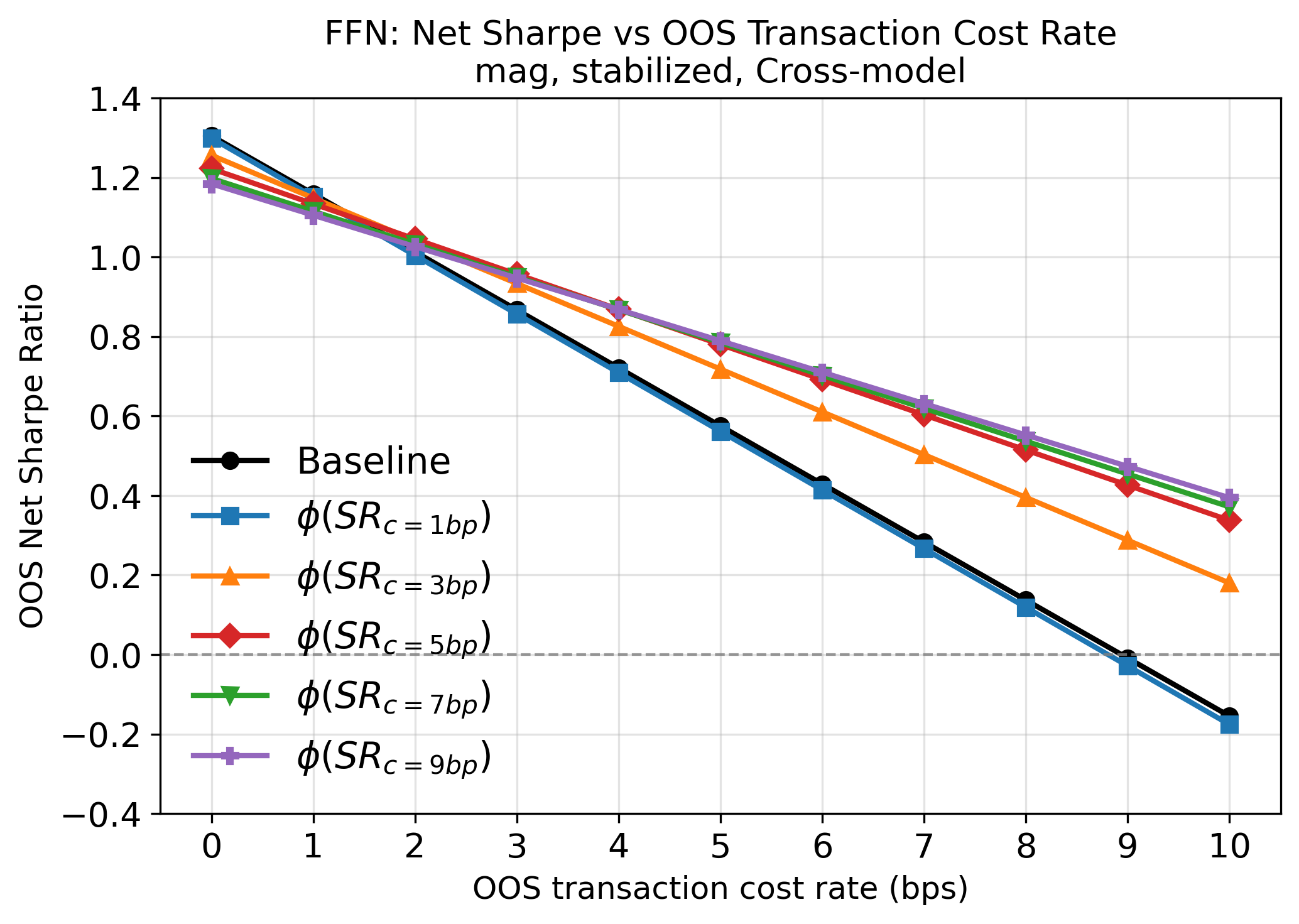}
			\small (h) FFN: Stabilised, cross-model
			\label{fig:tc_ffn_cross_stabilised}
		\end{minipage}
		
		\caption{
			%			Out-of-sample net Sharpe ratios under different transaction cost rates.
			%			The horizontal axis reports the out-of-sample transaction cost rate
			%			$c^{\mathrm{OOS}}$, while the vertical axis reports the corresponding
			%			out-of-sample net Sharpe ratio. The baseline portfolio is constructed
			%			without feature-exposure adjustment, while the remaining curves use
			%			feature contributions derived from $SR_{c^{\mathrm{IS}}}$
			%			under different transaction cost assumptions with $c^{\mathrm{IS}}
			%			\in
			%			\{1,3, 5, 7, 9\}
			%			\ \text{bps}$. 
			%			Results are shown separately for XGB and FFN under unstabilised and stabilised feature-exposure update rules, using model-specific and cross-model attribution sources with attribution magnitude information.
			%			The confidence parameter is $p=0.5$.
			Out-of-sample net Sharpe ratios under alternative transaction-cost
			assumptions. The horizontal axis reports the out-of-sample transaction-cost
			rate $c^{\mathrm{OOS}}$, and the vertical axis reports the corresponding
			net Sharpe ratio. The baseline uses no feature-exposure adjustment, while
			the intervention strategies use net Sharpe ratio attributions derived under
			$c^{\mathrm{IS}}\in\{1,3,5,7,9\}$ bps. Results are reported for XGB and
			FFN under model-specific and cross-model attribution rules, with and
			without stabilisation. The confidence parameter is $\eta=0.5$.
		}
		\label{fig:tc_sensitivity}
	\end{figure}
	
	Net portfolio returns are computed as
	$
		R_{c,t}
		=
		R_{0, t}
		-
		c
		\sum_{n=1}^{N}
		\left|
		w^R_{n,t-1}
		-
		w^R_{n,t-2}
		\right|,
	$
	where $c$ denotes the one-way proportional transaction cost. For each transaction cost $c^{\mathrm{IS}}$ in net SR as attribution target metric, feature contributions are learnt from $SR_{c^{\mathrm{IS}}}$, and the resulting feature exposure multipliers are used to construct out-of-sample portfolios. We consider $
	c^{\mathrm{IS}}
	\in
	\{1,3, 5, 7, 9\}
	\ \text{bps}.
	$
	The out-of-sample portfolios are subsequently evaluated under transaction costs
	$
	c^{\mathrm{OOS}}
	\in
	\{0,1,\ldots,10\}
	\ \text{bps},
	$
	allowing the sensitivity of the proposed attribution framework to be examined across different combinations of  transaction costs in learning attribution and evaluation trading performance.

	Figure~\ref{fig:tc_sensitivity} reports out-of-sample net Sharpe ratios
	under alternative transaction-cost assumptions. Overall, the
	attribution-guided feature-exposure interventions remain effective across
	a broad range of transaction costs, indicating that their economic value
	is not confined to a particular cost specification. Moreover, attributions
	estimated under higher in-sample transaction-cost rates generally perform
	better when evaluated in higher-cost out-of-sample environments. As the
	transaction cost incorporated into the attribution increases, the resulting
	exposure adjustments place greater emphasis on trading efficiency and tend
	to favour lower-turnover allocations. This pattern is particularly
	pronounced for FFN.
	
	For XGB, improvements in net Sharpe ratio are driven primarily by higher
	gross risk-adjusted performance, as the baseline portfolio already exhibits
	relatively low turnover. Consequently, the benefits of the feature-exposure
	intervention are less sensitive to the out-of-sample transaction-cost
	assumption. The intervention and baseline Sharpe-ratio curves are
	approximately parallel as transaction costs increase, indicating that the
	performance improvement remains relatively stable across different cost
	environments.
	
	For FFN, by contrast, improvements in net Sharpe ratio are primarily
	associated with lower portfolio turnover and the resulting reduction in
	transaction costs. As the in-sample transaction-cost rate
	$c^{\mathrm{IS}}$ increases, net Sharpe-ratio attributions place greater
	emphasis on the turnover-cost channel, leading to more conservative
	feature-exposure adjustments and lower-turnover portfolios. 
	The benefits therefore become more pronounced in higher-cost environments.
	Attributions estimated under very low transaction-cost assumptions, such as
	$c^{\mathrm{IS}}=1$ bp, result in weaker performance, whereas specifications
	with $c^{\mathrm{IS}}\geq5$ bps perform substantially better as
	out-of-sample transaction costs increase. Under the stabilised rules, for
	example, the feature-exposure interventions maintain a positive net Sharpe
	ratio of approximately 0.4 at $c^{\mathrm{OOS}}=10$ bps when
	$c^{\mathrm{IS}}\geq5$ bps, while the corresponding baseline net Sharpe
	ratio becomes negative.

	\section{Conclusion}

	To translate model explainability into actionable portfolio decisions, we develop a local first-order attribution-propagation framework that measures the sensitivity of sequential portfolio metrics to feature exposures in the allocation function. By explicitly linking feature-level contributions to economically interpretable exposure adjustments, the framework enables attribution-guided post-training adaptation without requiring retraining of the underlying allocation model. Empirically, these feature-exposure interventions generate statistically significant improvements in out-of-sample performance when the net Sharpe ratio is used as the target portfolio objective.
	
	A central implication of our analysis is that model-level attribution to allocation weights does not map directly into attribution to economically relevant portfolio outcomes. Features that are influential for portfolio weights need not be those that contribute most positively to risk-adjusted performance after turnover and transaction costs are taken into account. For both gross and net Sharpe ratios, the proposed local first-order attributions exhibit strong agreement with the corresponding estimated Shapley values, while Sharpe-ratio attributions are statistically significant for a non-negligible subset of feature-year observations.
	
	To improve the robustness of attribution-guided exposure adjustment, temporal stabilisation and cross-model averaging can serve as regularisation mechanisms. The results suggest that cumulative attribution structures exhibit greater persistence over time, particularly for turnover and net Sharpe ratio. Cross-model attribution and exposure rules also remain effective, suggesting the presence of model-agnostic economic components in feature contributions that depend more fundamentally on signal characteristics and market conditions. Together, these findings demonstrate how portfolio-level attribution can move beyond explaining model decisions toward economically meaningful and actionable post-training portfolio adaptation.

	\paragraph{\textbf{Limitations and future research}}

	Our analysis also suggests several directions for future research. First, the proposed local first-order objective attribution focuses on the marginal contribution of individual features and does not explicitly account for interaction effects when multiple feature exposures are adjusted simultaneously around the baseline trading strategy. For the feature universe considered in this study, the first-order attribution captures the dominant contribution structure and provides useful guidance for feature-exposure adjustment, leading to significant improvements in out-of-sample net Sharpe ratios. However, with hundreds of potentially correlated features, interactions may become economically important, and simultaneous exposure adjustments may no longer be well approximated by individual first-order effects. Extending the framework to incorporate higher-order attribution and interaction effects would therefore be valuable for high-dimensional feature spaces.

	Moreover, feature attribution and interaction structures may be regime dependent. 
	%	A feature that contributes positively to a portfolio objective during a bull market may become less informative or even detrimental during a bear or high-volatility regime, while other features may become more valuable as market dislocations and arbitrage opportunities increase. For example, short-horizon signals may generate higher portfolio turnover while exploiting transient arbitrage opportunities. Such signals may be particularly valuable during volatile periods but may reduce the net Sharpe ratio in more stable markets once transaction costs are taken into account.
	A feature that improves portfolio performance in one regime may become less informative or even detrimental in another. For example, signals that exploit transient arbitrage opportunities may generate higher turnover and become more valuable during volatile periods, but reduce the net Sharpe ratio in stable markets once transaction costs are considered.
	For interaction effects, complementarities among important signals may become stronger during periods of market stress.
	This motivates regime-dependent feature-exposure policies that condition exposure adjustments on prevailing market states. 
	%	Moreover, interaction effects may also vary across regimes. For example, complementarities among important signals may become stronger during periods of market stress. 
	Incorporating regime-dependent first- and higher-order attribution into feature-exposure calibration therefore could represent a promising extension of the proposed framework.

    \newpage
%	\appendix

	% ============================================================
%	\appendix

	\begin{appendices}
	\section{Regime-Switching Models}
	\label{sec:rs_ou_model}
	
	\subsection{Markov Regime-Switching Ornstein–Uhlenbeck Model}
	The Markov regime-switching (MRS) framework, originally proposed by \citet{Hamilton1989}, provides a natural extension of the Ornstein-Uhlenbeck (OU) process by allowing parameters to vary across a finite number of regimes. In this setting, the number of regimes must be specified in advance.
	
	A regime-switching OU process is given by
	\begin{equation}
		d X_t= \begin{cases}\kappa_1\left(\mu_1-X_t\right) d t+ \sigma_1 d W_t, & \text { for } S_t=1 \\ \quad \quad \quad  \quad \vdots & \quad  \quad \vdots \\ \kappa_r\left(\mu_r-X_t\right) d t+ \sigma_r d W_t, & \text { for } S_t=r\end{cases}
	\end{equation}
	%		$$
	%		d X_t=\kappa_{s_t}\left(\theta_{s_t}+X_{t}\right) d t+\sigma_{s_t} d W_t
	%		$$
	where $d W_t$ is a standard Brownian motion, and $S_t \in\{1, \ldots, r\}$ is a latent state variable. This regime switching behaviour is described by a continuous-time Markov chain $\left\{S_t\right\}_{t \geq 0}$, where the random variable $S_t$ denotes the state of the process at time $t$. For each state, the $d X_t$ follows a mean-reversion stochastic process characterised by a constant speed of reversion, $\kappa_{s_t}$, long-term mean level, $\mu_{s_t}$, and volatility $\sigma_{s_t}^2$. 
	
	The discretised version of the MRS-OU model, conditional on past observations $x_{1: t-1}=\left\{x_1, \ldots, x_{t-1}\right\}$, can be written as
	
	\begin{equation}
		\begin{aligned}
			&  
			X_{t}=\lambda_{s_t}+\beta_{s_t} x_{t-1}+ \varepsilon_t, \quad \varepsilon_t \sim N(0,\sigma_{e, s_t}^2) \\
			& S_t \mid s_{t-1} \sim \Pi_{s_{t-1}}\\
			& \text{where} \quad \lambda_{s_t} = \left(1-e^{-\kappa_{s_t} }\right) \mu_{s_t}, \quad \beta_{s_t} = e^{-\kappa_{s_t} }, \quad \sigma_{e, s_t}^2 = \sigma_{s_t}^2 \frac{1-e^{-2\kappa_{s_t} }}{2\kappa_{s_t}}
		\end{aligned}
		\label{eq:RS_OU}
	\end{equation}
	
	Each $\varepsilon_t$ is a normal, independent and identically distributed random variable. $\Pi_{s_{t-1}}$ denotes row $s_{t-1}$ of the $r \times r$ transition matrix $\Pi$ and is the discrete distribution governing the move from state $s_{t-1}$ to $s_t$. Note that each row in $\Pi$ sums to one. The process is well defined only if $\beta_{s_t}<1$ which corresponds to the mean-reversion restriction of the OU process.
	
	For instance, when $r = 2$, the two regimes can be regarded as representations of bull and bear markets intuitively. Since the discretisation of the OU process is a kind of AR(1) model, there is a standard linear regression in each regime. In practice, parameter estimation proceeds by Maximum Likelihood Estimation (MLE), yielding state-dependent parameters $\kappa_{s_t}$, $\mu_{s_t}$ and $\sigma_{s_t}$ ($s_t = 1, ..., r$) for each regime, along with the transition matrix $P$. Out-of-sample regime probabilities and parameters are then used to inform dynamic portfolio allocation.
	
	\subsection{Hierarchical Dirichlet Process Hidden Markov Model}
	\label{sec:HDPHMM}
	
	While the MRS framework is useful, it requires the number of regimes to be fixed a priori. This limitation motivates the use of more flexible approaches that allow the number of regimes to be inferred from the data. The Hierarchical Dirichlet Process Hidden Markov Model (HDP-HMM) provides such flexibility. It can adaptively determine the appropriate number of regimes and is capable of identifying extreme or rare market conditions that may not be anticipated beforehand.
	
	The HDP-HMM builds on the Infinite Hidden Markov Model (IHMM) by placing a Hierarchical Dirichlet Process (HDP) prior on the transition distributions. A further refinement, the sticky HDP-HMM, introduces a self-persistence parameter that encourages regimes to exhibit temporal persistence, making it particularly well-suited for financial time series.
	
	\textbf{Hierarchical Dirichlet Process}
	
	The hierarchical Dirichlet process (HDP) prior is introduced by \citet{TehJordanBealBlei2006} as an extension to the DP prior. The HDP has a hierarchical structure constructed by two DPs,
	\begin{equation}
		\begin{aligned}
			& G_0 \mid \eta, H \sim D P(\eta, H) \\
			& G_j \mid \alpha, G_0 \stackrel{i i d}{\sim} D P\left(\alpha, G_0\right), \quad j=1, \ldots, \infty
		\end{aligned}
	\end{equation}
	where the process defines group-specific probability measures $G_j$ conditional on a global probability measure $G_0$. The scalars $\eta$ and $\alpha$ are concentration parameters and $H$ is the base measure. 
	
	Using the stick-breaking construction, we have the following representations of the HDP.
	\begin{equation}
		\begin{array}{ll}
			G_0=\sum_{i=1}^{\infty} \gamma_i \delta_{\theta_i}, & \Gamma=\left\{\gamma_i\right\}_{i=1}^{\infty} \sim \operatorname{Stick}(\eta), \quad \theta_i \stackrel{i i d}{\sim} H \\
			G_j=\sum_{i=1}^{\infty} \pi_{j i} \delta_{\theta_i}, & \Pi_j=\left\{\pi_{j i}\right\}_{i=1}^{\infty} \stackrel{i i d}{\sim} \operatorname{Stick} 2(\alpha, \Gamma),\quad j=1, \ldots, \infty
		\end{array}
	\end{equation}
	where $\left\{\gamma_i\right\}_{i=1}^{\infty} \sim \operatorname{Stick}(\eta)$ represents
	\begin{equation}
		\gamma_i=v_i \prod_{l=1}^{i-1}\left(1-v_l\right), \quad v_i \stackrel{i i d}{\sim} \operatorname{Beta}(1, \eta), \quad i=1,2, \ldots
	\end{equation}
	and the weights of the latter measure $G_j$ denoted as $\operatorname{Stick} 2(\alpha, \Gamma)$ mean
	\begin{equation}
		\pi_{j i}=\hat{\pi}_{j i} \prod_{l=1}^{i-1}\left(1-\hat{\pi}_{j l}\right), \quad \hat{\pi}_{j i} \stackrel{i i d}{\sim} \operatorname{Beta}\left(\alpha \gamma_i, \alpha\left(1-\sum_{l=1}^i \gamma_l\right)\right), \quad i=1,2, \ldots
	\end{equation}
	
	Note that all $G_j$ share the same atoms $\theta_i$ ($ i=1,2, \ldots$) but have different weights, $\pi_{j i}$. If $s_t \in\{1,2, \ldots\}$, then row $j$ of the transition matrix of an infinite Markov chain is $\Pi_j$ and directs the possible moves of $s_t=j$ to $s_{t+1}$. 
	
	\textbf{Infinite Hidden Markov Model}
	
	If $s_t \in\{1,2, \ldots\}$ is the unobserved state variable and $\Pi_j=\left(\pi_{j 1}, \pi_{j 2}, \ldots\right)$ is the $j$th row of the transition matrix, then $\pi_{j i}$ becomes the prior probability of $s_t$ moving from state $j$ to $i$ at time $t+1$. The infinite hidden Markov model (IHMM) can be written as
	\begin{equation}
		\begin{aligned}
			& \Gamma \mid \eta \sim \operatorname{Stick}(\eta), \quad 
			\theta_i \stackrel{iid}{\sim} H, \quad i=1,2,\ldots \\
			& \Pi_j \mid \alpha, \Gamma \stackrel{iid}{\sim} \operatorname{Stick}2(\alpha, \Gamma), 
			\quad j=1,2,\ldots \\
			& S_t \mid s_{t-1}, \Pi_{s_{t-1}} \sim \Pi_{s_{t-1}}, 
			\quad t=1,\ldots,T \\
			& X_t \mid s_t, \Theta \sim F\!\left(X_t \mid \theta_{s_t}\right)
		\end{aligned}
	\end{equation}
	%		\begin{figure}[htbp]
		%			\centering
		%			\begin{minipage}{0.45\textwidth}
			%				\small
			%				\[
			%				\begin{aligned}
				%					& \Gamma \mid \eta \sim \operatorname{Stick}(\eta), \quad 
				%					\theta_i \stackrel{iid}{\sim} H, \quad i=1,2,\ldots \\
				%					& \Pi_j \mid \alpha, \Gamma \stackrel{iid}{\sim} \operatorname{Stick}2(\alpha, \Gamma), 
				%					\quad j=1,2,\ldots \\
				%					& s_t \mid s_{t-1}, \Pi_{s_{t-1}} \sim \Pi_{s_{t-1}}, 
				%					\quad t=1,\ldots,T \\
				%					& y_t \mid s_t, \Theta \sim F\!\left(y_t \mid \theta_{s_t}\right)
				%				\end{aligned}
			%				\]
			%			\end{minipage}\hfill
		%			\begin{minipage}{0.45\textwidth}
			%				\centering
			%%				\includegraphics[width=0.9\textwidth]{Sticky_HDP_HMM}
			%			\end{minipage}
		%			\caption{Graphical representation and generative process of the Sticky HDP-HMM.}
		%			\label{fig:sticky_hdp_hmm}
		%		\end{figure}
	where $\Theta = \left\{\theta_i\right\}_{i=1}^{\infty}$, $F\!\left(\cdot \mid \cdot\right)$ is the distribution of $X_t$ given parameters $\theta_{s_t}$. According to \citet{MaheuYang2016}, the extension of the OU process to the IHMM is given by
	\begin{equation}
		X_t\mid x_{t-1}, s_t, \Theta \sim N\left(\lambda_{s_t}+\beta_{s_t} x_{t-1}, \sigma_{e, s_t}^2 \right)
	\end{equation}
	where $\theta_i=\left(\lambda_i, \beta_i, \sigma_{e, i}\right)$, and $s_t$ following a first-order Markov chain. To allow the prior to be centred and concentrated in empirically important regions of the parameter space, the hierarchical prior for $\theta_i$ is given by $H(\xi)$, with $\xi=(\phi, B, \chi, \nu)$:
	\begin{equation}
		\vartheta_i = \left(\lambda_i, \beta_i\right) \sim N(\phi, B) \quad \sigma_{e, i}^{-2} \sim \operatorname{Gamma}(\chi, \nu) 
	\end{equation}
	
	Different combinations of $\eta$ and $\alpha$ can be used to enforce various prior beliefs about the Markov chain.  Larger values of $\eta$ favour more active states, whereas larger values of $\alpha$ increase the likelihood of introducing new states. Each row $\Pi_j$, of the transition matrix $\Pi$ is assumed to be drawn from a DP prior with a common base measure $\Gamma=\left(\gamma_1, \gamma_2, \ldots\right)$ and precision $\alpha$, i.e. $\Pi_j \mid \alpha, \Gamma \sim \operatorname{DP}(\alpha, \Gamma)$. As a result, it can be shown that
	the expectation of state-specific transition distributions $\mathbb{E}\left[\pi_{j i}\mid \alpha, \Gamma\right]=\gamma_i$. In other words, the HDP prior on the transition matrix encourages states to have similar transition distributions.
	
	\textbf{Sticky Hierarchical Dirichlet Process Hidden Markov Model}
	
	A limitation of the standard HDP-HMM is that it does not distinguish between self-transitions and transitions across different states. This flexibility often results in posterior sequences with unrealistically rapid switching dynamics. To mitigate this issue, \citet{FoxSudderthJordanWillsky2011} propose the sticky HDP-HMM, in which $\Pi_j$ is specified as
	\begin{equation}
		\Pi_j \mid \alpha, \Gamma, k \sim \operatorname{DP}\left(\alpha+k, \frac{\alpha \Gamma+k \delta_j}{\alpha+k}\right) 
	\end{equation}
	
	Here, $\left(\alpha \Gamma+k \delta_j\right)$ indicates that an amount $k>0$ adds extra weight to the $j^{\text {th }}$ component of $\alpha \Gamma$. This construction increases the expected probability of self-transition by an amount proportional to $k$. Specifically, the expected set of weights for transition distribution $\Pi_j$ is a convex combination of the global base measure $\Gamma$ and a state-specific spike defined by $k$ :
	\begin{equation}
		\mathbb{E}\left[\pi_{j i} \mid \alpha, \Gamma, k \right]=\frac{\alpha}{\alpha+k} \gamma_i+\frac{k}{\alpha+k} \delta(j, i) 
	\end{equation}
	
	Here, $\delta(j, i)$ denotes the discrete Kronecker delta, i.e. $\delta(j,i) =
	\begin{cases}
		1, & \text{if } i = j, \\
		0, & \text{if } i \neq j.
	\end{cases}$. 
	When $k=0$, the original HDP-HMM is recovered. 
	
	\textbf{Posterior Sampling for Inference and Predictive Density}
	
	Posterior inference for the state sequence is carried out using the Forward Filtering Backward Sampling (FFBS) algorithm of \citet{Chib1996}.
	
	During the filtering step, states follow
	\begin{equation}
		p\left(s_t \mid y_{1: t}, \Pi \right) \propto p\left(y_t \mid y_{1: t-1}, s_t\right) \sum_{s_{t-1}} \pi_{s_{t-1} s_t} p\left(s_{t-1} \mid y_{1: t-1}, \Pi\right) 
	\end{equation}
	
	Once the forward pass is computed for $t=1, \ldots, T$, the backward step follows from
	\begin{equation}
		p\left(s_t \mid s_{t+1}, y_{1: T}\right) \propto p\left(s_t \mid y_{1: t}\right) \pi_{s_t, s_{t+1}}, \quad t=T-1, \ldots, 1 
	\end{equation}
	
	This is initiated with a draw of $s_T$ from the last value of the filter $p\left(s_T \mid y_{1: T},\Pi\right)$.
	
	After initialising the parameters, the MCMC routine proceeds as follows:
	\begin{enumerate}[leftmargin=2cm]
		\item sample $s_{1:T} \mid y_{1:T}, \Pi$
		\item sample $\Theta \mid s_{1:T}, y_{1:T}, \xi$
		\item sample $\Gamma \mid s_{1:T}, \eta$
		\item sample $\Pi \mid s_{1:T}, \alpha, \Gamma, k$
	\end{enumerate}
	
	Iterating over these sampling steps yields draws from the posterior denoted as $\Omega=\left\{S_{1: T}, \Theta, \Gamma, \Pi\right\}$. We adopt an empirical Bayes strategy for hyperparameters $\left\{\eta, \alpha, k, \xi\right\}$, while performing MCMC inference for state sequences and emission parameters. After a suitable burn-in period, a large number of draws are collected from which features of the posterior and predictive densities can be estimated.
	
	Given the model estimation, the predictive density for $X_{T+1}$ is then approximated by
	\begin{equation}
		p\left(X_{T+1} \mid x_{1: T}\right) \approx \sum_{s_{T+1}} N\left(X_{T+1} \mid \lambda_{s_{T+1}}+\beta_{s_{T+1}} x_T, \sigma^{2}_{e, s_{T+1}}\right) \pi_{s_{T}, s_{T+1}} 
	\end{equation}
	where $N(\cdot \mid \cdot, \cdot)$ is the normal probability density function. Similarly, the predictive mean of $X_{T+1}$ can be estimated as
	\begin{equation}
		\mathbb{E}\left[X_{T+1} \mid x_{1: T}\right] \approx \sum_{s_{T+1}} \left(\lambda_{s_{T+1}}+\beta_{s_{T+1}} x_T\right) \pi_{s_{T}, s_{T+1}} 
	\end{equation}

	\section{Robustness Analysis}

	\subsection{Glass-box validation of the objective-propagation framework}
	\label{sec:linear_model}
	
	The linear benchmark provides a glass-box validation of the proposed
	objective-propagation framework. We consider a linear portfolio allocation
	model trained by directly maximising the net Sharpe ratio, or equivalently
	by minimising $-\mathrm{SR}^{\mathrm{net}}$, under proportional transaction
	costs of $c=0.0005$.
	
	The portfolio allocation function is given by
	\begin{equation}
		\boldsymbol{w}_{t-1}^{\epsilon \mid \mathrm{LM}}
		\left(z_{t-1}\right)
		=
		g_{\boldsymbol{\theta}}^{\mathrm{LM}}(z_{t-1})
		=
		z_{t-1}\boldsymbol{\theta}
		=
		\sum_{i=1}^{p} z_{i,t-1}\theta_i,
	\end{equation}
	where $z_{t-1}=[z_{1,t-1},\ldots,z_{p,t-1}]$ denotes the input features
	and $\boldsymbol{\theta}$ is the vector of model parameters. Because the
	allocation function is linear, the contribution of feature $i$ to the
	allocation weight is directly given by
	\begin{equation}
		\phi_i^{\mathrm{LM}}
		\left(w_{t-1}^{\epsilon}\right)
		=
		z_{i,t-1}\theta_i.
	\end{equation}
	For given features $z_{t-1}$ and a trained allocation function
	$g_{\boldsymbol{\theta}}^{\mathrm{LM}}(\cdot)$, these contributions satisfy
	the exact additive decomposition
	$
		w_{t-1}^{\epsilon \mid \mathrm{LM}}
		(\boldsymbol{\lambda}=\mathbf{1}_p)
		=
		\sum_{i=1}^{p}
		\phi_i^{\mathrm{LM}}
		\left(w_{t-1}^{\epsilon}\right).
	$
	We set the reference allocation in the absence of feature signals to
	$
		w_{t-1}^{\epsilon \mid \mathrm{LM}}
		(\boldsymbol{\lambda}=\mathbf{0}_p)
		=
		\phi_0^{\mathrm{LM}}
		\left(w_{t-1}^{\epsilon}\right)
		=
		0.
	$
	
	Portfolio-level attributions are then obtained by propagating
	$\phi_i^{\mathrm{LM}}(w_{t-1}^{\epsilon})$ through the first-order
	objective-propagation framework introduced in
	Section~\ref{sec:model_portfolio_explainability}. The propagation procedure
	is identical to that applied to the nonlinear XGB and FFN allocation models.
	Although feature contributions to allocation weights are directly separable
	under the linear allocation function, their contributions to downstream
	portfolio metrics, including turnover and the Sharpe ratio, become
	nontrivial after portfolio construction. The proposed framework therefore
	remains useful for quantifying the economic contributions of individual
	features to portfolio-level objectives.

\begin{figure}[htbp]
		\centering
		
		\begin{minipage}[t]{0.49\textwidth}
			\centering
			\includegraphics[width=\linewidth]{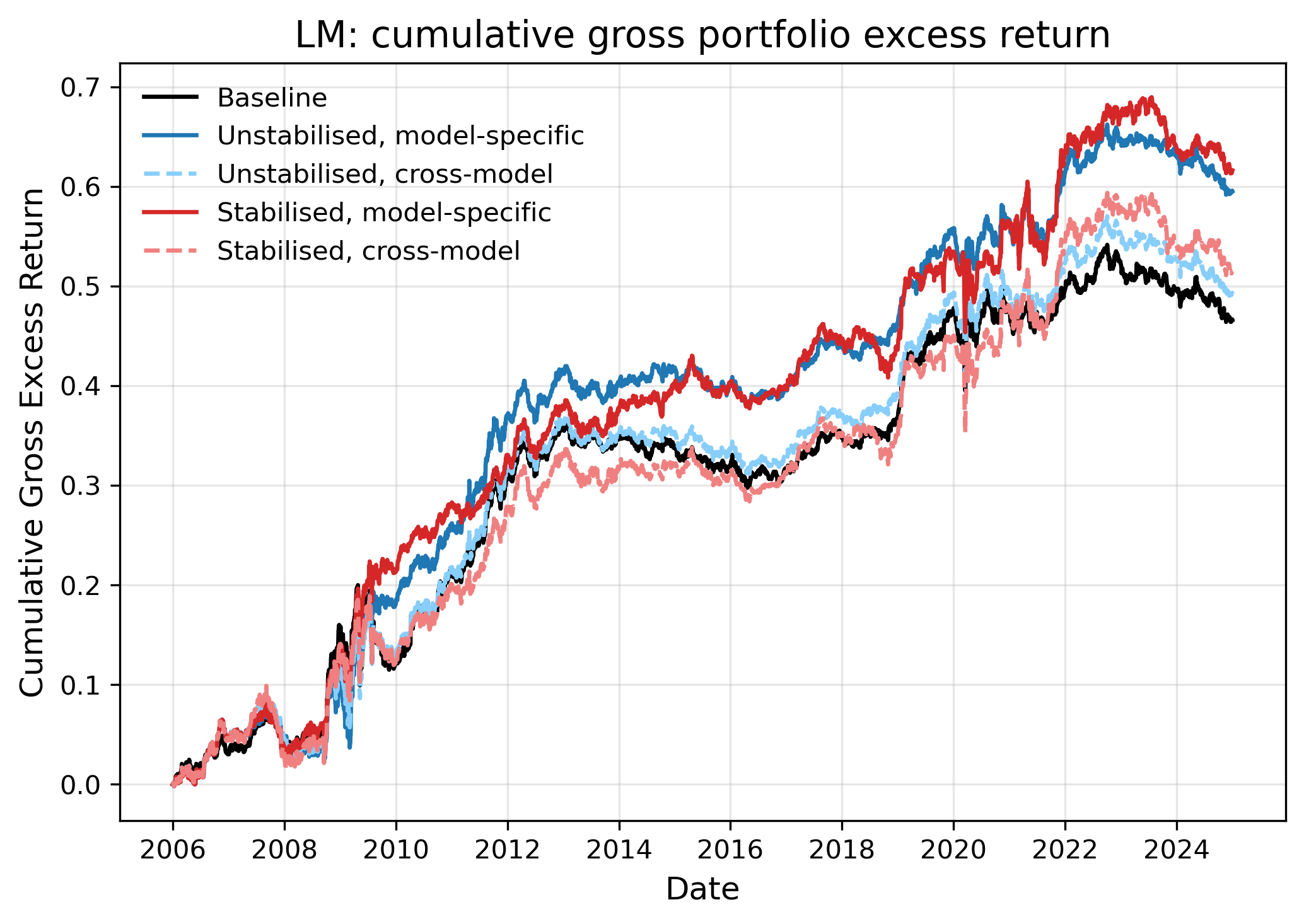}
			\small (a) LM: Cumulative gross returns.
			\label{fig:lm_gross}
		\end{minipage}
		\hfill
		\begin{minipage}[t]{0.49\textwidth}
			\centering
			\includegraphics[width=\linewidth]{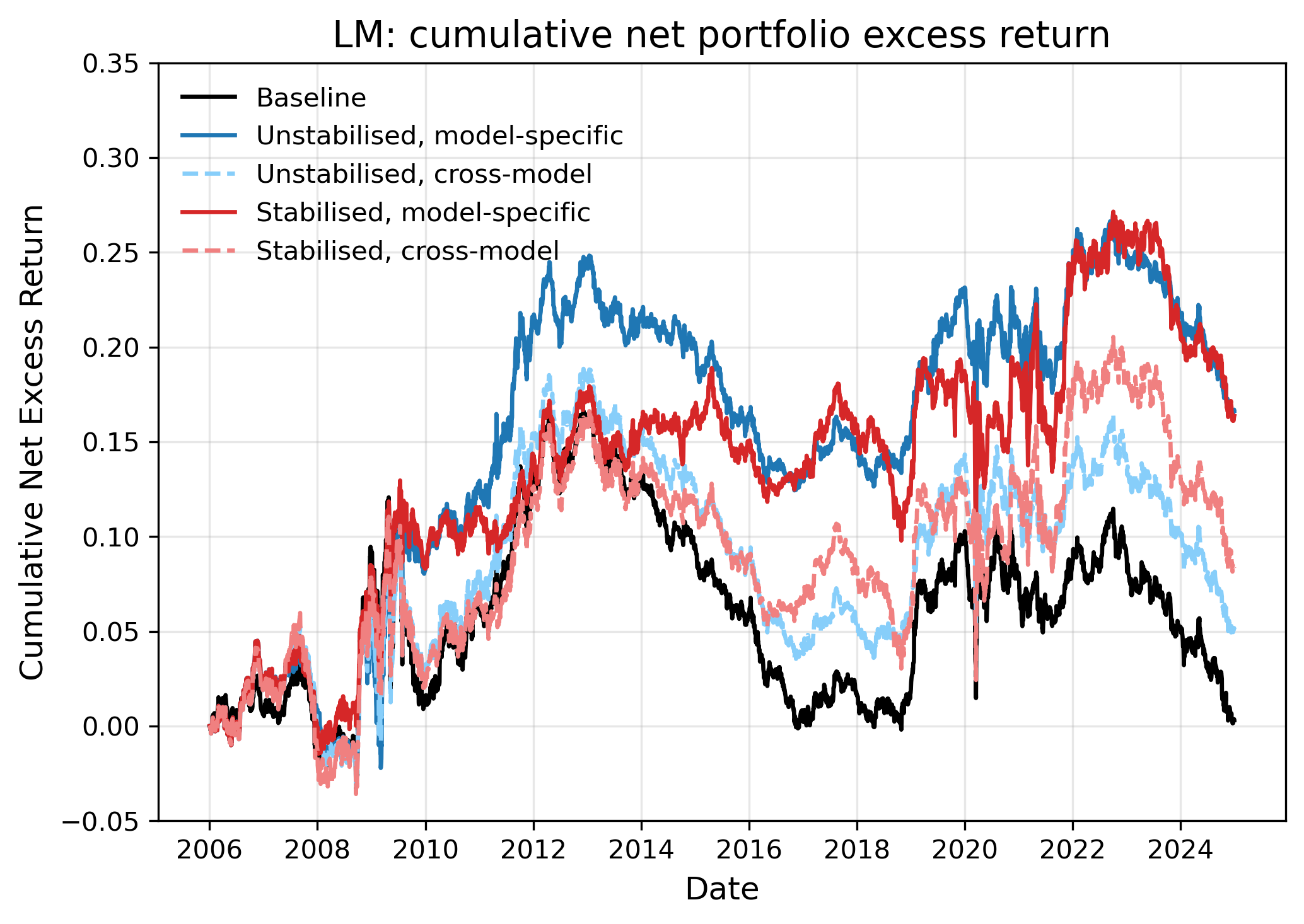}
			\small (b) LM: Cumulative net returns.
			\label{fig:lm_net}
		\end{minipage}
		
		\caption{Cumulative gross and net portfolio excess returns with feature exposure interventions for linear allocation model.}
		\label{fig:lm_cumulative_returns}
	\end{figure}

	Table~\ref{tab:LM_attribution_persistence} reports the
	validation-to-out-of-sample persistence of the resulting portfolio-level
	feature attributions. We subsequently use the validation-period attributions
	to guide feature-exposure interventions for the linear allocation model.
	The corresponding out-of-sample trading performance is reported in
	Table~\ref{tab:feature_scaling_LM_performance}, while
	Figure~\ref{fig:lm_cumulative_returns} presents the cumulative gross and net
	portfolio excess returns.
	
		\begin{table}[htbp]
		\centering
		\caption{Attribution persistence for the linear allocation model}
		\label{tab:LM_attribution_persistence}
		\small
		\setlength{\tabcolsep}{5pt}
		
		\resizebox{\textwidth}{!}{
			\begin{tabular}{ccccccc}
				\toprule
				&
				\multicolumn{3}{c}{Correlation}
				&
				\multicolumn{3}{c}{Sign Consistency} \\
				\cmidrule(lr){2-4}
				\cmidrule(lr){5-7}
				
				Metric
				& Mean
				& Median
				& $\Pr(\rho>0)$
				& Mean
				& Median
				& $\Pr(SC>50\%)$ \\
				
				\midrule
				
				Portfolio Turnover
				& 0.9827
				& 0.9927$^{***}$
				& 100.00\%
				& 84.80\%
				& 83.33\%$^{***}$
				& 100.00\% \\
				
				Gross Sharpe Ratio
				& 0.2529
				& 0.4313$^{**}$
				& 63.16\%
				& 55.26\%
				& 50.00\%$^{**}$
				& 47.37\% \\
				
				Net Sharpe Ratio
				& 0.2823
				& 0.4345$^{**}$
				& 68.42\%
				& 55.85\%
				& 55.56\%$^{*}$
				& 52.63\% \\
				
				\bottomrule
			\end{tabular}
		}
		\vspace{2mm}
		
		\begin{minipage}{0.95\textwidth}
			\footnotesize
			\textit{Notes:}
			The table reports validation-to-out-of-sample persistence of feature
			attributions for the linear allocation model. For each trading
			period, correlation is the cross-feature Pearson correlation between
			the attribution vectors estimated over the validation period and the
			subsequent out-of-sample period. Sign consistency is the proportion of
			features with the same nonzero attribution sign across the two periods.
			The reported statistics are computed across the 19 validation-to-OOS
			comparisons. $\Pr(\rho>0)$ denotes the proportion of periods with
			positive correlation, while $\Pr(SC>50\%)$ denotes the proportion with
			sign consistency exceeding 50\%. Significance stars attached to the
			medians are based on one-sided Wilcoxon signed-rank tests against zero
			for correlation and 50\% for sign consistency.
			$^{***}$, $^{**}$, and $^{*}$ denote significance at the 1\%, 5\%,
			and 10\% levels, respectively.
		\end{minipage}
		
	\end{table}

	The attribution-persistence results show that turnover attributions are highly persistent across time windows, with validation-to-OOS correlations exceeding 0.9 in almost all periods, consistent with the findings for XGB and FFN. For the linear allocation model, the validation-to-OOS correlations of both gross and net Sharpe-ratio attributions are significantly positive, providing evidence of temporal persistence in their feature-contribution structures. The portfolio using unstabilised model-specific feature exposures significantly outperforms the baseline in terms of the net Sharpe ratio, with an improvement of approximately 0.20, driven primarily by an increase in the gross Sharpe ratio. Stabilised and cross-model interventions, where cross-model exposures average attribution information across LM, XGB, and FFN, also yield positive improvements in the net Sharpe ratio relative to the baseline.

	\begin{table}[htbp]
		\centering
		\caption{Portfolio performance under attribution-guided
			feature-exposure interventions for the linear allocation model}
		\label{tab:feature_scaling_LM_performance}
		\small
		
		\resizebox{\textwidth}{!}{
			\begin{tabular}{ccccccccccl}
				\toprule
				Model &
				Update Rule &
				$SR^{gross}$ &
				$\mu^{gross}$ (\%) &
				$\sigma^{gross}$ (\%) &
				$SR^{net}$ &
				$\mu^{net}$ (\%) &
				$\sigma^{net}$ (\%) &
				Avg.\ Turnover (\%) &
				$\Delta SR^{gross}$ &
				$\Delta SR^{net}$ \\
				\midrule
				
				% =====================================================
				% Baseline
				% =====================================================
				
				\textbf{LM}
				& Baseline
				& 0.5577
				& 2.46
				& 4.40
				& 0.0036
				& 0.02
				& 4.40
				& 19.37
				& --
				& -- \\
				
				\midrule
				
				% =====================================================
				% Panel A: Model-specific
				% =====================================================
				
				\multicolumn{11}{c}{\textbf{Panel A: Model-specific}} \\
				\midrule
				
				\multirow{2}{*}{\textbf{LM}}
				& Unstabilised
				& 0.7429
				& 3.14
				& 4.22
				& 0.2075
				& 0.88
				& 4.22
				& 17.94
				& 0.1852
				& $0.2039^{**}$ \\
				
				& Stabilised
				& 0.7474
				& 3.25
				& 4.35
				& 0.1986
				& 0.86
				& 4.34
				& 18.93
				& 0.1897
				& 0.1950 \\
				
				\midrule
				
				% =====================================================
				% Panel B: Cross-model
				% =====================================================
				
				\multicolumn{11}{c}{\textbf{Panel B: Cross-model}} \\
				\midrule
				
				\multirow{2}{*}{\textbf{LM}}
				& Unstabilised
				& 0.6116
				& 2.60
				& 4.25
				& 0.0633
				& 0.27
				& 4.25
				& 18.47
				& 0.0539
				& 0.0597 \\
				
				& Stabilised
				& 0.5733
				& 2.71
				& 4.73
				& 0.0935
				& 0.44
				& 4.73
				& 18.03
				& 0.0156
				& 0.0899 \\
				
				\bottomrule
			\end{tabular}
		}
		
		\vspace{2mm}
		
		\begin{minipage}{0.95\textwidth}
			\footnotesize
			\textit{Notes:}
			The table reports out-of-sample portfolio performance for the baseline
			and attribution-guided feature-exposure strategies under the linear
			allocation model. The baseline portfolio is reported first as the common
			benchmark. Panel A uses model-specific attribution information, while
			Panel B uses cross-model averaged attribution information from LM, XGB and FFN. The
			unstabilised rules use the attribution signal from the preceding
			validation period directly, whereas the stabilised rules recursively
			update feature exposures using information from previous periods.
			Feature-exposure adjustments use the magnitude of the attribution signal
			with intervention-strength parameter $\eta=0.5$.
			$SR^{gross}$ and $SR^{net}$ denote annualised gross and net Sharpe
			ratios, respectively. $\mu$ and $\sigma$ denote annualised mean return
			and volatility, respectively. Net returns incorporate transaction costs
			of 5 basis points per unit of portfolio turnover.
			$\Delta SR^{gross}$ and $\Delta SR^{net}$ denote changes in gross and
			net Sharpe ratios relative to the baseline linear-model portfolio.
			Statistical significance reported for $\Delta SR^{net}$ is based on the
			one-sided Sharpe-ratio improvement test.
			$^{***}$, $^{**}$, and $^{*}$ denote significance at the 1\%, 5\%,
			and 10\% levels, respectively.
		\end{minipage}
		
	\end{table}

%	\begin{table}[H]
%		\centering
%		\caption{Portfolio performance after feature scaling.}
%		\begin{tabular}{lccc}
%			\toprule
%			Feature scaling & Gross SR & Net SR & Avg. turnover \\
%			\midrule
%			Baseline & 0.5578 & 0.0036 & 0.1934\\
%			LM (unstabilised) & \textbf{0.7202} & \textbf{0.1775} & \textbf{0.1764}\\
%			LM (stabilised) & 0.5686 & 0.0690 & 0.1727\\
%			AVG(FFN,XGB) (unstabilised) & 0.5590 & 0.0117 & 0.1880\\
%			AVG(FFN,XGB) (stabilised) & \textbf{0.7632} & \textbf{0.2135} & \textbf{0.1821}\\
%			\bottomrule
%		\end{tabular}
%		\label{tab:linear_performance}
%	\end{table}
	
	%	\begin{figure}[H]
		%		\centering
		%		\includegraphics[width=0.8\textwidth]{figures/linear_model/LM_cumulative_gross_returns.png}
		%		\caption{Cumulative gross returns.}
		%		\label{fig:lm_gross}
		%	\end{figure}
	%	
	%	\begin{figure}[H]
		%		\centering
		%		\includegraphics[width=0.8\textwidth]{figures/linear_model/LM_cumulative_net_returns.png}
		%		\caption{Cumulative net returns.}
		%		\label{fig:lm_net}
		%	\end{figure}

%	\begin{figure}[H]
%		\centering
%		
%		\begin{subfigure}[t]{0.49\textwidth}
%			\centering
%			\includegraphics[width=\linewidth]{figures/linear_model/LM_gross_cumulative_portfolio_excess_return.png}
%			\caption{LM: Cumulative gross returns.}
%			\label{fig:lm_gross}
%		\end{subfigure}
%		\hfill
%		\begin{subfigure}[t]{0.49\textwidth}
%			\centering
%			\includegraphics[width=\linewidth]{figures/linear_model/LM_net_cumulative_portfolio_excess_return.png}
%			\caption{LM: Cumulative net returns.}
%			\label{fig:lm_net}
%		\end{subfigure}
%		
%		\caption{Cumulative gross and net portfolio excess returns with feature exposure interventions for linear allocation model.}
%		\label{fig:lm_cumulative_returns}
%	\end{figure}

	\subsection{Robustness to volatility scaling}
	\label{sec:volatility_scaling}
	
	To examine whether the proposed feature exposure adjustment provides information beyond conventional risk management, we further evaluate the proposed method after applying a standard volatility scaling procedure to the factor-neutral stock portfolio. Let $w_{t-1}^{\epsilon}$ denote the residual portfolio weights and let
	$
	\tilde{w}_{t-1}^{R}
	=
	{w_{t-1}^{\epsilon}}^{\top}\Phi_{t-1}
	$
	be the corresponding factor-neutral stock portfolio, where $\Phi$ denotes the factor-neutral projection matrix introduced in Section~\ref{sec:factor_neutral_portfolios}. Before constructing the final portfolio, each stock weight is scaled by the inverse of its historical volatility,
	\begin{equation}
		w_{n,t-1}^{R}
		=
		\frac{
			\tilde{w}_{n,t-1}^{R}
			/\hat{\sigma}_{n,t-1}
		}{
			\sum_{m=1}^{N}
			\left|
			\tilde{w}_{m,t-1}^{R}
			/\hat{\sigma}_{m,t-1}
			\right|
		},
		\label{eq:volatility_scaling}
	\end{equation}
	where
	$
	\hat{\sigma}_{n,t-1}
	=
	\operatorname{std}
	\left(
	r_{n,t-L},
	\ldots,
	r_{n,t-1}
	\right)
	$
	denotes the historical volatility of stock $n$ estimated from the previous $L$ daily simple returns. Throughout this section, we set $L=30$ trading days. 
	
	\begin{table}[htbp]
		\centering
		\caption{Portfolio performance under volatility-scaled
			attribution-guided feature-exposure interventions}
		\label{tab:feature_scaling_volatility_performance}
		\small
		
		\resizebox{\textwidth}{!}{
			\begin{tabular}{ccccccccccl}
				\toprule
				Model &
				Update Rule &
				$SR^{gross}$ &
				$\mu^{gross}$ (\%) &
				$\sigma^{gross}$ (\%) &
				$SR^{net}$ &
				$\mu^{net}$ (\%) &
				$\sigma^{net}$ (\%) &
				Avg.\ Turnover (\%) &
				$\Delta SR^{gross}$ &
				$\Delta SR^{net}$ \\
				\midrule
				
				% =====================================================
				% Baseline
				% =====================================================
				
				\textbf{XGB}
				& Baseline
				& 1.1175
				& 6.02
				& 5.39
				& 0.8078
				& 4.36
				& 5.39
				& 13.24
				& --
				& -- \\
				
				\textbf{FFN}
				& Baseline
				& 1.2301
				& 6.89
				& 5.60
				& 0.6760
				& 3.79
				& 5.60
				& 24.64
				& --
				& -- \\
				
				\midrule
				
				% =====================================================
				% Panel A: Model-specific
				% =====================================================
				
				\multicolumn{11}{c}{\textbf{Panel A: Model-specific}} \\
				\midrule
				
				\multirow{2}{*}{\textbf{XGB}}
				& Unstabilised
				& 1.1248
				& 6.09
				& 5.41
				& 0.8239
				& 4.46
				& 5.41
				& 12.92
				& 0.0074
				& 0.0161 \\
				
				& Stabilised
				& 1.1173
				& 6.34
				& 5.67
				& 0.8002
				& 4.54
				& 5.67
				& 14.26
				& -0.0002
				& -0.0076 \\
				
				\midrule
				
				\multirow{2}{*}{\textbf{FFN}}
				& Unstabilised
				& 1.2237
				& 6.81
				& 5.57
				& 0.7613
				& 4.24
				& 5.57
				& 20.43
				& -0.0064
				& $0.0853^{***}$ \\
				
				& Stabilised
				& 1.1792
				& 7.04
				& 5.97
				& 0.9070
				& 5.42
				& 5.97
				& 12.90
				& -0.0510
				& $0.2311^{***}$ \\
				
				\midrule
				
				% =====================================================
				% Panel B: Cross-model
				% =====================================================
				
				\multicolumn{11}{c}{\textbf{Panel B: Cross-model}} \\
				\midrule
				
				\multirow{2}{*}{\textbf{XGB}}
				& Unstabilised
				& 1.1406
				& 6.10
				& 5.35
				& 0.8335
				& 4.46
				& 5.35
				& 13.02
				& 0.0232
				& $0.0258^{*}$ \\
				
				& Stabilised
				& 1.2234
				& 6.67
				& 5.45
				& 0.8972
				& 4.89
				& 5.45
				& 14.11
				& 0.1060
				& $0.0894^{**}$ \\
				
				\midrule
				
				\multirow{2}{*}{\textbf{FFN}}
				& Unstabilised
				& 1.2348
				& 6.85
				& 5.55
				& 0.7414
				& 4.11
				& 5.55
				& 21.71
				& 0.0046
				& $0.0654^{***}$ \\
				
				& Stabilised
				& 1.2123
				& 7.03
				& 5.80
				& 0.8626
				& 5.01
				& 5.80
				& 16.10
				& -0.0178
				& $0.1867^{***}$ \\
				
				\bottomrule
			\end{tabular}
		}
		
		\vspace{2mm}
		
		\begin{minipage}{0.95\textwidth}
			\footnotesize
			\textit{Notes:}
			The table reports out-of-sample portfolio performance for the baseline
			and attribution-guided feature-exposure strategies under volatility
			scaling. The baseline portfolios are reported first as common benchmarks.
			Panel A uses attribution information from the corresponding allocation
			model, while Panel B uses cross-model averaged attribution information.
			The unstabilised rules use the attribution signal from the preceding
			validation period directly, whereas the stabilised rules recursively
			update feature exposures using information from previous periods.
			Feature-exposure adjustments use the magnitude of the attribution signal
			with intervention-strength parameter $\eta=0.5$.
			$SR^{gross}$ and $SR^{net}$ denote annualised gross and net Sharpe
			ratios, respectively. $\mu$ and $\sigma$ denote annualised mean return
			and volatility, respectively. Net returns incorporate transaction costs
			of 5 basis points per unit of portfolio turnover. Average turnover is
			reported in percentage terms. Volatility scaling is implemented using
			$L=30$ trading days.
			$\Delta SR^{gross}$ and $\Delta SR^{net}$ denote changes in gross and
			net Sharpe ratios relative to the corresponding volatility-scaled
			baseline portfolio. Statistical significance reported for
			$\Delta SR^{net}$ is based on the one-sided Sharpe-ratio improvement
			test using 4,999 bootstrap replications and five-trading-day blocks.
			$^{***}$, $^{**}$, and $^{*}$ denote significance at the 1\%, 5\%,
			and 10\% levels, respectively.
		\end{minipage}
		
	\end{table}
	
	We apply volatility scaling to both the baseline and exposure-adjusted portfolios to examine whether attribution-guided feature-exposure interventions remain effective after conventional volatility-based risk adjustment. Table~\ref{tab:feature_scaling_volatility_performance} reports the resulting out-of-sample portfolio performance. The results show that the feature-exposure interventions continue to improve net portfolio performance after volatility scaling. For FFN, all four attribution-guided portfolios achieve statistically significant improvements in the net Sharpe ratio relative to the volatility-scaled baseline at the 1\% level. For XGB, the cross-model interventions improve the net Sharpe ratio, with the stabilised specification significant at the 5\% level and the unstabilised specification at the 10\% level.
	
	Under volatility scaling, the baseline portfolios achieve net Sharpe ratios of 0.8078 for XGB and 0.6760 for FFN. Applying stabilised cross-model feature exposures increases these ratios to 0.8972 and 0.8626, respectively. The mechanisms underlying these improvements are consistent with those observed without volatility scaling. For FFN, the increase in the net Sharpe ratio is primarily associated with lower portfolio turnover and the resulting reduction in transaction costs. For XGB, the stabilised cross-model intervention improves both gross and net Sharpe ratios, suggesting that the attribution signal identifies feature exposures associated with stronger gross portfolio performance in addition to their effects on trading costs.
	
	Overall, these results provide evidence that attribution-guided feature-exposure interventions contain incremental economic information beyond conventional volatility-based risk adjustment. Portfolio-objective attribution therefore complements volatility scaling, generating additional improvements in out-of-sample net portfolio performance.

	\subsection{Information of feature contributions}
	\label{sec:phi_information}
	% ============================================================
	
	We investigate which aspects of feature-contribution information are most useful for feature-exposure interventions. While the main analysis uses the magnitude of feature contributions, we further consider interventions based on their sign and rank to assess the information content of alternative attribution signals. Tables~\ref{tab:model_cross_model_performance_sign_rank} reports the corresponding trading performance.

	\begin{table}[htbp]
		\centering
		\caption{Portfolio performance using sign- and rank-based feature contributions}
		\label{tab:model_cross_model_performance_sign_rank}
		\footnotesize
		
		\resizebox{\textwidth}{!}{
			\begin{tabular}{ccccccccccl}
				\toprule
				Model &
				Update Rule &
				$SR^{gross}$ &
				$\mu^{gross}$ (\%) &
				$\sigma^{gross}$ (\%) &
				$SR^{net}$ &
				$\mu^{net}$ (\%) &
				$\sigma^{net}$ (\%) &
				Avg.\ Turnover (\%) &
				$\Delta SR^{gross}$ &
				$\Delta SR^{net}$ \\
				\midrule
				
				% =====================================================
				% Panel A: Sign-based
				% =====================================================
				\multicolumn{11}{c}{\textbf{Panel A: Sign-based feature contributions}} \\
				\midrule
				
				\multicolumn{11}{c}{\textit{Model-specific}} \\
				
				\multirow{2}{*}{\textbf{XGB}}
				& Unstabilised
				& 1.0576
				& 4.37
				& 4.13
				& 0.6689
				& 2.76
				& 4.13
				& 12.73
				& 0.0369
				& $0.0514^{*}$ \\
				
				& Stabilised
				& 1.0404
				& 4.61
				& 4.42
				& 0.6730
				& 2.99
				& 4.42
				& 12.93
				& 0.0197
				& 0.0555 \\
				
				\addlinespace
				
				\multirow{2}{*}{\textbf{FFN}}
				& Unstabilised
				& 1.2922
				& 5.93
				& 4.59
				& 0.6750
				& 3.10
				& 4.59
				& 22.47
				& -0.0129
				& $0.1000^{***}$ \\
				
				& Stabilised
				& 1.1114
				& 5.68
				& 5.11
				& 0.7818
				& 4.00
				& 5.11
				& 13.37
				& -0.1937
				& $0.2068^{***}$ \\
				
				\midrule
				
				\multicolumn{11}{c}{\textit{Cross-model}} \\
				
				\multirow{2}{*}{\textbf{XGB}}
				& Unstabilised
				& 1.0668
				& 4.39
				& 4.11
				& 0.6692
				& 2.76
				& 4.11
				& 12.98
				& 0.0461
				& $0.0517^{**}$ \\
				
				& Stabilised
				& 1.1085
				& 4.81
				& 4.34
				& 0.7295
				& 3.17
				& 4.34
				& 13.05
				& 0.0878
				& $0.1120^{*}$ \\
				
				\addlinespace
				
				\multirow{2}{*}{\textbf{FFN}}
				& Unstabilised
				& 1.2966
				& 5.94
				& 4.48
				& 0.6335
				& 2.85
				& 4.48
				& 23.68
				& -0.0085
				& $0.0585^{**}$ \\
				
				& Stabilised
				& 1.1718
				& 5.79
				& 4.94
				& 0.7813
				& 3.86
				& 4.94
				& 15.31
				& -0.1333
				& $0.2063^{***}$ \\
				
				% =====================================================
				% Panel B: Rank-based
				% =====================================================
				\midrule
				\multicolumn{11}{c}{\textbf{Panel B: Rank-based feature contributions}} \\
				\midrule
				
				\multicolumn{11}{c}{\textit{Model-specific}} \\
				
				\multirow{2}{*}{\textbf{XGB}}
				& Unstabilised
				& 1.0689
				& 4.40
				& 4.12
				& 0.6799
				& 2.81
				& 4.12
				& 12.74
				& 0.0482
				& $0.0624^{*}$ \\
				
				& Stabilised
				& 1.0354
				& 4.52
				& 4.37
				& 0.6668
				& 2.91
				& 4.37
				& 12.96
				& 0.0147
				& 0.0493 \\
				
				\addlinespace
				
				\multirow{2}{*}{\textbf{FFN}}
				& Unstabilised
				& 1.2863
				& 5.90
				& 4.58
				& 0.6897
				& 3.13
				& 4.58
				& 21.75
				& -0.0188
				& $0.1147^{***}$ \\
				
				& Stabilised
				& 1.1113
				& 5.81
				& 5.03
				& 0.7985
				& 4.06
				& 5.03
				& 12.70
				& -0.1938
				& $0.2235^{***}$ \\
				
				\midrule
				
				\multicolumn{11}{c}{\textit{Cross-model}} \\
				
				\multirow{2}{*}{\textbf{XGB}}
				& Unstabilised
				& 1.0668
				& 4.39
				& 4.11
				& 0.6704
				& 2.76
				& 4.11
				& 12.93
				& 0.0461
				& $0.0529^{**}$ \\
				
				& Stabilised
				& 1.1271
				& 4.91
				& 4.28
				& 0.7389
				& 3.23
				& 4.28
				& 13.48
				& 0.1064
				& $0.1214^{*}$ \\
				
				\addlinespace
				
				\multirow{2}{*}{\textbf{FFN}}
				& Unstabilised
				& 1.2952
				& 5.94
				& 4.48
				& 0.6425
				& 2.87
				& 4.48
				& 23.29
				& -0.0099
				& $0.0675^{**}$ \\
				
				& Stabilised
				& 1.1619
				& 5.86
				& 4.98
				& 0.7930
				& 3.95
				& 4.98
				& 14.42
				& -0.1432
				& $0.2180^{***}$ \\
				
				\bottomrule
			\end{tabular}
		}
		
		\vspace{2mm}
		
		\begin{minipage}{0.95\textwidth}
			\footnotesize
			\textit{Notes:}
			The table reports out-of-sample portfolio performance under
			attribution-guided feature-exposure interventions using the sign
			(Panel A) and rank (Panel B) of feature contributions.
			The intervention-strength parameter is set to $\eta=0.5$, with feature exposures constrained to $0\leq\lambda\leq3$.
			Model-specific interventions use attribution information from the
			corresponding allocation model, while cross-model interventions use
			cross-model averaged attribution information. The unstabilised rules
			use the attribution signal from the preceding validation period directly,
			whereas the stabilised rules recursively update the signal.
			$\Delta SR^{gross}$ and $\Delta SR^{net}$ denote changes in gross and
			net Sharpe ratios relative to the corresponding baseline portfolios
			reported in Section~\ref{sec:baseline_portfolio}.
			Statistical significance reported for $\Delta SR^{net}$ is based on the
			one-sided Ledoit-Wolf Sharpe-ratio improvement test.
			Inference is based on 4,999 bootstrap replications using
			five-trading-day blocks to preserve short-run dependence in daily
			portfolio returns.
			$^{***}$, $^{**}$, and $^{*}$ denote significance at the 1\%, 5\%,
			and 10\% levels, respectively.
		\end{minipage}
		
	\end{table}
	The results show that interventions based on the sign, rank, and magnitude of feature contributions generate remarkably similar portfolio performance. This suggests that the direction of feature contributions constitutes an important component of the attribution signal for feature-exposure interventions, while information about their relative ranking and magnitude provides incremental benefits.
	
	Consistent with the magnitude-based results, the sources of performance improvement differ across allocation models. For XGB, cross-model averaging provides additional attribution information that primarily improves the gross Sharpe ratio, while portfolio turnover remains relatively stable. For FFN, the improvement is driven mainly by a substantial reduction in turnover, which raises the net Sharpe ratio despite a lower gross Sharpe ratio. Stabilised feature-exposure interventions consistently outperform their unstabilised counterparts for FFN.
	
	Finally, after combining temporal stabilisation with cross-model averaging, the XGB and FFN strategies exhibit similar portfolio characteristics in terms of gross Sharpe ratio, turnover, and net Sharpe ratio. This convergence provides further evidence that stabilised cross-model attributions capture common and economically meaningful information about feature efficiency.

	\subsection{Robustness to the number of selected feature contributions}
	
%	To examine the robustness of the proposed feature exposure intervention framework, features are ranked each validation year according to the absolute net Sharpe ratio attribution,
%	$|\phi_i(SR_c)|$ with $c=0.0005$,
%	and only the top-ranked features are used to construct the feature scaling coefficients.
%	
%	Figure~\ref{fig:top_features_sharpe_model_specific_cross_model} reports the out-of-sample gross and net Sharpe ratios obtained using different numbers of selected features for both XGB and FFN under sign-, rank-, and magnitude-based scaling.
%	
%	
%	The results demonstrate findings.
%	First, the improvement in the out-of-sample net Sharpe ratio is robust with respect to the number of selected feature contributions. Performance remains consistently above the baseline across a wide range of feature-selection thresholds.
%	Second, the highest-ranked features according to
%	the absolute value of net sharpe ratio attribution
%	carry important economically useful information, particularly for the XGBoost allocation model.
%%	Finally, although larger attribution magnitudes generally lead to larger performance improvements, the relationship is not strictly monotonic, suggesting that feature interactions to portfolio metrics impact the marginal value of individual feature contributions.

	To examine the robustness of the proposed feature-exposure intervention
	framework to the number of selected features, we rank features in each
	validation period according to the absolute magnitude of their net
	Sharpe-ratio attributions, $|\phi_i(SR_c)|$, with $c=0.0005$, and construct
	the exposure adjustments using only the top-ranked features.
	
\begin{figure}[htbp]
		\centering
		
		% ============================================================
		% Row 1: XGB Magnitude -- Gross and Net
		% ============================================================
		\begin{minipage}[b]{0.48\textwidth}
			\centering
			\includegraphics[width=\textwidth]{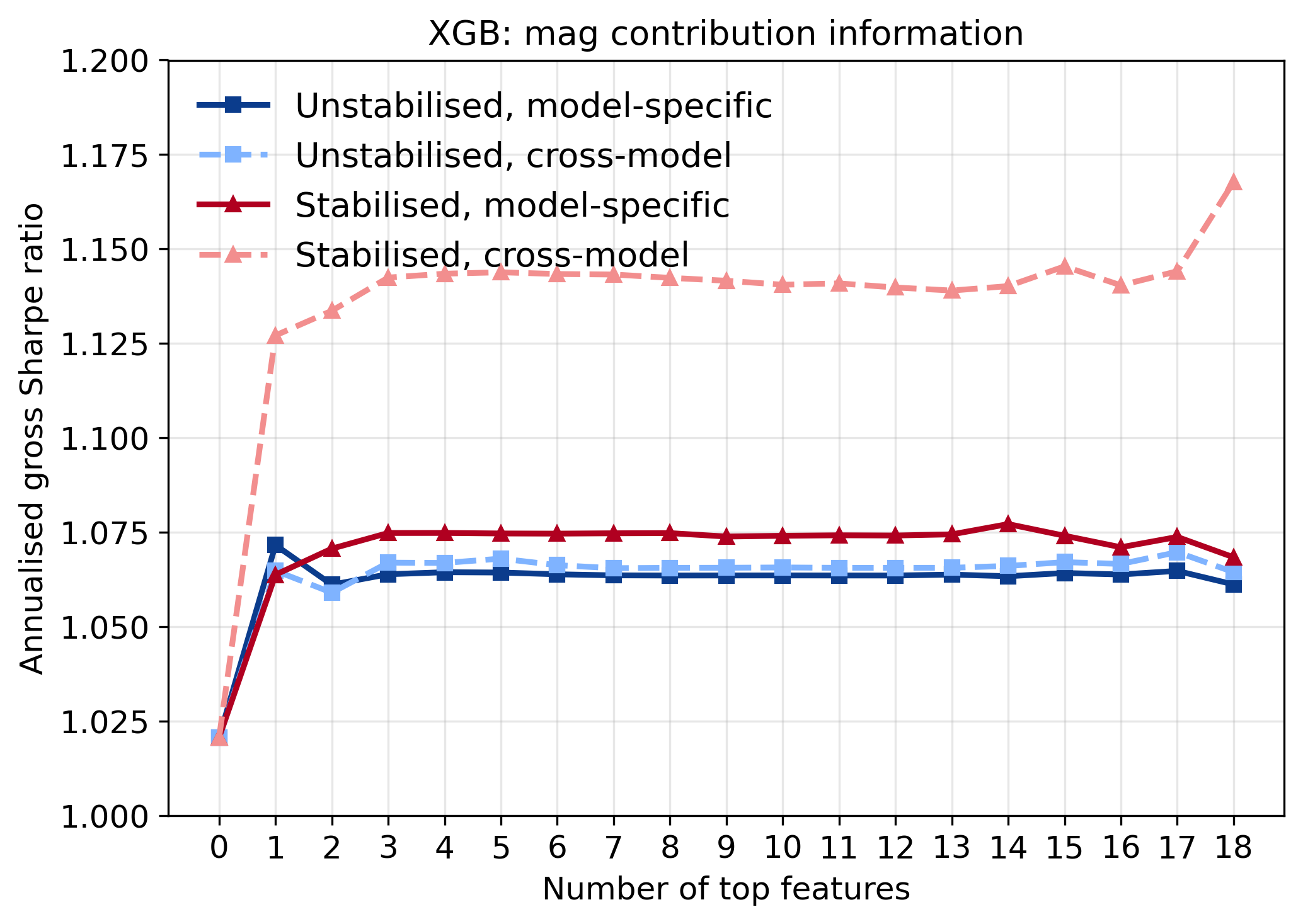}
			\small (a) XGB: Gross Sharpe ratio
		\end{minipage}
		\hfill
		\begin{minipage}[b]{0.48\textwidth}
			\centering
			\includegraphics[width=\textwidth]{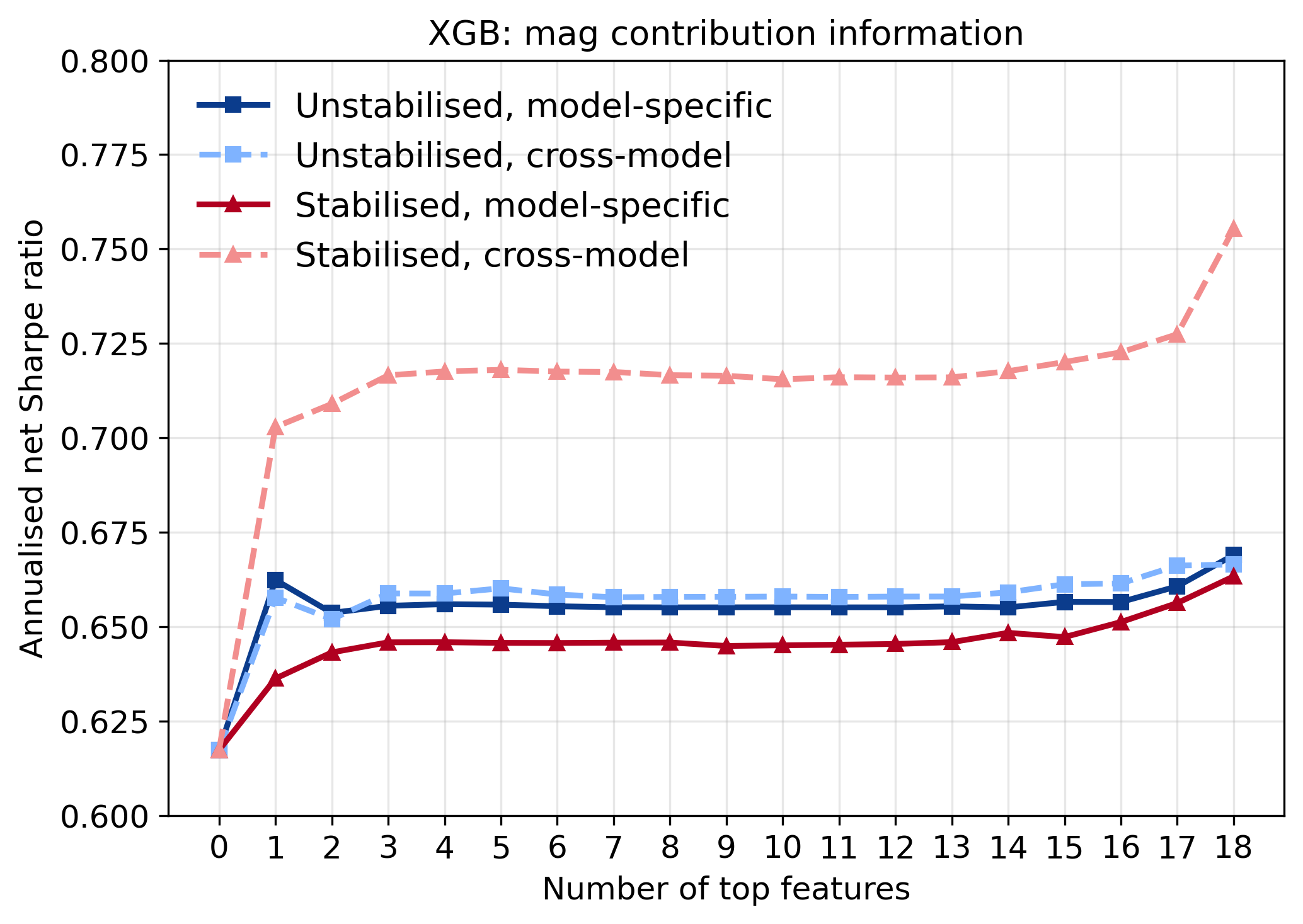}
			\small (b) XGB: Net Sharpe ratio
		\end{minipage}
		
		\medskip
		
		% ============================================================
		% Row 2: FFN Magnitude -- Gross and Net
		% ============================================================
		\begin{minipage}[b]{0.48\textwidth}
			\centering
			\includegraphics[width=\textwidth]{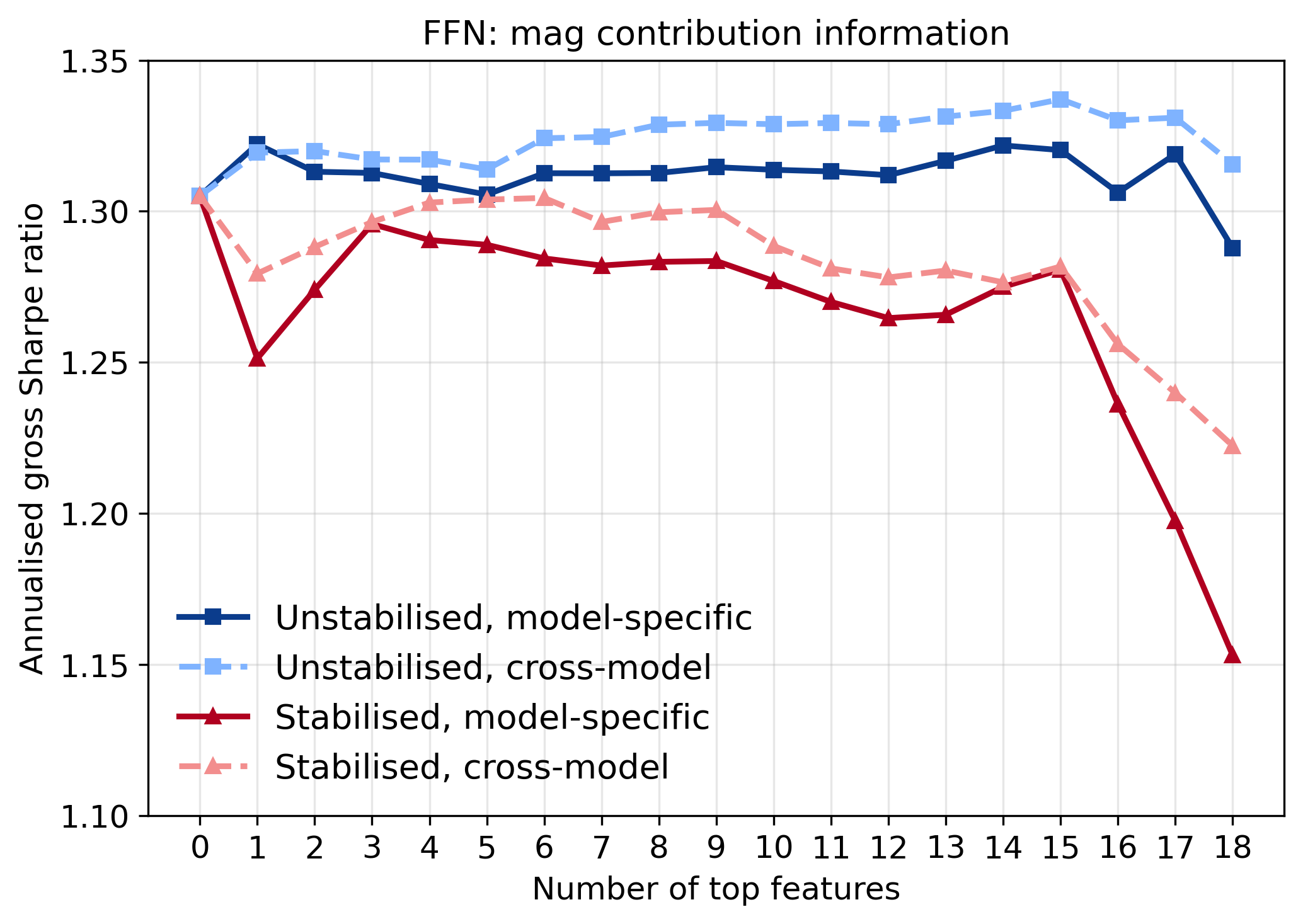}
			\small (c) FFN: Gross Sharpe ratio
		\end{minipage}
		\hfill
		\begin{minipage}[b]{0.48\textwidth}
			\centering
			\includegraphics[width=\textwidth]{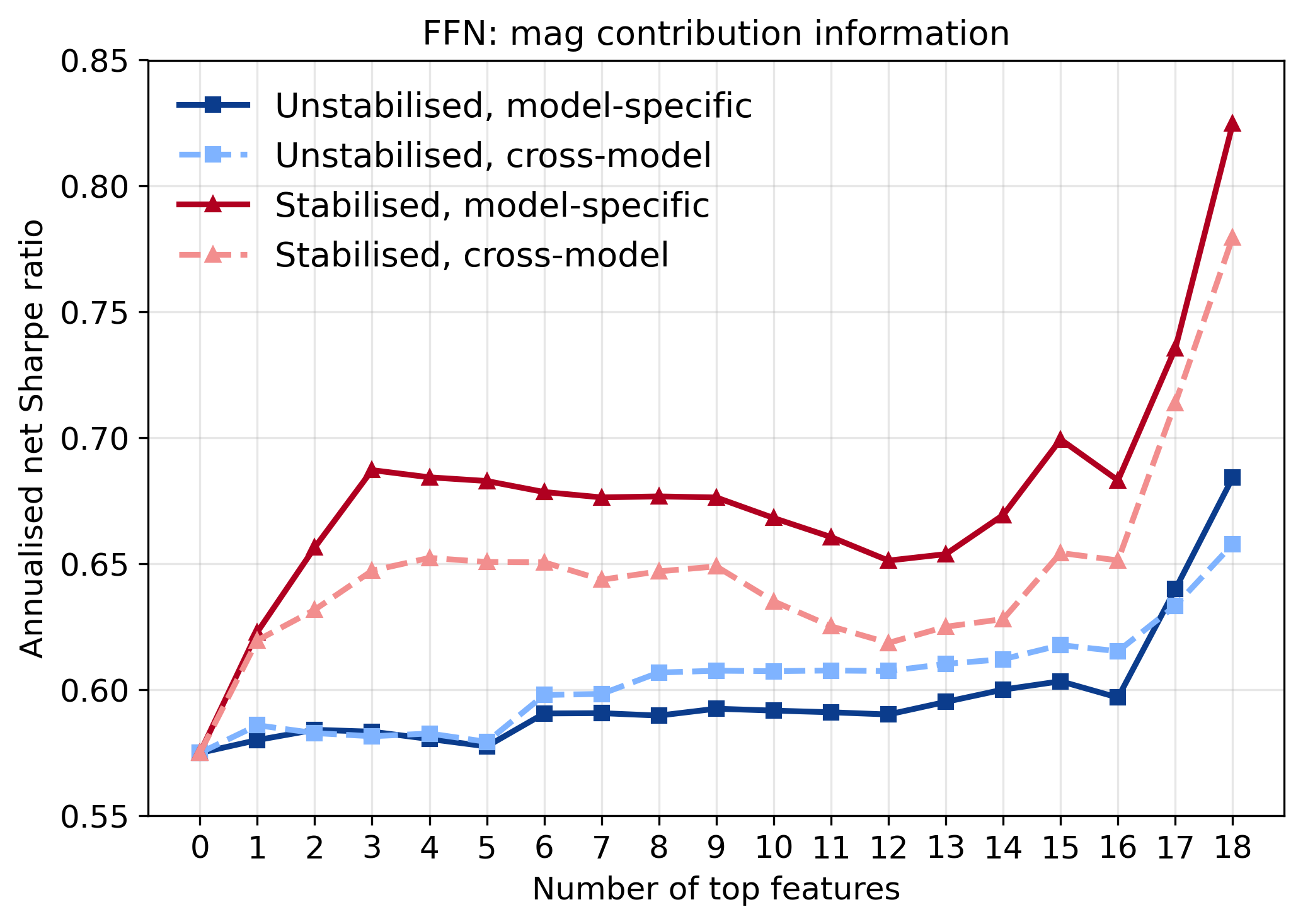}
			\small (d) FFN: Net Sharpe ratio
		\end{minipage}
		
		\caption{
			Portfolio gross and net Sharpe ratios under attribution-guided feature-exposure
			interventions using increasing numbers of top-ranked features.
			Results are reported for magnitude-based attribution ($\eta=0.5$) under both unstabilised and stabilised feature-exposure rules, using both model-specific and cross-model specifications.
		}
		\label{fig:top_features_magnitude_sharpe}
	\end{figure}

	Figure~\ref{fig:top_features_magnitude_sharpe} reports the out-of-sample gross and net Sharpe ratios as the number of selected features varies for both XGB and FFN under magnitude-based interventions. The results show that improvements in the out-of-sample net Sharpe ratio are robust to the number of selected features, with performance remaining above the corresponding baseline across a broad range of feature-selection thresholds. Moreover, interventions based on only the highest-ranked features can already generate economically meaningful performance improvements, particularly for XGB.

	\subsection{Quantile portfolio analysis}
	\label{sec:quantile_portfolio}
	
	To examine whether the proposed feature exposure adjustment remains effective for portfolios with different concentration levels, we further construct top-quantile portfolios based on the absolute stock portfolio weights. Let $w_{t-1}^{R}$ denote the factor-neutral stock weights obtained from the residual portfolio weights $w_{t-1}^{\epsilon}$. At each rebalancing date, stocks are ranked according to the absolute portfolio weights $|w_{t-1}^{R}|$, and only the top $q\%$ of stocks are retained, where
	$
	q\in\{0.1,0.2,\ldots,0.9,1.0\}.
	$
	The remaining portfolio weights are subsequently normalized to satisfy the $L^1$ exposure constraint.
	
%	Feature exposure multipliers are then constructed using the proposed feature contribution information. Specifically, three feature contribution signals are considered, namely the sign, rank, and magnitude signals described in Section~\ref{sec:feature_contribution_signals}. 
%%	The corresponding exposure multiplier is given by
%%	$
%%	f(T_i)
%%	=
%%	1
%%	+
%%	p
%%	\tanh
%%	\left(
%%	\frac{T_i}
%%	{\operatorname{sd}(T)}
%%	\right),
%%	\quad
%%	p\in[0,1],
%%	$
%%	with attribution information confidence $p\in[0,1]$. 
%	We consider both unstabilized and stabilized exposure updates, as well as model-specific and cross-model feature exposures.
%	
%	Figures~\ref{fig:quantile_mag_p05} reports the out-of-sample portfolio performance under different quantile portfolios.
	
	\begin{figure}[htbp]
		\centering
		
		% ============================================================
		% Row 1: XGB Gross
		% ============================================================
		\begin{minipage}[t]{\textwidth}
			\centering
			\includegraphics[width=\textwidth]
			{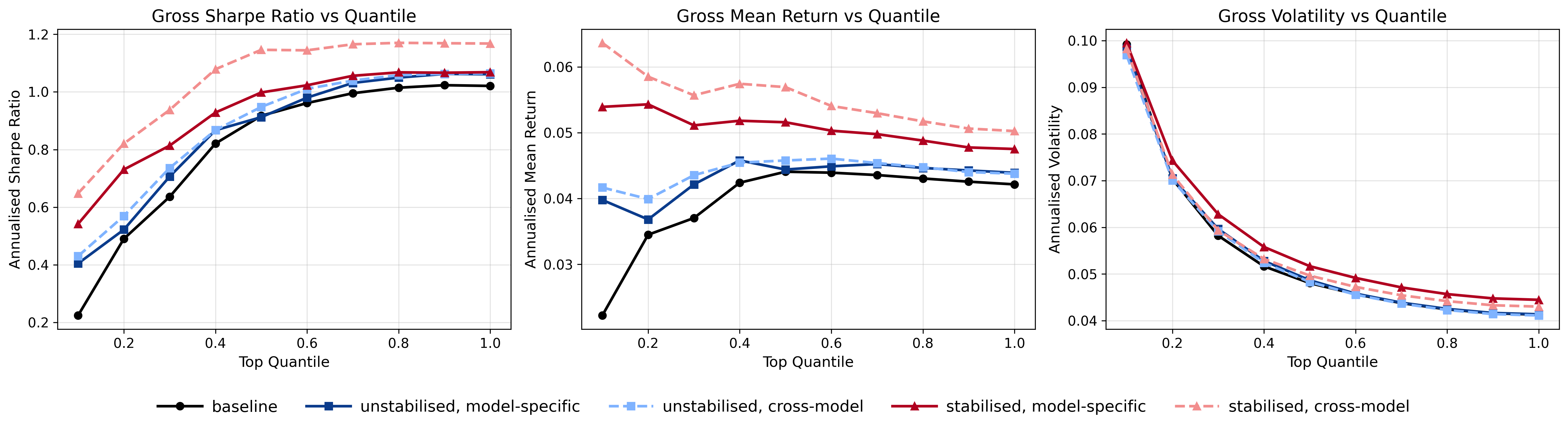}
			\small (a) XGB: Gross portfolio performance
		\end{minipage}
		
		\vspace{2mm}
		
		% ============================================================
		% Row 2: XGB Net
		% ============================================================
		\begin{minipage}[t]{\textwidth}
			\centering
			\includegraphics[width=\textwidth]
			{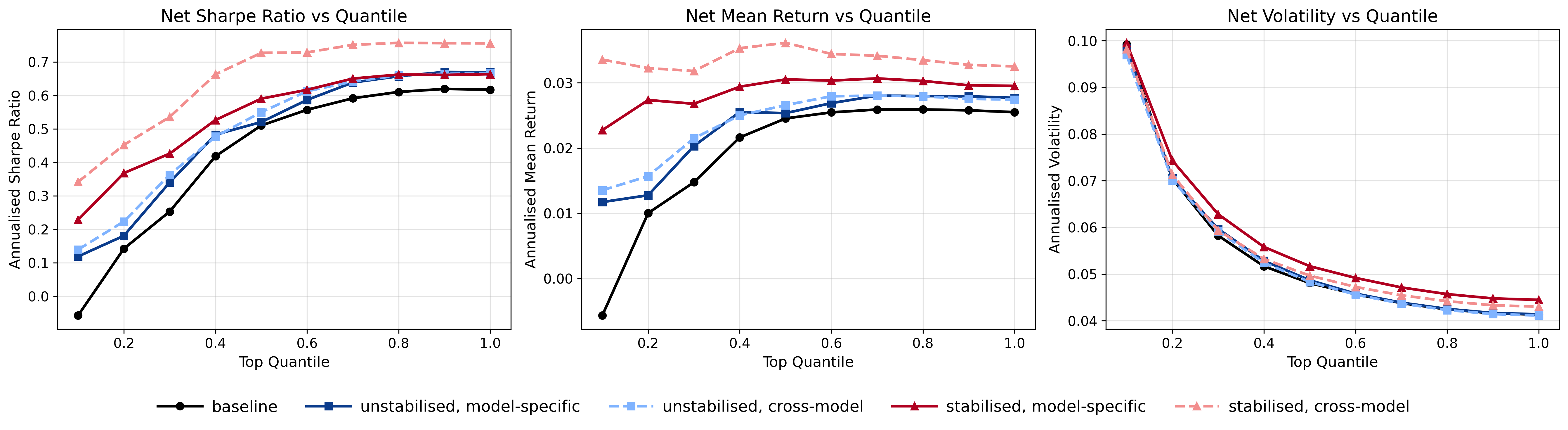}
			\small (b) XGB: Net portfolio performance
		\end{minipage}
		
		\vspace{2mm}
		
		% ============================================================
		% Row 3: FFN Gross
		% ============================================================
		\begin{minipage}[t]{\textwidth}
			\centering
			\includegraphics[width=\textwidth]
			{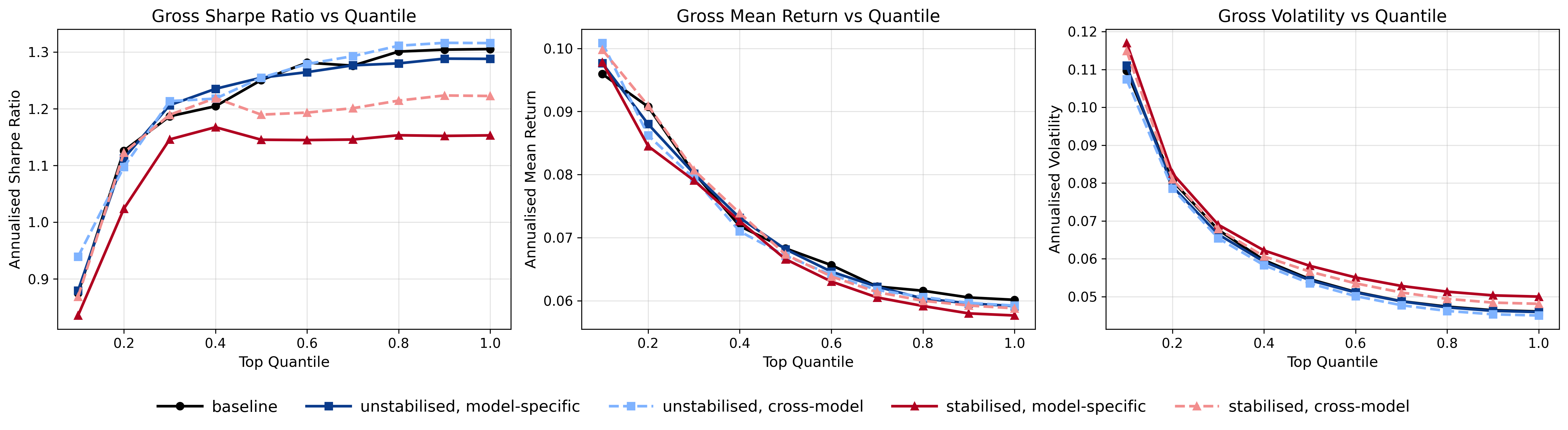}
			\small (c) FFN: Gross portfolio performance
		\end{minipage}
		
		\vspace{2mm}
		
		% ============================================================
		% Row 4: FFN Net
		% ============================================================
		\begin{minipage}[t]{\textwidth}
			\centering
			\includegraphics[width=\textwidth]
			{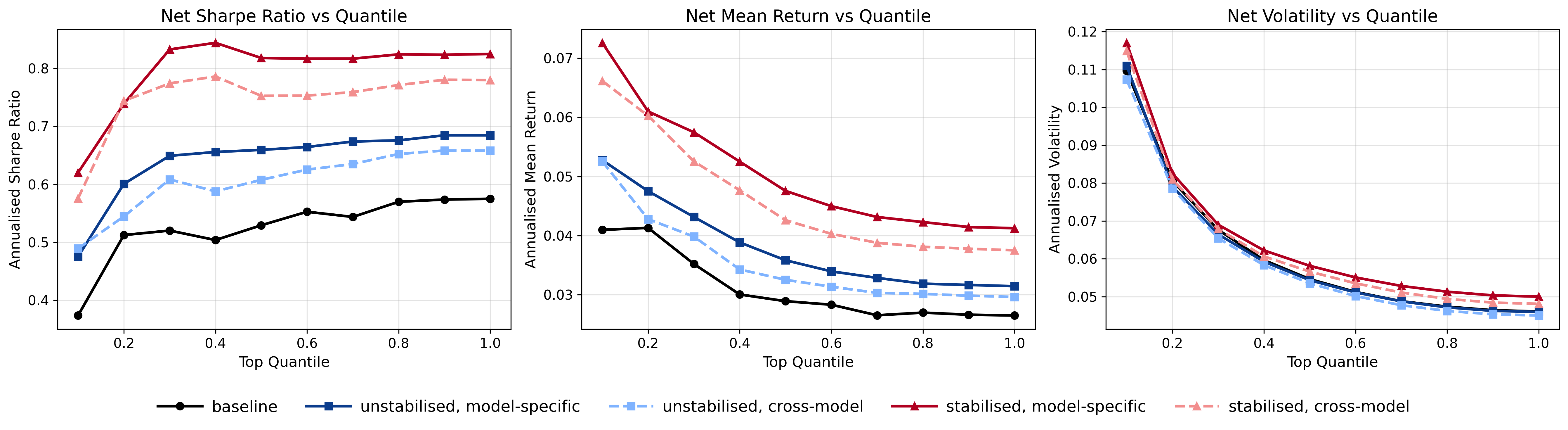}
			\small (d) FFN: Net portfolio performance
		\end{minipage}
		
		\caption{Out-of-sample portfolio performance across different stock quantiles using magnitude-based feature-contribution signals ($\eta=0.5$). From left to right, the panels report the Sharpe ratio, mean return, and volatility, respectively. }
		\label{fig:quantile_mag_p05}
		
	\end{figure}

	Figures~\ref{fig:quantile_mag_p05} report the out-of-sample portfolio performance across different portfolio quantiles. The proposed feature-exposure adjustment consistently improves the out-of-sample net Sharpe ratio across different levels of portfolio concentration. The improvement is particularly pronounced under the stabilised exposure update, suggesting that recursively accumulated feature exposures provide a more robust basis for portfolio adjustment. Moreover, the cross-model feature exposures generally deliver performance comparable to their model-specific counterparts, indicating that the feature contributions capture economically meaningful information that is relatively robust to the choice of allocation model.
	
	For the FFN, the improvement in the net Sharpe ratio is primarily attributable to higher annualised net mean returns and substantially lower portfolio turnover, suggesting that the feature-exposure adjustment enhances trading efficiency. In contrast, for XGBoost, the feature-exposure adjustment consistently improves both gross and net mean returns across different portfolio quantiles, indicating a more direct enhancement of portfolio return performance. Moreover, more concentrated portfolios generally exhibit larger improvements in the net Sharpe ratio, with this pattern being particularly pronounced for XGBoost.

	\subsection{Performance under different market regimes}
	\label{sec:regimes}
	% ============================================================

%	We classify each trading day into regime groups based on rolling market conditions computed from the daily average universe return. Regime variables are defined using a 252-day lookback window. We consider five regime dimensions:
%	
%	\begin{itemize}
%		\item volatility regime: HighVol, LowVol, NormalVol;
%		\item drawdown regime: DeepDD, ShallowDD;
%		\item momentum regime: HighMom, LowMom, NormalMom;
%		\item volume-surprise regime: HighVolSurp, LowVolSurp, NormalVolSurp;
%		\item turnover regime: HighTurnover, LowTurnover, NormalTurnover.
%	\end{itemize}
%	
%	Volatility, momentum, volume-surprise, and turnover regimes are
%	formed using rolling 252-day 25th and 75th percentiles of lagged state
%	variables, while drawdown is based on lagged market drawdown with a threshold
%	of $-5\%$. All regimes contain more than 1,000 daily observations. We evaluate the baseline and feature-weighted portfolios using magnitude-based feature-contribution signals ($\eta=0.5$).

	To examine the robustness of portfolio performance across different market conditions, we classify each trading day into market regimes based on five market-wide indicators: volatility, drawdown, momentum, volume surprise, and turnover. 
	The volatility, momentum, volume-surprise, and turnover regimes are defined relative to their respective trailing 252-day distributions, whereas the drawdown regime is based on a fixed $-10\%$ threshold. The rolling classification scheme allows market regimes to adapt to changes in the underlying distribution of market conditions while avoiding the use of future information.

	Let the cross-sectional average market return on day $t$ be defined as $\bar{R}_t = \frac{1}{N}\sum_{n=1}^{N} R^{gross}_{n,t}$, where $R^{gross}_{n,t}$ denotes the daily simple return of stock $n$ on day $t$, and $N$ is the number of stocks in the investment universe.
	
	\paragraph{\textbf{Volatility regime.}}
	Market volatility is measured as the trailing 21-trading-day standard deviation of the cross-sectional average market return, $\sigma_t^{(21)} = \operatorname{Std}\left(\bar{R}_{t-20},\ldots,\bar{R}_t\right)$. Let $Q_{0.25,t}^{\sigma}$ and $Q_{0.75,t}^{\sigma}$ denote the 25th and 75th percentiles, respectively, of $\sigma^{(21)}$ over the preceding 252 trading days. A trading day is classified as \texttt{HighVol} if $\sigma_t^{(21)} > Q_{0.75,t}^{\sigma}$, as \texttt{LowVol} if $\sigma_t^{(21)} < Q_{0.25,t}^{\sigma}$, and as \texttt{NormalVol} otherwise.
	
	\paragraph{\textbf{Drawdown regime.}}
	We first construct a compounded market wealth index from the cross-sectional average market return as $W_t = \prod_{s=t_0}^{t}(1+\bar{R}_s)$, where $t_0$ denotes the beginning of the sample period in January 2000. The running maximum of the wealth index is defined as $W_t^{\max} = \max_{t_0\leq s\leq t} W_s$, and the current drawdown is calculated as $DD_t = \frac{W_t}{W_t^{\max}}-1$. Using a fixed threshold of $-10\%$, a trading day is classified as \texttt{DeepDD} if $DD_t < -0.1$ and as \texttt{ShallowDD} if $DD_t \geq -0.1$.
	
	\paragraph{\textbf{Momentum regime.}}
	Market momentum is measured by the trailing 21-trading-day average of the cross-sectional average market return, $M_t^{(21)} = \frac{1}{21}\sum_{s=t-20}^{t}\bar{R}_s$. Let $Q_{0.25,t}^{M}$ and $Q_{0.75,t}^{M}$ denote the 25th and 75th percentiles, respectively, of $M^{(21)}$ over the preceding 252 trading days. A trading day is classified as \texttt{HighMom} if $M_t^{(21)} > Q_{0.75,t}^{M}$, as \texttt{LowMom} if $M_t^{(21)} < Q_{0.25,t}^{M}$, and as \texttt{NormalMom} otherwise.
	
	\paragraph{\textbf{Volume-surprise regime.}}
	Let $VS_{n,t}^{(21)}$ denote the 21-day volume-surprise measure for stock $n$ on day $t$. We construct the market-wide volume-surprise indicator as the cross-sectional average $\overline{VS}_t^{(21)} = \frac{1}{N}\sum_{n=1}^{N}VS_{n,t}^{(21)}$. Let $Q_{0.25,t}^{VS}$ and $Q_{0.75,t}^{VS}$ denote the 25th and 75th percentiles, respectively, of $\overline{VS}^{(21)}$ over the preceding 252 trading days. A trading day is classified as \texttt{HighVolSurp} if $\overline{VS}_t^{(21)} > Q_{0.75,t}^{VS}$, as \texttt{LowVolSurp} if $\overline{VS}_t^{(21)} < Q_{0.25,t}^{VS}$, and as \texttt{NormalVolSurp} otherwise.
	
	\paragraph{\textbf{Turnover regime.}}
	Let $\operatorname{TO}_{n,t}$ denote the turnover of stock $n$ on day $t$. The market-wide turnover indicator is defined as the cross-sectional average $\overline{\operatorname{TO}}_t = \frac{1}{N}\sum_{n=1}^{N}\operatorname{TO}_{n,t}$. Let $Q_{0.25,t}^{TO}$ and $Q_{0.75,t}^{TO}$ denote the 25th and 75th percentiles, respectively, of $\overline{\operatorname{TO}}$ over the preceding 252 trading days. A trading day is classified as \texttt{HighTurnover} if $\overline{\operatorname{TO}}_t > Q_{0.75,t}^{TO}$, as \texttt{LowTurnover} if $\overline{\operatorname{TO}}_t < Q_{0.25,t}^{TO}$, and as \texttt{NormalTurnover} otherwise.

	We evaluate portfolios with and without feature-exposure adjustments using magnitude-based feature-contribution signals ($\eta=0.5$). Table~\ref{tab:regime_conditional_net_sr} shows that the feature-exposure interventions improve net Sharpe ratios across a broad range of market regimes. The regime-conditional results indicate that the improvements tend to be particularly pronounced under market conditions in which the baseline strategies perform relatively poorly, most notably during low-volatility and deep-drawdown periods. In the deep-drawdown regime, the XGB net Sharpe ratio increases from 0.1867 to 0.8334, while the FFN net Sharpe ratio rises from 0.3355 to 0.6434 under the stabilised cross-model exposure rule.
	Stabilised feature-contribution signals also tend to produce significant improvements in net Sharpe ratios across several market regimes. For example, under the stabilised cross-model exposure rule, the XGB net Sharpe ratio increases from 0.0392 to 0.4333 in the high-momentum regime, while the FFN net Sharpe ratio increases from 0.7137 to 1.0395 in the low-turnover regime.

	\begin{table}[htbp]
		\centering
		\caption{Regime-conditional net Sharpe ratios under
			attribution-guided feature-exposure interventions}
		\label{tab:regime_conditional_net_sr}
		\small
		\setlength{\tabcolsep}{4pt}
		
		\resizebox{\textwidth}{!}{%
			\begin{tabular}{cccllll}
				\toprule
				Regime
				& $n_{\mathrm{days}}$
				& $SR^{net}_{Baseline}$
				& $SR^{net}_{MS,U}$
				& $SR^{net}_{MS,S}$
				& $SR^{net}_{CM,U}$
				& $SR^{net}_{CM,S}$ \\
				\midrule
				
				\multicolumn{7}{c}{\textbf{Panel A: XGB}} \\
				\midrule
				
				HighVol
				& 1187
				& 1.0766
				& 1.0200\,(-0.0566)
				& 1.0796\,(0.0030)
				& 1.0886\,(0.0120)
				& 1.3329\,(0.2563)
				\\
				
				LowVol
				& 1391
				& -0.4027
				& -0.3125\,(0.0902$^{***}$)
				& -0.3896\,(0.0130)
				& -0.3650\,(0.0376$^{*}$)
				& -0.3463\,(0.0564)
				\\
				
				NormalVol
				& 2203
				& 0.9715
				& 1.0664\,(0.0950$^{**}$)
				& 1.0320\,(0.0606)
				& 1.0518\,(0.0804$^{**}$)
				& 1.0544\,(0.0830)
				\\
				
				\midrule
				
%				DeepDD
%				& 1399
%				& 0.4711
%				& 0.5741\,(0.1030)
%				& 0.7695\,(0.2984)
%				& 0.5882\,(0.1171$^{**}$)
%				& 0.9498\,(0.4787$^{**}$)
%				\\
%				
%				ShallowDD
%				& 3382
%				& 0.6947
%				& 0.7232\,(0.0286)
%				& 0.6152\,(-0.0794)
%				& 0.7107\,(0.0160)
%				& 0.6618\,(-0.0329)
%				\\
				
				DeepDD
				& 622
				& 0.1867
				& 0.3390\,(0.1523)
				& 0.4263\,(0.2396)
				& 0.4025\,(0.2159$^{**}$)
				& 0.8334\,(0.6467$^{**}$)
				\\
				
				ShallowDD
				& 4159
				& 0.7080
				& 0.7432\,(0.0352)
				& 0.7254\,(0.0174)
				& 0.7254\,(0.0174)
				& 0.7476\,(0.0396)
				\\
				
				\midrule
				
				HighMom
				& 1106
				& 0.0392
				& 0.0899\,(0.0507)
				& 0.3279\,(0.2887)
				& 0.0793\,(0.0401)
				& 0.4333\,(0.3941$^{**}$)
				\\
				
				LowMom
				& 1271
				& 0.6639
				& 0.7336\,(0.0697)
				& 0.5666\,(-0.0973)
				& 0.7566\,(0.0927$^{*}$)
				& 0.8077\,(0.1437)
				\\
				
				NormalMom
				& 2404
				& 0.8707
				& 0.9111\,(0.0404)
				& 0.9136\,(0.0429)
				& 0.8940\,(0.0233)
				& 0.8917\,(0.0210)
				\\
				
				\midrule
				
				HighVolSurp
				& 1224
				& -0.0897
				& -0.0544\,(0.0353)
				& -0.2306\,(-0.1409)
				& -0.0070\,(0.0827$^{**}$)
				& -0.0629\,(0.0268)
				\\
				
				LowVolSurp
				& 1190
				& 1.1419
				& 1.1375\,(-0.0044)
				& 0.8170\,(-0.3249)
				& 1.1237\,(-0.0181)
				& 0.9813\,(-0.1606)
				\\
				
				NormalVolSurp
				& 2367
				& 0.8286
				& 0.9127\,(0.0842)
				& 1.1451\,(0.3165$^{***}$)
				& 0.8886\,(0.0600$^{*}$)
				& 1.1624\,(0.3339$^{***}$)
				\\
				
				\midrule
				
				HighTurnover
				& 1232
				& 0.3287
				& 0.2997\,(-0.0290)
				& 0.3128\,(-0.0158)
				& 0.4003\,(0.0716)
				& 0.5127\,(0.1840)
				\\
				
				LowTurnover
				& 1334
				& 0.8646
				& 0.8786\,(0.0140)
				& 0.9226\,(0.0580)
				& 0.8290\,(-0.0356)
				& 0.9761\,(0.1115)
				\\
				
				NormalTurnover
				& 2215
				& 0.7062
				& 0.8403\,(0.1341$^{***}$)
				& 0.8140\,(0.1078)
				& 0.7875\,(0.0813$^{***}$)
				& 0.8397\,(0.1335$^{*}$)
				\\
				
				\midrule
				\multicolumn{7}{c}{\textbf{Panel B: FFN}} \\
				\midrule
				
				HighVol
				& 1187
				& 0.9332
				& 1.0259\,(0.0927)
				& 0.9776\,(0.0444)
				& 1.0522\,(0.1189$^{**}$)
				& 0.9927\,(0.0595)
				\\
				
				LowVol
				& 1391
				& -0.2146
				& -0.1033\,(0.1113$^{**}$)
				& 0.0922\,(0.3068$^{***}$)
				& -0.1301\,(0.0845$^{***}$)
				& 0.0348\,(0.2495$^{***}$)
				\\
				
				NormalVol
				& 2203
				& 0.8017
				& 0.9285\,(0.1268$^{**}$)
				& 1.1877\,(0.3860$^{***}$)
				& 0.8699\,(0.0682$^{**}$)
				& 1.0968\,(0.2951$^{***}$)
				\\
				
				\midrule
				
%				DeepDD
%				& 1399
%				& 0.6586
%				& 0.7776\,(0.1190)
%				& 0.8674\,(0.2088)
%				& 0.7589\,(0.1003$^{*}$)
%				& 0.8139\,(0.1553)
%				\\
%				
%				ShallowDD
%				& 3382
%				& 0.5391
%				& 0.6461\,(0.1070$^{***}$)
%				& 0.8443\,(0.3051$^{***}$)
%				& 0.6131\,(0.0740$^{***}$)
%				& 0.7911\,(0.2520$^{***}$)
%				\\
				
				DeepDD
				& 622
				& 0.3355
				& 0.6080\,(0.2725$^{***}$)
				& 0.7977\,(0.4623$^{*}$)
				& 0.5488\,(0.2133$^{**}$)
				& 0.6434\,(0.3079)
				\\
				
				ShallowDD
				& 4159
				& 0.6441
				& 0.7196\,(0.0755$^{**}$)
				& 0.8809\,(0.2368$^{***}$)
				& 0.6950\,(0.0508$^{**}$)
				& 0.8504\,(0.2063$^{***}$)
				\\
				
				\midrule
				
				HighMom
				& 1106
				& 0.9178
				& 0.9715\,(0.0537)
				& 1.0353\,(0.1175)
				& 0.9244\,(0.0066)
				& 1.0242\,(0.1063)
				\\
				
				LowMom
				& 1271
				& 0.1647
				& 0.2620\,(0.0973)
				& 0.3521\,(0.1873)
				& 0.2614\,(0.0967$^{*}$)
				& 0.2992\,(0.1345)
				\\
				
				NormalMom
				& 2404
				& 0.7320
				& 0.8768\,(0.1449$^{***}$)
				& 1.1331\,(0.4012$^{***}$)
				& 0.8386\,(0.1066$^{***}$)
				& 1.0633\,(0.3313$^{***}$)
				\\
				
				\midrule
				
				HighVolSurp
				& 1224
				& 0.0153
				& 0.0868\,(0.0715)
				& 0.1943\,(0.1789)
				& 0.0207\,(0.0053)
				& 0.1307\,(0.1154)
				\\
				
				LowVolSurp
				& 1190
				& 0.9208
				& 0.9731\,(0.0523)
				& 1.0880\,(0.1672)
				& 1.0965\,(0.1757$^{***}$)
				& 1.0491\,(0.1283)
				\\
				
				NormalVolSurp
				& 2367
				& 0.8061
				& 0.9653\,(0.1592$^{***}$)
				& 1.1581\,(0.3520$^{**}$)
				& 0.8969\,(0.0907$^{***}$)
				& 1.1201\,(0.3139$^{***}$)
				\\
				
				\midrule
				
				HighTurnover
				& 1232
				& 0.5196
				& 0.5923\,(0.0727)
				& 0.7627\,(0.2431)
				& 0.5126\,(-0.0070)
				& 0.6826\,(0.1630)
				\\
				
				LowTurnover
				& 1334
				& 0.7137
				& 0.8200\,(0.1064$^{*}$)
				& 1.1181\,(0.4044$^{***}$)
				& 0.8017\,(0.0880$^{**}$)
				& 1.0395\,(0.3258$^{***}$)
				\\
				
				NormalTurnover
				& 2215
				& 0.5782
				& 0.7243\,(0.1460$^{***}$)
				& 0.7937\,(0.2154)
				& 0.7299\,(0.1516$^{***}$)
				& 0.7792\,(0.2009$^{**}$)
				\\
				
				\bottomrule
			\end{tabular}%
		}
		
		\vspace{2mm}
		
		\begin{minipage}{0.98\textwidth}
			\footnotesize
			\textit{Notes:}
			The table reports net Sharpe ratios conditional on observable market-state
			regimes. Regimes are defined using only information available prior to each
			trading day. Volatility, momentum, volume-surprise, and turnover regimes are
			formed using rolling 252-day 25th and 75th percentiles of lagged state
			variables, while drawdown is based on lagged market drawdown with a threshold
			of $-10\%$. Each row uses the same regime-specific trading days for the
			corresponding baseline and attribution-guided strategies.
			For each intervention, the reported value is the regime-specific net Sharpe
			ratio, with the change relative to the corresponding baseline reported in
			parentheses. $MS$ denotes model-specific attribution, $CM$ denotes
			cross-model averaged attribution, $U$ denotes the unstabilised rule, and
			$S$ denotes the stabilised rule. Statistical significance of the
			Sharpe-ratio improvement is assessed using a one-sided circular block-bootstrap
			test with 4,999 replications and five-trading-day blocks.
			$^{***}$, $^{**}$, and $^{*}$ denote significance at the 1\%, 5\%, and 10\%
			levels, respectively.
		\end{minipage}
		
	\end{table}

	\end{appendices}

\end{document}